Distributional Schwarzschild Geometry from nonsmooth regularization via Horizon. Colombeau disributional Rindler space-time with disributional Levi-Cività connection induced vacuum dominance.Unruh effect revisited.

Jaykov Foukzon

Center for Mathematical Sciences, Israel Institute of Technology, Haifa, Israel
Email: jaykovfoukzon@list.ru

## Abstract

**The vacuum energy density of free scalar quantum field $\Phi$ in a Rindler distributional space-time with distributional Levi-Cività connection is considered. It has been widely believed that, except in very extreme situations, the influence of acceleration on quantum fields should amount to just small, sub-dominant contributions. Here we argue that this belief is wrong by showing that in a Rindler distributional background spacetime with distributional Levi-Cività connection the vacuum energy of free quantum fields is forced, by the very same background distributional space-time such a Rindler distributional background space-time, to become dominant over any classical energy density component. This semiclassical gravity effect finds its roots in the singular behavior of quantum fields on a Rindler distributional space-times with distributional Levi-Cività connection. In particular we obtain that the vacuum fluctuations $\langle\Phi^2\rangle$ have a singular behavior at a Rindler horizon $\delta_R = 0 : \langle\Phi^2(\delta)\rangle \asymp \delta^{-4}, \delta \asymp c^2/a, a \to \infty$. Therefore sufficiently strongly accelerated observer burns up near the Rindler horizon. Thus Polchinski's account doesn't violate of the Einstein equivalence principle.**

## 6mm . Introduction 3mm

In March 2012, Joseph Polchinski claimed that the following three statements cannot all be true cite: AlmheiriMarolfPolchinskiSully13[1]: (i) Hawking radiation is in a pure state, (ii) the information carried by the radiation is emitted from the region near the horizon, with low energy effective field theory valid beyond some microscopic distance from the horizon, (iii) the infalling observer encounters nothing unusual at the horizon. Joseph Polchinski argue that the most conservative resolution is: the infalling observer burns up at the horizon. In Polchinski's account, quantum effects would turn the event horizon into a seething maelstrom of particles. Anyone who fell into it would hit a wall of fire and be burned to a crisp in an instant. As pointed out by physics community such firewalls would violate a foundational tenet of contemporary physics known as the equivalence principle, it states in part that an observer falling in a gravitational field — even the powerful one inside a black hole — will see exactly the same phenomena as an accelerated observer floating in empty space.

In this paper we argue that Polchinski was not wrong, but Unruh effect revision is needed.

### 6mm . 1.1.What is Colombeau distributional semi-Riemannian

**geometry**? 3mm

Recall that the classical Cartan's structural equations show in a compact way the relation between a connection and its curvature, and reveal their geometric interpretation in terms of moving frames. In order to study the mathematical properties of singularities, we need to study the geometry of manifolds endowed on the tangent bundle with a symmetric bilinear form it is allowed to become degenerate (singular).

**Remark 1.1.1.**But if the fundamental tensor is allowed to be degenerate (singular), there are some obstructions in constructing the geometric objects normally associated to the fundamental tensor. Also,local orthonormal frames and coframes no longer exist, as well as the metric connection and its curvature operator cite: Kupeli96[2].

**Remark 1.1.2.**"Singular Semi-Riemannian Geometry"- the main brunch of contemporary semi-Riemannian geometry in which have been studied a smooth manifolds $M$ furnished with a degenerate (singular) on a smooth submanifold $M' \subsetneq M$ metric tensor of arbitrary signature cite: Kupeli96[2].

**Remark 1.1.3.**In order to solve problems of the gravitational singularity in classical general relativity the singular semi-Riemannian geometry based on Colombeau calculas and Colombeau generalized functions was mach developed,see
cite: Colombeau84Parker79VickersWilson98VickersWilson99Vickers99GerochTraschen87
cite: BalasinNachbagauer93BalasinNachbagauer94KawaiSakane97PantojaRago97
cite: PantojaRago00KunzingerSteinbauer02KunzingerSteinbauer01
cite: GrosserFarkasKunzingerSteinbauer01HeinzleSteinbauer02Foukzon15
cite: FoukzonPotapovMenkova16Vickers12Steinbauer00 cite: GolubevKelner05 [3]-[22].

**Remark 1.1.4.**Let $\mathbf{G}(M')$be algebra of Colombeau generalized functions on $M' \subset M$ , let $\widetilde{\mathbb{R}}$ be the ring of Colombeau generalized numbers cite: Colombeau84Parker79VickersWilson98 [3]-[5]. Let $(g_\varepsilon)_\varepsilon$ be Colombeau generalized metric tensor on $M$ and let $\mathbf{Ric}_{M'}(p)$ be generalized Ricci tensor of the metric $(g_\varepsilon(p))_\varepsilon|_{M'}$ cite: Vickers12Steinbauer00 [20]-[21].The main properties of such nonclassical manifolds with a degenerate (singular) metric tensor that is
$\mathbf{Ric}_{M'}(p) \in \mathbf{G}(M')\backslash C^\infty(M'),$ i.e. for all $p \in M'$ : $\mathbf{Ric}_{M'}(p) \in \widetilde{\mathbb{R}}\backslash\mathbb{R}$.

**Definition 1.1.1.**Let $\mathbf{G}(M')$ be algebra of Colombeau generalized functions on $M' \subset M,$ and let $(g_\varepsilon(p))_\varepsilon$ be Colombeau generalized metric tensor on $M$ such that $(g_\varepsilon(p))_\varepsilon$ is the Colombeau solution of the generalized Einstein field equations (1.3.19),(see Remark 1.3.7). We define now the Colombeau distributional scalar curvature $\mathbf{R}_M(p) = [(\mathbf{R}_{\varepsilon,M}(p))_\varepsilon]$ (or distributional Ricci cite: Vickers12Steinbauer00[20]-[21] scalar) as the trace of
$\mathbf{Ric}_M(p) : \mathbf{R}_M(p) = \mathbf{tr}(\mathbf{Ric}_M(p)).$ Assume that $\mathbf{R}_{M'}(p) \in \mathbf{G}(M')\backslash C^\infty(M')$.

Then we say that: (i) gravitational field $(g_\varepsilon(p))_\varepsilon$ (or corresponding distributional spacetime) has a gravitational singularity on a smooth compact submanifold $M_c \subset M$ if
$\mathbf{R}_{M_c}(p) \in \mathbf{G}(M_c)\backslash C^\infty(M_c);$ (ii) gravitational field $(g_\varepsilon(p))_\varepsilon$ has a gravitational singularity with compact support iff $\mathbf{R}_{M_c}(p) \in D'(\mathbb{R}^3)$.

**Remark 1.1.5.**It turns out that the distributional Schwarzschild spacetime has a gravitational singularity with compact support at origin $\{r = 0\}$
cite: VickersWilson99Vickers99GerochTraschen87
cite: BalasinNachbagauer93BalasinNachbagauer94KawaiSakane97[6]-[11] and at Schwarzschild horizon $\mathbf{S}^2 \times \{r = 2m\}$ cite: Foukzon15FoukzonPotapovMenkova16 [18]-[19].

**Definition 1.1.2.**(i) Let $\mathbf{G}(M)$ be algebra of Colombeau generalized functions on $M,$ and let $(g_\varepsilon(p))_\varepsilon$ be Colombeau generalized metric tensor on $M$ such that $(g_\varepsilon(p))_\varepsilon$ is the Colombeau solution of the generalized Einstein field equations (1.3.19).The generalized point value of $(g_\varepsilon(p))_\varepsilon$ at generalized point $((p_\varepsilon))_\varepsilon$ is $(g_\varepsilon(p_\varepsilon))_\varepsilon$.(ii)We define now the generalized point value of the distributional scalar curvature $\mathbf{R}_M(p)$ at generalized point $\mathbf{p} = [((p_\varepsilon))_\varepsilon]$ by formula $\mathbf{R}_M(\mathbf{p}) = [(\mathbf{R}_{\varepsilon,M}(p_\varepsilon))_\varepsilon]$.

## 6mm . 1.2.Distributional Møller's geometry as Colombeau extension of the classical Møller's spacetime 3mm

As important example of Colombeau extension of the singular semi-Riemannian geometry mentioned above, we consider now Møller's uniformly accelerated frame given by Møller's line element cite: Moller43[23]:

$$ds^2 = -(a+gx)^2dt^2 + dx^2 + dy^2 + dz^2. \qquad (1.2.1)$$

Of couse Møller's metric (1.2.1) degenerate at Møller horizon $x_{hor}^{M\phi l} = -(a/g)^{-1}$. Note that formally corresponding to the metric (1.2.1) classical Levi-Civitá connection is cite: Moller43 [23]

$$\Gamma_{44}^{1}(x) = (a+gx), \Gamma_{14}^{4}(x) = \Gamma_{41}^{4}(x) = \frac{1}{a+gx} \qquad (1.2.2)$$

and therefore classical Levi-Civit'a connection (1.2.2) of couse is not available at Møller horizon $x_{hor}^{M\phi l} = -a \cdot g^{-1}$. Recall that fundamental tensor corresponding to the metric (1.2.1) was obtained in Møller's paper cite: Moller43 [23] as a vacuum solution of the classical Einstein's field equations

$$G_i^k = R_i^k - \frac{1}{2}\delta_i^k R = 0, \qquad (1.2.3)$$

where $R_i^k$ is the contracted Riemann-Christoffel tensor formally calculated by canonical way by using classical Levi-Civitá connection (1.2.2) and $R = R_i^i$. Using Dingle's formula cite: Moller43 [23] in case of the metric (1.2.1) we get

$$\begin{gathered}G_2^2(x) = G_3^3(x) = -\frac{1}{2\Delta(x)}\left\{\Delta''(x) - \frac{[\Delta'(x)]^2}{2\Delta(x)}\right\}, \\ \Delta(x) = (a+gx)^2,\end{gathered} \qquad (1.2.4)$$

where $\Delta'(x) = \partial\Delta(x)/\partial x$ and all other components of $G_i^k$ vanishes identically. Note that

$$\Delta'(x) = 2g(a+gx), \Delta''(x) = 2g^2. \qquad (1.2.5)$$

Thus for any $x \neq -a \cdot g^{-1}$ we get a classical result

$$G_2^2(x) = G_3^3(x) = -\frac{1}{2\Delta(x)}\left\{2g^2 - \frac{4g^2(a+gx)^2}{2\Delta(x)}\right\} \equiv 0. \qquad (1.2.6)$$

Let $\{x_n\}_{n\in\mathbb{N}}$ be a sequence such that $\lim_{n\to\infty} x_n = -a \cdot g^{-1}, x_n \neq -a \cdot g^{-1}, n \in \mathbb{N}$. Then for any $n \in \mathbb{N}$ we get

$$\Im(x_n) = G_2^2(x_n) = G_3^3(x_n) = -\frac{1}{2\Delta(x_n)}\left\{2g^2 - \frac{4g^2(a+gx_n)^2}{2\Delta(x_n)}\right\} \equiv 0, \qquad (1.2.7)$$

and therefore $\lim_{n\to\infty} \Im(x_n) \equiv 0$. However

$$\lim_{n\to\infty} \Gamma_{14}^{4}(x_n) = \lim_{n\to\infty} \Gamma_{41}^{4}(x_n) = \lim_{n\to\infty} \frac{1}{a+gx_n} = \infty, \qquad (1.2.8)$$

i.e. classical Levi-Civit'a connection given by (1.2.2) unavailable at Møller horizon.

**Remark 1.2.1**. In order to avoid difficultness mentioned above, we consider now the regularized Møller's metric

$$\begin{gathered}d_\varepsilon s^2 = -\Delta_\varepsilon(x)dt^2 + dx^2 + dy^2 + dz^2, \\ \Delta_\varepsilon(x) = \left[(a+gx)^2 + \varepsilon^2\right], \varepsilon \in (0,1].\end{gathered} \qquad (1.2.9)$$

Using now Dingle's formula cite: Moller43 [23] for the case of (1.2.9) we get

$$\begin{gathered}\Im(x;\varepsilon) = G_2^2(x;\varepsilon) = G_3^3(x;\varepsilon) = -\frac{1}{2\Delta_\varepsilon(x)}\left\{\Delta_\varepsilon''(x) - \frac{[\Delta_\varepsilon'(x)]^2}{2\Delta_\varepsilon(x)}\right\}, \\ \Delta_\varepsilon(x) = \left[(a+gx)^2 + \varepsilon^2\right].\end{gathered} \qquad (1.2.10)$$

Note that

$$\Delta'_{\varepsilon} = 2g(1+gx), \Delta''_{\varepsilon} = 2g^2 \quad (1.2.11)$$

and therefore

$$\Im(x;\varepsilon) = -\frac{1}{2\Delta_{\varepsilon}(x)}\left\{2g^2 - \frac{2g^2(a+gx)^2}{\Delta_{\varepsilon}(x)}\right\} =$$
$$-\frac{1}{2\Delta_{\varepsilon}(x)}\left\{2g^2 - \frac{2g^2\left[(a+gx)^2+\varepsilon^2\right]-2g^2\varepsilon^2}{\Delta_{\varepsilon}(x)}\right\} = \quad (1.2.12)$$
$$= -\frac{g^2\varepsilon^2}{\Delta^2_{\varepsilon}(x)}.$$

**Remark 1.2.2.**(i)Note that $(\Im(x;\varepsilon))_{\varepsilon}, \varepsilon \in (0,1]$ is Colombeau generalized function such that $\mathbf{cl}[(\Im(x;\varepsilon))_{\varepsilon}] \in \mathbf{G}(\mathbb{R})$ and $\mathbf{cl}[(\Im(-a \cdot g^{-1};\varepsilon))_{\varepsilon}] = \mathbf{cl}[(\varepsilon^{-2})_{\varepsilon}] \in \widetilde{\mathbb{R}}$.
(ii)Note that $\mathbf{cl}[(\Im(x;\varepsilon))_{\varepsilon}] \sim \dfrac{\delta(a+gx)}{a+gx} \notin D'(\mathbb{R})$.

**Remark 1.2.3**. Note that: (i) at any point $\widetilde{x} \in \widetilde{\mathbb{R}}$ such that $x \in \mathbb{R}$ and $x \neq -a \cdot g^{-1}$ one obtains $(\Im(\widetilde{x};\varepsilon))_{\varepsilon} \approx_{\widetilde{\mathbb{R}}} \widetilde{0}$ (see Definition 1.5.0 (i)) and therefore the Ricci tensor as well as the Ricci scalar are infinite small beyond Møller horizon $x^M_{hor} = \widetilde{-a \cdot g^{-1}}$. Thus at any point $\widetilde{x}$ such that $x \in \mathbb{R}$ and $x \neq -a \cdot g^{-1}$we obtain the disered result in a good agriment with formall canonical calculation (see for example [24],subsect.2.1.6),
(ii) obviously at any finite point $x_{fin} \in \widetilde{\mathbb{R}}$ (see Definition 1.5.0 (iii)) one obtains again $(\Im(\widetilde{x};\varepsilon))_{\varepsilon} \approx_{\widetilde{\mathbb{R}}} \widetilde{0}$.

**Remark 1.2.4.**(I)Thus Colombeau generalized fundamental tensor $(g_{ik}(\varepsilon))_{\varepsilon}$ corresponding to Colombeau metric

$$(d_{\varepsilon}s^2) = -(\Delta_{\varepsilon}(x)dt^2)_{\varepsilon} + dx^2 + dy^2 + dz^2,$$
$$(\Delta_{\varepsilon}(x))_{\varepsilon} = \left(\left[(a+gx)^2+\varepsilon^2\right]\right)_{\varepsilon}, \varepsilon \in (0,1] \quad (1.2.13)$$

that is non vacuum Colombeau solution (see cite: Foukzon15 [18] section 6 and cite: FoukzonPotapovMenkova16 [19] subsection 2.3 Distributional general relativity) of the Einstein's field equations

$$(G^k_i(\varepsilon))_{\varepsilon} = (R^k_i(\varepsilon))_{\varepsilon} - \frac{1}{2}\delta^k_i(R(\varepsilon))_{\varepsilon} = -g^2\left(\frac{\varepsilon^2}{\Delta^2_{\varepsilon}(x)}\right)_{\varepsilon}. \quad (1.2.14)$$

For Rindler metric $a = 0, g = 1$ and we get

$$(G^k_i(\varepsilon))_{\varepsilon} = (R^k_i(\varepsilon))_{\varepsilon} - \frac{1}{2}\delta^k_i(R(\varepsilon))_{\varepsilon} = -\left(\frac{\varepsilon^2}{(x^2+\varepsilon^2)^2}\right)_{\varepsilon} \in \mathbf{G}(\mathbb{R}). \quad (1.2.15)$$

**Definition 1.2.1.**Distributional Møller's geometry that is Colombeau extension of the classical Møller's spacetime given by Colombeau generalized fundamental tensor (1.2.13).

## 6mm . 1.3.Distributional Schwarzschild geometry as Colombeau extension of the classical singular Schwarzschild spacetime 3mm

### 5mm . 1.3.1.Colombeau extension of the classical singular Schwarzschild spacetime furnished with a degenerate and singular Schwarzschild metric 2mm

As another important example of Colombeau extension of the singular semi-Riemannian geometry we consider now classical singular Schwarzschild spacetime furnished with a degenerate and singular Schwarzschild metric

$$ds^2 = -\left(1 - \frac{2m}{r}\right)dt^2 + \left(1 - \frac{2m}{r}\right)^{-1}dr^2 + r^2d\Omega^2 \tag{1.3.1}$$

**Remark 1.3.1.**Note that formally corresponding to the metric (1.3.1) classical Levi-Civitá connection given by canonical Christoffel symbols are cite: MullerGrave10 [24]:

$$\begin{gathered}
\Gamma^1_{00}(r)|_{r=2m} = \lim_{r\to 2m} \frac{m(r-2m)}{r^3} = 0, \Gamma^1_{11}(r)|_{r=2m} = \lim_{r\to 2m} \frac{-m}{r(r-2m)} = \infty, \\
\Gamma^0_{01}(r)|_{r=2m} = \lim_{r\to 2m} \frac{m}{r(r-2m)} = \infty, \\
\Gamma^2_{12}(r)|_{r=2m} = \lim_{r\to 2m} \frac{1}{r} = 2^{-1}m^{-1}, \Gamma^1_{22}|_{r=2m} = -\lim_{r\to 2m}(r-2m) = 0, \\
\Gamma^3_{13}|_{r=2m} = \lim_{r\to 2m} \frac{1}{r} = 2^{-1}m^{-1}, \Gamma^1_{33}|_{r=2m} = -\lim_{r\to 2m}(r-2m)\sin^2\theta = 0, \\
\Gamma^1_{00}(r)|_{r=0} = \lim_{r\to 0} \frac{m(r-2m)}{r^3} = \infty \Gamma^1_{11}(r)|_{r=0} = \lim_{r\to 0} \frac{-m}{r(r-2m)} = \infty, \\
\ldots\ldots\ldots\ldots\ldots\ldots\ldots\ldots\ldots\ldots\ldots\ldots\ldots\ldots\ldots\ldots\ldots\ldots\ldots\ldots \\
\Gamma^2_{33} = -\sin\theta\cos\theta, \Gamma^3_{23} = \frac{\cos\theta}{\sin\theta}.
\end{gathered} \tag{1.3.2}$$

i.e. classical Levi-Civita connection given by Eq.(1.3.2) unavaluble at Schwarzschild horizon.

**Remark 1.3.2.**Newertheles in classical handbooks cite: MullerGrave10Reall12Hooft98Choquet-Bruhat09FeliceClarke10MisnerThorneWheeler73 [24-29] mistakenly it is assumed that classical semi-Riemannian geometry holds on whole Schwarzschild manifold and therefore canonical formal calculation gives

$$R_{abcd}(r)R^{abcd}(r) = \frac{16m^2}{r^6}. \tag{1.3.3}$$

By Eq. (1.3.2) it is mistakenly pointed out that the Schwarzschild metric has only a coordinate singularity at $r = 2m$ and there is no gravitational singularity at Schwarzschild horizon.

**Remark 1.3.3.**Note that canonical formal calculation gives

$$\begin{gathered}
R_{abcd}(r)R^{abcd}(r) = \frac{16m^2}{r^6} + 4\left[-\frac{m}{r\left(1 - \frac{2m}{r}\right)}\right]\left[-\frac{m}{r^5}\left(1 - \frac{2m}{r}\right)\right] \\
+4\left[-\frac{m}{r\left(1 - \frac{2m}{r}\right)}\sin^2\theta\right]\left[-\frac{m}{r^5\sin^2\theta}\left(1 - \frac{2m}{r}\right)\right] + \\
\frac{4m}{r}\left(1 - \frac{2m}{r}\right)\frac{m}{r^5\left(1 - \frac{2m}{r}\right)} + 8mr\sin^2\theta\frac{2m}{r^7\sin^2\theta} + \\
\frac{4m}{r}\left(1 - \frac{2m}{r}\right)\sin^2\theta\frac{m}{r^5\sin^2\theta\left(1 - \frac{2m}{r}\right)}
\end{gathered} \tag{1.3.4}$$

Assume that $r \neq 0$ and $r \neq 2m$, i.e. $1 - \frac{2m}{r} \neq 0, \sin^2\theta \neq 0$, then from Eq.(1.3.4) one obtains directly

$$\begin{gathered}
R_{abcd}(r)R^{abcd}(r) = \frac{16m^2}{r^6} + 4\left[-\frac{m}{r}\right]\left[-\frac{m}{r^5}\right] \\
+4\left[-\frac{m}{r}\right]\left[-\frac{m}{r^5}\right] + \frac{4m^2}{r^6} + \frac{16m^2}{r^6} + \frac{4m^2}{r^6} = \frac{48m^2}{r^6} = \frac{12r_s^2}{r^6}.
\end{gathered} \tag{1.3.5}$$

**Remark 1.3.4.**Notice that: if $r = 2m$ then RHS of the Eq.(1.3.4) become uncertainty

$$
\begin{gathered}
R_{abcd}(r)R^{abcd}(r) = \\
\frac{16m^2}{r^6} + 4\left[-\frac{m}{r0}\right]\left[-\frac{m}{r^5}0\right] + 4\left[-\frac{m}{r0}\sin^2\theta\right]\left[-\frac{m}{r^5\sin^2\theta}0\right] + \\
\frac{4m}{r}0\frac{m}{r^5 0} + 8mr\sin^2\theta\frac{2m}{r^7\sin^2\theta} + \frac{4m}{r}0\sin^2\theta\frac{m}{(r^5\sin^2\theta)0} = \\
\frac{16m^2}{r^6} + \frac{0}{0}.
\end{gathered}
\tag{1.3.6}
$$

In order to avoid this difficulty mentioned above one defines $R_{abcd}(r)R^{abcd}(r)$ at $r = 2m$ by the limit

$$
\begin{gathered}
\lim_{r\to 2m} R_{abcd}(r)R^{abcd}(r) = \\
\lim_{r\to 2m}\left\{\frac{16m^2}{r^6} + 4\left[-\frac{m}{r\left(1-\frac{2m}{r}\right)}\right]\left[-\frac{m}{r^5}\left(1-\frac{2m}{r}\right)\right]\right. \\
+4\left[-\frac{m}{r\left(1-\frac{2m}{r}\right)}\sin^2\theta\right]\left[-\frac{m}{r^5\sin^2\theta}\left(1-\frac{2m}{r}\right)\right] + \\
\frac{4m}{r}\left(1-\frac{2m}{r}\right)\frac{m}{r^5\left(1-\frac{2m}{r}\right)} + 8mr\sin^2\theta\frac{2m}{r^7\sin^2\theta} + \\
\left.\frac{4m}{r}\left(1-\frac{2m}{r}\right)\sin^2\theta\frac{m}{r^5\sin^2\theta\left(1-\frac{2m}{r}\right)}\right\} = \frac{48m^2}{r^6}
\end{gathered}
\tag{1.3.7}
$$

However Eq. (1.3.7) doesn't holds because classical Levi-Civitá connection (1.3.2) of couse is not available at Schwarzschild horizon, see Remark 1.3.1.

**Remark 1.3.5**. Thus from Eq.(1.3.4) for $r \neq 0$ and $r \neq 2m$ we get

$$
R_{abcd}(r)R^{abcd}(r) = \frac{16m^2}{r^6} \Leftrightarrow (r \neq 0) \wedge (r \neq 2m), \tag{1.3.8}
$$

and we get nothing at Schwarzschild horizon.Therefore semi-Riemannian geometry break down at Schwarzschild horizon cite: Foukzon15FoukzonPotapovMenkova16 [18]-[19].

**Remark 1.3.6**.Recall that canonical derivation of the canonical singular Schwarzschild metric in classical handbooks is always based on assumption that:

**Assumption 1.3.1**.Classical semi-Riemannian geometry holds on the whole semi-Riemannian manifold, see for example cite: Hooft98 [26].

Let $ds^2$ be the metric

$$
ds^2 = -A(r)dt^2 + B(r)dr^2 + r^2 d\Omega^2, \tag{1.3.9}
$$

where $A, B \to 1$ as $r \to \infty$. Then under Assumption 1.3.1 one obtains cite: Hooft98 [26]:

(i) all $\Gamma^1_{\mu\nu}$ are zero except

$$
\Gamma^1_{00} = A'/2B, \Gamma^1_{11} = B'/2B, \Gamma^1_{22} = -r/B, \Gamma^1_{33} = -(r/B)\sin^2\theta, \tag{1.3.10}
$$

The equations $R_{\mu\nu} = 0, \mu, \nu = 0,1,2,3$ are

$$R_{00}=\Gamma^{1}_{00,1}-2\Gamma^{1}_{00}\Gamma^{0}_{01}+\Gamma^{1}_{00}\left(\log\sqrt{-g}\right)_{,1}=$$
$$\left(\frac{A'}{2B}\right)'-\frac{A'^{2}}{2AB}+\frac{A'}{2B}\left(\frac{A'}{2A}+\frac{B'}{2B}+\frac{2}{r}\right)= \qquad (1.3.11)$$
$$\frac{1}{2B}\left(A''-\frac{A'B'}{2B}-\frac{A'^{2}}{2A}+\frac{2A'}{r}\right)=0,$$

and

$$R_{11}=-\left(\log\sqrt{-g}\right)_{,1,1}+\Gamma^{1}_{11,1}-(\Gamma^{0}_{10})^{2}-(\Gamma^{1}_{11})^{2}-(\Gamma^{2}_{21})^{2}-$$
$$-(\Gamma^{3}_{31})^{2}+\Gamma^{1}_{11}\left(\log\sqrt{-g}\right)_{,1}= \qquad (1.3.12)$$
$$\frac{1}{2A}\left(-A''+\frac{A'B'}{2B}+\frac{A'^{2}}{2A}+\frac{2AB'}{rB}\right)=0,$$

and

$$R_{22}=-\left(\log\sqrt{-g}\right)_{,2,2}+\Gamma^{1}_{22,1}-2\Gamma^{1}_{22}\Gamma^{2}_{21}-(\Gamma^{3}_{23})^{2}+\Gamma^{1}_{22}\left(\log\sqrt{-g}\right)_{,1}=$$
$$-\frac{d\cot\theta}{d\theta}-\left(\frac{r}{B}\right)'+\frac{2}{B}-\cot^{2}\theta-\frac{r}{B}\left(\frac{2}{r}+\frac{(AB)'}{2AB}\right)=0. \qquad (1.3.13)$$

From Eq.(1.3.11)-Eq.(1.3.12) one obtains

$$\frac{2(AB)'}{rB}=0. \qquad (1.3.14)$$

Therefore $AB$ = constant. Since at $r\to\infty$ we have $A$ and $B\to1$ one obtains $B=A^{-1}$. From Eq.(1.3.13)-Eq.(1.3.14) one obtains

$$\left(\frac{r}{B}\right)'=1, \qquad (1.3.15)$$

and by integration Eq.(1.3.15) one obtains $r/B=r-2m$, where $2m$ is an integration constant. Finally one obtains well known classical result

$$A(r)=1-\frac{2m}{r},B(r)=\left(1-\frac{2m}{r}\right)^{-1}. \qquad (1.3.16)$$

From Eq.(1.3.16) and consideration above (see Remark1.3.4) Assumption 1.3.1 wrong, otherwise one obtains the contradiction.

**Remark 1.3.7.**In order to avoid this difficulty:

(i) we have introduced instead a classical Einstein field equations

$$R_{\mu\nu}-\tfrac{1}{2}Rg_{\mu\nu}=-8\pi GT_{\mu\nu}, \qquad (1.3.17)$$

[where the sign of the energy-momentum tensor is defined by ($\rho$ is the energy density)]

$$T_{44}=-T_{00}=T^{0}_{0}=\rho, \qquad (1.3.18)$$

apropriate Colombeau generalization of the Eq.(1.3.17)-Eq.(1.3.18) such that

$$(R_{\mu\nu}(\varepsilon))_{\varepsilon}-\tfrac{1}{2}(R(\varepsilon)g_{\mu\nu}(\varepsilon))_{\varepsilon}=-8\pi G(T_{\mu\nu}(\varepsilon))_{\varepsilon}, \qquad (1.3.19)$$

where the sign of the distributional energy-momentum tensor is defined by

$$T_{44}(\varepsilon)=-T_{00}(\varepsilon)=T^{0}_{0}(\varepsilon)=\rho(\varepsilon)\in\mathbf{G}(M), \qquad (1.3.20)$$

see cite: Foukzon15FoukzonPotapovMenkova16 [18]-[19].

(ii) we have introduced instead of Assumption 1.3.1 the following assumption.

**Assumption 1.3.2.**Distributional semi-Riemannian geometry holds on whole distributional semi-Riemannian manifold.

**Definition 1.3.1.**Let $A^{\pm}_{\varepsilon}(r),\varepsilon\in[0,1]$ and $B^{\pm}_{\varepsilon}(r),\varepsilon\in[0,1]$ the regularization of the functions $A^{\pm}(r)$ and $B^{\pm}(r)$ [defined above by Eq.(1.3.16)] such that the following conditions are satisfied:

(i) $(A_\varepsilon^\pm(r))_\varepsilon \in \mathbf{G}(\mathbb{R}_+)$ and $(B_\varepsilon^\pm(r))_\varepsilon \in \mathbf{G}(\mathbb{R}_+), \varepsilon \in (0,1]$ are Colombeau generalized functions;

(ii)
$$A_0^\pm(r) = 1 - \frac{2m}{r}, B_0^\pm(r) = \left(1 - \frac{2m}{r}\right)^{-1}; \qquad (1.3.21)$$

(iii) $(A_\varepsilon^\pm(2m))_\varepsilon = \mp(\varepsilon)_\varepsilon \in \widetilde{\mathbb{R}}, (B_\varepsilon^\pm(2m))_\varepsilon = \pm(\varepsilon^{-1})_\varepsilon \in \widetilde{\mathbb{R}}$;

(iv) $(A_\varepsilon^\pm(0))_\varepsilon = 1 - \frac{2m}{(\varepsilon)_\varepsilon} \in \widetilde{\mathbb{R}}, (B_\varepsilon^\pm(0))_\varepsilon = \left(1 - \frac{2m}{(\varepsilon)_\varepsilon}\right)^{-1} \in \widetilde{\mathbb{R}}$.

Let $ds_\varepsilon^2$ be the Colombeau metric

$$(ds_\varepsilon^2)_\varepsilon = -(A_\varepsilon^\pm(r)dt^2)_\varepsilon + (B_\varepsilon^\pm(r)dr^2)_\varepsilon + r^2 d\Omega^2, \qquad (1.3.22)$$

and let $(\Gamma_{\mu\nu}^\eta(\varepsilon))_\varepsilon$ be the distributional Levi-Civita connection cite: Foukzon15FoukzonPotapovMenkova16 [18]-[19] corresponding to Colombeau metric (1.3.22). Then under Assumption 1.3.2 one obtains:

(i) all $(\Gamma_{\mu\nu}^1(\varepsilon))_\varepsilon$ are zero except

$$\begin{gathered}(\Gamma_{00}^1(\varepsilon))_\varepsilon = (A_\varepsilon')_\varepsilon / 2(B_\varepsilon)_\varepsilon, (\Gamma_{11}^1(\varepsilon))_\varepsilon = (B_\varepsilon')_\varepsilon / 2(B_\varepsilon)_\varepsilon, \\ (\Gamma_{22}^1(\varepsilon))_\varepsilon = -r/(B_\varepsilon), (\Gamma_{33}^1(\varepsilon))_\varepsilon = -(r/B_\varepsilon)_\varepsilon \sin^2\theta,\end{gathered} \qquad (1.3.23)$$

$$\begin{gathered}(R_{00}(\varepsilon))_\varepsilon = (\Gamma_{00,1}^1(\varepsilon))_\varepsilon - 2(\Gamma_{00}^1(\varepsilon))_\varepsilon(\Gamma_{01}^0(\varepsilon))_\varepsilon + \\ (\Gamma_{00}^1(\varepsilon))_\varepsilon\left(\log\sqrt{-(g^\pm(\varepsilon))_\varepsilon}\right)_{,1} = \\ \left(\frac{(A_\varepsilon')_\varepsilon}{2(B_\varepsilon)_\varepsilon}\right)' - \frac{(A_\varepsilon'^2)_\varepsilon}{2(A_\varepsilon)_\varepsilon(B_\varepsilon)_\varepsilon} + \frac{(A_\varepsilon')_\varepsilon}{2(B_\varepsilon)_\varepsilon}\left(\frac{(A_\varepsilon')_\varepsilon}{2(A_\varepsilon)_\varepsilon} + \frac{(B_\varepsilon')_\varepsilon}{2(B_\varepsilon)_\varepsilon} + \frac{2}{r}\right) = \\ \frac{1}{2(B_\varepsilon)_\varepsilon}\left((A'') - \frac{(A')_\varepsilon(B')_\varepsilon}{2(B_\varepsilon)_\varepsilon} - \frac{(A_\varepsilon'^2)_\varepsilon}{2(A_\varepsilon)_\varepsilon} + \frac{2(A_\varepsilon')_\varepsilon}{r}\right),\end{gathered} \qquad (1.3.24)$$

and

$$\begin{gathered}(R_{11}(\varepsilon))_\varepsilon = -\left(\log\sqrt{-(g^\pm(\varepsilon))_\varepsilon}\right)_{,1,1} + (\Gamma_{11,1}^1(\varepsilon))_\varepsilon - \\ (\Gamma_{10}^0(\varepsilon))^2 - \left((\Gamma_{11}^1(\varepsilon))^2\right)_\varepsilon - \left((\Gamma_{21}^2(\varepsilon))^2\right)_\varepsilon - \\ -\left((\Gamma_{31}^3(\varepsilon))^2\right)_\varepsilon + (\Gamma_{11}^1(\varepsilon))_\varepsilon\left(\log\sqrt{-(g^\pm(\varepsilon))_\varepsilon}\right)_{,1} = \\ \frac{1}{2(A_\varepsilon)_\varepsilon}\left(-(A_\varepsilon'') + \frac{(A_\varepsilon')_\varepsilon(B_\varepsilon')_\varepsilon}{2(B_\varepsilon)_\varepsilon} + \frac{(A_\varepsilon'^2)_\varepsilon}{2(A_\varepsilon)_\varepsilon} + \frac{2(A_\varepsilon)_\varepsilon(B_\varepsilon')_\varepsilon}{r(B_\varepsilon)_\varepsilon}\right),\end{gathered} \qquad (1.3.25)$$

and

$$\begin{gathered}(R_{22}(\varepsilon))_\varepsilon = -\left(\log\sqrt{-(g^\pm(\varepsilon))_\varepsilon}\right)_{,2,2} + ((\Gamma_{22,1}^1(\varepsilon)))_\varepsilon - \\ 2((\Gamma_{22}^1(\varepsilon)))_\varepsilon((\Gamma_{21}^2(\varepsilon)))_\varepsilon - \left((\Gamma_{23}^3(\varepsilon))^2\right)_\varepsilon + \Gamma_{22}^1(\varepsilon)\left(\log\sqrt{-(g^\pm(\varepsilon))_\varepsilon}\right)_{,1} = \\ -\frac{d\cot\theta}{d\theta} - \left(\frac{r}{(B_\varepsilon)_\varepsilon}\right)' + \frac{2}{(B_\varepsilon)_\varepsilon} - \cot^2\theta - \frac{r}{(B_\varepsilon)_\varepsilon}\left(\frac{2}{r} + \frac{((A_\varepsilon)_\varepsilon(B_\varepsilon)_\varepsilon)'}{2(A_\varepsilon)_\varepsilon(B_\varepsilon)_\varepsilon}\right).\end{gathered} \qquad (1.3.26)$$

Weak distributional limit in $D'(\mathbb{R}^3)$ of the RHS of the Eq.(1.3.18), i.e. $w\text{-}\lim_{\varepsilon\to 0} T_{\mu\nu}(\varepsilon)$ is calculated in our papers cite: Foukzon15FoukzonPotapovMenkova16 [18]-[19], see also Appendix B.

**Remark 1.3.8.** It turns out that the distributional Schwarzschild metric (1.3.22) has a gravitational singularity with compact support at origin $\{r = 0\}$ cite: VickersWilson99Vickers99GerochTraschen87
cite: BalasinNachbagauer93BalasinNachbagauer94KawaiSakane97 [6]-[11] and at Schwarzschild horizon $\mathbf{S}^2 \times \{r = 2m\}$ cite: Foukzon15FoukzonPotapovMenkova16 [18]-[19].

5mm . **1.3.2.Colombeau extension of the Schwarzschild spacetime in isotropic coordinates** 2mm

Let us consider now nonclassical spacetime furnished with a degenerate at horizon $\overline{\rho} = r_s/4$ but nonsingular (at horizon) metric and known in physical literature as Schwarzschild spacetime in isotropic coordinates cite: MullerGrave10 [24]:

$$ds^2 = -\left(\frac{1 - r_s/4\rho}{1 + r_s/4\rho}\right)^2 c^2dt^2 + \left(1 + \frac{r_s}{4\rho}\right)^4 [d\rho^2 + \rho^2(d\theta^2 + \sin^2\theta d\varphi)]. \qquad (1.3.27)$$

Nonsingular metric (1.3.27) is obtained by the coordinate transformation: $r = \rho(1 + r_s/4\rho)^2$, between the Schwarzschild radial coordinate $r$ and the isotropic radial coordinate $\rho$. Under formal calculation one obtains cite: MullerGrave10 [24]:

$$\Gamma^t_{t\rho} = \frac{8r_s}{16\rho^2 - r_s}, \qquad (1.3.28)$$

i.e. classical Levi-Civitá connection given by (1.3.28) of course unavaluble at horizon $\overline{\rho} = r_s/4$. However in physical literature under ubnormal calculation it was mistakenly pointed out that the Ricci tensor and the Ricci scalar vanish identically and Kretschman scalar is

$$R_{abcd}(\rho)R^{abcd}(\rho) = \frac{3 \cdot 4^{13}\rho^6 r_s^2}{(4\rho + r_s)^{12}}. \qquad (1.3.29)$$

**Remark 1.3.9.** In order to avoid difficultness whis the degeneracy of the metric (1.3.27) mentioned above, we consider now the corresponding distributional Colombeau metric which reads

$$(ds^2_\varepsilon)_\varepsilon = -\frac{\left((\rho - \overline{\rho})^2 + \varepsilon^2\right)_\varepsilon}{(\rho + \overline{\rho})^2} c^2dt^2 + \left(1 + \frac{\overline{\rho}}{\rho}\right)^4 [d\rho^2 + \rho^2(d\theta^2 + \sin^2\theta d\varphi)], \qquad (1.3.30)$$

where $\varepsilon \in (0,1]$.

**Definition 1.3.2.** Distributional Schwarzschild geometry in isotropic coordinates which is Colombeau extension of the classical spacetime (1.3.27), given by Colombeau generalized fundamental tensor (1.3.30).

Colombeau generalized metric (1.3.30) nondegenerate at horizon in Colombeau sence and distributional Levi-Civitá connection now available on the whole distributional Schwarzschild spacetime in isotropic coordinates. Notice that generalized metric (1.3.30) has the form given by Eq. (A.1) (see apendix A) with ,

$A_\varepsilon(\rho) = \left((\rho - \overline{\rho})^2 + \varepsilon^2\right)(\rho + \overline{\rho})^{-2}, B_\varepsilon(\rho) = \left(1 + \frac{\overline{\rho}}{\rho}\right)^4, D_\varepsilon(\rho) = 0, C_\varepsilon(\rho) = 0.$ Thus

$$\Delta_\varepsilon(\rho) = A_\varepsilon(\rho)B_\varepsilon(\rho) = \left((\rho - \overline{\rho})^2 + \varepsilon^2\right)(\rho + \overline{\rho})^2\rho^{-1}. \qquad (1.3.31)$$

From Eq. (A.2) (see apendix A) and Eq. (1.3.31) in the limit $|\rho - \overline{\rho}| \asymp 0$, we get

$$(\mathbf{R}(\rho,\varepsilon))_\varepsilon \asymp -\left(\frac{O(\overline{\rho})\varepsilon^2}{\left[(\rho - \overline{\rho})^2 + \varepsilon^2\right]^2}\right)_\varepsilon,$$
$$(\mathbf{R}^{\mu\nu}(\rho,\varepsilon)\mathbf{R}_{\mu\nu}(\rho,\varepsilon))_\varepsilon \asymp \left(\frac{O(\overline{\rho})\varepsilon^4}{\left[(\rho - \overline{\rho})^2 + \varepsilon^2\right]^4}\right)_\varepsilon, \qquad (1.3.32)$$
$$(\mathbf{R}^{\rho\sigma\mu\nu}(\rho,\varepsilon)\mathbf{R}_{\rho\sigma\mu\nu}(\rho,\varepsilon))_\varepsilon \asymp \left(\frac{O(\overline{\rho})\varepsilon^4}{\left[(\rho - \overline{\rho})^2 + \varepsilon^2\right]^4}\right)_\varepsilon.$$

Compare the equation $(\mathbf{R}(\rho,\varepsilon))_\varepsilon \asymp -O(\overline{\rho})\varepsilon^2\left[(\rho - \overline{\rho})^2 + \varepsilon^2\right]^{-2}$ with Eq.(1.2.12).

**Remark 1.3.10**. Notice that in contrast with result of naive formal calculation mentioned above (see Eq.(1.3.29)) we get: (i) $(\mathbf{R}^{\rho\sigma\mu\nu}(\rho,\varepsilon)\mathbf{R}_{\rho\sigma\mu\nu}(\rho,\varepsilon))_{\varepsilon} \in \mathbf{G}(\mathbb{R}^3)$

## 6mm . 1.4.On the near horizon Colombeau approximation for the classical singular Schwarzschild black hole geometry 3mm

Let us perform the following coordinate transformation

$$\bar{t} = \frac{t}{4m}, \quad \bar{r}_{\varepsilon} = \sqrt{8m(r-2m)+\varepsilon^2}, \varepsilon \in (0,1] \tag{1.4.1}$$

to the classical singular Schwarzschild metric

$$ds^2 = -\left(1-\frac{2m}{r}\right)dt^2 + \left(1-\frac{2m}{r}\right)^{-1}dr^2 + r^2 d\Omega^2 \tag{1.4.2}$$

we get

$$ds_{\varepsilon}^2 = -\bar{r}_{\varepsilon}^2\left(1+\frac{\bar{r}_{\varepsilon}^2}{16m^2}\right)^{-1}d\bar{t}^2 + \left(1+\frac{\bar{r}_{\varepsilon}^2}{16m^2}\right)d\bar{r}^2 + 4m^2\left(1+\frac{\bar{r}_{\varepsilon}^2}{16m^2}\right)^2 d\Omega^2. \tag{1.4.3}$$

In Eq.(1.4.2), $m$ is the central mass, $d\Omega^2 = d\theta^2 + \sin^2\theta d\phi^2$ and $G = c = 1$.Taking the limit $m \to \infty$ , the spherical horizon becomes planar and Eq. (1.4.3) leads to the Colombeau type metric

$$(ds_{\varepsilon}^2)_{\varepsilon} = -\{(\bar{r}_{\varepsilon}^2)_{\varepsilon}\}d\bar{t}^2 + (d\bar{r}_{\varepsilon}^2)_{\varepsilon} + 4m^2 d\Omega^2 \tag{1.4.4}$$

which is distributional Rindler’s spacetime if we neglect the angular contribution.The condition $m \to \infty$ is equivalent to the ”near horizon approximation” for the exterior geometry of a black hole : for $r \approx 2m$ $(r > 2m)$ the line element (1.4.2) appears, indeed, as

$$ds^2 = -\frac{r-2m}{2m}dt^2 + \frac{2m}{r-2m}dr^2 + 4m^2 d\Omega^2. \tag{1.4.5}$$

By using simple coordinate transformations it could be shown that (1.4.5) again becomes the distributional Rindler metric when we take $\theta, \phi = const.$ or $\Delta\theta$ and $\Delta\phi$ are negligible. We stress that the condition $r \approx 2m$ only is not enough to obtain Rindler’s spacetime which has no spherical symmetry as Schwarzschild.

**Remark 1.4.1**. At this stage of consideration, it is already clear that near horizon Schwarzschild black hole geometry has a gravitational singularity at horizon. Notice that in classical literature (see, for example, cite: MullerGrave10
cite: Reall12Hooft98Choquet-Bruhat09FeliceClarke10MisnerThorneWheeler73
cite: LandauLifshitz75Grant08Eddington24Finkelstein58Lemaitre33LoingerMarsico09
cite: HawkingEllis73 [24]-[36]) near horizon Schwarzschild black hole geometry mistakenly acepted as regular with the Ricci tensor and the Ricci scalar vanish identically.

## 6mm . 1.5. Colombeau distributional semi-Riemannian geometry.Preliminaries 3mm

### 5mm . 1.5.1.The ring of Colombeau generalized numbers $\widetilde{\mathbb{R}}$ 2mm

**Designation 1.5.1**. We denote by $\widetilde{\mathbb{R}}$ the ring of real, Colombeau generalized numbers. Recall that cite: Kupeli96Colombeau84 [2]-[3] by definition
$\widetilde{\mathbb{R}} = \mathbf{E}_{\mathbb{R}}(\mathbb{R})/\mathbf{N}(\mathbb{R})$ where

$$\begin{gathered}\mathbf{E}_{\mathbb{R}}(\mathbb{R}) = \{(x_{\varepsilon})_{\varepsilon} \in \mathbb{R}^{(0,1)} | (\exists a \in \mathbb{R}_{+})(\exists \varepsilon_0 \in (0,1))(\forall \varepsilon \leq \varepsilon_0)[|x_{\varepsilon}| \leq \varepsilon^{-a}]\}, \\ \mathbf{N}(\mathbb{R}) = \{(x_{\varepsilon})_{\varepsilon} \in \mathbb{R}^{(0,1)} | (\forall a \in \mathbb{R}_{+})(\exists \varepsilon_0 \in (0,1))(\forall \varepsilon \leq \varepsilon_0)[|x_{\varepsilon}| \leq \varepsilon^{a}]\}.\end{gathered} \tag{1.5.0}$$

**Designation 1.5.2**. In the sequel we denote by:
$\mathbf{E}[\Omega]$, where $\Omega$ an open subset of $\mathbb{R}^n$, the algebra of all the sequences $(u_{\varepsilon}(x))_{\varepsilon\in(0,1]}$ (for short, $(u_{\varepsilon}(x))_{\varepsilon}$ ) of smooth functions $u_{\varepsilon}(x) \in C^{\infty}(\Omega)$.
$\mathbf{E}_M[\Omega]$ is the differential subalgebra of the elements $(u_{\varepsilon}(x))_{\varepsilon} \in \mathbf{E}[\Omega]$ such that for all $K \Subset \Omega$, for all $\alpha \in \mathbb{N}^n$ there exists $N \in \mathbb{N}$ with the following property: $\sup_{x\in K}|\partial^{\alpha}u_{\varepsilon}(x)|$

$= O(\varepsilon^{-N})$ as $\varepsilon \to 0$.
$\mathbf{N}(\Omega)$ the differential subalgebra of the elements $(u_\varepsilon(x))_\varepsilon \in \mathbf{E}[\Omega]$ such that for all $K \Subset \Omega$, for all $\alpha \in \mathbb{N}^n$ and $q \in \mathbb{N}$ the following property holds: $\sup_{x\in K}|\partial^\alpha u_\varepsilon(x)| = O(\varepsilon^q)$ as $\varepsilon \to 0$.
**Definition 1.5.1**.The elements of $\mathbf{E}_M[\Omega]$ and $\mathbf{N}(\Omega)$ are called moderate and negligible, respectively.The factor algebra

$$\mathbf{G}(\Omega) = \mathbf{E}_M[\Omega]/\mathbf{N}(\Omega) \quad (1.5.1)$$

is the algebra of Colombeau generalized functions on $\Omega$.
**Remark 1.5.0**.Note that: (i) there exists natural embedding $\widetilde{r} : \mathbb{R} \hookrightarrow \widetilde{\mathbb{R}}$ such that for all $r \in \mathbb{R}$, $\widetilde{r} : r \to (r_\varepsilon)_\varepsilon$, $r_\varepsilon = r$, for all $\varepsilon \in [1,0)$, (ii) the ring $\widetilde{\mathbb{R}}$ can be endowed with the structure of a partially ordered ring: for $\mathbf{r},\mathbf{s} \in \widetilde{\mathbb{R}}$, $\mathbf{r} \leq_{\widetilde{\mathbb{R}}} \mathbf{s}$ if and only if there are representatives $(r_\varepsilon)_\varepsilon$ and $(s_\varepsilon)_\varepsilon$ with $r_\varepsilon \leq s_\varepsilon$ for all $\varepsilon \in [1,0)$.
**Definition 1.5.2**.(i) Let $\delta \in \widetilde{\mathbb{R}}$. We say that $\delta$ is infinite small Colombeau generalized number and abbraviate

$$\delta \approx_{\widetilde{\mathbb{R}}} \widetilde{0}$$

or $\delta \approx_{\widetilde{\mathbb{R}}} 0$, if there exists representative $(\delta_\varepsilon)_\varepsilon, \varepsilon \in (0,1]$ and some $q \in \mathbb{N}$ such that $|\delta_\varepsilon| = O(\varepsilon^q)$, as $\varepsilon \to 0$. We abbraviate $\alpha \approx_{\widetilde{\mathbb{R}}} \beta$ iff $\alpha - \beta \approx_{\widetilde{\mathbb{R}}} 0$.
(ii) We say that $\Delta \in \widetilde{\mathbb{R}}$ is infinite large Colombeau generalized number and abbraviate

$$\Delta \approx_{\widetilde{\mathbb{R}}} \infty$$

if $\Delta^{-1} \approx_{\widetilde{\mathbb{R}}} 0$.
(iii) We say that $\mathbf{r} \in \widetilde{\mathbb{R}}$ is finite Colombeau generalized number and abbraviate $\mathbf{r}_{fin}$ if there are $r_1, r_2 \in \mathbb{R}$ such that $\widetilde{r}_1 \leq \mathbf{r} \leq \widetilde{r}_2$.
**Definition 1.5.3**.Let $\Omega_{1,M}$ and $\Omega_{2,M}$ be a set

$$\Omega_{1,M} = \left\{(x_\varepsilon)_\varepsilon \in \Omega^{(0,1]} | \exists N \in \mathbb{N}, |x_\varepsilon| = O(\varepsilon^{-N}), \text{as } \varepsilon \to 0\right\},$$
$$\Omega_{2,M} = \left\{(x_\varepsilon)_\varepsilon \in \Omega^{(0,1]} | \exists N \in \mathbb{N}, |x_\varepsilon| = O(\varepsilon^{N}), \text{as } \varepsilon \to 0\right\}.$$

correspondingly. We introduce equivalence relation given by

$$(x_\varepsilon)_\varepsilon \sim (y_\varepsilon)_\varepsilon \Leftrightarrow \forall q \in \mathbb{N}, |x_\varepsilon - y_\varepsilon| = O(\varepsilon^q), \text{as } \varepsilon \to 0,$$

and denote by $\widetilde{\Omega} = (\Omega_{1,M} \cup \Omega_{2,M})/\sim$ the set of generalized points. Moreover, if $[(x_\varepsilon)_\varepsilon]$ is the class of $(x_\varepsilon)_\varepsilon$ in $\widetilde{\Omega}$ then the set of compactly generalized points is

$$\widetilde{\Omega}_c = \left\{x = [(x_\varepsilon)_\varepsilon] \in \widetilde{\Omega} \middle| \exists K(K \Subset \Omega)\exists \eta(\eta > 0)\forall \varepsilon(\varepsilon \in (0,\eta])[x_\varepsilon \in K]\right\}.$$

Note that if the $\widetilde{\Omega}_c$- property holds for one representative of $x \in \widetilde{\Omega}_c$ then it holds for every representative. Also, for $\Omega = \mathbb{R}$ we have that the factor $\mathbb{R}_M/\sim$ is the usual algebra of real generalized numbers.
**Definition 1.5.4**.We denote by $\widetilde{\Omega}_{IL} \triangleq \Omega_{1,M}/\sim$ and $\widetilde{\Omega}_{ISM} \triangleq \Omega_{2,M}/\sim$ the set of inite large and infite small generalized points correspondingly. It is clear that the generalized point value of $u(x)$ at $x \in \widetilde{\Omega}_{ISM}$ is

$$u(x) = (u_\varepsilon(x_\varepsilon))_\varepsilon + \mathbf{N}(\Omega).$$

**Definition 1.5.5**.

## 5mm . 1.5.2.A real Colombeau vector bundle 2mm

**Definition 1.5.6**.A real vector bundle consists of:
1.topological spaces $X$ (base space) and $E$ (total space)
2.a continuous surjection $\pi : E \to X$ (bundle projection)
3.for every $x$ in $X$, the structure of a finite-dimensional vector space over Colombeau ring $\widetilde{\mathbb{R}}$ on the fiber $\pi^{-1}(\{x\})$ where the following compatibility condition is satisfied: for every point in $X$, there is an open neighborhood $U$, a natural number $k$, and a homeomorphism $\varphi : U \times \widetilde{\mathbb{R}}^k \to \pi^{-1}(U)$ such that for all $x \in U, (\pi \circ \varphi)(x,v) = x$ for all vectors v in $\widetilde{\mathbb{R}}^k$, and the map $v \mapsto \varphi(x,v)$ is a linear isomorphism between the vector spaces $\widetilde{\mathbb{R}}^k$ and $\pi^{-1}(\{x\})$.

The open neighborhood $U$ together with the homeomorphism $\varphi$ is called a local trivialization of the Colombeau vector bundle. The local trivialization shows that locally the map $\pi$ "looks like" the projection of $U \times \widetilde{\mathbb{R}}^k$ on $U$.

The Cartesian product $X \times \widetilde{\mathbb{R}}^k$, equipped with the projection $X \times \widetilde{\mathbb{R}}^k \to X$, is called the trivial bundle of rank $k$ over $X$.

### 5mm . 1.5.3.The algebra of Colombeau generalized functions 2mm

The basic idea of Colombeau's theory of generalized functions is regularization by sequences (nets) of smooth functions and the use of asymptotic estimates in terms of a regularization parameter $\varepsilon \in (0,1]$. Let $(u_\varepsilon)_{\varepsilon\in(0,1]}$ with $u_\varepsilon \in C^\infty(M)$ for all $\varepsilon \in (0,1]$ ($M$ a separable, smooth orientable Hausdorff manifold of dimension $n$).The algebra of Colombeau generalized functions on $M$ is defined as the quotient

$$\mathbf{G}(M) = \mathbf{E}_M(M)/\mathbf{N}(M) \tag{1.5.2}$$

of the space $\mathbf{E}_M(M)$ of sequences of moderate growth modulo the space $\mathbf{N}(M)$ of negligible sequences. More precisely the notions of moderateness resp. negligibility are defined by the following asymptotic estimates ($\widetilde{\mathbf{X}}_{\widetilde{\mathbb{R}}}(M)$ or $\widetilde{\mathbf{X}}(M)$ denoting the space of smooth vector fields on $M$).

$$\begin{gathered}\mathbf{E}_M(M) = \Big\{(u_\varepsilon)_\varepsilon \in C^\infty(M)^{(0,1]} \,\Big|\, (\forall K \subset\subset M)(\forall k \in \mathbb{N}_0)(\exists n \in \mathbb{N}) \\ \Big(\forall \xi_1 \in \widetilde{\mathbf{X}}(M),\ldots,\forall \xi_k \in \widetilde{\mathbf{X}}(M)\Big)[\sup_{p\in K}|L_{\xi_1}\ldots L_{\xi_n}u_\varepsilon(p)| \leq O(\varepsilon^{-n})]\Big\}, \\ \mathbf{N}(M) = \{(u_\varepsilon)_\varepsilon \in M^{(0,1]} | (\forall K \subset\subset M)(\forall k,q,l \in \mathbb{N}_0) \\ \Big(\forall \xi_1 \in \widetilde{\mathbf{X}}(M),\ldots,\forall \xi_k \in \widetilde{\mathbf{X}}(M)\Big)[\sup_{p\in K}|L_{\xi_1}\ldots L_{\xi_n}u_\varepsilon(p)| \leq O(\varepsilon^q\delta^l)]\Big\}.\end{gathered} \tag{1.5.3}$$

Elements of $G(M)$ are denoted by

$$\bar{u} = \mathbf{cl}[(u_\varepsilon)_\varepsilon] = \overline{[(u_\varepsilon)_\varepsilon]} = (u_\varepsilon)_\varepsilon + \mathbf{N}(M). \tag{1.5.4}$$

With componentwise operations $G(M)$ is a fine sheaf of differential algebras with respect to the Lie derivative defined by

$$L_\xi \bar{u} := \mathbf{cl}[(L_\xi u_\varepsilon)_\varepsilon] = \overline{[(L_\xi u_\varepsilon)_\varepsilon]}. \tag{1.5.5}$$

The spaces of moderate resp. negligible sequences and hence the algebra itself may be characterized locally, i.e., $\bar{u} \in G(M)$ iff $\bar{u}\circ\psi_\alpha \in G(\psi_\alpha(V_\alpha))$ for all charts $(V_\alpha,\psi_\alpha)$, where on the open set $\psi_\alpha(V_\alpha) \subset \mathbb{R}^n$ in the respective estimates Lie derivatives are replaced by partial derivatives. Smooth functions are embedded into $G(M)$ simply by the "constant" embedding $\sigma$, i.e., $\sigma(f) := cl[(f)_\varepsilon]$, hence $C^\infty(M)$ is a faithful subalgebra of $G(M)$. On open sets of $\mathbb{R}^n$ compactly supported distributions are embedded into $G$ via convolution with a mollifier $\rho \in S(\mathbb{R}^n)$ with unit integral satisfying $\int \rho(x)x^\alpha dx = 0$ for all $|\alpha| \geq 1$; more precisely setting $\rho_\varepsilon(x) = (1/\varepsilon^n)\rho(x/\varepsilon)$ we have $\iota(w) := cl[(w * \rho_\varepsilon)_\varepsilon]$. In case $supp(w)$ is not compact one uses a sheaf-theoretical construction.

### 5mm . 1.5.4.Colombeau tangent vector 2mm

Let $\bar{f} = cl[(f_\varepsilon(\mathbf{x}))_\varepsilon] = \overline{[(f_\varepsilon(\mathbf{x}))_\varepsilon]} \in G(\mathbb{R}^n)$, where $f_\varepsilon(\mathbf{x}) : \mathbb{R}^n \to \mathbb{R}, \varepsilon \in (0,1)$ is a differentiable function and let $v$ be a vector in $\mathbb{R}^n$ . We define the Colombeau directional derivative in the $v$ direction at a point $x \in \mathbb{R}^n$ by

$$D_{\mathbf{v}}^{Col}[\bar{f}] = D_{\mathbf{v}}\left(\overline{[(f_\varepsilon(\mathbf{x}))_\varepsilon]}\right) = \overline{[(D_{\mathbf{v}}f_\varepsilon(\mathbf{x}))_\varepsilon]} =$$
$$\overline{\left[\left(\frac{d}{dt}f_\varepsilon(\mathbf{x}+t\mathbf{v})|_{t=0}\right)_\varepsilon\right]} = \overline{\left[\left(\sum_{i=1}^{n}\mathbf{v}_i\frac{\partial f_\varepsilon(\mathbf{x})}{\partial x_i}\right)_\varepsilon\right]}. \tag{1.5.6}$$

The Colombeau tangent vector at the point $x$ may then be defined as

$$\mathbf{v}^{Col}(\bar{f}) = \mathbf{v}\left(\overline{[(f_\varepsilon(\mathbf{x}))_\varepsilon]}\right) = \overline{[(D_{\mathbf{v}}f_\varepsilon(\mathbf{x}))_\varepsilon]}. \tag{1.5.7}$$

Let $\bar{f} = \overline{[(f_\varepsilon(\mathbf{x}))_\varepsilon]} \in G(\mathbb{R}^n), \bar{g} = \overline{[(g_\varepsilon(\mathbf{x}))_\varepsilon]} \in G(\mathbb{R}^n)$, where $f_\varepsilon, g_\varepsilon : \mathbb{R}^n \to \mathbb{R}, \varepsilon \in (0,1)$ be differentiable functions, let $v, w$ be tangent vectors in $\mathbb{R}^n$ at $x \in \mathbb{R}^n$ and let $a, b \in \widetilde{\mathbb{R}}$ . Then

1.$(a \cdot \mathbf{v} + b \cdot \mathbf{w})^{Col}(\bar{f}) = (a \cdot \mathbf{v} + b \cdot \mathbf{w})\left(\overline{[(f_\varepsilon)_\varepsilon]}\right) = a\mathbf{v}\left(\overline{[(f_\varepsilon)_\varepsilon]}\right) + b\mathbf{w}\left(\overline{[(f_\varepsilon)_\varepsilon]}\right) =$
$= a\mathbf{v}^{Col}(\bar{f}) + b\mathbf{w}^{Col}(\bar{f})$;

2.$\mathbf{v}^{Col}(a \cdot \bar{f} + b \cdot \bar{g}) = \mathbf{v}(a \cdot \overline{[(f_\varepsilon)_\varepsilon]} + b \cdot \overline{[(g_\varepsilon)_\varepsilon]}) = a \cdot \mathbf{v}\left(\overline{[(f_\varepsilon)_\varepsilon]}\right) + b \cdot \mathbf{v}\left(\overline{[(g_\varepsilon)_\varepsilon]}\right) =$
$a \cdot \mathbf{v}^{Col}(\bar{f}) + b \cdot \mathbf{v}^{Col}(\bar{g})$;

3.$\mathbf{v}^{Col}(\bar{f} \cdot \bar{g}) = \mathbf{v}(\overline{[(f_\varepsilon \cdot g_\varepsilon)_\varepsilon]}) = \overline{[(f_\varepsilon(\mathbf{x}))_\varepsilon]} \cdot \mathbf{v}\left(\overline{[(g_\varepsilon)_\varepsilon]}\right) + \overline{[(g_\varepsilon(\mathbf{x}))_\varepsilon]} \cdot \mathbf{v}([(f_\varepsilon)_\varepsilon]) =$
$\bar{f} \cdot \mathbf{v}^{Col}(\bar{g}) + \bar{g} \cdot \mathbf{v}^{Col}(\bar{f})$.

### 5mm . 1.5.5.Colombeau tangent vector to differentiable manifold $M$ 2mm

Let $M$ be a differentiable manifold and let $\mathbf{G}(M)$ be the algebra of real-valued Colombeau generalized functions on $M$. Then the tangent vector to $M$ at a point $x$ in the manifold is given by the derivation $D_{\mathbf{v}} : \mathbf{G}(M) \to \widetilde{\mathbb{R}}$ which shall be linear - i.e., for any
$\bar{f} = \overline{[(f_\varepsilon)_\varepsilon]}, \bar{g} = \overline{[(g_\varepsilon)_\varepsilon]} \in \mathbf{G}(M)$ and $a, b \in \widetilde{\mathbb{R}}$ we have

1.$D_{\mathbf{v}}^{Col}(a \cdot \bar{f} + b \cdot \bar{g}) = D_{\mathbf{v}}(a \cdot \overline{[(f_\varepsilon)_\varepsilon]} + b \cdot \overline{[(g_\varepsilon)_\varepsilon]}) = a \cdot D_{\mathbf{v}}(\overline{[(f_\varepsilon)_\varepsilon]}) + b \cdot D_{\mathbf{v}}(\overline{[(g_\varepsilon)_\varepsilon]}) =$
$= a \cdot D_{\mathbf{v}}^{Col}(\bar{f}) + b \cdot D_{\mathbf{v}}^{Col}(\bar{g})$.

Note that the derivation will by definition have the Leibniz property

2.$D_{\mathbf{v}}^{Col}(\bar{f} \cdot \bar{g}) = D_{\mathbf{v}}(\overline{[(f_\varepsilon \cdot g_\varepsilon)_\varepsilon]}) = D_{\mathbf{v}}(\overline{[(f_\varepsilon)_\varepsilon]}) \cdot \overline{[(g_\varepsilon(x))_\varepsilon]} + \overline{[(f_\varepsilon(x))_\varepsilon]} \cdot D_{\mathbf{v}}(\overline{[(g_\varepsilon)_\varepsilon]}) =$
$= D_{\mathbf{v}}^{Col}(\bar{f}) \cdot \bar{g} + \bar{f} \cdot D_{\mathbf{v}}^{Col}(\bar{g})$.

### 5mm . 1.5.6.Colombeau vector fields on distributional manifolds 2mm

Colombeau vector field $\widetilde{\mathbf{X}}_{\widetilde{\mathbb{R}}}$ (denoted often by $\widetilde{\mathbf{X}}$ ) on a manifold $M$ is a linear map $\widetilde{\mathbf{X}}_{\widetilde{\mathbb{R}}}$ :
$\mathbf{G}(M) \to \mathbf{G}(M)$ such that for all $\bar{f}, \bar{g} \in \mathbf{G}(M)$:

$$\widetilde{\mathbf{X}}_{\widetilde{\mathbb{R}}}(\bar{f} \cdot \bar{g}) = \bar{f} \cdot \widetilde{\mathbf{X}}_{\widetilde{\mathbb{R}}}(\bar{f} \cdot \bar{g}) + \widetilde{\mathbf{X}}_{\widetilde{\mathbb{R}}}(\bar{f}) \cdot \bar{g}. \tag{1.5.8}$$

### 5mm . 1.5.7.Colombeau tangent space 2mm

Suppose now that $M$ is a $C^\infty$ manifold. A real-valued Colombeau generalized function $(f_\varepsilon)_\varepsilon$ : $M \to \widetilde{\mathbb{R}}, \varepsilon \in (0,1]$ is said to belong to $\mathbf{G}(M)$ if and only if for every coordinate chart $\varphi : U \to \mathbb{R}^n$, the map $f_\varepsilon \circ \varphi^{-1} : \varphi[U] \subseteq \mathbb{R}^n \to \mathbb{R}$ is infinitely differentiable. Note that $\mathbf{G}(M)$ is a real associative algebra with respect to the pointwise product and sum of Colombeau generalized functions. Pick a point $x \in M$. A derivation at $x$ is defined as a linear map $D : \mathbf{G}(M) \to \widetilde{\mathbb{R}}$ that satisfies the Leibniz identity:

$\bar{f} = \overline{[(f_\varepsilon)_\varepsilon]}, \bar{g} = \overline{[(g_\varepsilon)_\varepsilon]} \in \mathbf{G}(M) : D(\bar{f} \cdot \bar{g}) = D(\bar{f}) \cdot \bar{g}(x) + \bar{f}(x) \cdot D(\bar{g})$,

which is modeled on the product rule of calculus.

If we define addition and scalar multiplication on the set of derivations at $x$ by
$(D_1 + D_2)(\bar{f}) = \bar{f} \cdot D_1(\bar{f}) + D_2(\bar{f})$ and
$(\lambda \cdot D)(\bar{f}) = \bar{f} \cdot \lambda \cdot D(\bar{f})$,
where $\lambda \in \widetilde{\mathbb{R}}$, then we obtain a real vector space over $\widetilde{\mathbb{R}}$, which we define as the Colombeau tangent space $T_x^{Col}M$ of $M$ at $x$.

### 5mm . 1.5.8.Colombeau dual space 2mm

Given any vector space $V_{\widetilde{\mathbb{R}}}$ over Colombeau algebra $\widetilde{\mathbb{R}}$, the (algebraic) Colombeau dual space

$V^*_{\widetilde{\mathbb{R}}}$ (also denoted for a short by $V^*$) is defined as the set of all linear maps $\varphi : V_{\widetilde{\mathbb{R}}} \to \widetilde{\mathbb{R}}$. Since linear maps are vector space homomorphisms, the Colombeau dual space is also sometimes denoted by $Hom(V,\widetilde{\mathbb{R}})$. The Colombeau dual space $V^*_{\widetilde{\mathbb{R}}}$ itself becomes a vector space over $\widetilde{\mathbb{R}}$ when equipped with an addition and scalar multiplication satisfying: (i) $(\varphi+\psi)(x) = \varphi(x)+\psi(x)$ and (ii) $\alpha\varphi(x) = \varphi(\alpha x)$, where $\varphi(x),\psi(x) \in V^*_{\widetilde{\mathbb{R}}}, \alpha \in \widetilde{\mathbb{R}}$.

5mm . **1.5.9.Colombeau cotangent space** 2mm

Let $M$ be a smooth manifold and let $x$ be a point in M. Let $T_xM$ be Colombeau tangent space at $x$. Then Colombeau cotangent space at $x$ is defined as the Colombeau dual space of $T_xM$ : $T^*_xM = (T_xM)^*_{\widetilde{\mathbb{R}}}$.

Suppose now that $M$ is a $C^\infty$ manifold and let $f \in \mathbf{G}(M)$. The differential of $f$ at a point $x$ is the map: $df_x(X_x) = X_x(f)$ where $X_x$ is a tangent vector at $x$, thought of as a derivation. In either case, $df_x$ is a linear map on $T_xM$ and hence it is a tangent covector at $x$.

We can then define the differential map $d : \mathbf{G}(M) \to T^*_xM$ at a point $x$ as the map which sends $f$ to $df_x$. Properties of the differential map include:

(i) $d(af+bg) = adf+bdg$ for $a,b \in \widetilde{\mathbb{R}}$, (ii) $d(fg) = f(x)dg_x + g(x)df_x$.

Let $\varphi_\varepsilon : M \to N$ for any $\varepsilon \in (0,1]$ be a smooth map of smooth manifolds .Given some $x \in M$, the Colombeau differential of $(\varphi_\varepsilon)_\varepsilon$ at $x$ is a linear map $(d\varphi_{\varepsilon,x})_\varepsilon : T_xM \to T_{(\varphi_\varepsilon(x))_\varepsilon}N$from Colombeau tangent space of $M$ at $x$ to Colombeau tangent space of $N$ at $(\varphi_\varepsilon(x))_\varepsilon$. The application of $(d\varphi_{\varepsilon,x})_\varepsilon$ to a tangent vector $X$ is called the pushforward of $X$ by $(\varphi_\varepsilon)_\varepsilon$.

5mm . **1.5.10.$\mathbf{G}(M)$-module of generalized sections $\mathbf{G}(M,E)$ of a vector bundle** $E \to M$ 2mm

The $\mathbf{G}(M)$-module of generalized sections $\mathbf{G}(M,E)$ of a vector bundle $E \to M$ and in particular the space of generalized tensor fields $\mathbf{G}^r_s(M)$ is defined along the same lines using analogous asymptotic estimates with respect to the norm induced by any Riemannian metric on the respective fibers. We denote generalized sections by $S = \mathbf{cl}[(s_\varepsilon)_\varepsilon] = (s_\varepsilon)_\varepsilon + N(M,E)$. Alternatively we may describe a section $S \in \mathbf{G}(M,E)$ by a family $(S_\alpha)_\alpha = ((S^i_\alpha))^N_{i=1}$, where $S_\alpha$ is called the local expression of $S$ with its components $S^i_\alpha \triangleq \Psi^i_\alpha \circ S \circ \psi^{-1}_\alpha \in \mathbf{G}(\psi_\alpha(V_\alpha))$ $((V_\alpha,\Psi_\alpha)_\alpha$ a vector bundle atlas and $i = 1,\ldots,N$, with $N$denoting the dimension of the fibers) satisfying $S^i_\alpha(x) = (\psi_{\alpha\beta})^i_j(\psi_\beta\circ\psi^{-1}_\alpha(x))S^j_\beta(\psi_\beta\circ\psi^{-1}_\alpha(x))$ for all $x \in (V \cap V)$, where $\psi_{\alpha\beta}$ denotes the transition functions of the bundle.

**Remark 1.5.1**. Smooth sections of $E \to M$ again may be embedded as constant nets, i.e.,

$$\Theta(s) : s \to \mathbf{cl}[(s)_\varepsilon].$$

Since $C^\infty(M)$ is a subring of $\mathbf{G}(M), \mathbf{G}(M,E)$ also may be viewed as $C^\infty(M)$-module and the two respective module structures are compatible with respect to the embeddings.

Moreover we have the following algebraic characterization of the space of generalized sections

$$\mathbf{G}(M,E) = \mathbf{G}(M) \otimes \Gamma(M,E), \tag{1.5.9}$$

where $\Gamma(M,E)$ denotes the space of smooth sections and the tensor product is taken over the module $C^\infty(M)$.Generalized tensor fields may be viewed likewise as $C^\infty$- resp. $\mathbf{G}$-multilinear mappings, i.e., as $C^\infty(M)$-resp.$\mathbf{G}(M)$-modules we have

$$\begin{aligned} \mathbf{G}^r_s(M) &\cong L_{C^\infty(M)}\left(\widetilde{X}(M)^r, \widetilde{X}^*(M)^s; \mathbf{G}(M)\right), \\ \mathbf{G}^r_s(M) &\cong L_{\mathbf{G}(M)}(\mathbf{G}^0_1(M)^r, \mathbf{G}^1_0(M)^s; \mathbf{G}(M)). \end{aligned} \tag{1.5.10}$$

Here $\widetilde{X}(M)$ resp. $\widetilde{X}^*(M)$ denotes the space of smooth vector resp. covector fields on $M$.

5mm . **1.5.11.Generalized pseudo-Riemannian manifold** 2mm

A generalized $(0,2)$ tensor field $\bar{g} \in \mathbf{G}^0_2(M)$ is called a generalized Pseudo-Riemannian metric if it has a representative $(g_\varepsilon)_\varepsilon$ satisfying

(i) $g_\varepsilon$ is a smooth Pseudo-Riemannian metric for all $\varepsilon \in (0,1]$, and
(ii) $(\det g_\varepsilon(p))_\varepsilon$ is strictly nonzero on compact sets, i.e.,
$\forall K(K \subset\subset X)\exists m(m \in N)[\inf_{p\in K}|\det g_\varepsilon(p)| \geq \varepsilon^m]$.

We call a separable, smooth Hausdorff manifold $M$ furnished with a generalized pseudo-Riemannian metric $\mathbf{cl}[(g_\varepsilon)_\varepsilon] \triangleq \bar{g}$ generalized pseudo-Riemannian manifold or generalized spacetime and denote it by $(M,\bar{g})$. The action of the metric on a pair $\{H_1,H_2\}$ of generalized vector fields will be denoted by $\bar{g}(H_1,H_2)$ or $\langle H_1,H_2\rangle$
cite: KunzingerSteinbauer02Vickers12Steinbauer00 [14], [20]-[21].

A generalized metric $\bar{g}$ is non-degenerate in the following sense:

$$[H_1 \in \mathbf{G}^1_0(M)] \wedge \forall H_2(H_2 \in \mathbf{G}^1_0(M))[\bar{g}(H_1,H_2) = 0] \Rightarrow H_1 = 0. \tag{1.5.11}$$

Note that condition (ii) above is precisely equivalent to invertibility of $\det(\bar{g})$ in the generalized sense.

The inverse metric $\bar{g}^{-1} \triangleq \mathbf{cl}[(g_\varepsilon^{-1})_\varepsilon]$ is a well defined element of $\mathbf{G}^2_0(M)$, depending exclusively on $\bar{g}$ (i.e., independent of the particular representative $(g_\varepsilon)_\varepsilon$ ).

Moreover if $\bar{g} \approx_k g$, where $g$ is a classical $C^k$-pseudo-Riemannian metric then $\bar{g}^{-1} \approx_k g^{-1}$.

From now on we denote the inverse metric by $\bar{g}^{ab}$, its components by $\bar{g}^{ij}$ and the components of a representative by $\left(g_\varepsilon^{ij}\right)_\varepsilon$. Also we shall denote the generalized metric $\bar{g}_{ab}$ by $\overline{ds}^2 = \mathbf{cl}[(ds^2_\varepsilon)_\varepsilon]$ and its representative by $(ds^2_\varepsilon)_\varepsilon = (g_{ij}(\varepsilon)dx^i dx^j)_\varepsilon$ and use summation convention.

Notice that $\bar{g}$ induces a $\mathbf{G}(M)$-linear isomorphism $\mathbf{G}^1_0(M) \to \mathbf{G}^0_1(M)$ by $\Phi \mapsto \bar{g}(\Phi,\cdot)$, which as in the classical context extends naturally to generalized tensor fields of all types.

### 5mm . 1.5.12. Colombeau isometric embedding 2mm

Let $(M,\bar{g})$ and $(N,\bar{h})$ be generalized pseudo-Riemannian manifolds. An isometric Colombeau embedding is a Colombeau generalized function $(f_\varepsilon)_\varepsilon : M \to N$ which preserves the metric in the sense that $(g_\varepsilon)_\varepsilon$ is equal to the pullback of $(h_\varepsilon)_\varepsilon$ by $(f_\varepsilon)_\varepsilon$, i.e. $(g_\varepsilon)_\varepsilon = (f^*_\varepsilon h_\varepsilon)_\varepsilon$. Explicitly, for any two tangent vectors $\mathbf{v},\mathbf{w} \in T_x(M)$ we have $\left(g_\varepsilon(\mathbf{v}\ ,\ \mathbf{w})\right)_\varepsilon = (h_\varepsilon(df_\varepsilon(\mathbf{v}),df_\varepsilon(\mathbf{w})))_\varepsilon$.

### 5mm . 1.5.13. Generalized connection on a generalized pseudo-Riemannian manifold 2mm

Generalized connection $\bar{D}_{H_1}H_2$ on a manifold $M$ is a map $\mathbf{G}^1_0(M) \times \mathbf{G}^1_0(M) \to \mathbf{G}^1_0(M)$ satisfying:
(D1) $\bar{D}_{H_1}(H_2)$ is $\widetilde{\mathbb{R}}$-linear in $H_2$.
(D2) $\bar{D}_{H_1}(H_2)$ is $\mathbf{G}(M)$-linear in $H_1$.
(D3) $\bar{D}_{H_1}\left(\bar{\mathbf{f}} \cdot H_2\right) = \bar{\mathbf{f}} \cdot \bar{D}_{H_1}(H_2) + H_1(\bar{\mathbf{f}})H_2$ for all $\bar{\mathbf{f}} \in \mathbf{G}(M)$.

Let $(V_\alpha,\psi_\alpha)$ be a chart on M with coordinates $x^i$. The generalized Christoffel symbols for this chart are given by the $[\dim(M)]^3$ generalized functions $\bar{\Gamma}^k_{ij} \in \mathbf{G}(V_\alpha)$defined by

$$\bar{D}_{\partial_i}(\partial_j) = \sum_k \bar{\Gamma}^k_{ij}\partial_k. \tag{1.5.12}$$

**Theorem**.cite: Steinbauer00 [21].**I**.Let $(M,\bar{g})$ be a generalized pseudo-Riemannian manifold. Then there exists a unique generalized connection $\bar{D}_{H_1}(H_2)$ such that
(D4) $[H_1,H_2] = \bar{D}_{H_1}(H_2) - \bar{D}_{H_2}(H_1)$ and
(D5) $H_1\langle H_2,H_3\rangle = \langle\bar{D}_{H_1}(H_2),H_3\rangle + \langle H_2,\bar{D}_{H_1}(H_3)\rangle$
hold for all $H_1,H_2,H_3$ in $\mathbf{G}^1_0(M)$.

$\bar{D}_{H_1}(H_2)$ is called generalized Levi-Civita connection of $M$ and characterized by the so-called Koszul formula

$$\begin{aligned} 2\langle\bar{D}_{H_1}(H_2),H_3\rangle &= H_1\langle H_2,H_3\rangle + H_2\langle H_3,H_1\rangle - H_3\langle H_1,H_2\rangle - \\ &-\langle H_1,[H_2,H_3]\rangle + \langle H_2,[H_3,H_1]\rangle + \langle H_3,[H_1,H_2]\rangle. \end{aligned} \tag{1.5.13}$$

**II**.On every chart $(V_\alpha,\psi_\alpha)$ we have for the generalized Levi-Civita connection $\bar{D}_{H_1}(H_2)$ of

$(M,\overline{g})$ and any vector field $H \in \mathbf{G}_0^1(M)$

$$\overline{D}_{\partial_i}(H^j\partial_j) = \left(\frac{\partial H^k}{\partial x^i} + \overline{\Gamma}^k_{ij}H^j\right)\partial_k. \tag{1.5.14}$$

The generalized Christoffel symbols are given by

$$\overline{\Gamma}^k_{ij} = \frac{1}{2}\overline{g}^{km}\left(\frac{\partial \overline{g}_{jm}}{\partial x^i} + \frac{\partial \overline{g}_{im}}{\partial x^j} + \frac{\partial \overline{g}_{ij}}{\partial x^m}\right), \tag{1.5.15}$$

or by using representative

$$\left(\overline{\Gamma}^k_{ij}(\varepsilon)\right)_\varepsilon = \frac{1}{2}\left[(\overline{g}^{km}(\varepsilon))_\varepsilon\right]\left[\left(\frac{\partial \overline{g}_{jm}(\varepsilon)}{\partial x^i}\right)_\varepsilon + \left(\frac{\partial \overline{g}_{im}(\varepsilon)}{\partial x^j}\right)_\varepsilon + \left(\frac{\partial \overline{g}_{ij}(\varepsilon)}{\partial x^m}\right)_\varepsilon\right]. \tag{1.5.16}$$

We define now the action of a classical (smooth) connection $D$ on generalized vector fields $H_1 = \mathbf{cl}[(\xi_\varepsilon)_\varepsilon]$ and $H_2 = \mathbf{cl}[(\eta_\varepsilon)_\varepsilon]$ by $D_{H_1}(H_2) = \mathbf{cl}[(D_{\xi_\varepsilon}(\eta_\varepsilon))_\varepsilon]$

**III**.Let $(M,\overline{g})$ be a generalized pseudo-Riemannian manifold.

(i) If $\overline{g}_{ab} = \Theta(g_{ab})$ where $g_{ab}$ is a classical smooth metric then we have,in any chart, $\overline{\Gamma}^i_{jk} = \Theta(\Gamma^i_{jk})$ (with $\Gamma^i_{jk}$ denoting the Christoffel Symbols of $g_{ab}$). Hence for all $H \in \mathbf{G}_0^1(M)$ : $\overline{D}_{H_1}(H_2) = D_{H_1}(H_2)$.

(ii) If $\overline{g}_{ab} \approx_\infty g_{ab}$, where $g_{ab}$ a classical smooth metric, $H_1,H_2 \in \mathbf{G}_0^1(M)$ and $H_1 \approx_\infty \xi \in \widetilde{X}(M)$, $H_2 \approx \eta \in D_0'^1(M)$ or $H_1 \approx \xi \in D_0'^1(M)$,$H_2 \approx_\infty \eta \in \widetilde{X}(M)$, then $\overline{D}_{H_1}(H_2) \approx D_\xi(\eta)$.

(iii) Let $g_{ab} \approx_k g_{ab}$, where $g_{ab}$ a classical $C^k$-metric, then, in any chart, $\overline{\Gamma}^i_{jk} \approx_{k-1} \Gamma^i_{jk}$. If in addition,
$H_1,H_2 \in \mathbf{G}_0^1(M)$, $H_1 \approx \xi_{k-1} \in \Gamma^{k-1}(M,TM)$ and $H_2 \approx_k \eta \in \Gamma^k(M,TM)$ then $\overline{D}_{H_1}(H_2) \approx_k D_\xi(\eta)$.

### 5mm . **1.5.14.The generalized Riemannian curvature tensor** 2mm

Let $(M,\overline{g})$ be a generalized pseudo-Riemannian manifold with a generalized Levi-Civita connection $\overline{D}$.

(i) The generalized Riemannian curvature tensor $\overline{R}_{abc}{}^d \in \mathbf{G}_3^1(M)$ is defined by

$$\overline{R}_{H_1,H_2}H_3 \triangleq \overline{D}_{[H_1,H_2]}H_3 - [\overline{D}_{H_1},\overline{D}_{H_2}]H_3. \tag{1.5.17}$$

(ii) The generalized Ricci curvature tensor is defined by

$$\overline{R}_{ab} \triangleq \overline{R}_{cab}{}^c. \tag{1.5.18}$$

(iii) The generalized Ricci scalar is defined by

$$R \triangleq R^a{}_a. \tag{1.5.19}$$

(iv) Finally we define the generalized Einstein tensor by

$$\overline{G}_{ab} \triangleq \overline{R}_{ab} - \tfrac{1}{2}\overline{R}\overline{g}_{ab}. \tag{1.5.20}$$

## 6mm . **1.6.Super generalized functions**. 3mm

### 5mm . **1.6.1.The nonsmooth regularization via horizon** 2mm

Examining now the Schwarzschild metric (1.3.1) (note that the origin is now excluded from our considerations, the space we are working on is $\mathbb{R}^3_+ \triangleq \mathbb{R}^3\backslash\{0\}$) in a neighborhood of the horizon, we see that, whereas $h(r)$ is smooth, $h^{-1}(r)$ is not even $L^1_{loc}$. Thus, regularizing the Schwarzschild metric amounts to embedding $h^{-1}$ into $\mathbf{G}(\mathbb{R}^3_+)$,for example as that given in paper cite: HeinzleSteinbauer02 [17], one obtains Colombeau generalized metric

$$
\begin{gathered}
(ds_\varepsilon^2)_\varepsilon = h(r)dt^2 - (h^{-1}(r,\varepsilon)dr^2)_\varepsilon + r^2 d\Omega^2, \\
h^{-1}(r,\varepsilon) = h_\varepsilon^{-1} = -1 - 2m\left(\mathbf{vp}\left(\frac{1}{r-2m}\right)\right) * \rho_\varepsilon(r) = \\
-1 - 2m \cdot \iota_\varepsilon\left(\mathbf{vp}\left(\frac{1}{r-2m}\right)\right).
\end{gathered} \tag{1.6.1}
$$

Here $\rho_\varepsilon(r) = \varepsilon^{-3}\rho(r\varepsilon^{-3})$ and $\rho(r)$ is a mollifier.

Obviously, (1.6.1) is degenerate at $r = 2m$, because $h(r)$ is zero at the horizon. Due to the degeneracy of (33), the Levi-Civitá connection is not available. In order avoid this difficultness in literature the following generalized pseudo-connection $\overline{\Gamma}_{kj}^i$ was considered $\overline{\Gamma}_{kj}^i = \mathbf{cl}\left[(\Gamma_{kj}^i(\varepsilon))_\varepsilon\right] \in \mathbf{G}(\mathbb{R}_+^3)$ cite: HeinzleSteinbauer02 [17]:

$$
(\Gamma_{kj}^i(\varepsilon))_\varepsilon = \frac{1}{2}\left(\iota_\varepsilon(g^{im})\left[\iota_\varepsilon(g)_{mk,j} + \iota_\varepsilon(g)_{mj,k} - \iota_\varepsilon(g)_{kj,m}\right]\right)_\varepsilon. \tag{1.6.2}
$$

Obviously the generalized pseudo-connection $\overline{\Gamma}_{kj}^i$ coincides with the classical Levi-Civitá connection $\Gamma_{kj}^i$ on $\mathbb{R}_+^3 \backslash \{r = 2m\}$ since $(\iota_\varepsilon(g^{im}))_\varepsilon = g^{im}$, $(\iota_\varepsilon(g_{im}))_\varepsilon = g_{im}$ there. However the generalized pseudo- connection $\overline{\Gamma}_{kj}^i$ is not a true generalized Levi-Civitá connection on $\mathbb{R}_+^3$ since $\overline{\Gamma}_{kj}^i$ does not respect the Colombeau generalized metric (1.6.1),i.e.,$\left(\iota_\varepsilon(g)_{ij,k}\right)_\varepsilon \neq 0$, e.g., $\left(\iota_\varepsilon(g)_{00,1}\right)_\varepsilon = (1 - h(h_\varepsilon^{-1})_\varepsilon)h'$. Compatibility with the metric $(\iota_\varepsilon(g))_\varepsilon$ is a priori ruled out by the following statement: there exists no connection whatsoever under which $(\iota_\varepsilon(g))_\varepsilon$ would be a parallel tensor.However in a weak sense, the connection (1.6.2) is metric compatible: $\left(\iota_\varepsilon(g)_{ij,k}\right)_\varepsilon \approx 0$. In additional cite: HeinzleSteinbauer02 [17]:

$$
(R_{ij}(\varepsilon))_\varepsilon \approx 0, \tag{1.6.3}
$$

where $(R_{ij}(\varepsilon))_\varepsilon$ is a Ricci tensor corresponding to generalized pseudo-connection (1.6.2), and therefore Colombeau object $(R_{ij}(\varepsilon))_\varepsilon$ viewed as a classical distribution on $\mathbb{R}^3 \backslash \{0\}$ gives

$$
R_{ij} = 0 \text{ on } \mathbb{R}^3 \backslash \{0\}. \tag{1.6.4}
$$

**Remark 1.6.1**. In paper cite: HeinzleSteinbauer02 [17] the equality (1.6.4) mistakenly considered as a proof that the metric singularity at the Schwarzschild horizon is only a coordinate singularity.

**Remark 1.6.2**.Due to the degeneracy of any smooth regularization of the metric (1.3.1) no canonical Levi-Civitá connection could be defined. In order to avoid such difficultnes in our papers cite: Foukzon15FoukzonPotapovMenkova16 [18]-[19] the nonsmooth regularization via horizon is considered, see section 2 below. However such regularization demands appropriate extension of the Colombeau algebra $\mathbf{G}(M)$.

5mm . **1.6.2.The super generalized functions** 2mm

The basic idea of the theory of super generalized functions is regularization by sequences (nets) of appropriate classes of non smooth and discontinuous functions or classical distributions and the use of asymptotic estimates in terms of a regularization parameter $\varepsilon \in (0,1]$. Let $(u_\varepsilon)_{\varepsilon\in(0,1]}$ with $u_\varepsilon \in D'(M)$ for all $\varepsilon \in (0,1]$ ($M$ a separable, smooth orientable Hausdorff manifold of dimension $n$).Such sequences are called super generalized functions. The algebra of super generalized functions on $M$ is defined as the quotient

$$
\mathbf{SG}(M) = \mathbf{SE}_M'(M)/\mathbf{SN}'(M) \tag{1.6.5}
$$

of the space $\mathbf{SE}_M'(M)$ of sequences of moderate growth modulo the space $\mathbf{N}(M)$ of negligible sequences. More precisely the notions of moderateness resp. negligibility are defined by the following asymptotic estimates ($\widetilde{\mathbf{X}}_{\widetilde{\mathbb{R}}}(M)$ or $\widetilde{\mathbf{X}}(M)$ denoting the space of smooth vector fields on $M$) cite: Foukzon15FoukzonPotapovMenkova16 [18]-[19]:

$$\mathbf{SE}'_M(M) = \Big\{(u_\varepsilon)_\varepsilon \in (D'(M))^{(0,1]}\Big|(\forall K \subset\subset M)(\forall k \in \mathbb{N}_0)(\exists n \in \mathbb{N})$$
$$\Big(\forall \xi_1 \in \widetilde{\mathbf{X}}(M),\ldots,\forall \xi_k \in \widetilde{\mathbf{X}}(M)\Big)(\forall f \in C_0^\infty(K))[|L_{\xi_1}\ldots L_{\xi_n}u_\varepsilon(f)| \le O_f(\varepsilon^{-n})]\Big\},$$
$$\mathbf{SN}'(M) = \Big\{(u_\varepsilon)_\varepsilon \in (D'(M))^{(0,1]}\Big|(\forall K \subset\subset M)(\forall k,q \in \mathbb{N}_0)(\exists n \in \mathbb{N})$$
$$\Big(\forall \xi_1 \in \widetilde{\mathbf{X}}(M),\ldots,\forall \xi_k \in \widetilde{\mathbf{X}}(M)\Big)(\forall f \in C_0^\infty(K))[|L_{\xi_1}\ldots L_{\xi_n}u_\varepsilon(f)| \le O_f(\varepsilon^{q})]\Big\}. \tag{1.6.6}$$

Here $L^w_{\xi_k}, k = 1,2,\ldots,n$ denoting the weak Lie derivative in L.Schwartz sense and where Landau symbol $a_\varepsilon = O_f(\psi(\varepsilon))$ appears, having the following meaning:
$\forall f \exists C_f\ (C_f > 0)\exists\varepsilon_0(\varepsilon_0 \in (0,1])\forall\varepsilon(\varepsilon < \varepsilon_0)[a_\varepsilon \le C_f\ \psi(\varepsilon)]$.

We denote by $\mathbf{S}\widetilde{\mathbb{R}}$ the ring of real, Colombeau super generalized numbers.
$\mathbf{S}\widetilde{\mathbb{R}} = \mathbf{SE}_{\mathbb{R}}(\mathbb{R})/\mathbf{SN}(\mathbb{R})$, where

$$\mathbf{SE}_{\mathbb{R}}(\mathbb{R}) =$$
$$\Big\{(x_{\varepsilon,\delta})_{\varepsilon,\delta} \in \mathbb{R}^{(0,1]\times(0,1]}\Big|(\exists a,b \in \mathbb{R}_+)(\exists\varepsilon_0,\delta_0 \in (0,1])$$
$$(\forall\varepsilon \le \varepsilon_0)(\forall\delta \le \delta_0)[|x_{\varepsilon,\delta}| \le \varepsilon^{-a}\delta^{-b}]\},$$
$$\mathbf{SN}(\mathbb{R}) = \Big\{(x_{\varepsilon,\delta})_{\varepsilon,\delta} \in \mathbb{R}^{(0,1]\times(0,1]}\Big|(\forall a,b \in \mathbb{R}_+)(\exists\varepsilon_0,\delta_0 \in (0,1))$$
$$(\forall\varepsilon \le \varepsilon_0)(\forall\delta \le \delta_0)[|x_\varepsilon| \le \varepsilon^{a}\delta^{b}]\}. \tag{1.6.7}$$

Let $(u_{\varepsilon,\delta})_{\varepsilon,\delta\in(0,1]}$ with $u_{\varepsilon,\delta} \in C^\infty(M)$ for all $\varepsilon,\delta \in (0,1]$ ($M$ a separable, smooth orientable Hausdorff manifold of dimension $n$). By using canonical imbeding $D'(M) \hookrightarrow \mathbf{G}(M)$ the algebra $\mathbf{SG}(M)$ of Colombeau super generalized functions on $M$ is defined also in the equivalent form as the quotient

$$\mathbf{SG}(M) = \mathbf{SE}_M(M)/\mathbf{SN}(M) \tag{1.6.8}$$

of the space $\mathbf{SE}_M(M)$ of sequences of moderate growth modulo the space $\mathbf{SN}(M)$ of negligible sequences. More precisely the notions of moderateness resp. negligibility are defined by the following asymptotic estimates ($\widetilde{\mathbf{X}}_{\widetilde{\mathbb{R}}}(M)$ or $\widetilde{\mathbf{X}}(M)$ denoting the space of smooth vector fields on $M$).

$$\mathbf{SE}_M(M) = \Big\{(u_{\varepsilon,\delta})_{\varepsilon,\delta} \in C^\infty(M)^{(0,1]\times(0,1]}\Big|(\forall K \subset\subset M)(\forall k \in \mathbb{N}_0)(\exists n,m \in \mathbb{N})$$
$$\Big(\forall \xi_1 \in \widetilde{\mathbf{X}}(M),\ldots,\forall \xi_k \in \widetilde{\mathbf{X}}(M)\Big)[\sup_{p\in K}|L_{\xi_1}\ldots L_{\xi_n}u_{\varepsilon,\delta}(p)| \le O(\varepsilon^{-n}\delta^{-m})]\Big\},$$
$$\mathbf{SN}(M) = \{(u_{\varepsilon,\delta})_{\varepsilon,\delta} \in M^{(0,1\times(0,1]]}|(\forall K \subset\subset M)(\forall k,q,l \in \mathbb{N}_0)$$
$$\Big(\forall \xi_1 \in \widetilde{\mathbf{X}}(M),\ldots,\forall \xi_k \in \widetilde{\mathbf{X}}(M)\Big)[\sup_{p\in K}|L_{\xi_1}\ldots L_{\xi_n}u_\varepsilon(p)| \le O(\varepsilon^{q}\delta^{l})]\Big\}. \tag{1.6.9}$$

The $\mathbf{SG}(M)$-module of super generalized sections $\mathbf{SG}(M,E)$ of a vector bundle $E \to M$ and in particular the space of super generalized tensor fields $\mathbf{SG}^r_s(M)$ is defined along the same lines using analogous asymptotic estimates with respect to the norm induced by any Riemannian metric on the respective fibers. We denote super generalized sections by
$S = \mathbf{cl}[(s_{\varepsilon,\delta})_{\varepsilon,\delta}] = (s_{\varepsilon,\delta})_{\varepsilon,\delta} + \mathbf{SN}(M,E)$. Alternatively we may describe a section $S \in \mathbf{SG}(M,E)$by a family $(S_\alpha)_\alpha = ((S^i_\alpha))^N_{i=1}$, where $S_\alpha$ is called the local expression of $S$ with its components $S^i_\alpha \triangleq \Psi^i_\alpha \circ S \circ \psi^{-1}_\alpha \in \mathbf{SG}(\psi_\alpha(V_\alpha))$ $((V_\alpha,\Psi_\alpha)_\alpha$ a vector bundle atlas and $i = 1,\ldots,N$, with $N$ denoting the dimension of the fibers) satisfying $S^i_\alpha(x) = (\psi_{\alpha\beta})^i_j(\psi_\beta \circ \psi^{-1}_\alpha(x))$
$S^j_\beta(\psi_\beta \circ \psi^{-1}_\alpha(x))$ for all $x \in (V \cap V)$, where $\psi_{\alpha\beta}$ denotes the transition functions of the bundle.

**Remark 1.6.3**. Smooth sections of $E \to M$ again may be embedded as constant nets, i.e.,

$$\Theta(s) : s \to \mathbf{cl}[(s)_{\varepsilon,\delta}].$$

Since $C^\infty(M)$ is a subring of $\mathbf{SG}(M)$, $\mathbf{SG}(M,E)$ also may be viewed as $C^\infty(M)$-module and the two respective module structures are compatible with respect to the embeddings.

Moreover we have the following algebraic characterization of the space of super generalized sections

$$\mathbf{SG}(M,E) = \mathbf{SG}(M) \otimes \Gamma(M,E), \qquad (1.6.10)$$

where $\Gamma(M,E)$ denotes the space of smooth sections and the tensor product is taken over the module $C^\infty(M)$.Generalized tensor fields may be viewed likewise as $C^\infty$- resp. $\mathbf{SG}$-multilinear mappings, i.e., as $C^\infty(M)$-resp. $\mathbf{SG}(M)$-modules we have

$$\begin{gathered}\mathbf{SG}^r_s(M) \cong L_{C^\infty(M)}\left(\widetilde{X}(M)^r, \widetilde{X}^*(M)^s; \mathbf{SG}(M)\right),\\ \mathbf{SG}^r_s(M) \cong L_{\mathbf{SG}(M)}(\mathbf{SG}^0_1(M)^r, \mathbf{SG}^1_0(M)^s; \mathbf{SG}(M)).\end{gathered} \qquad (1.6.11)$$

Here $\widetilde{X}(M)$ resp. $\widetilde{X}^*(M)$ denotes the space of smooth vector resp. covector fields on $M$.

5mm . **1.6.3.Super generalized pseudo-Riemannian manifold** 2mm

A super generalized $(0,2)$ tensor field $\bar{g} \in \mathbf{SG}^0_2(M)$ is called a super generalized Pseudo-Riemannian metric if it has a representative $(g_{\varepsilon,\delta})_{\varepsilon,\delta}$ satisfying:

(i) $g_{\varepsilon,\delta}$ is a smooth Pseudo-Riemannian metric for all $\varepsilon,\delta \in (0,1]$, and

(ii) $(\det g_{\varepsilon,\delta}(p))_{\varepsilon,\delta}$ is strictly nonzero on compact sets, i.e.,

$\forall K(K \subset\subset X)\exists m(m \in N)\exists l(l \in N)[\inf_{p\in K}|\det g_{\varepsilon,\delta}(p)| \geq \varepsilon^m \delta^l]$.

We call a separable, smooth Hausdorff manifold $M$ furnished with a super generalized pseudo-Riemannian metric $\mathbf{cl}\left[(g_{\varepsilon,\delta})_{\varepsilon,\delta}\right] \triangleq \bar{g}$ super generalized pseudo-Riemannian manifold or super generalized spacetime and denote it by $(M,\bar{g})$. The action of the metric on a pair $\{H_1,H_2\}$ of super generalized vector fields will be denoted by $\bar{g}(H_1,H_2)$ or $\langle H_1,H_2\rangle$.

A super generalized metric $\bar{g}$ is non-degenerate in the following sense:

$$[H_1 \in \mathbf{SG}^1_0(M)] \wedge \forall H_2(H_2 \in \mathbf{SG}^1_0(M))[\bar{g}(H_1,H_2) = 0] \Rightarrow H_1 = 0. \qquad (1.6.12)$$

Note that condition (ii) above is precisely equivalent to invertibility of $\det(\bar{g})$ in the super generalized sense. The inverse metric

$\bar{g}^{-1} \triangleq \mathbf{cl}[(g^{-1}_{\varepsilon,\delta})_{\varepsilon,\delta}]$ is a well defined element of $\mathbf{SG}^2_0(M)$, depending exclusively on $\bar{g}$ (i.e., independent of the particular representative

$(g_{\varepsilon,\delta})_{\varepsilon,\delta}$ ). Moreover if $\bar{g} \approx_k g$, where $g$ is a classical $C^k$-pseudo-Riemannian metric then $\bar{g}^{-1} \approx_k g^{-1}$. From now on we denote the inverse metric by $\bar{g}^{ab}$, its components by $\bar{g}^{ij}$ and the components of a representative by $\left(g^{ij}_\varepsilon\right)_{\varepsilon,\delta}$. Also we shall denote the super generalized metric $\bar{g}_{ab}$ by $\overline{ds}^2 = \mathbf{cl}[(ds^2_{\varepsilon,\delta})_{\varepsilon,\delta}]$ and its representative by $(ds^2_{\varepsilon,\delta})_{\varepsilon,\delta} = (g_{ij}(\varepsilon,\delta)dx^i dx^j)_{\varepsilon,\delta}$ and use summation convention.Notice that $\bar{g}$ induces a $\mathbf{SG}(M)$-linear isomorphism

$\mathbf{SG}^1_0(M) \to \mathbf{SG}^0_1(M)$ by $\Phi \mapsto \bar{g}(\Phi,\cdot)$, which as in the classical context extends naturally to generalized tensor fields of all types.

Let $(M,\bar{g})$ and $(N,\bar{h})$ be super generalized pseudo-Riemannian manifolds. An isometric Colombeau embedding is a Colombeau super generalized function $(f_{\varepsilon,\delta})_{\varepsilon,\delta} : M \to N$ which preserves the metric in the sense that $(g_{\varepsilon,\delta})_{\varepsilon,\delta}$ is equal to the pullback of $(h_{\varepsilon,\delta})_{\varepsilon,\delta}$ by $(f_{\varepsilon,\delta})_{\varepsilon,\delta}$, i.e. $(g_{\varepsilon,\delta})_{\varepsilon,\delta} = (f^*_{\varepsilon,\delta} h_{\varepsilon,\delta})_{\varepsilon,\delta}$.Explicitly, for any two tangent vectors $\mathbf{v},\mathbf{w} \in T_x(M)$ we have $\left(g_{\varepsilon,\delta}(\mathbf{v}\,,\,\mathbf{w})\right)_{\varepsilon,\delta} = (h_{\varepsilon,\delta}(df_{\varepsilon,\delta}(\mathbf{v}), df_{\varepsilon,\delta}(\mathbf{w})))_{\varepsilon,\delta}$.

5mm . **1.6.4.Super generalized connection on a super generalized pseudo-Riemannian manifold** 2mm

Super generalized connection $\bar{D}_{H_1}H_2$ on a manifold $M$ is a map

$\mathbf{SG}^1_0(M) \times \mathbf{SG}^1_0(M) \to \mathbf{SG}^1_0(M)$ satisfying:

(D1) $\bar{D}_{H_1}(H_2)$ is $\mathbf{S}\widetilde{\mathbb{R}}$-linear in $H_2$.

(D2) $\bar{D}_{H_1}(H_2)$ is $\mathbf{SG}(M)$-linear in $H_1$.

(D3) $\bar{D}_{H_1}\left(\bar{\mathbf{f}} \cdot H_2\right) = \bar{\mathbf{f}} \cdot \bar{D}_{H_1}(H_2) + H_1(\bar{\mathbf{f}})H_2$ for all $\bar{\mathbf{f}} \in \mathbf{SG}(M)$.

Let $(V_\alpha,\psi_\alpha)$ be a chart on M with coordinates $x^i$. The super generalized Christoffel symbols for this chart are given by the $[\dim(M)]^3$ super generalized functions $\bar{\Gamma}^k_{ij} \in \mathbf{SG}(V_\alpha)$ defined by

$$\overline{D}_{\partial_i}(\partial_j) = \sum_k \overline{\Gamma}_{ij}^k \partial_k. \qquad (1.6.13)$$

**Theorem**.**I**.Let $(M,\overline{g})$ be a super generalized pseudo-Riemannian manifold. Then there exists a unique super generalized connection $\overline{D}_{H_1}(H_2)$ such that

(D4) $[H_1,H_2] = \overline{D}_{H_1}(H_2) - \overline{D}_{H_2}(H_1)$ and

(D5) $H_1\langle H_2,H_3\rangle = \langle \overline{D}_{H_1}(H_2),H_3\rangle + \langle H_2,\overline{D}_{H_1}(H_3)\rangle$

hold for all $H_1,H_2,H_3$ in $\mathbf{SG}_0^1(M)$.

$\overline{D}_{H_1}(H_2)$ is called super generalized Levi-Civita connection of $M$ and characterized by the so-called Koszul formula

$$\begin{gathered} 2\langle \overline{D}_{H_1}(H_2),H_3\rangle = H_1\langle H_2,H_3\rangle + H_2\langle H_3,H_1\rangle - H_3\langle H_1,H_2\rangle - \\ -\langle H_1,[H_2,H_3]\rangle + \langle H_2,[H_3,H_1]\rangle + \langle H_3,[H_1,H_2]\rangle. \end{gathered} \qquad (1.6.14)$$

**II**.On every chart $(V_\alpha,\psi_\alpha)$ we have for the super generalized Levi-Civita connection $\overline{D}_{H_1}(H_2)$ of $(M,\overline{g})$ and any vector field $H \in \mathbf{SG}_0^1(M)$

$$\overline{D}_{\partial_i}(H^j\partial_j) = \left(\frac{\partial H^k}{\partial x^i} + \overline{\Gamma}_{ij}^k H^j\right)\partial_k. \qquad (1.6.15)$$

The super generalized Christoffel symbols are given by

$$\overline{\Gamma}_{ij}^k = \frac{1}{2}\overline{g}^{km}\left(\frac{\partial \overline{g}_{jm}}{\partial x^i} + \frac{\partial \overline{g}_{im}}{\partial x^j} + \frac{\partial \overline{g}_{ij}}{\partial x^m}\right), \qquad (1.6.16)$$

or by using representative

$$\begin{gathered} \left(\overline{\Gamma}_{ij}^k(\varepsilon,\delta)\right)_{\varepsilon,\delta} = \\ \frac{1}{2}\left[\left(\overline{g}^{km}(\varepsilon,\delta)\right)_{\varepsilon,\delta}\right]\left[\left(\frac{\partial \overline{g}_{jm}(\varepsilon,\delta)}{\partial x^i}\right)_{\varepsilon,\delta} + \left(\frac{\partial \overline{g}_{im}(\varepsilon,\delta)}{\partial x^j}\right)_{\varepsilon,\delta} + \left(\frac{\partial \overline{g}_{ij}(\varepsilon,\delta)}{\partial x^m}\right)_{\varepsilon,\delta}\right]. \end{gathered} \qquad (1.6.17)$$

We define now the action of a classical (smooth) connection $D$ on super generalized vector fields $H_1 = \mathbf{cl}\left[(\xi_{\varepsilon,\delta})_{\varepsilon,\delta}\right]$ and $H_2 = \mathbf{cl}\left[(\eta_{\varepsilon,\delta})_{\varepsilon,\delta}\right]$ by $D_{H_1}(H_2) = \mathbf{cl}\left[(D_{\xi_{\varepsilon,\delta}}(\eta_{\varepsilon,\delta}))_{\varepsilon,\delta}\right]$

**III**.Let $(M,\overline{g})$ be a super generalized pseudo-Riemannian manifold.

(i) If $\overline{g}_{ab} = \Theta(g_{ab})$ where $g_{ab}$ is a classical smooth metric then we have,in any chart, $\overline{\Gamma}_{jk}^i = \Theta(\Gamma_{jk}^i)$ (with $\Gamma_{jk}^i$ denoting the Christoffel Symbols of $g_{ab}$). Hence for all $H \in \mathbf{SG}_0^1(M) : \overline{D}_{H_1}(H_2) = D_{H_1}(H_2)$.

(ii) If $\overline{g}_{ab} \approx_\infty g_{ab}$, where $g_{ab}$ a classical smooth metric, $H_1,H_2 \in \mathbf{SG}_0^1(M)$ and $H_1 \approx_\infty \xi \in \widetilde{X}(M)$, $H_2 \approx \eta \in D_0'^1(M)$ or $H_1 \approx \xi \in D_0'^1(M)$,$H_2 \approx_\infty \eta \in \widetilde{X}(M)$, then $\overline{D}_{H_1}(H_2) \approx D_\xi(\eta)$.

(iii) Let $g_{ab} \approx_k g_{ab}$, where $g_{ab}$ a classical $C^k$-metric, then, in any chart, $\overline{\Gamma}_{jk}^i \approx_{k-1} \Gamma_{jk}^i$.

If in addition , $H_1,H_2 \in \mathbf{SG}_0^1(M)$, $H_1 \approx \xi_{k-1} \in \Gamma^{k-1}(M,TM)$ and $H_2 \approx_k \eta \in \Gamma^k(M,TM)$ then $\overline{D}_{H_1}(H_2) \approx_k D_\xi(\eta)$.

## 5mm . **1.6.5.The super generalized Riemannian curvature tensor** 2mm

Let $(M,\overline{g})$ be a super generalized pseudo-Riemannian manifold with a super generalized Levi-Civita connection $\overline{D}$.

(i) The super generalized Riemannian curvature tensor $\overline{R}_{abc}{}^d \in \mathbf{SG}_3^1(M)$ is defined by

$$\overline{R}_{H_1,H_2}H_3 \triangleq \overline{D}_{[H_1,H_2]}H_3 - [\overline{D}_{H_1},\overline{D}_{H_2}]H_3. \qquad (1.6.18)$$

(ii) The super generalized Ricci curvature tensor is defined by

$$\overline{R}_{ab} \triangleq \overline{R}_{cab}{}^c. \qquad (1.6.19)$$

(iii) The super generalized Ricci scalar is defined by

$$R \triangleq R^a{}_a. \quad (1.6.20)$$

(iv) Finally we define the super generalized Einstein tensor by

$$\overline{G}_{ab} \triangleq \overline{R}_{ab} - \tfrac{1}{2}\overline{R}\overline{g}_{ab}. \quad (1.6.21)$$

6mm .**1.7.The generalized gravitational field equations**.

.

# 1.7. Colombeau generalized curvilinear coordinates

Let us consider the Colombeau generalized transformation from one generalized coordinate system, $(x^0_\varepsilon)_\varepsilon, (x^1_\varepsilon)_\varepsilon, (x^2_\varepsilon)_\varepsilon, (x^3_\varepsilon)_\varepsilon$, to another generalized coordinate system $(x'^0_\varepsilon)_\varepsilon, (x'^1_\varepsilon)_\varepsilon, (x'^2_\varepsilon)_\varepsilon, (x'^3_\varepsilon)_\varepsilon$ : transform according to the relation

$$(x^i_\varepsilon)_\varepsilon = \left(f^i_\varepsilon(x'^0_\varepsilon, x'^1_\varepsilon, x'^2_\varepsilon, x'^3_\varepsilon)\right)_\varepsilon, \quad (1.7.1)$$

where the $\left(f^i_\varepsilon\right)_\varepsilon$ are certain Colombeau generalized functions and where $(\mathbf{J}_\varepsilon(x'^0, x'^1, x'^2, x'^3))_\varepsilon$

$$(\mathbf{J}_\varepsilon(x'^0_\varepsilon, x'^1_\varepsilon, x'^2_\varepsilon, x'^3_\varepsilon))_\varepsilon = \left(\frac{\partial(x^0_\varepsilon, x^1_\varepsilon, x^2_\varepsilon, x^3_\varepsilon)}{\partial(x'^0_\varepsilon, x'^1_\varepsilon, x'^2_\varepsilon, x'^3_\varepsilon)}\right)_\varepsilon \neq 0_{\widetilde{\mathbb{R}}} \quad (1.7.1.a)$$

is the Jacobian of the Colombeau generalized transformation (1.7.1).

**Remark 1.7.1.**When we transform the coordinates, their Colombeau differentials $(dx^i_\varepsilon)_\varepsilon$ transform according to the relation

$$(dx^i_\varepsilon)_\varepsilon = \left(\frac{\partial x^i_\varepsilon}{\partial x'^k_\varepsilon} dx'^k_\varepsilon\right)_\varepsilon = \left(\frac{\partial x^i_\varepsilon}{\partial x'^k_\varepsilon}\right)_\varepsilon (dx'^k_\varepsilon)_\varepsilon. \quad (1.7.2)$$

**Definition 1.7.1.**Every tuple of four Colombeau quantities $(A^i_\varepsilon)_\varepsilon, i = 0,1,2,3$, which under a transformation (1.7.1) of coordinates, transform like the Colombeau coordinate differentials (1.7.2), is called Colombeau contravariant four-vector:

$$(A^i_\varepsilon)_\varepsilon = \left(\frac{\partial x^i_\varepsilon}{\partial x'^k_\varepsilon} A'^k_\varepsilon\right)_\varepsilon = \left(\frac{\partial x^i_\varepsilon}{\partial x'^k_\varepsilon}\right)_\varepsilon (A'^k_\varepsilon)_\varepsilon. \quad (1.7.3)$$

Let $(\varphi_\varepsilon)_\varepsilon$ be the Colombeau scalar. Under a coordinate transformation (1.7.1), the four Colombeau quantities $\left(\frac{\partial \varphi_\varepsilon}{\partial x^i_\varepsilon}\right)_\varepsilon, i = 0,1,2,3$ transform according to the formula

$$\left(\frac{\partial \varphi_\varepsilon}{\partial x^i_\varepsilon}\right)_\varepsilon = \left(\frac{\partial \varphi_\varepsilon}{\partial x'^k_\varepsilon}\frac{\partial x'^k_\varepsilon}{\partial x^i_\varepsilon}\right)_\varepsilon = \left(\frac{\partial \varphi_\varepsilon}{\partial x'^k_\varepsilon}\right)_\varepsilon \left(\frac{\partial x'^k_\varepsilon}{\partial x^i_\varepsilon}{}_\varepsilon\right). \quad (1.7.4)$$

**Definition 1.7.2.**Every tuple of four Colombeau generalized functions $(A_{i,\varepsilon})_\varepsilon$ which, under a coordinate transformation (1.7.1), transform like the Colombeau derivatives of a scalar, is called Colombeau generalized covariant four-vector

$$(A_{i,\varepsilon})_\varepsilon = \left(\frac{\partial x'^k_\varepsilon}{\partial x^i_\varepsilon} A'_{k,\varepsilon}\right)_\varepsilon = \left(\frac{\partial x'^k_\varepsilon}{\partial x^i_\varepsilon}\right)_\varepsilon (A'_{k,\varepsilon})_\varepsilon. \quad (1.7.5)$$

**Definition 1.7.3.**We call the Colombeau generalized contravariant tensor of the second rank, $(A^{ik}_\varepsilon)_\varepsilon$,any tuple of sixteen Colombeau generalized functions which transform like the products of the components of two Colombeau generalized contravariant vectors, i.e.

according to the law

$$(A_\varepsilon^{ik})_\varepsilon = \left(\frac{\partial x_\varepsilon'^k}{\partial x_\varepsilon^i}\frac{\partial x_\varepsilon'^m}{\partial x_\varepsilon^i}A'_{im,\varepsilon}\right)_\varepsilon = \left(\frac{\partial x_\varepsilon'^k}{\partial x_\varepsilon^i}\frac{\partial x_\varepsilon'^m}{\partial x_\varepsilon^i}\right)_\varepsilon (A'_{im,\varepsilon})_\varepsilon \quad (1.7.6)$$

and a mixed Colombeau generalized tensor transforms as follows

$$(A^i_{k,\varepsilon})_\varepsilon = \left(\frac{\partial x_\varepsilon^i}{\partial x_\varepsilon'^l}\frac{\partial x_\varepsilon'^m}{\partial x_\varepsilon^k}A'^l_{m,\varepsilon}\right)_\varepsilon = \left(\frac{\partial x_\varepsilon^i}{\partial x_\varepsilon'^l}\frac{\partial x_\varepsilon'^m}{\partial x_\varepsilon^k}\right)_\varepsilon (A'^l_{m,\varepsilon})_\varepsilon. \quad (1.7.7)$$

**Remark 1.7.2.** Note that the scalar product of two four-vectors $(A^i_\varepsilon B_{i,\varepsilon})$ is invariant since

$$(A^i_\varepsilon B_{i,\varepsilon})_\varepsilon = \left(\frac{\partial x_\varepsilon^i}{\partial x_\varepsilon'^l}\frac{\partial x_\varepsilon'^m}{\partial x_\varepsilon^i}A'^l_\varepsilon B'_{m,\varepsilon}\right)_\varepsilon = \left(\frac{\partial x_\varepsilon'^m}{\partial x_\varepsilon'^l}A'^l_\varepsilon B'_{m,\varepsilon}\right)_\varepsilon = (A'^l_\varepsilon B'_{l,\varepsilon})_\varepsilon. \quad (1.7.8)$$

The unit four-tensor $\delta^i_k$ is defined the same as in classical case: $\delta^i_k = 0$ for $i \neq k$ and $\delta^i_k = 1$ for $i = k$. If $(A^k_\varepsilon)_\varepsilon$ is a Colombeau generalized four-vector,then multiplying by $\delta^i_k$ we obtain

$$(A^k_\varepsilon \delta^i_k)_\varepsilon = (A^i_\varepsilon)_\varepsilon, \quad (1.7.9)$$

i.e. again Colombeau generalized four-vector; this proves that $\delta^i_k$ is a tensor.

**Remark 1.7.3.** The square of the Colombeau generalized line element $(ds^2_\varepsilon)_\varepsilon$ in curvilinear coordinates is a quadratic form in the differentials $dx^i, i = 0,1,2,3$ :

$$(ds^2_\varepsilon)_\varepsilon = (g_{ik,\varepsilon}dx^i dx^k)_\varepsilon = \{(g_{ik,\varepsilon})_\varepsilon\}dx^i dx^k. \quad (1.7.10)$$

where the $(g_{ik,\varepsilon})_\varepsilon$ are Colombeau generalized functions of the coordinates; $(g_{ik,\varepsilon})_\varepsilon$ is symmetric in the indices $i$ and $k$ :

$$(g_{ik,\varepsilon})_\varepsilon = (g_{ki,\varepsilon})_\varepsilon. \quad (1.7.11)$$

**Definition 1.7.4.** Since the (contracted) product of $(g_{ik,\varepsilon})_\varepsilon$ and the contravariant tensor $dx^i dx^k$ is a scalar, the $(g_{ik,\varepsilon})_\varepsilon$ form a covariant tensor; it is called the Colombeau generalized metric tensor.

**Definition 1.7.5.** Two tensors $(A_{ik,\varepsilon})_\varepsilon$ and $(B^{ik}_\varepsilon)_\varepsilon$ are said to be reciprocal to each other if

$$(A_{ik,\varepsilon}B^{ik}_\varepsilon)_\varepsilon = \{(A_{ik,\varepsilon})_\varepsilon\}\left\{(B^{ik}_\varepsilon)_\varepsilon\right\} = \delta^i_k. \quad (1.7.12)$$

In particular the contravariant metric tensor is the tensor $(g_{ik,\varepsilon})_\varepsilon$ reciprocal to the tensor $(g^{ik}_\varepsilon)_\varepsilon$ ,that is,

$$\{(g_{ik,\varepsilon})_\varepsilon\}\left\{(g^{ik}_\varepsilon)_\varepsilon\right\} = \delta^i_k. \quad (1.7.13)$$

The same physical quantity can be represented in contravariant or covariant components. It is obvious that the only quantities that can determine the connection between the different forms are the components of the metric tensor. This connection is given by the formulas:

$$(A^i_\varepsilon)_\varepsilon = (g^{ik}_\varepsilon A_{k,\varepsilon})_\varepsilon, (A_{i,\varepsilon})_\varepsilon = (g_{ik,\varepsilon}A^k_\varepsilon)_\varepsilon. \quad (1.7.14)$$

These remarks also apply to Colombeau generalized tensors. The transition between the different forms of a given physical generalized tensor is accomplished by using the metric tensor according to the formulas:

$$(A^i_{k,\varepsilon})_\varepsilon = (g^{il}_\varepsilon A_{lk,\varepsilon})_\varepsilon, (A^{ik}_\varepsilon)_\varepsilon = (g^{il}_\varepsilon g^{km}_\varepsilon A_{lm,\varepsilon})_\varepsilon, \text{etc.} \quad (1.7.15)$$

The completely antisymmetric unit pseudotensor in galilean coordinates we denote by $e^{iklm}$. Let us transform it to an arbitrary system of Colombeau generalized coordinates, and now denote it by $(E^{iklm}_\varepsilon)_\varepsilon$. We keep the notation $e^{iklm}$ for the quantities defined as before by $e^{0123} = 1$ (or $e_{0123} = -1$).Let the $x'^i, i = 0,1,2,3$ be galilean, and the $(x^i_\varepsilon)_\varepsilon, i = 0,1,2,3$ be arbitrary Colombeau generalized curvilinear coordinates. According to the general rules for transformation of Colombeau generalized tensors, we have

$$(E^{iklm}_\varepsilon)_\varepsilon = \left\{\left(\frac{\partial x^i_\varepsilon}{\partial x'^p}\frac{\partial x^k_\varepsilon}{\partial x'^r}\frac{\partial x^l_\varepsilon}{\partial x'^s}\frac{\partial x^m_\varepsilon}{\partial x'^t}\right)_\varepsilon\right\}e^{prst}, \quad (1.7.16)$$

or

$$(E_\varepsilon^{iklm})_\varepsilon = \{(\mathbf{J}_\varepsilon(x'^0,x'^1,x'^2,x'^3))_\varepsilon\}e^{prst}, \tag{1.7.17}$$

where $(\mathbf{J}_\varepsilon(x'^0,x'^1,x'^2,x'^3))_\varepsilon \neq 0_{\widetilde{\mathbb{R}}}$ is the determinant formed from the derivatives $(\partial x_\varepsilon^i/\partial x'^p)_\varepsilon$, i.e. it is just the Jacobian of the Colombeau generalized transformation from the galilean to the Colombeau generalized curvilinear coordinates:

$$(\mathbf{J}_\varepsilon(x'^0,x'^1,x'^2,x'^3))_\varepsilon = \left(\frac{\partial x_\varepsilon^i}{\partial x'^p}\frac{\partial x_\varepsilon^k}{\partial x'^r}\frac{\partial x_\varepsilon^l}{\partial x'^s}\frac{\partial x_\varepsilon^m}{\partial x'^t}\right)_\varepsilon = \left(\frac{\partial(x_\varepsilon^0,x_\varepsilon^1,x_\varepsilon^2,x_\varepsilon^3)}{\partial(x'^0,x'^1,x'^2,x'^3)}\right)_\varepsilon. \tag{1.7.18}$$

This Jacobian can be expressed in terms of the determinant of the Colombeau generalized metric tensor $(g_{ik,\varepsilon})_\varepsilon$ (in the system $(x_\varepsilon^i)_\varepsilon$). To do this we write the formula for the transformation of the metric tensor:

$$(g_\varepsilon^{ik})_\varepsilon = \left\{\left(\frac{\partial x_\varepsilon^i}{\partial x'^i}\frac{\partial x_\varepsilon^k}{\partial x'^m}\right)_\varepsilon\right\}g^{(0)im}, \tag{1.7.19}$$

where

$$g^{(0)im} = g_{im}^{(0)} = \left\{\begin{matrix} 1 & 1 & 1 & 1 \\ 1 & -1 & 1 & 1 \\ 1 & 1 & -1 & 1 \\ 1 & 1 & 1 & -1 \end{matrix}\right\}, \tag{1.7.20}$$

and equate the determinants of the two sides of this equation. The determinant of the reciprocal tensor $\det\left|(g_\varepsilon^{ik})_\varepsilon\right| = 1/(g_\varepsilon)_\varepsilon$. The determinant $\det|g^{(0)im}| = -1$. Thus we have $1/(g_\varepsilon)_\varepsilon = -(\mathbf{J}_\varepsilon^2(x'^0,x'^1,x'^2,x'^3))_\varepsilon$, and so

$$(\mathbf{J}_\varepsilon^2(x'^0,x'^1,x'^2,x'^3))_\varepsilon = 1/\sqrt{(g_\varepsilon)_\varepsilon}. \tag{1.7.21}$$

Thus, in curvilinear coordinates the antisymmetric unit tensor of rank four must bedefined as

$$(E_\varepsilon^{iklm})_\varepsilon = \frac{1}{\sqrt{-(g_\varepsilon)_\varepsilon}}e^{iklm} \tag{1.7.22}$$

and its covariant components are

$$(E_{iklm,\varepsilon})_\varepsilon = \sqrt{-(g_\varepsilon)_\varepsilon}\,e_{iklm}. \tag{1.7.23}$$

In a galilean coordinate system $x'^i, i = 0,1,2,3$ the integral of a scalar with respect to $d\Omega' = dx'^0dx'^1dx'^2dx'^3$ is also a scalar, i.e. the element $d\Omega'$ behaves like a scalar in the integration. On transforming to Colombeau generalized curvilinear coordinates $(x_\varepsilon^i)_\varepsilon, i = 0,1,2,3$, the element of integration $d\Omega'$ goes over into

$$d\Omega' := \left\{(\mathbf{J}_\varepsilon^{-1})_\varepsilon\right\}d\Omega = \sqrt{-(g_\varepsilon)_\varepsilon}\,(d\Omega_\varepsilon)_\varepsilon, \tag{1.7.24}$$

where $(d\Omega_\varepsilon)_\varepsilon = \{(dx_\varepsilon^0)_\varepsilon\}\{(dx_\varepsilon^1)_\varepsilon\}\{(dx_\varepsilon^2)_\varepsilon\}\{(dx_\varepsilon^3)_\varepsilon\}$.

Thus, in Colombeau generalized curvilinear coordinates, when integrating over a four-volume the quantity $\sqrt{-(g_\varepsilon)_\varepsilon}\,(d\Omega_\varepsilon)_\varepsilon$ behaves like an invariant.

**Remark 1.7.4.** The element of "area" of the Colombeau generalized hypersurface spanned by three infinitesimal Colombeau generalized displacements is the contravariant antisymmetric Colombeau generalized tensor $(dS_\varepsilon^{ikl})_\varepsilon$; the vector dual to it is gotten by multiplying by the tensor $\sqrt{-(g_\varepsilon)_\varepsilon}\,e_{iklm}$, so it is equal to

$$\sqrt{-(g_\varepsilon)_\varepsilon}\,(dS_{\varepsilon,i})_\varepsilon = -\tfrac{1}{6}\sqrt{-(g_\varepsilon)_\varepsilon}\,e_{iklm}(dS_\varepsilon^{kim})_\varepsilon. \tag{1.7.25}$$

**Remark 1.7.5.** Let $(df_\varepsilon^{ik})_\varepsilon$ be the element of two-dimensional Colombeau generalized surface spanned by two infinitesimal Colombeau generalized displacements, the dual Colombeau generalized tensor is defined as

$$\sqrt{-(g_\varepsilon)_\varepsilon}\left(df^*_{ik,\varepsilon}\right)_\varepsilon = \tfrac{1}{2}\sqrt{-(g_\varepsilon)_\varepsilon}\, e_{iklm}\left(df^{lm}_\varepsilon\right)_\varepsilon. \quad (1.7.26)$$

We will be use the designations $(dS_{\varepsilon,i})$ and $\left(df^*_{ki,\varepsilon}\right)_\varepsilon$ for $e_{iklm}(dS^{kim}_\varepsilon)_\varepsilon$ and $e_{iklm}\left(df^{lm}_\varepsilon\right)_\varepsilon$ (and not for their products by $\sqrt{-(g_\varepsilon)_\varepsilon}$).

**Remark 1.7.6.**Note that the canonical rules for transforming the various integrals into one another remain the same, since their derivation was formal in character and not related to the tensor properties of the different quantities. Of particular importance is the rule for transforming the integral over a hypersurface into an integral over a four-volume (Gauss' theorem), which is accomplished by the substitution

$$(dS_{i,\varepsilon})_\varepsilon := \{(d\Omega_\varepsilon)_\varepsilon\}\left(\frac{\partial}{\partial x^i_\varepsilon}\right)_\varepsilon. \quad (1.7.27)$$

**Remark 1.7.7.**Note that for the integral of Colombeau generalized vector $(A^i_\varepsilon)_\varepsilon$ we have

$$\left(\oint A^i_\varepsilon dS_{i,\varepsilon}\right)_\varepsilon = \left(\int \frac{\partial A^i_\varepsilon}{\partial x^i_\varepsilon} d\Omega_\varepsilon\right)_\varepsilon = \int\left\{\left(\frac{\partial A^i_\varepsilon}{\partial x^i_\varepsilon}\right)_\varepsilon\right\}\{(d\Omega_\varepsilon)_\varepsilon\}. \quad (1.7.28)$$

This formula (2.3.28) is the generalization of Gauss' theorem.
An integral over a two-dimensional surface is transformed into an integral over the hypersurface "spanning" it by replacing the Colombeau generalized element of integration $\left(df^*_{ik,\varepsilon}\right)_\varepsilon$ by the operator

$$\left(df^*_{ik,\varepsilon}\right)_\varepsilon := \left(dS_{i,\varepsilon}\frac{\partial}{\partial x^k_\varepsilon}\right)_\varepsilon - \left(dS_{k,\varepsilon}\frac{\partial}{\partial x^i_\varepsilon}\right)_\varepsilon. \quad (1.7.29)$$

For example, for the integral of an antisymmetric Colombeau generalized tensor $(A^{ik}_\varepsilon)_\varepsilon$ we have

$$\left(\int A^{ik}_\varepsilon df^*_{ik,\varepsilon}\right)_\varepsilon = \left(\int dS_{i,\varepsilon}\frac{\partial A^{ik}_\varepsilon}{\partial x^k_\varepsilon}\right)_\varepsilon - \left(\int dS_{k,\varepsilon}\frac{\partial A^{ik}_\varepsilon}{\partial x^i_\varepsilon}\right)_\varepsilon = 2\left(\int dS_{i,\varepsilon}\frac{\partial A^{ik}_\varepsilon}{\partial x^k_\varepsilon}\right)_\varepsilon. \quad (1.7.30)$$

The integral over a four-dimensional Colombeau generalized closed curve is transformed into an integral over the surface spanning it by the substitution

$$(dx^i_\varepsilon)_\varepsilon := \left\{\left(df^{ki}_\varepsilon\right)_\varepsilon\right\}\left(\frac{\partial}{\partial x^i_\varepsilon}\right)_\varepsilon. \quad (1.7.31)$$

Thus for the integral of Colombeau generalized vector, we have:

$$\left(\oint A_{i,\varepsilon}dx^i_\varepsilon\right)_\varepsilon = \left(\int df^{ki}_\varepsilon\frac{\partial A_{i,\varepsilon}}{\partial x^i_\varepsilon}\right)_\varepsilon = \frac{1}{2}\int\left(df^{ik}_\varepsilon\frac{\partial A_{k,\varepsilon}}{\partial x^i_\varepsilon} - df^{ik}_\varepsilon\frac{\partial A_{i,\varepsilon}}{\partial x^k_\varepsilon}\right)_\varepsilon, \quad (1.7.32)$$

which is the generalization of Stokes' theorem.

# 1.8. Colombeau Generalized covariant derivatives

In galilean coordinates the Colombeau generalized differentials $(dA_{i,\varepsilon})_\varepsilon$ of a vector $(A_{i,\varepsilon})_\varepsilon$ form the Colombeau generalized vector, and the derivatives $(\partial A_{i,\varepsilon}/\partial x^k_\varepsilon)_\varepsilon$ of the components of a vector with respect to the coordinates form the Colombeau generalized tensor. In Colombeau generalized curvilinear coordinates this is not so; $(dA_{i,\varepsilon})_\varepsilon$ is not a vector, and $(\partial A_{i,\varepsilon}/\partial x^k_\varepsilon)_\varepsilon$ is not the Colombeau generalized tensor.This is due to the fact that $(dA_{i,\varepsilon})_\varepsilon$ is the difference of vectors located at different (infinitesimally separated) points of space; at different points in space vectors transform differently, since the coefficients in the transformation formulas (1.7.3), (1.7.4) are Colombeau generalized functions of the generalized coordinates.Thus in order to compare two infmitesimally separated generalized vectors we must subject one of them to a parallel translation to the point where the second is located. Let us consider an arbitrary generalized contravariant vector ; if its value at the point $x^i$ is $(A^i_\varepsilon)_\varepsilon$, then at the neighboring point $x^i + dx^i$ it is equal to $(A^i_\varepsilon)_\varepsilon + (dA^i_\varepsilon)_\varepsilon = (A^i_\varepsilon + dA^i_\varepsilon)_\varepsilon$. We subject the vector $(A^i_\varepsilon)_\varepsilon$ to an infinitesimal parallel displacement to the point $x^i + dx^i$; the change in the vector which results from this we denote by $(\delta A^i_\varepsilon)_\varepsilon$ Then the difference $(DA^i_\varepsilon)_\varepsilon$ between the two Colombeau generalized vectors which are now located at the same point is

$$(DA^i_\varepsilon)_\varepsilon = (dA^i_\varepsilon)_\varepsilon - (\delta A^i_\varepsilon)_\varepsilon. \qquad (1.8.1)$$

The change $(\delta A^i_\varepsilon)_\varepsilon$in the components of Colombeau generalized vector under an infinitesimal parallel displacement depends on the values of the components themselves, where the dependence must clearly be linear. This follows directly from the fact that the sum of two Colombeau generalized vectors must transform according to the same law as each of the constituents. Thus $(\delta A^i_\varepsilon)_\varepsilon$ has the form

$$(\delta A^i_\varepsilon)_\varepsilon = -(\Gamma^i_{kl,\varepsilon} A^k_\varepsilon dx^l)_\varepsilon, \qquad (1.8.2)$$

where $(\Gamma^i_{kl,\varepsilon})_\varepsilon$ the are certain Colombeau generalized functions of the coordinates. Their form depends, of course, on the coordinate system; for a galilean coordinate system $(\Gamma^i_{kl,\varepsilon})_\varepsilon = 0_{\widetilde{\mathbb{R}}}$. From this it is already clear that the quantities $(\Gamma^i_{kl,\varepsilon})_\varepsilon$ do not form Colombeau generalized tensor, since a tensor which is equal to zero in one coordinate system is equal to zero in every other one. In a curvilinear space it is, of course, impossible to make all the $(\Gamma^i_{kl,\varepsilon})_\varepsilon$ vanish over all of space. But we can choose a coordinate system for which the $(\Gamma^i_{kl,\varepsilon})_\varepsilon$become $0_{\widetilde{\mathbb{R}}}$ over a given infinitesimal region. The quantities $(\Gamma^i_{kl,\varepsilon})_\varepsilon$, are called generalized Christoffel symbols. In addition to the quantities $(\Gamma^i_{kl,\varepsilon})_\varepsilon$ we shall later also use Colombeau generalized quantities $(\Gamma_{i,kl,\varepsilon})_\varepsilon$ defined as follows

$$(\Gamma_{i,kl,\varepsilon})_\varepsilon = (g_{im,\varepsilon} \Gamma^m_{km,\varepsilon})_\varepsilon. \qquad (1.8.3)$$

Conversely,

$$(\Gamma^i_{kl,\varepsilon})_\varepsilon = (g^{im}_\varepsilon \Gamma_{m,kl,\varepsilon})_\varepsilon. \qquad (1.8.4)$$

It is also easy to relate the change in the components of a covariant vector under a parallel displacement to the Christoffel symbols. To do this we note that under a parallel displacement, a scalar is unchanged. In particular, the scalar product of two vectors does not change under a parallel displacement.

Let $(A_{i,\varepsilon})_\varepsilon$ and $(B^i_\varepsilon)_\varepsilon$ be any covariant and contravariant vectors. Then from $\delta(A_{i,\varepsilon} B^i_\varepsilon)_\varepsilon = 0_{\widetilde{\mathbb{R}}}$, we have

$$(B^i_\varepsilon \delta A_{i,\varepsilon}) = -(A_{i,\varepsilon} \delta B^i_\varepsilon)_\varepsilon = (\Gamma^i_{kl,\varepsilon} B^k_\varepsilon A_{i,\varepsilon} dx^l)_\varepsilon \qquad (1.8.5)$$

or, changing the indices,

$$(B^i_\varepsilon \delta A_{i,\varepsilon}) = (\Gamma^k_{il,\varepsilon} B^i_\varepsilon A_{k,\varepsilon} dx^l)_\varepsilon \qquad (1.8.6)$$

From this, by the arbitrariness of the $(B^i_\varepsilon)_\varepsilon$ one obtains

$$(\delta A_{i,\varepsilon})_\varepsilon = \left((\Gamma^k_{il,\varepsilon} A_{k,\varepsilon})_\varepsilon\right) dx^l \qquad (1.8.7)$$

which determines the change in a covariant vector under a parallel displacement. Substituting (1.8.2) and $(dA^i_\varepsilon)_\varepsilon = ((\partial A^i_\varepsilon/\partial x^l)_\varepsilon) dx^l$ in (1.8.1), we have

$$(DA^i_\varepsilon)_\varepsilon = \left[\left(\frac{\partial A^i_\varepsilon}{\partial x^l}\right)_\varepsilon + (\Gamma^i_{kl,\varepsilon} A^k_\varepsilon)_\varepsilon\right] dx^l. \qquad (1.8.8)$$

Similarly, one finds for a covariant Colombeau generalized vector

$$(DA_{i,\varepsilon})_\varepsilon = \left[\left(\frac{\partial A_{i,\varepsilon}}{\partial x^l}\right)_\varepsilon + (\Gamma^k_{il,\varepsilon} A_{k,\varepsilon})_\varepsilon\right] dx^l. \qquad (1.8.9)$$

The expressions in parentheses in (1.8.8) and (1.8.9) are Colombeau generalized tensors, since when multiplied by the vector $dx^l$ they give a vector. Clearly, these are the generalized tensors which give the desired generalization of the concept of a derivative to Colombeau generalized curvilinear coordinates. These Colombeau generalized tensors are called the Colombeau generalized covariant derivatives of the vectors $(A^i_\varepsilon)_\varepsilon$ and $(A_{i,\varepsilon})_\varepsilon$ respectively. We shall denote them by $A^i_{;k,\varepsilon}$ and $A_{i;k,\varepsilon}$. Thus,

$$(DA^i_\varepsilon)_\varepsilon = (A^i_{;l,\varepsilon}dx^l)_\varepsilon, (DA_{i,\varepsilon})_\varepsilon = (A_{i;l,\varepsilon}dx^l)_\varepsilon, \tag{1.8.10}$$

while the Colombeau generalized covariant derivatives themselves are

$$\left(A^i_{;\,l,\varepsilon}\right)_\varepsilon = \left(\frac{\partial A^i_\varepsilon}{\partial x^l}\right)_\varepsilon + (\Gamma^i_{kl,\varepsilon}A^k_\varepsilon)_\varepsilon \tag{1.8.11}$$

and

$$\left(A_{i;\,l,\varepsilon}\right)_\varepsilon = \left(\frac{\partial A_{i,\varepsilon}}{\partial x^l}\right)_\varepsilon + (\Gamma^k_{il,\varepsilon}A_{k,\varepsilon})_\varepsilon. \tag{1.8.12}$$

It is also easy to calculate the covariant derivative of the Colombeau generalized tensor. One gets the covariant derivative of the Colombeau generalized tensor $(A^{ik}_\varepsilon)_\varepsilon$ in the form

$$\left(A^{ik}_{;\,l,\varepsilon}\right)_\varepsilon = \left(\frac{\partial A^{ik}_\varepsilon}{\partial x^l}\right)_\varepsilon + (\Gamma^i_{ml,\varepsilon}A^{mk}_\varepsilon)_\varepsilon + (\Gamma^k_{ml,\varepsilon}A^{im}_\varepsilon)_\varepsilon. \tag{1.8.13}$$

The covariant derivative of the Colombeau generalized mixed tensor $(A^i_{k,\varepsilon})_\varepsilon$ and the Colombeau generalized covariant tensor $(A_{ik,\varepsilon})_\varepsilon$ has the form

$$\left(A^i_{k;\,l,\varepsilon}\right)_\varepsilon = \left(\frac{\partial A^i_{k,\varepsilon}}{\partial x^l}\right)_\varepsilon + (\Gamma^m_{kl,\varepsilon}A^i_{m,\varepsilon})_\varepsilon + (\Gamma^i_{ml,\varepsilon}A^m_{k,\varepsilon})_\varepsilon \tag{1.8.14}$$

and

$$\left(A_{ik;\,l,\varepsilon}\right)_\varepsilon = \left(\frac{\partial A_{ik,\varepsilon}}{\partial x^l}\right)_\varepsilon - (\Gamma^m_{il,\varepsilon}A_{mk,\varepsilon})_\varepsilon - (\Gamma^m_{kl,\varepsilon}A_{im,\varepsilon})_\varepsilon. \tag{1.8.15}$$

correspondingly.

If in a covariant derivative we raise the index signifying the differentiation, we obtain the so-called contravariant derivative. Thus,

$$\left(A^{;k}_{i,\varepsilon}\right)_\varepsilon = (g^{kl}_\varepsilon A_{i;l,\varepsilon})_\varepsilon; (A^{i;k}_\varepsilon)_\varepsilon = (g^{kl}_\varepsilon A^i_{;l,\varepsilon})_\varepsilon. \tag{1.8.16}$$

# 1.9. The relation of the Colombeau Generalized Christoffel symbols to the Colombeau Generalized metric tensor

Let us show that the Colombeau generalized covariant derivative of the Colombeau Generalized metric tensor $(g_{ik,\varepsilon})_\varepsilon$ is $0_{\widetilde{\mathbb{R}}}$. To do this we note that the relation

$$(DA_{i,\varepsilon})_\varepsilon = (g_{ik,\varepsilon}DA^k_\varepsilon)_\varepsilon \tag{1.9.1}$$

is valid for the generalized vector $(DA_{i,\varepsilon})_\varepsilon$ as for any generalized vector. On the other hand, $(A_{i,\varepsilon})_\varepsilon = (g_{ik,\varepsilon}A^k_\varepsilon)_\varepsilon$, so that

$$(DA_{i,\varepsilon})_\varepsilon = D(g_{ik,\varepsilon}A^k_\varepsilon)_\varepsilon = (g_{ik,\varepsilon}DA^k_\varepsilon)_\varepsilon + (A^k_\varepsilon Dg_{ik,\varepsilon})_\varepsilon. \tag{1.9.2}$$

Comparing with $(DA_{i,\varepsilon})_\varepsilon = (g_{ik,\varepsilon}DA^k_\varepsilon)_\varepsilon$, and remembering that the vector $(A^k_\varepsilon)_\varepsilon$ is arbitrary,

$$(Dg_{ik,\varepsilon})_\varepsilon = 0_{\widetilde{\mathbb{R}}}. \tag{1.9.3}$$

Therefore the covariant derivative

$$(g_{ik;l,\varepsilon})_\varepsilon = 0_{\widetilde{\mathbb{R}}}. \tag{1.9.4}$$

Thus $(g_{ik,\varepsilon})_\varepsilon$ may be considered as a constant under covariant differentiation.The equation $(g_{ik;l,\varepsilon})_\varepsilon = 0_{\widetilde{\mathbb{R}}}$ can be used to express the Colombeau generalized Christoffel symbols $(\Gamma^i_{kl,\varepsilon})_\varepsilon$ in terms of the Colombeau generalized metric tensor $(g_{ik,\varepsilon})_\varepsilon$. To do this we write in accordance with the general definition (1.8.9):

$$(g_{ik;l,\varepsilon})_\varepsilon = \left(\frac{\partial g_{ik,\varepsilon}}{\partial x^l}\right)_\varepsilon - (g_{mk,\varepsilon}\Gamma^m_{il,\varepsilon})_\varepsilon - (g_{im,\varepsilon}\Gamma^m_{kl,\varepsilon})_\varepsilon = \left(\frac{\partial g_{ik,\varepsilon}}{\partial x^l}\right)_\varepsilon - (\Gamma_{k,il,\varepsilon})_\varepsilon - (\Gamma_{i,kl,\varepsilon})_\varepsilon = 0_{\widetilde{\mathbb{R}}}. \tag{1.9.5}$$

Thus the derivatives of $(g_{ik,\varepsilon})_\varepsilon$ are expressed in terms of the Colombeau generalized Christoffel symbols.f We write the values of the derivatives of $(g_{ik,\varepsilon})_\varepsilon$ , permuting the indices $i,k,l$ :

$$\left(\frac{\partial g_{ik,\varepsilon}}{\partial x^l}\right)_\varepsilon = (\Gamma_{k,il,\varepsilon})_\varepsilon + (\Gamma_{i,kl,\varepsilon})_\varepsilon, \left(\frac{\partial g_{li,\varepsilon}}{\partial x^k}\right)_\varepsilon = (\Gamma_{i,kl,\varepsilon})_\varepsilon + (\Gamma_{l,ik,\varepsilon})_\varepsilon, \left(\frac{\partial g_{kl,\varepsilon}}{\partial x^i}\right)_\varepsilon = -(\Gamma_{l,ki,\varepsilon})_\varepsilon + (\Gamma_{k,li,\varepsilon})_\varepsilon. \tag{1.9.5'}$$

Taking half the sum of these equations, we find

$$(\Gamma_{i,kl,\varepsilon})_\varepsilon = \frac{1}{2}\left[\left(\frac{\partial g_{ik,\varepsilon}}{\partial x^l}\right)_\varepsilon + \left(\frac{\partial g_{il,\varepsilon}}{\partial x^k}\right)_\varepsilon - \left(\frac{\partial g_{kl,\varepsilon}}{\partial x^i}\right)_\varepsilon\right], \tag{1.9.6}$$

since $(\Gamma_{i,kl,\varepsilon})_\varepsilon = (\Gamma_{i,lk,\varepsilon})_\varepsilon$.From (2.5.6) we have for the symbols $(\Gamma^i_{kl,\varepsilon})_\varepsilon = (g^{im}_\varepsilon\Gamma_{m,kl,\varepsilon})_\varepsilon$

$$(\Gamma^i_{kl,\varepsilon})_\varepsilon = \frac{1}{2}((g^{im}_\varepsilon)_\varepsilon)\left[\left(\frac{\partial g_{mk,\varepsilon}}{\partial x^l}\right)_\varepsilon + \left(\frac{\partial g_{ml,\varepsilon}}{\partial x^k}\right)_\varepsilon - \left(\frac{\partial g_{kl,\varepsilon}}{\partial x^m}\right)_\varepsilon\right]. \tag{1.9.7}$$

# 1.10. The Colombeau Generalized Curvature Tensor

In this subsection we derive the general formula for the change in a vector after parallel displacement around any infinitesimal closed contour $\gamma$. This generalized change $(\Delta A_{k,\varepsilon})_\varepsilon \in \widetilde{\mathbb{R}}$ can clearly be written in the form $\left(\oint \delta A_{k,\varepsilon}\right)_\varepsilon$, where the Colombeau integral is taken over the given regular contour $\gamma$. Substituting in place of $(\delta A_{k,\varepsilon})_\varepsilon$the expression (1.8.7), we have

$$(\Delta A_{k,\varepsilon})_\varepsilon = \left(\oint_\gamma \Gamma^i_{kl,\varepsilon}(x)A_i(x)dx^l\right)_\varepsilon \in \widetilde{\mathbb{R}}, \tag{1.10.1}$$

where for any $i,k,l = 0,1,2,3$ : $(\Gamma^i_{kl,\varepsilon}(x))_\varepsilon \in \mathcal{G}(\mathbb{R}^4), x = (x^0,x^1,x^2,x^3)$, $A_i(x) \in \mathcal{D}(G)$, $G \subseteq \mathbb{R}^4$. Note that the vector $A_i$ which appears in the integrand obviously changes as we move along the contour $\gamma$.

**Definition 1.10.1**.We will say that generalized change $(\Delta A_{k,\varepsilon})_\varepsilon$ exists in the sense of the Schwartz distributions if for any $A_i(x) \in \mathcal{D}(G)$ the limit: $\lim_{\varepsilon\to 0} \Delta A_{k,\varepsilon}$ exists. Of course in this case obviously $\mathbf{cl}\left[(\Gamma^i_{kl,\varepsilon}(x))_\varepsilon\right] \in \mathcal{D}'(G)$ and $\mathbf{cl}[(\Delta A_{k,\varepsilon})_\varepsilon] \in \mathbb{R}$.

For the further transformation of this Colombeau integral, we must note the following. The values of the vector $A_i$ at points inside the contour are not unique; they depend on the path along which we approach the particular point. However, as we shall see from the result obtained below, this non-uniqueness is related to terms of second order. We may therefore, with the first-order accuracy which is sufficient for the transformation, regard the components of the vector $A_i$ at points inside the infinitesimal contour $\gamma$ as being uniquely determined by their values on the contour itself by the formulas $(\delta A_i(x))_\varepsilon = (\Gamma^n_{il,\varepsilon}(x)A_{n,\varepsilon}(x)dx^l)_\varepsilon$ , i.e., by the derivatives

$$\frac{\partial A_{i,\varepsilon}(x)}{\partial x^l} = (\Gamma^n_{il,\varepsilon}(x)A_{n,\varepsilon}(x))_\varepsilon. \tag{1.10.2}$$

Now applying generalized Stokes' theorem (Theorem 1.10.1) to the integral (1.10.1) and considering that the area enclosed by the contour has the infinitesimal value $(\Delta f^{im}_\varepsilon)_\varepsilon$, we get:

$$(\Delta A_{k,\varepsilon})_\varepsilon = \frac{1}{2}\left[\left(\frac{\partial(\Gamma^i_{km,\varepsilon}(x)A_i(x))}{\partial x^l}\right)_\varepsilon - \left(\frac{\partial(\Gamma^i_{kl,\varepsilon}(x)A_i(x))}{\partial x^m}\right)_\varepsilon\right](\Delta f^{im}_\varepsilon)_\varepsilon =$$

$$= \frac{1}{2}\left[A_i(x)\left(\frac{\partial(\Gamma^i_{km,\varepsilon}(x))}{\partial x^l}\right)_\varepsilon - A_i(x)\left(\frac{\partial(\Gamma^i_{kl,\varepsilon}(x))}{\partial x^m}\right)_\varepsilon + \left(\frac{\partial A_i(x)}{\partial x^l}\right)(\Gamma^i_{km,\varepsilon}(x))_\varepsilon - \left(\frac{\partial A_i(x)}{\partial x^m}\right)(\Gamma^i_{kl,\varepsilon}(x))_\varepsilon\right](\Delta f^{im}_\varepsilon)_\varepsilon. \quad (1.10.3)$$

**Definition 1.10.2.** Colombeau generalized $k$-form $(\omega_\varepsilon)_\varepsilon$ on a differentiable manifold $M$ is a smooth section of the bundle of alternating Colombeau generalized $k$-tensors on $M$. Equivalently, $(\omega_\varepsilon)_\varepsilon$ associates to each $x \in M$ an alternating Colombeau generalized $k$-tensor $(\omega_{x,\varepsilon})_\varepsilon$, in such a way that in any chart for $M$, the coefficients $(\omega_{i_1\ldots i_k,\varepsilon})_\varepsilon$ are Colombeau generalized functions.

**Theorem 1.10.1.** (Generalized Stokes' Theorem) Let $(\omega_\varepsilon)_\varepsilon$ be Colombeau generalized differential form. Then the Colombeau integral of a differential form $(\omega_\varepsilon)_\varepsilon$ over the boundary of some orientable manifold $\Sigma \subset M$ is equal to the integral of its exterior Colombeau derivative $(d\omega_\varepsilon)_\varepsilon$ over the whole of $\Sigma$, i.e.,

$$\int_{\partial\Sigma}(\omega_\varepsilon)_\varepsilon = \left(\int_{\partial\Sigma}\omega_\varepsilon\right)_\varepsilon = \left(\int_\Sigma d\omega_\varepsilon\right)_\varepsilon = \int_\Sigma (d\omega_\varepsilon)_\varepsilon. \quad (1.10.4)$$

Proof. Immediately from the classical Stokes' Theorem and definitions.

**Example 1.10.1.** For example, for the integral of Colombeau generalized vector $(A_{i,\varepsilon}(x))_\varepsilon$ we have

$$\left(\oint_\Gamma A_{i,\varepsilon}dx^i\right)_\varepsilon = \left(\int_\Sigma df^{ki}\frac{\partial A_{i,\varepsilon}}{\partial x^k}\right)_\varepsilon = \frac{1}{2}\left(\int\left[\left(df^{ki}_\varepsilon\right)_\varepsilon\right]\left(\frac{\partial A_{k,\varepsilon}}{\partial x^i} - \frac{\partial A_{i,\varepsilon}}{\partial x^k}\right)\right)_\varepsilon =$$

$$\frac{1}{2}\int\left[\left(df^{ki}_\varepsilon\right)_\varepsilon\right]\left(\frac{\partial A_{k,\varepsilon}}{\partial x^i} - \frac{\partial A_{i,\varepsilon}}{\partial x^k}\right)_\varepsilon = \quad (1.10.5)$$

$$\frac{1}{2}\int\left[\left(df^{ki}_\varepsilon\right)_\varepsilon\right]\left[\left(\frac{\partial A_{k,\varepsilon}}{\partial x^i}\right)_\varepsilon - \left(\frac{\partial A_{i,\varepsilon}}{\partial x^k}\right)_\varepsilon\right],$$

where $\Gamma = \partial\Sigma$ and $\left(df^{ki}_\varepsilon\right)_\varepsilon = (dx^i_\varepsilon dx'^k_\varepsilon)_\varepsilon - (dx^k_\varepsilon dx'^i_\varepsilon)_\varepsilon$ is the infinitesimal element of surface which is given by the antisymmetric tensor of second rank $\left(df^{ki}_\varepsilon\right)_\varepsilon$.

Substituting the values of the derivatives (1.10.2) into Eq.(1.10.3), we get

$$(\Delta A_{k,\varepsilon})_\varepsilon = \frac{1}{2}(R^i_{klm,\varepsilon}(x)A_{i,\varepsilon}(x)\Delta f^{im}_\varepsilon)_\varepsilon, \quad (1.10.6)$$

where $(R^i_{klm,\varepsilon}(x))_\varepsilon$ is a Colombeau generalized tensor field of the fourth rank:

$$(R^i_{klm,\varepsilon}(x))_\varepsilon = \left(\frac{\partial(\Gamma^i_{km,\varepsilon}(x))}{\partial x^l}\right)_\varepsilon - \left(\frac{\partial(\Gamma^i_{kl,\varepsilon}(x))}{\partial x^m}\right)_\varepsilon + (\Gamma^i_{ni,\varepsilon}(x)\Gamma^n_{km,\varepsilon}(x))_\varepsilon - (\Gamma^i_{nm,\varepsilon}(x)\Gamma^n_{kl,\varepsilon}(x))_\varepsilon. \quad (1.10.7)$$

**Definition 1.10.3.** The tensor field $(R^l_{kim,\varepsilon}(x))_\varepsilon$ is called the distributional curvature tensor or the distributional Riemann tensor.

**Remark 1.10.1.** Note that in general case for any $i,k,l = 0,1,2,3$ : $\mathbf{cl}\left[(R^i_{klm,\varepsilon}(x))_\varepsilon\right] \in \mathcal{G}(\mathbb{R}^4)$.

**Definition 1.10.4.** We will say that the distributional Riemann tensor $(R^i_{klm,\varepsilon}(x))_\varepsilon$ exists in the sense of the Schwartz distributions if for any $i,k,l = 0,1,2,3$ and for any $A_i(x) \in \mathcal{D}(G)$ the limit:

$$\lim_{\varepsilon\to 0}\int_G R^i_{klm,\varepsilon}(x)A_i(x)d^4x \tag{1.10.8}$$

exists.

**Definition 1.10.5.** We will say that the distributional Riemann tensor $(R^i_{klm,\varepsilon}(x))_\varepsilon$ exists in the classical sense at point $\breve{x} \in \mathbb{R}$ if there exists standard part of point value of Colombeau generalized function $(R^i_{klm,\varepsilon}(x))_\varepsilon$ at point $\breve{x} \in \mathbb{R}$, i.e. $\mathbf{st}\left(\mathbf{cl}\left[(R^i_{klm,\varepsilon}(\breve{x}))_\varepsilon\right]\right) \in \mathbb{R}$.

**Remark 1.10.2.** It is easy to obtain a similar formula to (1.10.6) for a generalized contravariant vector $(A^k_\varepsilon)_\varepsilon$. To do this we note, since under parallel displacement a scalar does not change, that $(\Delta(A^k_\varepsilon B_{k,\varepsilon}))_\varepsilon = 0_{\widetilde{\mathbb{R}}}$, where $(B_{k,\varepsilon})_\varepsilon$ is any Colombeau generalized covariant vector. With the help of (1.10.6), we obtain

$$\begin{gathered}(\Delta(A^k_\varepsilon B_{k,\varepsilon}))_\varepsilon = (\Delta A^k_\varepsilon B_{k,\varepsilon})_\varepsilon + (A^k_\varepsilon \Delta B_{k,\varepsilon})_\varepsilon = \\ \frac{1}{2}\left(A^k_\varepsilon B_{i,\varepsilon} R^i_{klm,\varepsilon}\Delta f^{lm}_\varepsilon\right)_\varepsilon + (B_{k,\varepsilon}\Delta A^k_\varepsilon)_\varepsilon = \\ \{(B_{k,\varepsilon})_\varepsilon\}\left((\Delta A^k_\varepsilon)_\varepsilon + \frac{1}{2}\left(A^l_\varepsilon R^k_{ilm,\varepsilon}\Delta f^{lm}_\varepsilon\right)_\varepsilon\right) = 0_{\widetilde{\mathbb{R}}}\end{gathered} \tag{1.10.9}$$

or, in view of the arbitrariness of the vector $(B_{k,\varepsilon})_\varepsilon$ we get

$$(\Delta A^k_\varepsilon)_\varepsilon = -\frac{1}{2}\left(R^k_{ilm,\varepsilon} A^k_\varepsilon f^{lm}_\varepsilon\right)_\varepsilon. \tag{1.10.10}$$

If we now twice differentiate Colombeau generalized vector $(A_{i,\varepsilon})_\varepsilon$ covariantly with respect to $x^k$ and $x^l$, then the result obviously depends on the order of differentiation, contrary to the situation for ordinary differentiation. It turns out that the difference $(A_{i;k;l,\varepsilon})_\varepsilon - (A_{i;l;k,\varepsilon})_\varepsilon$ is given by the same Colombeau generalized curvature tensor which we introduced above. Namely, one finds the formula

$$(A_{i;k;l,\varepsilon})_\varepsilon - (A_{i;l;k,\varepsilon})_\varepsilon = (A_{m,\varepsilon} R^m_{ikl,\varepsilon})_\varepsilon, \tag{1.10.11}$$

which is easily verified by direct calculation in the locally-geodesic coordinate system

Similarly, for a contravariant Colombeau generalized vector,one gets

$$(A^i_{;k;l,\varepsilon})_\varepsilon - (A^i_{;l;k,\varepsilon})_\varepsilon = (A^m_\varepsilon R^i_{mkl,\varepsilon})_\varepsilon, \tag{1.10.12}$$

Finally, it is easy to obtain similar formulas for the second derivatives of Colombeau generalized tensors for which this is done most easily by considering, for example, Colombeau generalized tensor of the form $(A_{i,\varepsilon}B_{k,\varepsilon})_\varepsilon$, and by using formulas (1.10.11) and (1.10.12); because of the linearity, the formulas thus obtained must be valid for an arbitrary Colombeau generalized tensor $(A_{ik,\varepsilon})_\varepsilon$. Thus one gets

$$(A_{ik;l;m,\varepsilon})_\varepsilon - (A_{ik;m;l,\varepsilon})_\varepsilon =_{\widetilde{\mathbb{R}}} (A_{in,\varepsilon} R^n_{klm,\varepsilon})_\varepsilon + (A_{nk,\varepsilon} R^n_{ilm,\varepsilon})_\varepsilon, \tag{1.10.13}$$

# 1.11. Main properties of the distributional Curvature Tensor

From the expression (1.10.7) it follows directly that $\forall x \in \mathbb{R}$ the distributional curvature tensor is antisymmetric in the indices $l$ and $m$ :

$$(R^i_{klm,\varepsilon}(x))_\varepsilon =_{\widetilde{\mathbb{R}}} -(R^i_{kml,\varepsilon}(x))_\varepsilon \tag{1.11.1}$$

and therefore $\forall z = (x_\varepsilon)_\varepsilon \in \widetilde{\mathbb{R}}_{\mathbf{fin}}$ the following identity holds

$$(R^i_{klm,\varepsilon}(x_\varepsilon))_\varepsilon =_{\widetilde{\mathbb{R}}} -(R^i_{kml,\varepsilon}(x_\varepsilon))_\varepsilon. \tag{1.11.1.a}$$

Obviously the following identity holds

$$(R^l_{kim,\varepsilon}(x))_\varepsilon + (R^i_{mkl,\varepsilon}(x))_\varepsilon + (R^i_{lmk,\varepsilon}(x)) =_{\widetilde{\mathbb{R}}} 0_{\widetilde{\mathbb{R}}} \tag{1.11.2}$$

and therefore $\forall z = (x_\varepsilon)_\varepsilon \in \widetilde{\mathbb{R}}_{\mathbf{fin}}$ the following identity holds

$$(R^l_{kim,\varepsilon}(x_\varepsilon))_\varepsilon + (R^i_{mkl,\varepsilon}(x_\varepsilon))_\varepsilon + (R^i_{lmk,\varepsilon}(x_\varepsilon)) =_{\widetilde{\mathbb{R}}} 0_{\widetilde{\mathbb{R}}} \quad (1.11.2.a)$$

In addition to the mixed distributional curvature tensor $(R^i_{klm,\varepsilon}(x))_\varepsilon$, one also uses the covariant distributional curvature tensor

$$(R_{iklm,\varepsilon}(x))_\varepsilon = (g_{in,\varepsilon}(x)R^n_{klm,\varepsilon}(x))_\varepsilon =_{\widetilde{\mathbb{R}}} ((g_{in,\varepsilon}(x))_\varepsilon)\Big((R^n_{klm,\varepsilon}(x))_\varepsilon\Big) \quad (1.11.3)$$

and therefore $\forall z = (x_\varepsilon)_\varepsilon \in \widetilde{\mathbb{R}}_{\mathbf{fin}}$ the following identity holds

$$(R_{iklm,\varepsilon}(x_\varepsilon))_\varepsilon = (g_{in,\varepsilon}(x_\varepsilon)R^n_{klm,\varepsilon}(x_\varepsilon))_\varepsilon =_{\widetilde{\mathbb{R}}} ((g_{in,\varepsilon}(x_\varepsilon))_\varepsilon)\Big((R^n_{klm,\varepsilon}(x_\varepsilon))_\varepsilon\Big) \quad (1.11.3.a)$$

Obviously by means of simple calculation the following expressions for $(R_{iklm,\varepsilon}(x))_\varepsilon$ holds

$$(R_{iklm,\varepsilon}(x))_\varepsilon =_{\widetilde{\mathbb{R}}}$$
$$\frac{1}{2}\left(\left(\frac{\partial^2 g_{im,\varepsilon}(x)}{\partial x^k \partial x^l}\right)_\varepsilon + \left(\frac{\partial^2 g_{kl,\varepsilon}(x)}{\partial x^i \partial x^m}\right)_\varepsilon - \left(\frac{\partial^2 g_{il,\varepsilon}(x)}{\partial x^k \partial x^m}\right)_\varepsilon - \left(\frac{\partial^2 g_{km,\varepsilon}(x)}{\partial x^i \partial x^l}\right)_\varepsilon\right) + \quad (1.11.4)$$
$$+((g_{np,\varepsilon})_\varepsilon)\Big[(\Gamma^n_{kl,\varepsilon}\Gamma^p_{im,\varepsilon})_\varepsilon - (\Gamma^n_{km,\varepsilon}\Gamma^p_{il,\varepsilon})_\varepsilon\Big]$$

and therefore $\forall z = (x_\varepsilon)_\varepsilon \in \widetilde{\mathbb{R}}_{\mathbf{fin}}$ the following identity holds

$$(R_{iklm,\varepsilon}(x_\varepsilon))_\varepsilon =_{\widetilde{\mathbb{R}}}$$
$$\frac{1}{2}\left(\left(\frac{\partial^2 g_{im,\varepsilon}(x_\varepsilon)}{\partial x^k \partial x^l}\right)_\varepsilon + \left(\frac{\partial^2 g_{kl,\varepsilon}(x_\varepsilon)}{\partial x^i \partial x^m}\right)_\varepsilon - \left(\frac{\partial^2 g_{il,\varepsilon}(x_\varepsilon)}{\partial x^k \partial x^m}\right)_\varepsilon\right.$$
$$\left. - \left(\frac{\partial^2 g_{km,\varepsilon}(x_\varepsilon)}{\partial x^i \partial x^l}\right)_\varepsilon\right) + \quad (1.11.4.a)$$
$$((g_{np,\varepsilon}(x_\varepsilon))_\varepsilon)\Big[(\Gamma^n_{kl,\varepsilon}(x_\varepsilon)\Gamma^p_{im,\varepsilon}(x_\varepsilon))_\varepsilon - (\Gamma^n_{km,\varepsilon}(x_\varepsilon)\Gamma^p_{il,\varepsilon}(x_\varepsilon))_\varepsilon\Big]$$

From this expressions (1.11.4)-(1.11.4.a) it follows

$$(R_{iklm,\varepsilon}(x))_\varepsilon =_{\widetilde{\mathbb{R}}} -(R_{kilm,\varepsilon}(x))_\varepsilon =_{\widetilde{\mathbb{R}}} -(R_{ikml,\varepsilon}(x))_\varepsilon,$$
$$(R_{iklm,\varepsilon}(x))_\varepsilon =_{\widetilde{\mathbb{R}}} (R_{lmik,\varepsilon}(x))_\varepsilon. \quad (1.11.5)$$

and

$$(R_{iklm,\varepsilon}(x_\varepsilon))_\varepsilon =_{\widetilde{\mathbb{R}}} -(R_{kilm,\varepsilon}(x_\varepsilon))_\varepsilon =_{\widetilde{\mathbb{R}}} -(R_{ikml,\varepsilon}(x_\varepsilon))_\varepsilon,$$
$$(R_{iklm,\varepsilon}(x_\varepsilon))_\varepsilon =_{\widetilde{\mathbb{R}}} (R_{lmik,\varepsilon}(x_\varepsilon))_\varepsilon. \quad (1.11.5.a)$$

For $(R_{iklm,\varepsilon}(x))_\varepsilon$ the following identities holds

$$(R_{iklm,\varepsilon}(x))_\varepsilon =_{\widetilde{\mathbb{R}}} (R_{lmik,\varepsilon}(x))_\varepsilon \quad (1.11.6)$$

and $\forall z = (x_\varepsilon)_\varepsilon \in \widetilde{\mathbb{R}}_{\mathbf{fin}}$

$$(R_{iklm,\varepsilon}(x_\varepsilon))_\varepsilon =_{\widetilde{\mathbb{R}}} (R_{lmik,\varepsilon}(x_\varepsilon))_\varepsilon. \quad (1.11.6.a)$$

For $(R_{iklm,\varepsilon})_\varepsilon$ the following identities holds

$$(R_{iklm,\varepsilon}(x))_\varepsilon + (R_{imkl,\varepsilon}(x))_\varepsilon + (R_{ilmk,\varepsilon}(x)) =_{\widetilde{\mathbb{R}}} 0_{\widetilde{\mathbb{R}}} \quad (1.11.7)$$

and $\forall z = (x_\varepsilon)_\varepsilon \in \widetilde{\mathbb{R}}_{\mathbf{fin}}$

$$(R_{iklm,\varepsilon}(x_\varepsilon))_\varepsilon + (R_{imkl,\varepsilon}(x_\varepsilon))_\varepsilon + (R_{ilmk,\varepsilon}(x_\varepsilon)) =_{\widetilde{\mathbb{R}}} 0_{\widetilde{\mathbb{R}}}. \quad (1.11.7.a)$$

The Bianchi identity holds

$$(R^n_{ikl;m,\varepsilon}(x))_\varepsilon + (R^n_{imk;l,\varepsilon}(x))_\varepsilon + (R_{ilm;k,\varepsilon}(x)) =_{\widetilde{\mathbb{R}}} 0_{\widetilde{\mathbb{R}}} \quad (1.11.8)$$

and $\forall z = (x_\varepsilon)_\varepsilon \in \widetilde{\mathbb{R}}_{\mathbf{fin}}$

$$(R^n_{ikl;m,\varepsilon}(x_\varepsilon))_\varepsilon + (R^n_{imk;l,\varepsilon}(x_\varepsilon))_\varepsilon + (R_{ilm;k,\varepsilon}(x_\varepsilon)) =_{\widetilde{\mathbb{R}}} 0_{\widetilde{\mathbb{R}}}. \quad (1.11.8.a)$$

From the Colombeau generalized curvature tensor we can, by contraction, construct

Colombeau generalized tensor of the second rank. This contraction can be carried out in only one way: contraction of the Colombeau generalized tensor$(R_{iklm,\varepsilon})_\varepsilon$ on the indices $i$ and $k$ or $l$ and $m$ gives zero because of the antisymmetry in these indices, while contraction on any other pair always gives the same result, except for sign. We define the Colombeau generalized tensor $(R_{ik,\varepsilon})_\varepsilon$ (the generalized Ricci tensor) as

$$(R_{ik,\varepsilon})_\varepsilon = (g_\varepsilon^{lm} R_{ilmk,\varepsilon})_\varepsilon = (R^l_{ilk,\varepsilon})_\varepsilon. \quad (1.11.9)$$

From Eq.(1.11.7) we get

$$\begin{gathered}(R_{ik,\varepsilon}(x))_\varepsilon = \left(\frac{\partial(\Gamma^l_{ik,\varepsilon}(x))}{\partial x^l}\right)_\varepsilon - \left(\frac{\partial(\Gamma^l_{il,\varepsilon}(x))}{\partial x^m}\right)_\varepsilon + \\ (\Gamma^l_{ik,\varepsilon}(x)\Gamma^m_{lm,\varepsilon}(x))_\varepsilon - (\Gamma^m_{il,\varepsilon}(x)\Gamma^l_{km,\varepsilon}(x))_\varepsilon.\end{gathered} \quad (1.11.10)$$

This Colombeau generalized tensor $(R_{ik,\varepsilon}(x))_\varepsilon$ is symmetric:

$$(R_{ik,\varepsilon}(x))_\varepsilon = (R_{ki,\varepsilon}(x))_\varepsilon. \quad (1.11.11)$$

$\forall z = (x_\varepsilon)_\varepsilon \in \widetilde{\mathbb{R}}_{\mathbf{fin}}$ From Eq.(1.11.10)-Eq.(1.11.11) we get

$$\begin{gathered}(R_{ik,\varepsilon}(x_\varepsilon))_\varepsilon = \left(\frac{\partial(\Gamma^l_{ik,\varepsilon}(x_\varepsilon))}{\partial x^l_\varepsilon}\right)_\varepsilon - \left(\frac{\partial(\Gamma^l_{il,\varepsilon}(x_\varepsilon))}{\partial x^m_\varepsilon}\right)_\varepsilon + \\ (\Gamma^l_{ik,\varepsilon}(x_\varepsilon)\Gamma^m_{lm,\varepsilon}(x_\varepsilon))_\varepsilon - (\Gamma^m_{il,\varepsilon}(x_\varepsilon)\Gamma^l_{km,\varepsilon}(x_\varepsilon))_\varepsilon\end{gathered} \quad (1.11.12)$$

and

$$(R_{ik,\varepsilon}(x_\varepsilon))_\varepsilon = (R_{ki,\varepsilon}(x_\varepsilon))_\varepsilon. \quad (1.11.13)$$

Contracting $(R_{ik,\varepsilon}(x))_\varepsilon$ , we obtain the invariant

$$(R(x))_\varepsilon = (g_\varepsilon^{ik}(x) R_{ik,\varepsilon}(x))_\varepsilon \quad (1.11.14)$$

which is called the Colombeau generalized scalar curvature.Finally, contracting $(R_{ik,\varepsilon}(x_\varepsilon))_\varepsilon$, we obtain the invariant

$$(R(x_\varepsilon))_\varepsilon = (g_\varepsilon^{ik}(x_\varepsilon) R_{ik,\varepsilon}(x_\varepsilon))_\varepsilon \quad (1.11.15)$$

is called the point value at point $(x_\varepsilon)_\varepsilon \in \widetilde{\mathbb{R}}_{\mathbf{fin}}$ of the Colombeau generalized scalar curvature $(R(x))_\varepsilon$.

**Remark 1.11.1**.Note that $\forall z = (x_\varepsilon)_\varepsilon \in \widetilde{\mathbb{R}}_{\mathbf{fin}}$ the point value at point $(x_\varepsilon)_\varepsilon$ of the Colombeau generalized scalar curvature $(R(x))_\varepsilon$ is $\mathbf{cl}[(R(x_\varepsilon))_\varepsilon] \in \widetilde{\mathbb{R}}$.

# 1.12. The action functional for the singular gravitational field

The action $\mathbf{S}[(g_\varepsilon)_\varepsilon]$ obviously must be expressed in terms of an scalar Colombeau integral taken over all space and over the time coordinate $x^0$ between two given values,i.e.

$$\mathbf{S}[(g_\varepsilon)_\varepsilon] = \left(\int G_\varepsilon \sqrt{-g_\varepsilon}\, d\Omega\right)_\varepsilon = \int (G_\varepsilon)_\varepsilon \sqrt{-(g_\varepsilon)_\varepsilon}\, d\Omega. \quad (1.12.1)$$

In order to determine this Colombeau scalar we assume that the equations of the gravitational field must contain derivatives of the "potentials" no higher than the second order. Since the field equations are obtained by varying the action, it is necessary that the integrand $(G_\varepsilon)_\varepsilon$ contain derivatives of $(g_{ik,\varepsilon})_\varepsilon$ no higher than first order; thus $(G_\varepsilon)_\varepsilon$ must contain only the tensor $(g_{ik,\varepsilon})_\varepsilon$ and the quantities $(\Gamma^l_{kl,\varepsilon})_\varepsilon$.However, it is impossible to construct an invariant from the quantities $(g_{ik,\varepsilon})_\varepsilon$ and $(\Gamma^l_{kl,\varepsilon})_\varepsilon$ alone.This is clear from the fact that by a suitable choice of generalized coordinate system we can always make all the quantities $(\Gamma^l_{kl,\varepsilon})_\varepsilon$ zero at a given point. There is, however, the Colombeau scalar $(R_\varepsilon)_\varepsilon$(the generalized curvature of the four-space), which though it contains in addition to the $(g_{ik,\varepsilon})_\varepsilon$ and its first derivatives also the second derivatives of $(g_{ik,\varepsilon})_\varepsilon$, is linear in the second order Colombeau derivatives. The action functional for the gravitational field reads

$$\left(\int R_\varepsilon \sqrt{-g_\varepsilon}\, d\Omega\right)_\varepsilon. \qquad (1.12.2)$$

The invariant Colombeau integral (2.8.2) can be transformed by means of Gauss'theorem to the integral of an expression not containing the second derivatives. Thus Colombeau integral (1.12.2) can be presented in the following form

$$\left(\int R_\varepsilon \sqrt{-g_\varepsilon}\, d\Omega\right)_\varepsilon = \left(\int G_\varepsilon \sqrt{-g_\varepsilon}\, d\Omega\right)_\varepsilon + \left(\int \frac{\partial\left(\sqrt{-g_\varepsilon}\, w_\varepsilon^i\right)}{\partial x^i} d\Omega\right)_\varepsilon, \qquad (1.12.3)$$

where $(G_\varepsilon)_\varepsilon$ contains only the tensor $(g_{ik,\varepsilon})_\varepsilon$ and its first derivatives, and the integrand of the second integral has the form of a divergence of a certain quantity wl (the detailed calculation is given at the end of this section). According to Gauss' theorem, this second integral can be transformed into an integral over a hypersurface surrounding the four-volume over which the integration is carried out in the other two integrals. When we vary the action, the variation of the second term on the right vanishes, since in the principle of least action, the variations of the field at the limits of the region of integration are zero. Consequently, we may write

$$\delta\left(\int R_\varepsilon \sqrt{-g_\varepsilon}\, d\Omega\right)_\varepsilon = \left(\delta \int R_\varepsilon \sqrt{-g_\varepsilon}\, d\Omega\right)_\varepsilon = \left(\delta \int G_\varepsilon \sqrt{-g_\varepsilon}\, d\Omega\right)_\varepsilon. \qquad (1.12.4)$$

The left side is Colombeau scalar; therefore the expression on the right is also Colombeau scalar (the quantity $(G_\varepsilon)_\varepsilon$ itself is, of course, not Colombeau scalar). The quantity $(G_\varepsilon)_\varepsilon$ satisfies the condition imposed above, since it contains only the $(g_{ik,\varepsilon})_\varepsilon$ and its Colombeau derivatives. Thus finally we obtain

$$\delta\mathbf{S}[(g_\varepsilon)_\varepsilon] = -\frac{c^3}{16\pi\kappa}\left(\delta \int G_\varepsilon \sqrt{-g_\varepsilon}\, d\Omega\right)_\varepsilon = -\frac{c^3}{16\pi k}\left(\delta \int R_\varepsilon \sqrt{-g_\varepsilon}\, d\Omega\right)_\varepsilon. \qquad (1.12.5)$$

The constant $\kappa$ is called the gravitational constant. The dimensions of $\kappa$ follow from (1.12.5). Its numerical value is $\kappa = 6.67 \times 10^{-8}\mathrm{sm}^3 \times \mathrm{gr}^{-1} \times \sec^{-2}$.

Let us calculate now the quantity $(G_\varepsilon)_\varepsilon$ of (1.12.5). From the Eq.(1.11.12) for $(R_{ik,\varepsilon})_\varepsilon$,we get

$$\begin{gathered}\left(\sqrt{-g_\varepsilon}\, R_\varepsilon\right)_\varepsilon = \left(\sqrt{-g_\varepsilon}\, g_\varepsilon^{ik} R_{ik,\varepsilon}\right)_\varepsilon = \left\{\left(\sqrt{-g_\varepsilon}\right)_\varepsilon\right\} \times \\ \left\{\left(g_\varepsilon^{ik} \frac{\partial \Gamma_{ik,\varepsilon}^l}{\partial x^l}\right)_\varepsilon - \left(g_\varepsilon^{ik} \frac{\partial \Gamma_{il,\varepsilon}^l}{\partial x^k}\right)_\varepsilon + (g_\varepsilon^{ik}\Gamma_{ik,\varepsilon}^l \Gamma_{lm,\varepsilon}^m)_\varepsilon + (g_\varepsilon^{ik}\Gamma_{il,\varepsilon}^m \Gamma_{km,\varepsilon}^l)_\varepsilon\right\}.\end{gathered} \qquad (1.12.6)$$

In the first two terms on the rhs of (1.12.6), we get

$$\begin{gathered}\left(\sqrt{-g_\varepsilon}\, g_\varepsilon^{ik} \frac{\partial \Gamma_{ik,\varepsilon}^l}{\partial x^l}\right)_\varepsilon = \frac{\partial}{\partial x^l}\left(\sqrt{-g_\varepsilon}\, g_\varepsilon^{ik} \Gamma_{ik,\varepsilon}^l\right)_\varepsilon - \left\{(\Gamma_{ik,\varepsilon}^l)_\varepsilon\right\} \frac{\partial}{\partial x^l}\left(\sqrt{-g_\varepsilon}\, g_\varepsilon^{ik}\right)_\varepsilon \\ \left(\sqrt{-g_\varepsilon}\, g_\varepsilon^{ik} \frac{\partial \Gamma_{il,\varepsilon}^l}{\partial x^k}\right)_\varepsilon = \frac{\partial}{\partial x^k}\left(\sqrt{-g_\varepsilon}\, g_\varepsilon^{ik} \Gamma_{il,\varepsilon}^l\right)_\varepsilon - \left\{(\Gamma_{il,\varepsilon}^l)_\varepsilon\right\} \frac{\partial}{\partial x^k}\left(\sqrt{-g_\varepsilon}\, g_\varepsilon^{ik}\right)_\varepsilon.\end{gathered} \qquad (1.12.7)$$

Dropping the total derivatives in the Eqs.(1.12.7), we find

$$\left(\sqrt{-g_\varepsilon}\, G_\varepsilon\right)_\varepsilon = \{(\Gamma_{im,\varepsilon}^m)_\varepsilon\} \frac{\partial}{\partial x^k}\left(\sqrt{-g_\varepsilon}\, g_\varepsilon^{ik}\right)_\varepsilon - \left\{(\Gamma_{ik,\varepsilon}^l)_\varepsilon\right\} \frac{\partial}{\partial x^l}\left(\sqrt{-g_\varepsilon}\, g_\varepsilon^{ik}\right)_\varepsilon. \qquad (1.12.8)$$

By using now formulas

$$(\Gamma_{ki,\varepsilon}^i)_\varepsilon = \frac{1}{2(g_\varepsilon)_\varepsilon}\left(\frac{\partial g_\varepsilon}{\partial x^k}\right)_\varepsilon = \frac{\partial \ln \sqrt{-(g_\varepsilon)_\varepsilon}}{\partial x^k},$$

$$(g_\varepsilon^{kl}\Gamma_{kl,\varepsilon}^i)_\varepsilon = -\frac{1}{\sqrt{(g_\varepsilon)_\varepsilon}} \frac{\partial\left(\left[(g_\varepsilon^{ik})_\varepsilon\right]\sqrt{-(g_\varepsilon)_\varepsilon}\right)}{\partial x^k},$$

we find that the first two terms on the right are equal to $\left(\sqrt{-g_\varepsilon}\right)_\varepsilon$ multiplied by

$$\begin{gathered}2(\Gamma^l_{ik,\varepsilon}\Gamma^i_{lm,\varepsilon}g^{mk}_\varepsilon)_\varepsilon - (\Gamma^m_{im,\varepsilon}\Gamma^i_{kl,\varepsilon}g^{kl}_\varepsilon)_\varepsilon - (\Gamma^l_{ik,\varepsilon}\Gamma^m_{lm,\varepsilon}g^{ik}_\varepsilon)_\varepsilon = \\ 2\left\{(g^{ik}_\varepsilon)_\varepsilon\right\}\left[(\Gamma^m_{il,\varepsilon}\Gamma^l_{km,\varepsilon} - \Gamma^l_{ik,\varepsilon}\Gamma^m_{lm,\varepsilon})_\varepsilon\right].\end{gathered} \tag{1.12.9}$$

Finally, we get

$$(G_\varepsilon)_\varepsilon = \left\{(g^{ik}_\varepsilon)_\varepsilon\right\}\left[(\Gamma^m_{im,\varepsilon}\Gamma^l_{km,\varepsilon})_\varepsilon - (\Gamma^l_{ik,\varepsilon}\Gamma^m_{lm,\varepsilon})_\varepsilon\right]. \tag{1.12.10}$$

**Remark 1.12.1.**Note that when the principle of least action is applied to a gravitational field, we can assert only that we can impose auxiliary conditions on the $(g_{ik,\varepsilon})_\varepsilon$, such that when they are fulfilled the action has a minimum with respect to variations of the $(g_{ik,\varepsilon})_\varepsilon$.

# 1.13.The distributional energy-momentum tensor.

In this subsection we consider the general rule for calculatingthe energy-momentum tensor of any physical system whose action is given in the form of an Colombeau integral over four dimensional distributional spase time.This integral reads

$$(\mathbf{S}_{m,\varepsilon})_\varepsilon = \frac{1}{c}\left(\int \Lambda_\varepsilon(x)\sqrt{-g_\varepsilon(x)}\,d\Omega\right)_\varepsilon. \tag{1.13.1}$$

Using generalized Gauss' theorem, and setting $(\delta g^{ik}_\varepsilon)_\varepsilon = 0_{\widetilde{\mathbb{R}}}$ at the integration limits, one finds $\delta(S_{m,\varepsilon})_\varepsilon$ in the form,(see Remark 1.13.2):

$$\begin{gathered}\delta(\mathbf{S}_{m,\varepsilon})_\varepsilon = \\ \frac{1}{c}\int\left\{\left(\frac{\partial\sqrt{-g_\varepsilon}\,\Lambda_\varepsilon}{\partial g^{ik}_\varepsilon}\delta g^{ik}_\varepsilon\right)_\varepsilon + \left(\frac{\partial}{\partial x^l}\frac{\partial\sqrt{-g_\varepsilon}\,\Lambda_\varepsilon}{\partial\frac{\partial g^{ik}_\varepsilon}{\partial x^l}}\delta\frac{\partial g^{ik}_\varepsilon}{\partial x^l}\right)_\varepsilon\right\}d\Omega = \\ \frac{1}{c}\int\left\{\left(\frac{\partial\sqrt{-g_\varepsilon}\,\Lambda_\varepsilon}{\partial g^{ik}_\varepsilon}\right)_\varepsilon - \left(\frac{\partial}{\partial x^l}\frac{\partial\sqrt{-g_\varepsilon}\,\Lambda_\varepsilon}{\partial\frac{\partial g^{ik}_\varepsilon}{\partial x^l}}\delta\frac{\partial g^{ik}_\varepsilon}{\partial x^l}\right)_\varepsilon\right\}\left\{(\delta g^{ik}_\varepsilon)_\varepsilon\right\}d\Omega\end{gathered} \tag{1.13.2}$$

We introduce now the notation

$$\frac{1}{2}\left(\sqrt{-g_\varepsilon}\,T_{ik,\varepsilon}\right)_\varepsilon = \left(\frac{\partial\sqrt{-g_\varepsilon}\,\Lambda_\varepsilon}{\partial g^{ik}_\varepsilon}\right)_\varepsilon - \left(\frac{\partial}{\partial x^l}\frac{\partial\sqrt{-g_\varepsilon}\,\Lambda_\varepsilon}{\partial\frac{\partial g^{ik}_\varepsilon}{\partial x^l}}\right)_\varepsilon, \tag{1.13.3}$$

then $\delta(S_{m,\varepsilon})_\varepsilon$ reads

$$\delta(S_{m,\varepsilon})_\varepsilon = \frac{1}{2c}\left(\int T_{ik,\varepsilon}\sqrt{-g_\varepsilon}\,\delta g^{ik}_\varepsilon d\Omega\right)_\varepsilon = -\frac{1}{2c}\left(\int T^{ik}_\varepsilon\sqrt{-g_\varepsilon}\,\delta g_{ik,\varepsilon}d\Omega\right)_\varepsilon, \tag{1.13.4}$$

note that $(T_{ik,\varepsilon}\delta g^{ik}_\varepsilon)_\varepsilon = -(T^{ik}_\varepsilon\delta g_{ik,\varepsilon})_\varepsilon$ since $(g^{ik}_\varepsilon\delta g_{ik,\varepsilon})_\varepsilon = -(g_{ik,\varepsilon}\delta g^{ik}_\varepsilon)_\varepsilon$.

**Remark 1.13.1.**In (1.13.1) we carry out now a transformation from the coordinates $x^i$ to the coordinates $(x'^i_\varepsilon)_\varepsilon = x^i + (\xi^i_\varepsilon)_\varepsilon$, where the $(\xi^i_\varepsilon)_\varepsilon \in \widetilde{\mathbb{R}}$ are small quantities. Under this transformation the $(g^{ik}_\varepsilon)_\varepsilon$ are transformed according to the formulas

$$(g_\varepsilon'^{ik}(x_\varepsilon'^{i}))_\varepsilon = \left(g_\varepsilon^{im}(x_\varepsilon^i)\frac{\partial x_\varepsilon'^i}{\partial x^l}\frac{\partial x_\varepsilon'^k}{\partial x^m}\right)_\varepsilon =$$

$$\{(g_\varepsilon^{im})_\varepsilon\}\left(\delta_l^i + \left(\frac{\partial \xi_\varepsilon^i}{\partial x^l}\right)_\varepsilon\right)\left(\delta_m^k + \left(\frac{\partial \xi_\varepsilon^k}{\partial x^m}\right)_\varepsilon\right) \simeq \tag{1.13.5}$$

$$(g_\varepsilon^{ik}(x^l))_\varepsilon + \{(g_\varepsilon^{im})_\varepsilon\}\left(\frac{\partial \xi_\varepsilon^k}{\partial x^m}\right)_\varepsilon + \left\{(g_\varepsilon^{kl})_\varepsilon\right\}\left(\frac{\partial \xi_\varepsilon^i}{\partial x^l}\right)_\varepsilon,$$

were the tensor $(g_\varepsilon'^{ik}(x_\varepsilon'^{i}))_\varepsilon$is a function of the $(x_\varepsilon'^{l})_\varepsilon$, while the tensor $(g_\varepsilon^{ik})_\varepsilon$ is a function of the original coordinates $x^l$. In order to represent all terms as functions of one and the same variables, we expand $(g_\varepsilon'^{ik}(x^l + \xi_\varepsilon^l))_\varepsilon$ in powers of $(\xi_\varepsilon^l)_\varepsilon$. Furthermore, if we neglect terms of higher order in $(\xi_\varepsilon^l)_\varepsilon$,we can in all terms containing $(\xi_\varepsilon^l)_\varepsilon$, replace $(g_\varepsilon'^{ik})_\varepsilon$ by $(g_\varepsilon^{ik})_\varepsilon$. Thus we obtain

$$(g_\varepsilon'^{ik}(x^l))_\varepsilon = (g_\varepsilon^{ik}(x^l))_\varepsilon - \left\{(\xi_\varepsilon^l)_\varepsilon\right\}\left(\frac{\partial g_\varepsilon^{ik}}{\partial x^l}\right)_\varepsilon + \left\{(g_\varepsilon^{il})_\varepsilon\right\}\left(\frac{\partial \xi_\varepsilon^k}{\partial x^l}\right)_\varepsilon + \left\{(g_\varepsilon^{kl})_\varepsilon\right\}\left(\frac{\partial \xi_\varepsilon^i}{\partial x^l}\right)_\varepsilon. \tag{1.13.6}$$

It is easy to verify by direct trial that the last three terms on the right can be written as a sum $(\xi_\varepsilon^{i;k})_\varepsilon + (\xi_\varepsilon^{k;i})_\varepsilon$ of contravariant derivatives of the $(\xi_\varepsilon^i)_\varepsilon$. Thus we finally obtain the transformation of the $(g_\varepsilon^{ik})_\varepsilon$ in the form

$$(g_\varepsilon'^{ik})_\varepsilon = (g_\varepsilon^{ik})_\varepsilon + (\delta g_\varepsilon^{ik})_\varepsilon, (\delta g_\varepsilon^{ik})_\varepsilon = (\xi_\varepsilon^{i;k})_\varepsilon + (\xi_\varepsilon^{k;i})_\varepsilon. \tag{1.13.7}$$

For the covariant components, we obtain

$$(g'_{ik,\varepsilon})_\varepsilon = (g_{ik,\varepsilon})_\varepsilon + (\delta g_{ik,\varepsilon})_\varepsilon, (\delta g_{ik,\varepsilon})_\varepsilon = -(\xi_{i;k,\varepsilon})_\varepsilon - (\xi_{k;i,\varepsilon})_\varepsilon, \tag{1.13.8}$$

so that, to terms of first order we satisfy the condition $(g_{il,\varepsilon}g_\varepsilon'^{kl})_\varepsilon = \delta_i^k$.

**Remark 1.13.2.**Since the action $(\mathbf{S}_{m,\varepsilon})_\varepsilon$ is a Colombeau scalar, it does not change under a transformation of coordinates. On the other hand, the change $\delta(\mathbf{S}_{m,\varepsilon})_\varepsilon$ in the action under a transformation of coordinates can be written in the following form:let $(q_\varepsilon)_\varepsilon$ denote the quantities defining the physical system to which the action $(\mathbf{S}_{m,\varepsilon})_\varepsilon$ applies. Under coordinate transformation the quantities $(q_\varepsilon)_\varepsilon$ change by $(\delta q_\varepsilon)_\varepsilon$. In calculating $\delta(\mathbf{S}_{m,\varepsilon})_\varepsilon = (\delta\mathbf{S}_{m,\varepsilon})_\varepsilon$ we need not write terms containing the changes in $(q_\varepsilon)_\varepsilon$. All such terms must cancel each other by virtue of the "equations of motion" of the physical system, since these equations are obtained by equating to zero the variation of $(\mathbf{S}_{m,\varepsilon})_\varepsilon$ with respect to the quantities $(q_\varepsilon)_\varepsilon$. Therefore it is sufficient to write the terms associated with changes in the $(g_{ik,\varepsilon})_\varepsilon$ . Using Gauss' theorem, and setting $(\delta g_\varepsilon^{ik})_\varepsilon = 0_{\widetilde{\mathbb{R}}}$ at the integration limits, one finds $\delta(S_{m,\varepsilon})_\varepsilon$ in the form of the Eq.(1.13.2).

**Remark 1.13.3.**We note that the equations

$$(\xi_\varepsilon^{i;k})_\varepsilon + (\xi_\varepsilon^{k;i})_\varepsilon = 0 \tag{1.13.9}$$

determine the infinitesimal generalized coordinate transformations that do not change the Colombeau generalized metric. These are called the distributional Killing equations.

# 1.14. The generalized gravitational field equations.

We now proceed to the derivation of the equations of the gravitational field. These equations are obtained from the principle of least action $\delta((S_{m,\varepsilon})_\varepsilon + (S_{g_\varepsilon})_\varepsilon) = 0_{\widetilde{\mathbb{R}}}$, where $(S_{m,\varepsilon})_\varepsilon$ and $(S_{g_\varepsilon})_\varepsilon$are the distributional actions of the gravitational field and matter respectively. We now subject the gravitational Colombeau metric field, that is,the quantities $g_{ik}$, to variation.

Calculating the variation $\delta(S_{g_\varepsilon})_\varepsilon$, we get

$$\delta\left(\int R_\varepsilon\sqrt{-g_\varepsilon}\,d\Omega\right)_\varepsilon = \left(\delta\int R_\varepsilon\sqrt{-g_\varepsilon}\,d\Omega\right)_\varepsilon = \left(\delta\int g_\varepsilon^{ik}R_{ik,\varepsilon}\sqrt{-g_\varepsilon}\,d\Omega\right)_\varepsilon =$$
$$\left\{\left(\int R_{ik,\varepsilon}\sqrt{-g_\varepsilon}\,\delta g_\varepsilon^{ik}d\Omega\right)_\varepsilon + \left(\int R_{ik,\varepsilon}g_\varepsilon^{ik}\delta\sqrt{-g_\varepsilon}\,d\Omega\right)_\varepsilon + \left(\int g_\varepsilon^{ik}\sqrt{-g_\varepsilon}\,\delta R_{ik,\varepsilon}d\Omega\right)_\varepsilon\right\} \quad (1.14.1)$$
$$\int\left\{\left(R_{ik,\varepsilon}\sqrt{-g_\varepsilon}\,\delta g_\varepsilon^{ik}\right)_\varepsilon + \left(R_{ik,\varepsilon}g_\varepsilon^{ik}\delta\sqrt{-g_\varepsilon}\right)_\varepsilon + \left(g_\varepsilon^{ik}\sqrt{-g_\varepsilon}\,\delta R_{ik,\varepsilon}\right)_\varepsilon\right\}d\Omega.$$

From formula $d(g_\varepsilon)_\varepsilon = \left[(g_\varepsilon g_\varepsilon^{ik})_\varepsilon\right]d(g_{ik,\varepsilon})_\varepsilon$, we obtain

$$\left(\delta\sqrt{-g_\varepsilon}\right)_\varepsilon = -\left\{\left(\frac{1}{2\sqrt{-g_\varepsilon}}\right)_\varepsilon\right\}(\delta g_\varepsilon)_\varepsilon = -\frac{1}{2}\left\{\left(\sqrt{-g_\varepsilon}\right)_\varepsilon\right\}(g_{ik,\varepsilon}\delta g_\varepsilon^{ik})_\varepsilon. \quad (1.14.2)$$

Substituting Eq.(1.14.2) into Eq.(1.14.1), we obtain

$$\delta\left(\int R_\varepsilon\sqrt{-g_\varepsilon}\,d\Omega\right)_\varepsilon = \left(\int\left\{R_{ik,\varepsilon} - \frac{1}{2}g_{ik,\varepsilon}R_\varepsilon\right\}\sqrt{-g_\varepsilon}\,\delta g_\varepsilon^{ik}d\Omega\right)_\varepsilon + $$
$$+\left(\int g_\varepsilon^{ik}\sqrt{-g_\varepsilon}\,\delta R_{ik,\varepsilon}d\Omega\right)_\varepsilon. \quad (1.14.3)$$

**Remark 1.14.1.** Note that although the quantities $(\Gamma^i_{kl,\varepsilon})_\varepsilon$ don't constitute a tensor, their variations $\delta(\Gamma^i_{kl,\varepsilon})_\varepsilon$ do form a generalized tensor, for $(\Gamma^i_{kl,\varepsilon}A_k dx^l)_\varepsilon$ is the change in a vector under parallel displacement from some point $P$ to an infinitesimally separated point $P'$ and therefore $(\Gamma^i_{kl,\varepsilon}A_k dx^l)_\varepsilon$ is the difference between the two vectors, obtained as the result of two parallel displacements (one with the unvaried, the other with the varied $(\Gamma^i_{kl,\varepsilon})_\varepsilon$ from the point $P$ to one and the same point $P'$. The difference between two vectors at the same point is a vector, and therefore $\delta(\Gamma^i_{kl,\varepsilon})_\varepsilon = (\delta\Gamma^i_{kl,\varepsilon})_\varepsilon$ is a generalized tensor.
Let us use now a locally geodesic system of a generalized coordinates. Then at that point all the $(\Gamma^i_{kl,\varepsilon})_\varepsilon = 0_{\widetilde{\mathbb{R}}}$. With the help ofexpression (92.10) for the $(R_{ik,\varepsilon})_\varepsilon$, we obtain (we remind that the first derivatives of the $(g_\varepsilon^{ik})_\varepsilon$ are now equal to $0_{\widetilde{\mathbb{R}}}$)

$$(g_\varepsilon^{ik}\delta R_{ik,\varepsilon})_\varepsilon = \left\{(g_\varepsilon^{ik})_\varepsilon\right\}\left\{\left(\frac{\partial}{\partial x^l}\delta\Gamma^l_{ik,\varepsilon}\right)_\varepsilon - \left(\frac{\partial}{\partial x^k}\delta\Gamma^l_{il,\varepsilon}\right)_\varepsilon\right\} =$$
$$\left(g_\varepsilon^{ik}\frac{\partial}{\partial x^l}\delta\Gamma^l_{ik,\varepsilon}\right)_\varepsilon - \left(g_\varepsilon^{il}\frac{\partial}{\partial x^l}\delta\Gamma^k_{ik,\varepsilon}\right)_\varepsilon = \left(\frac{\partial w_\varepsilon^l}{\partial x^l}\right)_\varepsilon \quad (1.14.4)$$

where

$$(w_\varepsilon^l)_\varepsilon = (g_\varepsilon^{ik}\delta\Gamma^l_{ik,\varepsilon})_\varepsilon - (g_\varepsilon^{il}\delta\Gamma^k_{ik,\varepsilon})_\varepsilon. \quad (1.14.5)$$

Since $(w_\varepsilon^l)_\varepsilon$ is a generalized vector, replacing $(dw_\varepsilon^l/dx^l)_\varepsilon$ by $\left(w^l_{;\,l,\varepsilon}\right)_\varepsilon$ and using formula

$$A^i_{;i,\varepsilon} = \left[\left(\frac{1}{\sqrt{-g_\varepsilon}}\right)_\varepsilon\right]\frac{\partial}{\partial x^i}\left(\sqrt{-g_\varepsilon}\,A_\varepsilon^i\right)_\varepsilon,$$

we may write the relation we have obtained above, in any generalized coordinate system, in the form

$$(g_\varepsilon^{ik}\delta R_{ik,\varepsilon})_\varepsilon = \left\{\left(\frac{1}{\sqrt{-g_\varepsilon}}\right)_\varepsilon\right\}\frac{\partial}{\partial x^l}\left(\sqrt{-g_\varepsilon}\,w_\varepsilon^l\right)_\varepsilon. \quad (1.14.6)$$

Therefore the second Colombeau integral on the right side of (1.14.3) is equal to

$$\left(\int g_\varepsilon^{ik}\sqrt{-g_\varepsilon}\,\delta R_{ik,\varepsilon}d\Omega\right)_\varepsilon = \left(\int\frac{\partial\sqrt{-g_\varepsilon}\,w_\varepsilon^l}{\partial x^l}d\Omega\right)_\varepsilon \quad (1.14.7)$$

and by generalized Gauss' theorem can be transformed into an Colombeau integral of $(w_\varepsilon^l)_\varepsilon$ over the hypersurface surrounding the whole four-volume. Since the variations of the field are zero at the integration limits, this term drops out. Thus, the variation $\mathbf{S}[(g_\varepsilon)_\varepsilon]$ is equal to

$$\mathbf{S}[(g_\varepsilon)_\varepsilon] = -\frac{c^3}{16\pi\kappa}\left(\int\left\{R_{ik,\varepsilon} - \frac{1}{2}g_{ik,\varepsilon}R_\varepsilon\right\}\sqrt{-g_\varepsilon}\,\delta g_\varepsilon^{ik}d\Omega\right)_\varepsilon. \tag{1.14.8}$$

**Remark 1.14.2.** We note that if we had started from the expression

$$\delta\mathbf{S}_g[(g_\varepsilon)_\varepsilon] = -\frac{c^3}{16\pi\kappa}\left(\delta\int G_\varepsilon\sqrt{-g_\varepsilon}\,d\Omega\right)_\varepsilon \tag{1.14.9}$$

for the action of the field, then we get

$$\delta\mathbf{S}[(g_\varepsilon)_\varepsilon] =$$

$$-\frac{c^3}{16\pi\kappa}\int\delta(g_\varepsilon^{ik})_\varepsilon d\Omega\left\{\left(\frac{\partial\left\{G_\varepsilon\sqrt{-g_\varepsilon}\right\}}{\partial g_\varepsilon^{ik}}\right)_\varepsilon - \left(\frac{\partial}{\partial x^l}\frac{\partial\left\{G_\varepsilon\sqrt{-g_\varepsilon}\right\}}{\partial\frac{\partial g_\varepsilon^{ik}}{\partial x^l}}\right)_\varepsilon\right\}. \tag{1.14.10}$$

Comparing Eq.(1.14.10) with Eq.(1.14.8), we get

$$(R_{ik,\varepsilon})_\varepsilon - \frac{1}{2}(g_{ik,\varepsilon}R_\varepsilon)_\varepsilon =$$

$$\left\{\left(\frac{1}{\sqrt{-g_\varepsilon}}\right)_\varepsilon\right\}\left\{\left(\frac{\partial\left\{G_\varepsilon\sqrt{-g_\varepsilon}\right\}}{\partial g_\varepsilon^{ik}}\right)_\varepsilon - \left(\frac{\partial}{\partial x^l}\frac{\partial\left\{G_\varepsilon\sqrt{-g_\varepsilon}\right\}}{\partial\frac{\partial g_\varepsilon^{ik}}{\partial x^l}}\right)_\varepsilon\right\}. \tag{1.14.11}$$

For the variation of the action of the matter we can write immediately from (1.13.4):

$$(\delta\mathbf{S}_{m,\varepsilon})_\varepsilon = \frac{1}{2c}\left(\int T_{ik,\varepsilon}\sqrt{-g_\varepsilon}\,\delta g_\varepsilon^{ik}d\Omega\right)_\varepsilon, \tag{1.14.12}$$

where $(T_{ik,\varepsilon})_\varepsilon \in \mathcal{G}(\mathbb{R}^4)$ is the generalized energy-momentum tensor of the matter fields.
Thus, from the principle of least action

$$\delta\{\mathbf{S}[(g_\varepsilon)_\varepsilon] + (\mathbf{S}_{m,\varepsilon})_\varepsilon\} = 0_{\widetilde{\mathbb{R}}} \tag{1.14.13}$$

one obtains

$$-\frac{c^3}{16\pi\kappa}\left(\int\left\{R_{ik,\varepsilon} - \frac{1}{2}g_{ik,\varepsilon}R_\varepsilon - \frac{8\pi\kappa}{c^4}T_{ik,\varepsilon}\right\}\sqrt{-g_\varepsilon}\,\delta g_\varepsilon^{ik}d\Omega\right)_\varepsilon = 0_{\widetilde{\mathbb{R}}}. \tag{1.14.14}$$

From Eq.(1.14.14), since of the arbitrariness of the $(\delta g_\varepsilon^{ik})_\varepsilon \in \mathcal{G}(\mathbb{R}^4)$ finally we get

$$(R_{ik,\varepsilon})_\varepsilon - \frac{1}{2}(g_{ik,\varepsilon}R_\varepsilon)_\varepsilon = \frac{8\pi\kappa}{c^4}(T_{ik,\varepsilon})_\varepsilon \tag{1.14.15}$$

or, in mixed components,

$$(R_{i,\varepsilon}^k)_\varepsilon - \frac{1}{2}\delta_i^k(R_\varepsilon)_\varepsilon = \frac{8\pi\kappa}{c^4}(T_{i,\varepsilon}^k)_\varepsilon. \tag{1.14.16}$$

They are called the generalized Einstein equations.
Contracting (1.14.16) on the indices $i$ and $k$, we get

$$(R_\varepsilon)_\varepsilon = -\frac{8\pi\kappa}{c^4}(T_{i,\varepsilon}^i)_\varepsilon = -\frac{8\pi\kappa}{c^4}(T_\varepsilon)_\varepsilon. \tag{1.14.17}$$

Therefore the generalized Einstein equations of the field can also be written in the form

$$(R_{ik,\varepsilon})_\varepsilon = \frac{8\pi\kappa}{c^4}\left\{(T_{ik,\varepsilon})_\varepsilon - \frac{1}{2}(g_{ik,\varepsilon}T_\varepsilon)_\varepsilon\right\}. \tag{1.14.18}$$

Note that the generalized Einstein field equations of the gravitational field are nonlinear Colombeau type equations.

# 1.15. Colombeau solutions of the

# generalized Einstein field equations

This paper dealing with Colombeau extension of the Einstein field equations using apparatus of the Colombeau generalized function [3]-[5] and contemporary generalization of the classical Lorentzian geometry named in literature Colombeau distributional geometry. The regularizations of singularities present in some solutions of the Einstein equations is an important part of this approach. Any singularities present in some solutions such that Schwarzschild solution etc. of the Einstein field equations recognized only in the sense of Colombeau generalized functions [3]-[5] and not classically. Note that in physical literature these singular solutions many years were mistakenly considered as vacuum solutions of the Einstein field equations, see for example [26],[30]. During last 30 years the applications classical linear distributional geometry in general relativity was many developed [4]-[17].

**Remark 1.15.1**. Let $(R^a_{bcd,\varepsilon})_\varepsilon \in \mathcal{G}_\delta(\mathbb{R}^4)$ be Colombeau generalized function obtained using the standard definition of the Riemann curvature in a coordinate basis,i.e.

$$(R^a_{bcd,\varepsilon})_\varepsilon = (\Gamma^a_{db,c,\varepsilon})_\varepsilon - (\Gamma^a_{cb,d,\varepsilon})_\varepsilon + \left(\Gamma^a_{cf,\varepsilon}\Gamma^f_{ab,\varepsilon}\right)_\varepsilon - \left(\Gamma^a_{df,\varepsilon}\Gamma^f_{cb,\varepsilon}\right)_\varepsilon, \qquad (1.15.1)$$

where $(\Gamma^a_{bc,\varepsilon})_\varepsilon \in \mathcal{G}_\delta(\mathbb{R}^n)$ and $\Gamma^a_{bc,\varepsilon}, \varepsilon \in (0,\delta], \delta \leq 1$ is the regularized Levi-Civita connection coefficients in terms of the regularized metric $g_{ab,\varepsilon}, \varepsilon \in (0,1]$ such that $(g_{ab,\varepsilon})_\varepsilon, (g^{ab}_\varepsilon)_\varepsilon \in \mathcal{G}(\mathbb{R}^4), (\det(g_{ab,\varepsilon}))_\varepsilon \neq 0_{\widetilde{\mathbb{R}}}$ It has been shown by many authors (see for example [9],[11]) that under apropriate regularization using the Eq.(1.15.1) one can defines the curvature scalar as a classical Schwartz distribution in $\mathcal{D}'(\mathbb{R}^n)$, [7].

**Remark 1.15.2.**This is the case even for the well-known Schwarzschild spacetime, which is given in the Schwarzschild coordinates $(\hat{x}^0, \hat{r}, \theta, \phi)$, by the metric

$$ds^2 = -\left(1 - \frac{a}{\hat{r}}\right)(d\hat{x}^0)^2 + \left(1 - \frac{a}{\hat{r}}\right)^{-1}(d\hat{r})^2 + \hat{r}^2[(d\theta)^2 + \sin^2\theta(d\phi)^2]. \qquad (1.15.2)$$

Here, $a$ is the Schwarzschild radius $a = 2GM/c^2$ with $G, M$ and $c$ being the Newton gravitational constant, mass of the source, and the light velocity in vacuum Minkowski space-time, respectively. Obviously the fundamental tensor corresponding to $ds^2$ has the components which is degenerate or singular: (i) at $\hat{r} = 0$ and (ii) at $\hat{r} = a$.

**Remark 1.15.3.**Note that in classical papers [7]-[11],etc.

(i) the Colombeau distributional metric tensor $(g_{ab,\varepsilon})_\varepsilon \in \mathcal{G}_\delta(\mathbb{R}^4)$ related to $(R^a_{bcd,\varepsilon})_\varepsilon \in \mathcal{G}_\delta(\mathbb{R}^4)$ by Eq.(1.15.2) never is considered as the Colombeau solution of the Einstein field equations,

(ii) Colombeau nonlinear distributional geometry never is considered as the rigorous mathematical model related to really physical spacetime but only as useful purely mathematical tools in order to obtain related to $(R^a_{bcd,\varepsilon})_\varepsilon \in \mathcal{G}_\delta(\mathbb{R}^4)$ classical Schwartz distributions in $\mathcal{D}'(\mathbb{R}^n)$,

(iii) there is no any important physical applications of the classical linear distributional geometry were obtained.

**Remark 1.15.4.**Originally fundamental physical applications of the Colombeau nonlinear distributional geometry has been obtained in authors paper [33]-[37].

By using now the Cartesian coordinates $(\hat{x}^0, \hat{x}^1, \hat{x}^2, \hat{x}^3)$, which are related to $(\hat{x}^0, \hat{r}, \theta, \phi)$ through the canonical relations: $\hat{x}^1 = \hat{r}\cos\phi\sin\theta$ , $\hat{x}^2 = \hat{r}\sin\phi\sin\theta$ , $\hat{x}^3 = \hat{r}\cos\theta$ ,the metric (1.15.2) reads $ds^2 = \hat{g}_{\mu\nu}d\hat{x}^\mu d\hat{x}^\nu$,where at points $\hat{r} \neq 0, \hat{r} \neq a$ the metric $\hat{g}_{\mu\nu}$ is given by [11]:

$$\begin{aligned} &\hat{g}_{00} = -(1-h)\ , \ \ \hat{g}_{0\alpha} = 0\ , \\ &\hat{g}_{\alpha\beta} = \delta^{\alpha\beta} + h(1-h)^{-1}\frac{\hat{x}^\alpha\hat{x}^\beta}{\hat{r}^2}\ , \quad \alpha,\beta = 1,2,3 \end{aligned} \qquad (1.15.3)$$

with $h \triangleq a/\hat{r}$. Well known that at points $\hat{r} \neq 0, \hat{r} \neq a$ :

$$\kappa\hat{\widetilde{T}}_0^0 = -\frac{h'}{\hat{r}} - \frac{h}{\hat{r}^2}\ ,$$

$$\kappa\hat{\widetilde{T}}_0^\alpha = 0\ ,\quad \kappa\hat{\widetilde{T}}_\alpha^0 = 0\ , \tag{1.15.4}$$

$$\kappa\hat{\widetilde{T}}_\alpha^\beta = \delta_\alpha^\beta\left(-\frac{h''}{2} - \frac{h'}{\hat{r}}\right) + \frac{\hat{x}^\alpha\hat{x}^\beta}{\hat{r}^2}\left(\frac{h''}{2} - \frac{h}{\hat{r}^2}\right),$$

where the hatted symbols $\hat{\widetilde{T}}_\mu^\nu$ represent the quantity $\widetilde{T}_\mu^\nu$ in the coordinate system $\{\hat{x}^\mu;\mu = 0,1,2,3\}$. Also, we have defined $h' \triangleq dh/d\hat{r}$ and $h'' \triangleq d^2h/d\hat{r}^2$.

**Remark 1.15.5**. We extend now the quantity (1.15.3)-(1.15.4) in point $\hat{r} = 0$ as Colombeau generalized functions from Colombeau algebra $\mathcal{G}_\delta(\mathbb{R}^3)$.Regularizing now the function $h = a/\hat{r}$ as $(h_\varepsilon)_\varepsilon = a/\left(\sqrt{\hat{r}^2+\varepsilon^2}\right)_\varepsilon$ and the function $\frac{\hat{x}^\alpha\hat{x}^\beta}{\hat{r}^2}$ as $\frac{\hat{x}^\alpha\hat{x}^\beta}{(\hat{r}^2+\varepsilon^2)_\varepsilon}$ with $\varepsilon \in (0,1]$, we replace now the the singular metric (1.15.3) by the Colombeau generalized metric

$$(ds_\varepsilon^2)_\varepsilon = \left(\hat{g}_{\mu\nu,\varepsilon}d\hat{x}^\mu d\hat{x}^\nu\right)_\varepsilon\ , \tag{1.15.5}$$

where

$$\left(\hat{g}_{00,\varepsilon}\right)_\varepsilon = -(1-h_\varepsilon)\ ,\quad \left(\hat{g}_{0\alpha,\varepsilon}\right)_\varepsilon = 0_{\widetilde{\mathbb{R}}}\ ,$$

$$\left(\hat{g}_{\alpha\beta,\varepsilon}\right)_\varepsilon = \delta^{\alpha\beta} + ((h_\varepsilon(1-h_\varepsilon)^{-1})_\varepsilon)\left(\frac{\hat{x}^\alpha\hat{x}^\beta}{\hat{r}^2+\varepsilon^2}\right)_\varepsilon,\quad \alpha,\beta = 1,2,3 \tag{1.15.6}$$

and therefore

$$\kappa\left(\hat{\widetilde{T}}_{0,\varepsilon}^0(\hat{x})\right)_\varepsilon = -\left(\frac{a\varepsilon^2}{(\hat{r}^2+\varepsilon^2)^{5/2}}\right)_\varepsilon,$$

$$\kappa\left(\hat{\widetilde{T}}_0^\alpha(\hat{x};\varepsilon)\right)_\varepsilon = 0_{\widetilde{\mathbb{R}}},\quad \kappa\left(\hat{\widetilde{T}}_\alpha^0(\hat{x};\varepsilon)\right)_\varepsilon = 0_{\widetilde{\mathbb{R}}},$$

$$\kappa\left(\hat{\widetilde{T}}_\alpha^\beta(\hat{x};\varepsilon)\right)_\varepsilon = \delta_\alpha^\beta\left(\frac{3a\varepsilon^2}{2(\hat{r}^2+\varepsilon^2)^{5/2}}\right)_\varepsilon - \tag{1.15.7}$$

$$\left(\left(\frac{\hat{x}^\alpha\hat{x}^\beta}{\hat{r}^2+\varepsilon^2}\right)_\varepsilon\right)\left(\frac{a\varepsilon^2}{(\hat{r}^2+\varepsilon^2)^{5/2}}\right)_\varepsilon\left(\frac{5}{2}+\left(\frac{\varepsilon^2}{\hat{r}^2+\varepsilon^2}\right)_\varepsilon\right).$$

Note that from Eq. (1.15.7) one obtains [11]:

$$\hat{\widetilde{T}}_\mu^\nu(\hat{x}) \triangleq w\text{-}\lim_{\epsilon\to 0}\hat{\widetilde{T}}_{\mu,\varepsilon}^\nu(\hat{x}) \sim -Mc^2\delta_\mu^0\delta_0^\nu\delta^{(3)}(\hat{x}). \tag{1.15.8}$$

**Remark 1.15.6**.Note that $\left(\hat{\widetilde{T}}_0^0(\hat{x};\varepsilon)\right)_\varepsilon, \left(\hat{\widetilde{T}}_\alpha^\beta(\hat{x};\varepsilon)\right)_\varepsilon \in \mathcal{G}(\mathbb{R}^4)$.Thus the generalized Einstein equation (1.14.8) related to Eq.(1.15.6)-Eq.(1.15.7) in Colombeau notations reads:

$$\left(\hat{G}_{\mu,\varepsilon}{}^\nu\right)_\varepsilon = \left(\hat{R}_{\mu,\varepsilon}{}^\nu\right)_\varepsilon - \frac{1}{2}\delta_\mu{}^\nu\left(\hat{R}_\varepsilon\right)_\varepsilon = \kappa\left(\widetilde{T}_{\mu,\varepsilon}{}^\nu\right)_\varepsilon, \tag{1.15.9}$$

where

$$\left(\hat{R}_\varepsilon(\hat{r})\right)_\varepsilon = \left(\hat{R}^\mu{}_{\mu,\varepsilon}(\hat{r})\right)_\varepsilon =$$

$$-\left(\frac{3a\varepsilon^2}{(\hat{r}^2+\varepsilon^2)^{5/2}}\right)_\varepsilon + \left(\frac{2a\varepsilon^2}{(\hat{r}^2+\varepsilon^2)^{5/2}}\right)_\varepsilon = -\left(\frac{a\varepsilon^2}{(\hat{r}^2+\varepsilon^2)^{5/2}}\right)_\varepsilon. \tag{1.15.10}$$

# 6mm . 2.Distributional Schwarzschild geometry by using nonsmooth regularization via horizon. 3mm

## 5mm . 2.1.1.Distributional Schwarzschild spacetime as Colombeau extension

**of the Lorentzian manifold with nonregularity conditions on Schwarzschild horizon** 2mm

Singular space-times present one of the major challenges in general relativity. Originally it was believed that their singular nature is due to the high degree of symmetry of the well-known examples ranging from the Schwarzschild geometry to the Friedmann-Robertson-Walker cosmological models. However, Penrose and Hawking cite: HawkingEllis73 [36] have shown in their classical singularity theorems that singularities are a phenomenon which is inherent to general relativity. Since the standard approach allows only for smooth space-time metrics, one has to exclude the socalled singular regions from the space-time manifold.In a recent work many authors advocated the use Colombeau distributional techniques cite: VickersWilson98
cite: VickersWilson99Vickers99GerochTraschen87BalasinNachbagauer93
cite: BalasinNachbagauer94KawaiSakane97PantojaRago97
cite: PantojaRago00KunzingerSteinbauer02KunzingerSteinbauer01
cite: GrosserFarkasKunzingerSteinbauer01
cite: HeinzleSteinbauer02Foukzon15FoukzonPotapovMenkova16
cite: Vickers12Steinbauer00GolubevKelner05 [5]-[22] to calculate the energy-momentum tensor of the Schwarzschild geometry.It turns out that it is possible to include the singular region (i.e. the space-like line $r = 0$ with respect to Schwarzschild coordinates) in the space-time which now no longer is a vacuum geometry, and to identify it with the support of the energy-momentum tensor
cite: VickersWilson98BalasinNachbagauer93KawaiSakane97PantojaRago97PantojaRago00 [5], [9], [11]-[13]. The same “physically expected” result for the distributional energy momentum tensor of the Schwarzschild geometry was obtained in papers cite: PantojaRago97
cite: PantojaRago00KunzingerSteinbauer02KunzingerSteinbauer01
cite: GrosserFarkasKunzingerSteinbauer01
cite: HeinzleSteinbauer02Foukzon15FoukzonPotapovMenkova16
cite: Vickers12Steinbauer00[12]-[21], i.e.,

$$T_0^0 = 8\pi m\delta(\vec{x}), \tag{2.1.1}$$

in a conceptually satisfactory way.

**Remark 2.1.1.**The result (2.1.1) can be easily obtained by using apropriate nonsmooth regularization of the Schwarzschild singularity at the origin $r = 0$.

The nonsmooth regularization of the Schwarzschild singularity at the origin $r = 0$ originally considered by N. R. Pantoja and H. Rago in paper cite: PantojaRago97 [12]. Such non smooth regularization of the Schwarzschild singularity is

$$(h_\varepsilon(r))_\varepsilon = -1 + \left(\frac{r_s}{r}\Theta_\varepsilon(r-\varepsilon)\right)_\varepsilon, \varepsilon \in (0,1], r < r_s. \tag{2.1.2}$$

Here $(\Theta_\varepsilon(u))_\varepsilon$ is the generalized Heaviside function,where

$$\Theta_\varepsilon(u) = \begin{cases} (\varepsilon)_\varepsilon & u < 0 \\ \\ 1 & u \geq 0 \end{cases} \tag{2.1.3}$$

and the limit $\varepsilon \to 0$ is understood in a weak distributional sense.The equation

$$\begin{gathered}(ds_\varepsilon^2)_\varepsilon = (h_\varepsilon(r)(dt)^2)_\varepsilon - (h_\varepsilon^{-1}(r)(dr)^2)_\varepsilon + r^2[(d\theta)^2 + \sin^2\theta(d\phi)^2], \\ h_0(r) = -1 + \frac{r_s}{r},\end{gathered} \tag{2.1.4}$$

with $h_\varepsilon, \varepsilon \in (0,1],$ as given in (2.1.4) can be considered as Colombeau version of the Schwarzschild line element in curvature coordinates. From equation (2.1.2), the calculation of the distributional Einstein tensor $(G_t^t(r,\varepsilon))_\varepsilon, (G_r^r(r,\varepsilon))_\varepsilon, (G_\theta^\theta(r,\varepsilon))_\varepsilon, (G_\varphi^\varphi(r,\varepsilon))_\varepsilon$ proceeds in a straighforward manner. By simple calculation one obtains cite: PantojaRago97 [12]:

$$(G_t^t(r,\varepsilon))_\varepsilon = (G_r^r(r,\varepsilon))_\varepsilon = -\left(\frac{h'_\varepsilon(r)}{r}\right)_\varepsilon - \left(\frac{1+h_\varepsilon(r)}{r^2}\right)_\varepsilon = \\ = -r_s\left(\frac{\delta(r-\varepsilon)}{r^2}\right)_\varepsilon = -r_s\frac{\delta(r)}{r^2} \tag{2.1.5}$$

and

$$(G_\theta^\theta(r,\varepsilon))_\varepsilon = (G_\varphi^\varphi(r,\varepsilon))_\varepsilon = -\left(\frac{h''_\varepsilon(r)}{2}\right)_\varepsilon - \left(\frac{h_\varepsilon(r)}{r^2}\right)_\varepsilon = \\ r_s\left(\frac{\delta(r-\varepsilon)}{r^2}\right)_\varepsilon - r_s\left(\frac{\varepsilon}{r^2}\frac{d}{dr}\delta(r-\varepsilon)\right)_\varepsilon \approx -r_s\frac{\delta(r)}{r^2}. \tag{2.1.6}$$

In papers cite: BalasinNachbagauer94Choquet-Bruhat09 [10], [27] Colombeau distributional techniques were extended to the general axisymmetric, stationary Kerr and Newman space-time family.This family also contains the Schwarzschild geometry and its charged extension the Reissner-Nordstrøm solution as special cases of spherical symmetry.In the paper cite: GolubevKelner05 [22] was shown that the solutions will satisfy the Einstein equations everywhere if the energy-momentum tensor has an appropriate singular addition of nonelectromagnetic origin. When this addition term is included, the total energy turns out to be finite and equal to $mc^2$, while the angular momentum for the Kerr and Kerr-Newman solutions is $mc\mathbf{a}$.

**Remark 2.1.2**. The nonsmooth regularization of the Schwarzschild singularity above the horizon $r = r_s$ is

$$(h_\varepsilon^+(r))_\varepsilon = -1 + \left(\frac{r_s}{r}\Theta_\varepsilon((r-r_s)-\varepsilon)\right)_\varepsilon, \varepsilon \in (0,1], r \geq r_s. \tag{2.1.7}$$

Here $(\Theta_\varepsilon(u))_\varepsilon$ is the generalized Heaviside function and the limit $\varepsilon \to 0$ is understood in a weak distributional sense.The equation

$$(ds_\varepsilon^{+2})_\varepsilon = (h_\varepsilon^+(r)(dt)^2)_\varepsilon - \left([h_\varepsilon^+(r)]^{-1}(dr)^2\right)_\varepsilon + r^2[(d\theta)^2 + \sin^2\theta(d\phi)^2], \\ h_0(r) = -1 + \frac{r_s}{r}, \tag{2.1.8}$$

$h_\varepsilon, \varepsilon \in (0,1]$, as given in (2.1.8) can be considered as Colombeau version of the Schwarzschild line element in curvature coordinates above horizon. From equation (2.1.7), the calculation of the distributional Einstein tensor above horizon $(G_t^{+t}(r,\varepsilon))_\varepsilon, (G_r^{+r}(r,\varepsilon))_\varepsilon$, $(G_\theta^{+\theta}(r,\varepsilon))_\varepsilon$, $(G_\varphi^{+\varphi}(r,\varepsilon))_\varepsilon$ proceeds in a straighforward manner. By simple calculation one obtains

$$(G_t^{+t}(r,\varepsilon))_\varepsilon = (G_r^{+r}(r,\varepsilon))_\varepsilon = -\left(\frac{h'_\varepsilon((r-r_s)-\varepsilon)}{r}\right)_\varepsilon \\ -\left(\frac{1+h_\varepsilon((r-r_s)-\varepsilon)}{r^2}\right)_\varepsilon = \\ = -r_s\left(\frac{\delta((r-r_s)-\varepsilon)}{r^2}\right)_\varepsilon \approx -r_s\frac{\delta(r-r_s)}{r^2}. \tag{2.1.9}$$

## The truncated distributional Schwarzschild geometry.

There exist two different types of distributional Schwarzschild blackhole geometry corresponding to classical Schwarzschild solution. That is: (i) full distributional Schwarzschild blackhole geometry, given by Colombeau generalized object, for example by Eq.(1.3.30) and (ii) the truncated distributional Schwarzschild space-time given by Colombeau generalized object (2.1.7)-(2.1.8), i.e. in this case distributional spacetime ends just on the Schwarzschild horizon.

Figure

Fig

Fig.2.1.1.(a) The picture of a distributional Schwarzschild blackhole,

given by Colombeau generalized object (1.3.30).

Distributional spacetime ends just on the Schwarzschild singularity.

(b) The truncated Schwarzschild distributional geometry,

given by Colombeau generalized object (2.1.7)-(2.1.8).

Distributional spacetime ends just on the Schwarzschild horizon.

**Remark 2.1.3.** In a nutshell, there is a widespread but mistaken belief that there exist true gravitational singularities,for example at origin $r = 0$ of the Schwarzschild spacetime, and non principal non gravitational,i.e. purely coordinate singularities, for example at horizon $r = r_g$ of the Schwarzschild spacetime. A coordinate singularity or coordinate degeneracy occurs when an apparent singularity or degeneracy occurs in one coordinate frame, which can be removed by choosing a different frame. Classical example of such mistake is ubnormal deletion of the gravitational singularity,for example from Schwarzschild spacetime

$$\mathbf{Sch} = (S^2 \times \{r \geq 2m\}) \times \mathbb{R}, g_{ij}(r,\theta,\phi), \qquad (2.1.10)$$

originally defined by singular and degenerate Schwarzschild metric cite: LandauLifshitz75 [30],

$$ds^2 = -h(r)(dx^0)^2 + h^{-1}(r)(dr)^2 + r^2[(d\theta)^2 + \sin^2\theta(d\phi)^2], h(r) = 1 - \frac{r_g}{r}. \qquad (2.1.11)$$

by using apropriate singular coordinate change
cite: Choquet-Bruhat09FeliceClarke10MisnerThorneWheeler73cite: LandauLifshitz75
cite: Grant08Eddington24Finkelstein58Lemaitre33LoingerMarsico09 [27]-[35].

**Remark 2.1.4.** Note that: (i) metric (2.1.11) is singular and degenerate at Schwarzschild horizon $r = r_g$, and thus metric (2.1.11) beiond canonical rigorous semi-Riemannian geometry.

(ii) however in physical literature (see for example
cite: FeliceClarke10MisnerThorneWheeler73 cite: LandauLifshitz75[28]-[30]) singularity and degeneracy at Schwarzschild horizon $r = r_g$ acepted as coordinate singularity and coordinate degeneracy.

**Remark 2.1.5.** (see cite: LandauLifshitz75 [30] section 100, p.296)."In the Schwarzschild metric (97.14), $g_{00}$goes to zero and $g_{11}$ to infinity at $r = r_g$ (on the "Schwarzschild sphere"). This could give the basis for concluding that there must be a singularity of the space-time metric and that it is therefore impossible for bodies to exist that have a "radius" (for a given mass) that is less than the gravitational radius. Actually,however, this conclusion would be wrong. This is already evident from the fact that the determinant $g(r) = -r^4\sin^2\theta$ has no singularity at $r = r_g$, so that the condition $g < 0$ (82.3) is not violated. We shall see that in fact we are dealing simply with the impossibility of establishing a suitable reference system for $r < r_g$."

**Remark 2.1.6**. Notice that consideration above meant the following definition of the gravitational singularity.

**Definition 2.1.1**.There is no gravitational singularity at $r = \bar{r}$ iff the determinant $g(r,\theta) = \det(g_{ij}(r,\theta))$ has no singularity at $r = \bar{r}$.

**Remark 2.1.7**. Notice that at singular point $r = r_g$ the determinant $g(r_g)$ is well defined only by the limit

$$g(r_g) = \lim_{r\to r_g}\det(g_{ij}(r,\theta)) = -r_g^4\sin^2\theta. \tag{2.1.12}$$

however in the limit $r \to r_g = 2m$ the classical Levi-Civitá connection $\Gamma^l_{kj}$ becomes infinite

$$\Gamma^1_{11}(r)|_{r=2m} = \lim_{r\to 2m}\frac{-m}{r(r-2m)} = \infty, \Gamma^0_{01}(r)|_{r=2m} = \lim_{r\to 2m}\frac{m}{r(r-2m)} = \infty, \tag{2.1.13}$$

and therefore the Definition 2.1.1 is not sound and even does not any sense under canonical semi-Riemannian geometry.

**Remark 2.1.8**.Notice that:

(i) in order to fixin problem with singularity and degeneracy of the Schwarzschild metric (2.1.11) at Schwarzschild horizon $r = r_g$ , in physical literature cite: Choquet-Bruhat09FeliceClarke10MisnerThorneWheeler73cite: LandauLifshitz75 cite: Grant08Eddington24Finkelstein58Lemaitre33LoingerMarsico09 [27]-[35], many years oneconsiders the abnormal formal change of coordinates obtained by replacing the canonical Schwarzschild time by "retarded time" $v(t,r)$,i.e., Eddington–Finkelstein coordinates, given by

$$\begin{gathered} dv(t,r) = dt + [h(r)]^{-1}dr, \\ h(r) = 1 - \frac{r_g}{r}; \end{gathered} \tag{2.1.14}$$

(ii) the change (2.1.14) of Schwarzschild coordinates is singular at Schwarzschild horizon $r = r_g$,as at Schwarzschild horizon $h(r_g) = \infty$ and therefore the change (2.1.14) does not holds on Schwarzschild horizon cite: LoingerMarsico09[35];

(ii) under the singular change (2.1.14) Schwarzschild metric (2.1.11) becomes to well known regular and nondegenerate Eddington-Finkelstein metric cite: Choquet-Bruhat09FeliceClarke10MisnerThorneWheeler73cite: LandauLifshitz75 cite: Grant08Eddington24Finkelstein58Lemaitre33LoingerMarsico09[27]-[35]:

$$ds^2_{\mathbf{EF}} = -\left(1 - \frac{2m}{r}\right)dv^2 + 2drdv + r^2[(d\theta)^2 + \sin^2\theta(d\phi)^2]; \tag{2.1.15}$$

(iii) in physical literature many years exist abnormal belief that by formal singular change (2.1.15) the singular and degenerate Schwarzschild spacetime $(\mathbf{S}^2 \times \{r > 2m\}) \times \mathbb{R}$ was immersed in a larger Eddington-Finkelstein spacetime

$$\mathbf{EF}_{\geq} = (\mathbf{S}^2 \times \{r \geq 2m\} \cup \{0 < r \leq 2m\}) \times \mathbb{R}\ , g_{\mathbf{EF}_{\geq}}(r,\theta), \tag{2.1.16}$$

with regular and non degenerate metric tensor $g_{\mathbf{EF}_{\geq}}(r,\theta)$, and whose manifold is not covered by the canonical Schwarzschild coordinate with $r \leq 2m$,and therefore singularity and degeneracy on Schwarzschild horizon $r = r_g$ is only coordinate singularity and coordinate degeneracy;

(iv) from statement (iii) it was mistakenly assumed that there is no gravitational singularity at BH horizon.

We remind now canonical definitions.

**Definition 2.1.2**. Let $(M,g)$ and $(N,h)$ be semi-Riemannian manifolds. An isometric embedding is a smooth embedding $f : M \to N$ which preserves the metric in the sense that $g$ is equal to the pullback of $h$ by $f$, i.e. $g = f^{*}h$. Explicitly, for any two tangent vectors $\mathbf{v},\mathbf{w} \in T_x(M)$ we have

$$g(\mathbf{v},\mathbf{w}) = h(df(\mathbf{v}),df(\mathbf{w})). \qquad (2.1.17)$$

**Remark 2.1.9.**Notice that such isometric embedding is a mathematical definition only and does not meant the equivalence $(M,g) \equiv (f(M),h)$ in absolute sense. Thus it is not alwais apropriate as equivalence of the Lorentzian manifolds $(M,g)$ and $(N,h)$ corresponding to the physical frames $(M_{\mathbf{ph}},g_{\mathbf{ph}})$ and $(N_{\mathbf{ph}},h_{\mathbf{ph}})$.

**Definition 2.1.3.**cite: Grant08[31].In general,a Lorentzian manifold $\left(M^{'},h\right)$ is said to be an extension of a Lorentzian manifold $(M,g)$ if there exists an isometric embedding $i : M \hookrightarrow M^{'}$.

**Remark 2.1.10.**Notice that such extension is a mathematical definition only and therefore it is not alwais apropriate as extension of the Lorentzian manifolds $(M,g)$ and $(M^{'},h)$corresponding to the physical frames $(M_{\mathbf{ph}},g_{\mathbf{ph}})$ and $(M^{'}_{\mathbf{ph}},h_{\mathbf{ph}})$.

**Remark 2.1.11**. In order to obtain example for the statement mentioned and Remark 2.1.8 and Remark 2.1.9 we go to prove below that the geometry of Schwarzschild spacetime $\mathbf{Sch}_{>} \triangleq \{(\mathbf{S}^2 \times \{r > 2m\}) \times \mathbb{R}, g_{\mathbf{Sch}_{>}}\}$ above Schwarzschild horizon, essantially cardinally different in comparizon with the geometry of Eddington-Finkelstein spacetime $\mathbf{EF}_{>} \triangleq \{(\mathbf{S}^2 \times \{r > 2m\}) \times \mathbb{R}, g_{\mathbf{EF}_{>}}\}$ above Eddington-Finkelstein horizon.

We remind now canonical definitions.

Figure

Fig

Fig.2.1.2.Paralel displacement along a closed contour $\Gamma$ in a curved space.

**Definition 2.1.4.**Let $\Delta_{\Gamma}A_k$ be the change in a vector $A_i(\hat{x})$ after parallel displacement around closed contour $\Gamma$ located in BH spacetime as ploted in Fig.2.1.3. This change $\Delta_{\Gamma}A_k$ can clearly be written in the form $\oint\limits_{\Gamma} \delta A_k$. Substituting in place of $\delta A_k$ the canonical expression $\delta A_k = \Gamma^{i}_{kl}(\hat{x})A_k dx^l$ (see 31],Eq.(85.5)) one obtains

$$\Delta_{\Gamma}A_k = \oint\limits_{\Gamma} \delta A_k = \oint\limits_{\Gamma} \Gamma^{i}_{kl}(\hat{x})A_k dx^l \,. \qquad (2.1.18)$$

Figure

Fig

Fig.2.1.3.Paralel displacement along a closed
contour $\Gamma$ in BH spacetime.

**Definition 2.1.5.(I)** Let $\Sigma_{Sh}^{g}$ be Schwarzschild horizon, let $\Gamma_{\hat{x}}$ be a contour located in Schwarzschild spacetime as plotted in Fig.2.1.4 and such that (i) $\hat{x} \in \Gamma_{\hat{x}}$, (ii) $\Sigma_{\mathbf{Sch}}^{g} \cap \Gamma_{\hat{x}} = \hat{x}$, and let $\Gamma^{\hat{x}}$ be a curve $\Gamma^{\hat{x}} = \Gamma_{\hat{x}} \backslash \hat{x}$. Let $\Delta_{\Gamma^{\hat{x}}} A_k$ be the integral

$$\Delta_{\Gamma^{\hat{x}}} A_k = \oint\limits_{\Gamma_{\hat{x}} \backslash \hat{x}} \delta A_k = \oint\limits_{\Gamma_{\hat{x}} \backslash \hat{x}} \Gamma_{kl}^{i}(\hat{x}) A_k dx^l \,. \qquad (2.1.19)$$

**(II)** Let $\Sigma_{\mathbf{EF}}^{g}$ be Eddington-Finkelstein horizon, let $\Gamma_{\hat{x}}$ be a contour located in Eddington-Finkelstein spacetime as plotted in Fig.2.1.5 and such that (i) $\hat{x} \in \Gamma_{\hat{x}}$, (ii) $\Sigma_{\mathbf{EF}}^{g} \cap \Gamma_{\hat{x}} = \hat{x}$, and let $\Gamma^{\hat{x}}$ be a curve $\Gamma^{\hat{x}} = \Gamma_{\hat{x}} \backslash \hat{x}$. Let $\Delta_{\Gamma^{\hat{x}}} A_k$ be the integral

$$\Delta_{\Gamma^{\hat{x}}} A_k = \oint\limits_{\Gamma_{\hat{x}} \backslash \hat{x}} \delta A_k = \oint\limits_{\Gamma_{\hat{x}} \backslash \hat{x}} \Gamma_{kl}^{i}(\hat{x}) A_k dx^l \,. \qquad (2.1.20)$$

Figure

Fig

Figure

Fig

Fig.2.1.4.Paralel displacement $\Delta_{\Gamma^{\hat{x}}}A_k$ along a curve $\Gamma^{\hat{x}}$ in Schwarzschild spacetime such that $\Sigma^{g}_{\mathbf{Sch}} \cap \Gamma_{\hat{x}} = \hat{x},$ then alwais $\Delta_{\Gamma^{\hat{x}}}A_k = \infty.$

Fig.2.1.5.Paralel displacement along a curve $\Gamma^{\hat{x}}$ in Eddington-Finkelstein spacetime $\Sigma^{g}_{\mathbf{EF}} \cap \Gamma_{\hat{x}} = \hat{x},$ then alwais $\Delta_{\Gamma^{\hat{x}}}A_k < \infty.$

Figure

Fig

Fig.2.1.6.Paralel displacement along aclosed contour $\Gamma$ located in region of the classical semi-Riemannian geometry of the Schwarzschild spacetime such that $\Sigma^{g}_{\mathbf{Sh}} \cap \Gamma = \varnothing,$then alwais $\Delta_{\Gamma}A_k < \infty.$

**Remark 2.1.12**. (**I**) Note that the geometry of Schwarzschild spacetime $\mathbf{Sch}_{>}$

$$\mathbf{Sch}_{>} \triangleq \{(\mathbf{S}^2 \times \{r > 2m\}) \times \mathbb{R}, g_{\mathbf{Sch}_{>}}\} \qquad (2.1.21)$$

above Schwarzschild horizon $\Sigma^{g}_{\mathbf{Sch}}$, essantially cardinally different in comparizon with the geometry of Eddington-Finkelstein spacetime $\mathbf{EF}_{>}$

$$\mathbf{EF}_{>} \triangleq \{(\mathbf{S}^2 \times \{r > 2m\}) \times \mathbb{R}, g_{\mathbf{EF}_{>}}\} \qquad (2.1.22)$$

above Eddington-Finkelstein horizon $\Sigma^{g}_{\mathbf{EF}}$.

(**II**) Note that Schwarzschild spacetime $\mathbf{Sch}_{>}$ obviously satisfies a very strong nonregularity condition

$$\text{if } \Sigma^{g}_{\mathbf{Sch}} \cap \Gamma_{\hat{x}} = \hat{x}, \text{ then } \Delta_{\Gamma^{\hat{x}}}A_k = \infty. \qquad (2.1.23)$$

Thus the geometry of spacetime $\mathbf{Sch}_{>}$ that is nonclassical geometry beyond apparatus of the classical semi-Riemannian geometry. Ofcourse the geometry any part of spacetime $\mathbf{Sch}_{>}$ located above some neighborhood of Schwarzschild horizon as plotted in Fig.2.1.6 that is a classical semi-Riemannian geometry.

**Remark 2.1.13.**Note that from Remark 2.1.11 it follows that Eddington-Finkelstein spacetime does not holds in regorous mathematical sense as extension of the Schwarzschild spacetime $\mathbf{Sch}_{>} \triangleq \{(\mathbf{S}^2 \times \{r > 2m\}) \times \mathbb{R}, g_{\mathbf{Sch}_{>}}\}$ above Schwarzschild horizon.

**Remark 2.1.14.**It is clear that nonregularity condition (2.1.23) arises not only from singularity of the function $h^{-1}(r)$ at point $r = r_g$ but from degeneracy of the function $h(r)$at point $r = r_g$.

**Remark 2.1.15**. We remind now that the relations (see cite: LandauLifshitz75[30] p.234, Eq.(84.7))

$$\gamma_{\alpha\beta} = -g_{\alpha\beta} + \frac{g_{0\alpha}g_{0\beta}}{g_{00}} \tag{2.1.24}$$

give the connection between the metric of real space

$$dl^2 = \gamma_{\alpha\beta}dx^\alpha dx^\beta \tag{2.1.25}$$

and the metric of the four-dimensional space-time

$$ds^2 = g_{\alpha\beta}dx^\alpha dx^\beta + 2g_{0\alpha}dx^0 dx^\alpha + g_{00}(dx^0)^2. \tag{2.1.26}$$

For Eddington-Finkelstein metric (2.1.15) metric of the corresponding real space is

$$dl^2_{\mathbf{EF}} = \frac{dr^2}{1 - \frac{2m}{r}} + r^2[(d\theta)^2 + \sin^2\theta(d\phi)^2]. \tag{2.1.27}$$

**Remark 2.1.16**. Notice that the Eddington-Finkelstein metric (2.1.15) is regular at the horizon and therefore the infalling observer encounters nothing unusual at the horizon. However from Eq.(2.1.17) it follows that the infalling observer encounters singularity on horizon. But this is a contradiction.

**Remark 2.1.17.**Note that in order dealing with singular Schwarzschild metric (2.1.11) using mathematically and logically soundness approach, one applies contemporary distributional geometry based on Colombeau generalized functions cite: Kupeli96cite: Colombeau84Parker79[2]-[4].Distributional Schwarzschild geometry and distributional BHs geometry by using Colombeau generalized functions cite: Kupeli96cite: Colombeau84Parker79[2]-[4] many developed in papers cite: Parker79VickersWilson98VickersWilson99Vickers99GerochTraschen87 cite: BalasinNachbagauer93BalasinNachbagauer94KawaiSakane97PantojaRago97 cite: PantojaRago00KunzingerSteinbauer02KunzingerSteinbauer01 cite: GrosserFarkasKunzingerSteinbauer01HeinzleSteinbauer02Foukzon15 cite: FoukzonPotapovMenkova16Vickers12Steinbauer00 cite: GolubevKelner05 [4]-[22]. By aproporiate regularization $g_{ij,\varepsilon}(r,\theta,\phi), \varepsilon \in (0,1]$ of the singular Schwarzschild metric$g_{ij}(r,\theta,\phi)$ such that:

(i) $g_{ij,0}(r,\theta,\phi) = g_{ij}(r,\theta,\phi)$ and

(ii) for any $\varepsilon \in (0,1]$ metric tensor $g_{ij,\varepsilon}(r,\theta,\phi)$ is regular and nondegenerate,one obtains Colombeau generalized object $[(g_{ij,\varepsilon}(r,\theta,\phi))_\varepsilon] \in \mathbf{G}(\mathbb{R}^3)$ with an representative $(g_{ij,\varepsilon}(r,\theta,\phi))_\varepsilon$,for a more detailed explanation see cite: KawaiSakane97 cite: Foukzon15 cite: FoukzonPotapovMenkova16[11],[18],[19]. Using rigorous Colombeau approach one obtains mathematically and logically soundness notion of singularity in Distributional Schwarzschild spacetime.

**Remark 2.1.18**. Note that in the case of Schwarzschild spacetime the conditions (i) and (ii) mentioned above (see Remark 2.1.13) are satisfied only by using non smooth regularization of the singular and degenerate Schwarzschild metric $g_{ij}(r,\theta,\phi)$ via Schwarzschild horizon cite: Foukzon15 cite: FoukzonPotapovMenkova16[18]-[19].

By apriporiate nonsmooth regularization one obtain Colombeau generalized object modeling the singular Schwarzschild metric above and below horizon cite: Foukzon15 cite: FoukzonPotapovMenkova16[18]-[19]:

$$
\begin{aligned}
(ds_\varepsilon^{+2})_\varepsilon &= -(h_\varepsilon^+(r)dt^2)_\varepsilon + \left([h_\varepsilon^+(r)]^{-1}dr^2\right)_\varepsilon + r^2 d\Omega^2,\\
(ds_\varepsilon^{-2})_\varepsilon &= (h_\varepsilon^-(r)dt^2)_\varepsilon - \left([h_\varepsilon^-(r)]^{-1}dr^2\right)_\varepsilon + r^2 d\Omega^2,\\
h_\varepsilon^+(r) &= \frac{\Theta_\varepsilon((r-r_s)-\varepsilon)\sqrt{(r-r_g)^2+\varepsilon^2}}{r}, r \geq r_g,\\
&\varepsilon \in (0,1].
\end{aligned}
\tag{2.1.24}
$$

**Remark 2.1.19**. Let us rewrite now the metric (2.1.24) (above horizon) in the form

$$
\begin{aligned}
(ds_\varepsilon^{+2})_\varepsilon &= -(h_\varepsilon^+(r)dt^2)_\varepsilon + \left([h_\varepsilon^+(r)]^{-1}dr^2\right)_\varepsilon + r^2(d\theta^2+\sin^2\theta d\varphi^2) =\\
-(h_\varepsilon^+(r))_\varepsilon&\left(\left[dt-[h_\varepsilon^+(r)]^{-1}dr\right]\left[dt+[h_\varepsilon^+(r)]^{-1}dr\right]\right)_\varepsilon + r^2(d\theta^2+\sin^2\theta d\varphi^2),
\end{aligned}
\tag{2.1.25}
$$

and define a new generalized Colombeau coordinates $((\tau_\varepsilon)_\varepsilon, \bar{r}, \theta, \varphi)$, where $(\tau_\varepsilon(t,r))_\varepsilon \in \mathbf{G}(\mathbb{R}^2)$, by formula

$$
\begin{aligned}
(d\tau_\varepsilon(t,r))_\varepsilon &= dt + \left([h_\varepsilon^+(r)]^{-1}dr\right)_\varepsilon,\\
\bar{r} &= r.
\end{aligned}
\tag{2.1.26}
$$

**Remark 2.1.20**. Notice that:

(i) Colombeau generalized coordinates (2.1.26) are the Colombeau extension of the canonical Eddington-Finkelstein coordinates (2.1.14) by Colombeau generalized function.

(ii) In contrast with canonical Eddington-Finkelstein coordinates (2.1.14) (see Remark 2.1.7), Colombeau generalized coordinates (2.1.26) holds at Schwarzschild horizon $r = r_g$as at Schwarzschild horizon Colombeau generalized function $\left([h_\varepsilon^+(r)]^{-1}\right)_\varepsilon$ become well defined Colombeau generalized number $[h_\varepsilon^+(r_g)]^{-1} \in \widetilde{\mathbb{R}}$.

Rewriting now the metric (2.1.25) in terms of the Colombeau generalized coordinates $((\tau_\varepsilon)_\varepsilon, \bar{r}, \theta, \varphi)$, it then above horizon takes the form

$$
\begin{aligned}
&(ds_\varepsilon^{+2})_\varepsilon =\\
-((h_\varepsilon^+(\bar{r}))_\varepsilon)&\left(\left[d\tau_\varepsilon - 2[h_\varepsilon^+(\bar{r})]^{-1}d\bar{r}\right]d\tau_\varepsilon\right)_\varepsilon + \bar{r}^2(d\theta^2+\sin^2\theta d\varphi^2) =\\
-((h_\varepsilon^+(\bar{r}))_\varepsilon)&(d\tau_\varepsilon^2)_\varepsilon + 2d\bar{r}(d\tau_\varepsilon)_\varepsilon + \bar{r}^2(d\theta^2+\sin^2\theta d\varphi^2).
\end{aligned}
\tag{2.1.27}
$$

We rewrite now Colombeau metric (2.1.27) in the equivalent form

$$
\begin{aligned}
&(ds_\varepsilon^{+2})_\varepsilon =\\
-(h_\varepsilon^+(r)d\tau^2)_\varepsilon &+ 2drd\tau + r^2(d\theta^2+\sin^2\theta d\varphi^2).
\end{aligned}
\tag{2.1.28}
$$

Colombeau metric (2.1.28) define the distributional Eddington-Finkelstein space-time

$$
\mathbf{EF}_{\geq}^{+} \triangleq \left\{\left(\widetilde{\mathbf{S}}^2 \times \{\bar{r} \geq 2m\}\right) \times \widetilde{\mathbb{R}}, g_{\mathbf{EF}_{\geq}}^{+}\right\}
\tag{2.1.29}
$$

above the Eddington-Finkelstein horizon $r = 2m$.

**Remark 2.1.21**. Notice that

$$
\begin{aligned}
(h_\varepsilon^+(r))_\varepsilon|_{r=r_g} &= r_g^{-1}(\varepsilon)_\varepsilon, \left([h_\varepsilon^+(r)]^{-1}\right)_\varepsilon\Big|_{r=r_g} = r_g \cdot (\varepsilon^{-1})_\varepsilon \in \widetilde{\mathbb{R}},\\
(d\tau_\varepsilon)_\varepsilon|_{r=r_g} &= dt + ((\varepsilon^{-1})_\varepsilon) \cdot r_g dr,\\
(d\tau_\varepsilon^2)_\varepsilon\Big|_{r=r_g} &= dt^2 + 2((\varepsilon^{-1})_\varepsilon) \cdot r_g dt dr + ((\varepsilon^{-2})_\varepsilon) r_g^2 dr^2,\\
&\varepsilon \in (0,1].
\end{aligned}
\tag{2.1.30}
$$

Of course at horizon $(h_\varepsilon^+(t,r_g))_\varepsilon \approx 0$, because at horizon $h_0^+(t,r_g) = 0,$however it follows from (2.1.24) at horizon the quantities $((h_\varepsilon^+(r_g))_\varepsilon)(d\tau_\varepsilon^2(t,r_g))_\varepsilon \approx ((\varepsilon^{-1})_\varepsilon) r_g dr^2$ and $(d\tau_\varepsilon)_\varepsilon \approx$ $((\varepsilon^{-1})_\varepsilon)r_g dr$ are infinite large Colombeau quantities,i.e.,the differential $(d\tau_\varepsilon)_\varepsilon$ is not classical but it is Colombeau differential.

**Remark 2.1.22**. Note that:

(i) under coordinate change (2.1.26) distributional curvature scalars of the distributional Schwarzschild space-time given by metric (2.1.24), does not changes because these scalars depend only on variable $r = \bar{r}$,

(ii) in contrast with classical Eddington-Finkelstein space-time

$$\mathbf{EF}_{\geq}=(\mathbf{S}^2 \times \{r \geq 2m\} \cup \{0 < r \leq 2m\}) \times \mathbb{R}\ , g_{\mathbf{EF}_{\geq}}(r,\theta),$$

distributional Eddington-Finkelstein space-time has a gravitational singularity at horizon.

**Remark 2.1.23**. Note that for the case of the distributional space-time the relations (2.1.24) obviously takes the form

$$(\gamma_{\alpha\beta}(\varepsilon))_\varepsilon = \left(-g_{\alpha\beta}(\varepsilon) + \frac{g_{0\alpha}(\varepsilon)g_{0\beta}(\varepsilon)}{g_{00}(\varepsilon)}\right)_\varepsilon \qquad (2.1.31)$$

where (2.1.30) give the connection between the Colombeau metric of the distributional real space

$$(dl^2_\varepsilon)_\varepsilon = ((\gamma_{\alpha\beta}(\varepsilon)dx^\alpha dx^\beta))_\varepsilon \qquad (2.1.32)$$

and the Colombeau metric of the four-dimensional distributional space-time

$$\begin{gathered}(ds^2_\varepsilon)_\varepsilon = \\ (g_{\alpha\beta}(\varepsilon)dx^\alpha dx^\beta)_\varepsilon + 2(g_{0\alpha}(\varepsilon)dx^0 dx^\alpha)_\varepsilon + \left(g_{00}(\varepsilon)(dx^0)^2\right)_\varepsilon.\end{gathered} \qquad (2.1.33)$$

For distributional Eddington-Finkelstein metric (2.1.25) above horizon of the corresponding Colombeau metric of the distributional real space is

$$\left(dl^{+2}_{\varepsilon,\mathbf{EF}^+_{\geq}}\right)_\varepsilon = \left((h^+_\varepsilon(r))^{-1}dr^2\right)_\varepsilon + r^2[(d\theta)^2 + \sin^2\theta(d\phi)^2]. \qquad (2.1.34)$$

**Remark 2.1.24.**Notice since the distributional Eddington-Finkelstein space-time (2.1.29) has a gravitational singularity (see Definition 1.1.1) at horizon,there is no contradiction mentioned above for the case of the regular classical Eddington-Finkelstein metric (2.1.15) and corresponding singular metric (2.1.17), see Remark 2.1.15.

### 5mm . 2.1.2.Distributional Kruskal-Szekeres space-time 2mm

Recall that the classical Kruskal–Szekeres coordinates are defined, from the classical Schwarzschild coordinates $(t,r,\theta,\phi)$, by replacing $t$ and $r$ by a new time coordinate $T$ and a new spatial coordinate $X$ :

$$\begin{gathered} r > r_g : \\ T = \left(\frac{r}{r_g} - 1\right)^{1/2} \exp\left(\frac{r}{r_g}\right)\sinh\left(\frac{t}{2r_g}\right), \\ X = \left(\frac{r}{r_g} - 1\right)^{1/2} \exp\left(\frac{r}{r_g}\right)\cosh\left(\frac{t}{2r_g}\right), \\ 0 < r < r_g : \\ T = \left(1 - \frac{r}{r_g}\right)^{1/2} \exp\left(\frac{r}{r_g}\right)\sinh\left(\frac{t}{2r_g}\right), \\ X = \left(1 - \frac{r}{r_g}\right)^{1/2} \exp\left(\frac{r}{r_g}\right)\cosh\left(\frac{t}{2r_g}\right). \end{gathered} \qquad (2.1.35)$$

It follows that the Schwarzschild radius $r$, in terms of Kruskal–Szekeres coordinates, is implicitly given by

$$T^2 - X^2 = \left(1 - \frac{r}{r_g}\right)^{1/2} \exp\left(\frac{r}{r_g}\right) \tag{2.1.36}$$

for both interior and exterior regions,i.e. $r \in \mathbb{R}_+ \backslash \{0\}$ .In these new coordinates the metric of the Schwarzschild black hole manifold formally is given by

$$ds^2 = \frac{4r_g^3}{r} \exp\left(-\frac{r}{r_g}\right)[-dT^2 + dX^2] + r^2 d\Omega^2. \tag{2.1.37}$$

The location of the event horizon ($r_g = 2GM$) in these coordinates obviously is given by

$$T^2 - X^2 = 0 \Rightarrow \quad T = \pm X. \tag{2.1.38}$$

**Remark 2.1.25.**Note that the metric (2.1.37) of course is perfectly well defined and non-singular at the event horizon.The curvature singularity is located at $T^2 - X^2 = 1$. Kruskal-Szekeres spacetime is regular Lorentzian spacetime, except singular submanifold $\{(T,X)|T^2 - X^2 = 1\}$.

**Remark 2.1.26.**In contrast with Eddington-Finkelstein coordinates the classical Kruskal–Szekeres coordinates holds at Schwarzschild horizon, but however the differentials $dT, dX$ of the functions $T(r,t), X(r,t)$ are singular at Schwarzschild horizon $r = r_g$ and therfore Kruskal-Szegeres spacetime cannot be considered as Schwarzschild spacetime in Kruskal–Szekeres coordinates (2.1.35)-(2.1.36).

**Remark 2.1.27.**In order to avoid these difficultness one can apply instead the Kruskal–Szekeres coordinates (2.1.35)-(2.1.36) the following distributional Kruskal–Szekeres coordinates to Colombeau generalized metric (2.1.8)

$$\begin{gathered}
r \geq r_g, \varepsilon \in (0,1] : \\
(T_\varepsilon)_\varepsilon = [(\Theta_\varepsilon((r - r_g) - \varepsilon))_\varepsilon] \\
\left[\left(\left(\frac{r}{r_g} - 1 + \varepsilon\right)^{1/2}\right)_\varepsilon\right] \exp\left(\frac{r}{r_g}\right) \sinh\left(\frac{t}{2r_g}\right), \\
(X_\varepsilon)_\varepsilon = [(\Theta_\varepsilon((r - r_g) - \varepsilon))_\varepsilon] \\
\left[\left(\left(\frac{r}{r_g} - 1 + \varepsilon\right)^{1/2}\right)_\varepsilon\right] \exp\left(\frac{r}{r_g}\right) \cosh\left(\frac{t}{2r_g}\right), \\
0 < r < r_g : \\
(T_\varepsilon)_\varepsilon = [(\Theta_\varepsilon((r - r_g) - \varepsilon))_\varepsilon] \\
\left[\left(\left(1 - \frac{r}{r_g} + \varepsilon\right)^{1/2}\right)_\varepsilon\right] \exp\left(\frac{r}{r_g}\right) \sinh\left(\frac{t}{2r_g}\right), \\
(X_\varepsilon)_\varepsilon = [(\Theta_\varepsilon((r - r_g) - \varepsilon))_\varepsilon] \\
\left[\left(\left(1 - \frac{r}{r_g} + \varepsilon\right)^{1/2}\right)_\varepsilon\right] \exp\left(\frac{r}{r_g}\right) \cosh\left(\frac{t}{2r_g}\right).
\end{gathered} \tag{2.1.39}$$

Therefore for both interior and exterior regions we get

$$(T_\varepsilon^2)_\varepsilon - (X_\varepsilon^2)_\varepsilon = \left[\left(\left(1 - \frac{r}{r_g} + \varepsilon\right)^{1/2}\right)_\varepsilon\right] \exp\left(\frac{r}{r_g}\right). \tag{2.1.40}$$

**Remark 2.1.29**. Note that in contrast with (2.1.37) at horizon $r = r_g$ :

$$(T_\varepsilon^2)_\varepsilon - (X_\varepsilon^2)_\varepsilon = e\left[(\varepsilon^{1/2})_\varepsilon\right] \neq 0_{\widetilde{\mathbb{R}}}. \tag{2.1.41}$$

In these new distributional coordinates the Colombeau metric (2.1.8) of the distributional Schwarzschild black hole manifold above horizon is given by formula

$$(ds_\varepsilon^2)_\varepsilon = \left(\frac{4r_g^3 (\Theta_\varepsilon((r - r_g) - \varepsilon))_\varepsilon}{r} \exp\left(-\frac{r}{r_g}\right)[-dT_\varepsilon^2 + dX_\varepsilon^2]\right)_\varepsilon + r^2 d\Omega^2. \tag{2.1.42}$$

Here $(\Theta_\varepsilon(u))_\varepsilon$ is the generalized Heaviside function given by Eq.(2.1.3).

**Remark 2.1.30**. Note that in contrast with (2.1.36) Colombeau generalized metric (2.1.39) non degerate at horizon $r = r_g$ in Colombeau sense.

## 5mm . 2.2.Distributional Schwarzschild space-time and distributional Rindler space-time with distributional Levi-Cività connection. Generalized Einstein equivalence principle 2mm

### 5mm . 2.2.1.Distributional Schwarzschild space-time with distributional Levi-Cività connection 2mm

**Remark 2.2.1**.Note that due to the degeneracy of the metric (2.1.11) at Schwarzschild horizon, the classical Levi-Civit‘a connection on whole Schwarzschild spacetime is not available cite: Foukzon15 cite: FoukzonPotapovMenkova16[18],[19] as classical Levi-Civit‘a connection on Schwarzschild horizon becomes infinity

$$\Gamma^1_{11}(r)|_{r=2m} = \lim_{r\to 2m} \frac{-m}{r(r-2m)} = -\infty, \Gamma^0_{01}(r)|_{r=2m} = \lim_{r\to 2m} \frac{m}{r(r-2m)} = \infty, \qquad (2.2.1)$$

**Remark 2.2.2**. In order to avoid difficultness with classical Levi-Civit‘a connection mentioned above in Remark 2.2.1, in papers cite: Foukzon15 cite: FoukzonPotapovMenkova16[18],[19] we have applied the non smooth regularization via Schwarzschild horizon, see Remark 2.1.5 and Eq.(2.1.6).Corresponding Colombeau distributional connections $(\Gamma^{+l}_{kj}(\varepsilon))_\varepsilon$ and $(\Gamma^{-l}_{kj}(\varepsilon))$ above and below Schwarzschild horizon are cite: Foukzon15 cite: FoukzonPotapovMenkova16[18]-[19]:

$$\begin{aligned}(\Gamma^{+l}_{kj}(\varepsilon))_\varepsilon &= \frac{1}{2}((g_\varepsilon^{+lm})[(g^+_\varepsilon)_{mk,j} + (g^+_\varepsilon)_{mj,k} - (g^+_\varepsilon)_{kj,m}])_\varepsilon,\\ (\Gamma^{-l}_{kj}(\varepsilon))_\varepsilon &= \frac{1}{2}((g_\varepsilon^{-lm})[(g^-_\varepsilon)_{mk,j} + (g^-_\varepsilon)_{mj,k} - (g^-_\varepsilon)_{kj,m}])_\varepsilon.\end{aligned} \qquad (2.2.2)$$

Obviously distributional connections $(\Gamma^{+l}_{kj}[h^+_\varepsilon])_\varepsilon, (\Gamma^{-l}_{kj}[h^+_\varepsilon])_\varepsilon$ coincides, in distributional sense, with the corresponding classical Levi-Cività connections on $\mathbb{R}^3\backslash\{r = 2m\}$, since $(h^+_\varepsilon)_\varepsilon = h^+_0, (h^-_\varepsilon)_\varepsilon = h^-_0$, and $(g_\varepsilon^{+lm})_\varepsilon = g_0^{+lm}$, $(g_\varepsilon^{-lm})_\varepsilon = g_0^{-lm}$ there. Clearly, connections $\Gamma^{+l}_{kj}(\epsilon), \Gamma^{-l}_{kj}(\epsilon), \epsilon \in (0,1]$ in respect the regularized metric $g^\pm_\epsilon, \epsilon \in (0,1]$, i.e., $(g^\pm_\epsilon)_{ij;k} = 0$. Proceeding in this manner, we obtain the nonstandard result cite: GolubevKelner05Moller43[22]-[23] see also Apendix B:

$$\begin{aligned}\left[\left([R^+_\epsilon]^1_1\right)_\epsilon\right] &= \left[\left([R^+_\epsilon]^0_0\right)_\varepsilon\right] = -4\pi m \frac{\delta(r-2m)}{r^2},\\ \left[\left([R^-_\epsilon]^1_1\right)_\epsilon\right] &= \left[\left([R^-_\epsilon]^0_0\right)_\varepsilon\right] = 4\pi m \frac{\delta(r-2m)}{r^2}.\end{aligned} \qquad (2.2.3)$$

**Remark 2.2.3**. As axpected, the distributional Ricci tensor as well as the distributional Ricci scalar vanish identically on $\mathbb{R}^3\backslash\{r = 2m\}$, since $\operatorname{supp}(\delta(r-2m)) = \{r = 2m\}$. This result in a good agrement with canonical result cite: MullerGrave10 cite: Reall12Hooft98Choquet-Bruhat09FeliceClarke10MisnerThorneWheeler73 cite: LandauLifshitz75 [24]-[30] on $\mathbb{R}^3\backslash\{r = 2m\}$ since distributional connections (2.2.2) coincides with the corresponding classical Levi-Cività connections on $\mathbb{R}^3\backslash\{r = 2m\}$ at least in distributional sense.We obtain for $r \asymp 2m$ the nonstandard result cite: Foukzon15 cite: FoukzonPotapovMenkova16[18]-[19]:

$$
\begin{aligned}
&(\mathbf{R}^{\pm\mu\nu}(\varepsilon)\mathbf{R}^{\pm}_{\mu\nu}(\varepsilon))_{\varepsilon} \asymp \left(\frac{c_1}{16m^2\left[(r-2m)^2+\varepsilon^2\right]}\right)_{\varepsilon} + \ldots, \\
&(\mathbf{R}^{\pm\rho\sigma\mu\nu}(\varepsilon)\mathbf{R}^{\pm}_{\rho\sigma\mu\nu}(\varepsilon))_{\varepsilon} \asymp \left(\frac{c_2}{16m^2\left[(r-2m)^2+\varepsilon^2\right]}\right)_{\varepsilon} + \ldots,
\end{aligned}
\tag{2.2.4}
$$

where $c_1, c_2 = O(1)$.

## 5mm . **2.2.2.Distributional Rindler space-time with distributional Levi-Cività connection** 2mm

We remind now that $2D$ Rindler spacetime is a patch of Minkowski spacetime. In $2D,$ the Rindler metric is

$$ds^2 = dR^2 - R^2 d\eta^2. \tag{2.2.5}$$

Figure

Fig

Fig.2.2.1.Hyperbolic motion in the right Rindler wedge.

$$x^2 - c^2t^2 = (c^2/a)^2.$$

**Remark 2.2.4.**Due to the degeneracy of the metric (2.2.5) at Rindler gorizon $R = 0,$ the classical Levi-Cività connection is not available on whole $\mathbb{R}^2$,e.g.,

$$\Gamma^1_{44} = R, \Gamma^4_{14} = \Gamma^4_{41} = R^{-1}, \tag{2.2.6}$$

and all other components being zero.

**Remark 2.2.5.**Note that in order to avoid this difficultnes, the origin in classical consideration is always excluded from the space $\mathbb{R}^{3.1}$ and we are working on $\mathbb{R}^{3.1}\backslash\{0\}$ $\mathbb{R}^{3,1}\mathbb{R}^4\backslash\{R = 0\}$, and therefore for Einstein's tensor

$$\mathbf{G}^k_i \triangleq \mathbf{R}^k_i - \tfrac{1}{2}\delta^k_i\mathbf{R}, \mathbf{R} \triangleq \mathbf{R}^i_i \tag{2.2.7}$$

following Møller [24] we get

$$\mathbf{G}^2_2 = \mathbf{G}^3_3 = -\frac{1}{2g_{44}}\left[g''_{44} - \frac{(g'_{44})^2}{2g_{44}}\right] = -\frac{1}{2R^2}\left[2 - \frac{(2R)^2}{2R^2}\right] \equiv 0, \tag{2.2.8}$$

where the accents indicate differentiation with respect variable $R,$ and all other components of $\mathbf{G}^k_i$ vanish identically. Thus Rindler metrical tensor satisfy on $\mathbb{R}^{3.1}\backslash\{0\}$ the Einstein field $\mathbb{R}^{3,1}\backslash\{R$field equations

$$\mathbf{G}_i^k \triangleq \mathbf{R}_i^k - \tfrac{1}{2}\delta_i^k\mathbf{R} = 0. \quad (2.2.9)$$

**Remark 2.2.6**. By calculations mentioned above, from Møller's times until nowdays,Rindler metrical tensor was mistakenly considered in physical literature as an vacuum solution of the Einstein's field equations,e.g.,solution for empty space,see Møller cite: Moller43 [23].

**Remark 2.2.7**.Note that Levi-Cività connection on whole space $\mathbb{R}^{3,1}$ is available only in Colombeau sense under smooth regularization $R^2 \to R^2 + \varepsilon^2, \varepsilon \in (0,1]$ and therefore we forced to change metric (2.5) by Colombeau object

$$\begin{gathered} [(ds_\varepsilon^2)_\varepsilon] = dR^2 - [(R^2+\varepsilon^2)_\varepsilon]dt^2 = dR^2 - [(g_{44,\varepsilon})_\varepsilon]dt^2 \\ [(g_{44,\varepsilon})_\varepsilon] = [(R^2+\varepsilon^2)_\varepsilon], \varepsilon \in (0,1]. \end{gathered} \quad (2.2.10)$$

Then for Einstein distributional tensor cite: Foukzon15
cite: FoukzonPotapovMenkova16Vickers12 [18]-[19],[20]:

$$(\mathbf{G}_{i,\varepsilon}^k)_\varepsilon \triangleq (\mathbf{R}_{i,\varepsilon}^k)_\varepsilon - \tfrac{1}{2}\delta_i^k(\mathbf{R}_\varepsilon)_\varepsilon, (\mathbf{R}_\varepsilon)_\varepsilon \triangleq \mathbf{R}_{i,\varepsilon}^i \quad (2.2.11)$$

we get

$$\begin{gathered} (\mathbf{G}_{2,\varepsilon}^2(R))_\varepsilon = (\mathbf{G}_{3,\varepsilon}^3(R))_\varepsilon = \\ -\left(\frac{1}{2g_{44,\varepsilon}}\left[g''_{44,\varepsilon} - \frac{(g'_{44,\varepsilon})^2}{2g_{44,\varepsilon}}\right]\right)_\varepsilon = -\left(\frac{\varepsilon^2}{(R^2+\varepsilon^2)^2}\right)_\varepsilon. \end{gathered} \quad (2.2.12)$$

Thus,

$$\left[(\mathbf{G}_{2,\varepsilon}^2(0))_\varepsilon\right] = \left[(\mathbf{G}_{3,\varepsilon}^3(0))_\varepsilon\right] = [(\varepsilon^{-2})_\varepsilon], \quad (2.2.13)$$

where $[(\varepsilon^{-2})_\varepsilon] \in \widetilde{\mathbb{R}}$ is infinite Colombeau generalized numbers, and therefore $(\mathbf{G}_{2,\varepsilon}^2(R))_\varepsilon$ and $(\mathbf{G}_{3,\varepsilon}^3(R))_\varepsilon$ is nontrivial Colombeau generalized functions and distributional Rindler metric tensor given by (2.2.12) that is non vacuum Colombeau solution of the Einstein field equations.

### 5mm . 2.2.3.Generalized Einstein equivalence principle 2mm

We remind that originally Einstein's gravity was formulated by using classical pseudo Riemannian geometry with classical Levi-Civit'a connection.In classical pseudo Riemannian geometry, the Levi-Civita connection is a specific connection on the tangent bundle of a manifold. More specifically, it is the torsion-free metric connection, i.e., the torsion-free connection on the tangent bundle (an affine connection) preserving a given (pseudo-Riemannian) Riemannian metric.The fundamental theorem of classical Riemannian geometry states that there is a unique connection which satisfies these properties.

**Remark 2.3.1**.Note that classical Einstein "Equivalence Principle" asserts the equivalence between inertial and gravitational forces of acceleration. The classical Einstein equivalence principle is the heart and soul of gravitational theory, for it is possible to argue convincingly that if EEP is valid, then gravitation must be a "curved spacetime" phenomenon, in other words, gravity must be governed by a "metric theory of gravity", whose postulates are:

1. Spacetime is endowed with a symmetric Lorentzian metric.
2. The trajectories of freely falling test bodies are geodesics of that metric.
3. In local freely falling reference frames, the non-gravitational laws of physics are those written in the language of special relativity.

In order to obtain appropriate generalization of EEP based on distributional Colombeau geometry cite: Parker79VickersWilson98VickersWilson99Vickers99[4]-[7] we claim the following generalized equivalence principle (GEEP):

1. Spacetime in general case is endowed with a symmetric distributional Lorentzian metric.
2. The trajectories of freely falling test bodies are geodesics of that distributional metric.
3. In local freely falling distributional reference frames, the non-gravitational laws of physics are those written in the language of special relativity.

## 6mm . 3.Quantum scalar field in curved distributional spacetime. Unruh effect revisited 3mm

### 5mm . 3.1.Canonical quantization in curved distributional space-time 2mm

In a recent work cite: FoukzonPotapovMenkova16Vickers12 [19] the authors advocated the use De Witt-Schwinger approach cite: FrolovNovikov98DeWitt75DeWitt57BirrellDavies84 [37]-[40] in order to establish QFT in general ditributional curved spacetime. The vacuum energy density of free scalar quantum field $\Phi$ with a distributional background spacetime is considered successfully. It has been widely believed that, except in very extreme situations, the influence of gravity on quantum fields should amount to just small, sub-dominant contributions. Here we argue that this belief is false by showing that there exist well-behaved spacetime evolutions where the vacuum energy density of free quantum fields is forced, by the very same background distributional spacetime such BHs, to become dominant over any classical energydensity component. This semiclassical gravity effect finds its roots in the singular behavior of quantum fields on curved distributional spacetimes. In particular we obtain that the vacuum fluctuations $\langle\Phi^2\rangle$ has a singular behavior on BHs horizon $r_+$ : $\langle\Phi^2(r)\rangle \sim |r-r_+|^{-2}$.

Much of formalism can be explained with Colombeau generalized scalar field cite: FoukzonPotapovMenkova16Vickers12 [19].The basic concepts and methods extend straightforwardly to distributional tensor and distributional spinor fields. To being with let's take a spacetime of arbitrary dimension $D$, with a metric $g_{\mu\nu}$ of signature$(+-\ldots-)$. The action for the Colombeau generalized scalar field $(\varphi_\varepsilon)_\varepsilon \in \mathbf{G}(M)$ is

$$(S_\varepsilon)_\varepsilon = \left(\int_M d^D x \frac{1}{2}\sqrt{|g_\varepsilon|}\,(g_\varepsilon^{\mu\nu}\partial_\mu\varphi_\varepsilon\partial_\nu\varphi_\varepsilon) - (m^2+\xi R_\varepsilon)\varphi_\varepsilon^2\right)_\varepsilon. \quad (3.1.1)$$

Here $\xi$ is a coupling constant (see cite: BirrellDavies84 [40] chapter 3). The corresponding equation of motion is

$$([\Box_{\varepsilon,x} + m^2 + \xi R_\varepsilon]\varphi_\varepsilon)_\varepsilon, \varepsilon \in (0,1]. \quad (3.1.2)$$

Here

$$(\Box_{\varepsilon,x}\varphi_\varepsilon)_\varepsilon = \left(|g_\varepsilon|^{-1/2}\partial_\mu|g_\varepsilon|^{1/2}g_\varepsilon^{\mu\nu}\partial_\mu\varphi_\varepsilon\right)_\varepsilon. \quad (3.1.3)$$

With $\hbar$ explicit, the mass $m$ should be replaced by $m/\hbar$. Separating out a time coordinate $x^0$, $x^\mu = (x^0, x^i), i = 1,2,3$ we can write the action as

$$(S_\varepsilon)_\varepsilon = \left(\int dx^0 L_\varepsilon\right)_\varepsilon, (L_\varepsilon)_\varepsilon = \left(\int d^{D-1}x\mathcal{L}_\varepsilon\right)_\varepsilon. \quad (3.1.4)$$

The canonical momentum at a time $x^0$ is given by

$$(\pi_\varepsilon(\underline{x}))_\varepsilon = (\delta L_\varepsilon/\delta(\partial_0\varphi_\varepsilon(\underline{x})))_\varepsilon = \left(|h_\varepsilon|^{1/2}n^\mu\partial_\mu\varphi_\varepsilon(\underline{x})\right)_\varepsilon, \quad (3.1.5)$$

where $\underline{x}$ labels a point on a surface of constant $x^0$, the $x^0$ argument of $(\varphi_\varepsilon)_\varepsilon$ is suppressed, $n^\mu$ is the unit normal to the surface, and $(|h_\varepsilon|)_\varepsilon$ is the determinant of the induced spatial metric $(h_{ij}(\varepsilon))_\varepsilon$. In order to quantize, the Colombeau generalized field $(\varphi_\varepsilon)_\varepsilon$ and its conjugate momentum $(\pi_\varepsilon(\underline{x}))_\varepsilon$ are now promoted to hermitian operators and required to satisfy the canonical commutation relation,

$$\left([\varphi_\varepsilon(\underline{x}), \pi_\varepsilon(\underline{y})]\right)_\varepsilon = i\hbar\delta^{D-1}(\underline{x},\underline{y}), \varepsilon \in (0,1]. \quad (3.1.6)$$

Here $\int d^{D-1}y\,\delta^{D-1}(\underline{x},\underline{y})f(\underline{y}) = f(\underline{x})$ for any scalar function $f \in D(\mathbb{R}^3)$, without the use of a metric volume element. We form now a conserved bracket from two complex Colombeau solutions to the scalar wave equation (3.1.2) by cite: FoukzonPotapovMenkova16Vickers12 [19]:

$$(\langle\varphi_\varepsilon,\phi_\varepsilon\rangle)_\varepsilon = \left(\int_\Sigma d\Sigma_\mu j^\mu_\varepsilon\right)_\varepsilon, \varepsilon \in (0,1], \tag{3.1.7}$$

where

$$(j^\mu_\varepsilon(\varphi_\varepsilon,\phi_\varepsilon))_\varepsilon = (i/\hbar)\left(|g_\varepsilon|^{1/2} g^{\mu\nu}_\varepsilon(\overline{\varphi}_\varepsilon\partial_\nu\phi_\varepsilon - \varphi_\varepsilon\partial_\nu\overline{\phi}_\varepsilon)\right)_\varepsilon. \tag{3.1.8}$$

Using equation of motion Eq.(3.1.2) one obtains corresponding Colombeau generalization of the canonical Green functions equations. In particular for the Colombeau distributional propagator

$$i(G^\pm_\varepsilon(x,x'))_\varepsilon = (\langle 0|T(\varphi^\pm_\varepsilon(x)\varphi^\pm_\varepsilon(x'))|0\rangle)_\varepsilon, \varepsilon \in (0,1], \tag{3.1.9}$$

one obtains directly

$$([\Box_{\varepsilon,x} + m^2 + \xi\mathbf{R}^\pm(x,\varepsilon)]G^\pm_\varepsilon(x,x'))_\varepsilon = -\left([-g^\pm(x,\varepsilon)]^{-1/2}\right)_\varepsilon \delta^n(x-x'). \tag{3.1.10}$$

We obtan now an adiabatic expansion of $(G^\pm_\varepsilon(x,x'))_\varepsilon$
cite: FoukzonPotapovMenkova16Vickers12 [19]. Introducing Riemann normal coordinates $y^\mu$ for the point $x$, with origin at the point $x'$ one obtains

$$\begin{aligned}(g^\pm_{\mu\upsilon}(x,\varepsilon))_\varepsilon = \eta_{\mu\upsilon} + \frac{1}{3}\left[(\mathbf{R}^\pm_{\mu\alpha\upsilon\beta}(\varepsilon))_\varepsilon\right]y^\alpha y^\beta - \frac{1}{6}\left[(\mathbf{R}^\pm_{\mu\alpha\upsilon\beta;\gamma}(\varepsilon))_\varepsilon\right]y^\alpha y^\beta y^\gamma + \\ +\left[\frac{1}{20}(\mathbf{R}^\pm_{\mu\alpha\upsilon\beta;\gamma\delta}(\varepsilon))_\varepsilon + \frac{2}{45}\left[(\mathbf{R}^\pm_{\alpha\mu\beta\lambda}(\varepsilon))_\varepsilon\right](\mathbf{R}^{\pm\lambda}_{\gamma\upsilon\delta}(\varepsilon))_\varepsilon\right]y^\alpha y^\beta y^\gamma y^\delta + \ldots\end{aligned} \tag{3.1.11}$$

where $\eta_{\mu\upsilon}$ is the Minkowski metric tensor, and the coefficients are all evaluated at $y = 0$. Defining now

$$(\mathcal{L}^\pm_\varepsilon(x,x'))_\varepsilon = \left[\left((-g^\pm_{\mu\upsilon}(x,\varepsilon))^{1/4}\right)_\varepsilon\right](G^\pm_\varepsilon(x,x'))_\varepsilon \tag{3.1.12}$$

and its Colombeau-Fourier transform $(\mathcal{L}^\pm_\varepsilon(k))_\varepsilon$ by

$$(\mathcal{L}^\pm_\varepsilon(x,x'))_\varepsilon = (2\pi)^{-n}\left(\int d^nke^{-ik\bullet y}\mathcal{L}^\pm_\varepsilon(k)\right)_\varepsilon \tag{3.1.13}$$

where $k \bullet y = \eta^{\alpha\beta}k_\alpha y_\beta$, one can work in a sort of localized momentum space. Expanding (3.1.10) in normal coordinates and converting to $k$-space, $(\mathcal{L}^\pm_\varepsilon(k))_\varepsilon$ can readily be solved by iteration to any adiabatic order. The result to adiabatic order four (i.e., four derivatives of the metric) is

$$\begin{aligned}(\mathcal{L}^\pm_\varepsilon(k))_\varepsilon = (k^2-m^2)^{-1} - \left(\frac{1}{6}-\xi\right)(k^2-m^2)^{-2}(\mathbf{R}^\pm(\varepsilon))_\varepsilon + \\ +\frac{i}{2}\left(\frac{1}{6}-\xi\right)\partial^\alpha(k^2-m^2)^{-2}(\mathbf{R}^\pm_{;\alpha}(\varepsilon))_\varepsilon - \\ -\frac{1}{3}\left[(a^\pm_{\alpha\beta}(\varepsilon))_\varepsilon\right]\partial^\alpha\partial^\beta(k^2-m^2)^{-2} + \\ \left[\left(\frac{1}{6}-\xi\right)^2(\mathbf{R}^{\pm 2}(\varepsilon))_\varepsilon + \frac{2}{3}\left(a^{\pm\lambda}_{\lambda}(\varepsilon)\right)_\varepsilon\right](k^2-m^2)^{-3},\end{aligned} \tag{3.1.14}$$

where $\partial_\alpha = \partial/\partial k^\alpha$,

$$\begin{aligned}(a^\pm_{\alpha\beta}(\varepsilon))_\varepsilon \asymp \left(\frac{1}{2}-\xi\right)(\mathbf{R}^\pm_{;\alpha\beta}(\varepsilon))_\varepsilon + \frac{1}{120}(\mathbf{R}^\pm_{;\alpha\beta}(\varepsilon))_\varepsilon - \frac{1}{140}\left(\mathbf{R}^{\pm}_{\alpha\beta;\lambda}{}^{\lambda}(\varepsilon)\right)_\varepsilon - \\ -\frac{1}{30}\left[\left(\mathbf{R}^{\pm\lambda}_{\alpha}(\varepsilon)\right)_\varepsilon\right](\mathbf{R}^\pm_{\lambda\beta}(\varepsilon))_\varepsilon + \frac{1}{60}\left[\left(\mathbf{R}^{\pm\kappa}{}_{\alpha}{}^{\lambda}{}_{\beta}(\varepsilon)\right)_\varepsilon\right](\mathbf{R}^\pm_{\kappa\lambda}(\varepsilon))_\varepsilon + \\ +\frac{1}{60}\left[\left(\mathbf{R}^{\pm\lambda\mu\kappa}{}_{\alpha}(\varepsilon)\right)_\varepsilon\right](\mathbf{R}^\pm_{\lambda\mu\kappa\beta}(\varepsilon))_\varepsilon,\end{aligned} \tag{3.1.15}$$

and we are using the symbol $\asymp$ to indicate that this is an asymptotic expansion. One ensures that Eq.(3.1.13) represents a time-ordered product by performing the $k^0$ integral along the appropriate contour in Fig.3.1.1.This is equivalent to replacing $m^2$ by $m^2 - i\epsilon$. Similarly, the adiabatic expansions of other Green functions can be obtained by using the other contours in

Fig.3.1.1.Substituting Eq.(3.1.14) into Eq.(3.1.13) gives
cite: FoukzonPotapovMenkova16Vickers12 [19]

$$\begin{gathered}(\mathcal{L}_{\varepsilon}^{\pm}(x,x'))_{\varepsilon} = \\ (2\pi)^{-n} \times \left( \int d^n k e^{-iky} (k^2 - m^2)^{-1} \left[ a_0^{\pm}(x,x';\varepsilon) + a_1^{\pm}(x,x';\varepsilon) \left( -\frac{\partial}{\partial m^2} \right) + \right.\right. \\ \left.\left. a_2^{\pm}(x,x';\varepsilon) \left( \frac{\partial}{\partial m^2} \right)^2 \right] \right)_{\varepsilon}, \end{gathered} \quad (3.1.16)$$

where $(a_0^{\pm}(x,x';\varepsilon))_{\varepsilon} = 1$ and, to adiabatic order 4,

$$\begin{cases} (a_1^{\pm}(x,x';\varepsilon))_{\varepsilon} = \\ \left( \frac{1}{6} - \xi \right) (\mathbf{R}^{\pm}(\varepsilon))_{\varepsilon} - \frac{i}{2} \left( \frac{1}{6} - \xi \right) [(\mathbf{R}_{;\alpha}^{\pm}(\varepsilon))_{\varepsilon}] y^{\alpha} - \frac{1}{3} \left[ (a_{\alpha\beta}^{\pm}(\varepsilon))_{\varepsilon} \right] y^{\alpha} y^{\beta} \\ (a_2^{\pm}(x,x';\varepsilon))_{\varepsilon} = \frac{1}{2} \left( \frac{1}{6} - \xi \right) (\mathbf{R}^{\pm 2}(\varepsilon))_{\varepsilon} + \frac{1}{3} \left( a^{\pm\lambda}_{\ \ \lambda}(\varepsilon) \right)_{\varepsilon} \end{cases} \quad (3.1.17)$$

with all geometric quantities on the right-hand side of Eq.(3.1.17) evaluated at $x'$.

Figure

Fig

Fig.3.1.1.The contour in the complex $k^0$ plane $\mathbb{C}$
to be used in the evaluation of the integral
giving $\mathcal{L}^{+}$. The cross indicates the pole at
$k^0 = (|\mathbf{k}|^2 + m^2)^{1/2}$.

in Eq.(3.16), then the $d^n k$ integration may be interchanged with the $ds$ integration, and performed explicitly to yield (dropping the $i\epsilon$)

$$\begin{gathered}(\mathcal{L}_{\varepsilon}^{\pm}(x,x'))_{\varepsilon} = -i(4\pi)^{-n/2} \left( \int_0^{\infty} ids(is)^{-n/2} \exp\left[ -im^2 s + \frac{\sigma(x,x')}{2is} \right] \mathcal{F}_{\varepsilon}^{\pm}(x,x';is) \right)_{\varepsilon} \\ \sigma(x,x') = \frac{1}{2} y_{\alpha} y^{\alpha}. \end{gathered} \quad (3.1.18)$$

The function $\sigma(x,x')$ which is one-half of the square of the proper distance between $x$ and $x'$, while the function $(\mathcal{F}_{\varepsilon}(x,x';is))_{\varepsilon}$ has the following asymptotic adiabatic expansion

$$(\mathcal{F}_{\varepsilon}^{\pm}(x,x';is))_{\varepsilon} \asymp (a_0^{\pm}(x,x';\varepsilon))_{\varepsilon} + is(a_1^{\pm}(x,x';\varepsilon))_{\varepsilon} + (is)^2 (a_2^{\pm}(x,x';\varepsilon))_{\varepsilon} + \ldots \quad (3.1.19)$$

Using Eq.(3.1.12), equation (3.1.18) gives a representation of $(G_{\varepsilon}^{\pm}(x,x'))_{\varepsilon}$ :

$$(G_\varepsilon^{\pm}(x,x'))_\varepsilon = -i(4\pi)_\varepsilon^{-n/2}$$

$$\left(\left[\left(\Delta_{\pm}^{1/2}(x,x';\varepsilon)\right)_\varepsilon\right]\int_0^\infty ids(is)^{-n/2}\exp\left[-im^2s+\frac{\sigma(x,x')}{2is}\right]\mathcal{F}_\varepsilon(x,x';is)\right) \quad (3.1.20)$$

7where $(\Delta_{\pm}(x,x';\varepsilon))_\varepsilon$ is the distributional Van Vleck determinant

$$(\Delta_{\pm}(x,x';\varepsilon))_\varepsilon = -\det[\partial_\mu\partial_\nu\sigma(x,x')]\left([g^{\pm}(x,\varepsilon)g^{\pm}(x',\varepsilon)]^{-1/2}\right)_\varepsilon. \quad (3.1.21)$$

In the normal coordinates about $x'$ that we are currently using, $(\Delta_{\pm}(x,x';\varepsilon))_\varepsilon$ reduces to $\left([-g^{\pm}(x,\varepsilon)]^{-1/2}\right)_\varepsilon$. The full asymptotic expansion of $(\mathcal{F}_\varepsilon^{\pm}(x,x';is))_\varepsilon$ to all adiabatic orders are

$$(\mathcal{F}_\varepsilon^{\pm}(x,x';is))_\varepsilon \asymp \sum_{j=0}^{\infty}(is)^j(a_2^{\pm}(x,x';\varepsilon))_\varepsilon \quad (3.1.22)$$

with $(a_0^{\pm}(x,x';\varepsilon))_\varepsilon = 1$, the other $(a_j^{\pm}(x,x';\varepsilon))_\varepsilon$ being given by canonical recursion relations which enable their adiabatic expansions to be obtained.

**Remark 3.1.1.** Note that the expansions (3.1.19) and (3.1.22) are, however, only asymptotic approximations in the limit of large adiabatic parameter $T$.

If (3.1.22) is substituted into (3.1.20) the integral can be performed to give the adiabatic expansion of the Feynman propagator in coordinate space:

$$(G_\varepsilon^{\pm}(x,x'))_\varepsilon \asymp -(4\pi i)^{-n/2}\left(\Delta_{\pm}^{1/2}(x,x';\varepsilon)\sum_{j=0}^{\infty}a_j^{\pm}(x,x';\varepsilon)\left(-\frac{\partial}{\partial m^2}\right)^j\times\right.$$
$$\left.\times\left[\left(-\frac{2m^2}{\sigma}\right)^{\frac{n-2}{4}}H_{(n-2)/2}^{(2)}\left((2m^2\sigma)^{\frac{1}{2}}\right)\right]\right)_\varepsilon \quad (3.1.23)$$

which, strictly, a small imaginary part $i\epsilon$ should be subtracted from $\sigma$.

**Remark 3.1.2.** Since we have not imposed global boundary conditions on the distributional Green function Colombeau solution of (3.1.10), the expansion (3.1.23) does not determine the particular vacuum state in (3.1.9). In particular, the "$i\epsilon$" in the expansion of $(G_\varepsilon^{\pm}(x,x'))_\varepsilon$ only ensures that (3.1.23) represents the expectation value, in some set of states, of a time-ordered product of fields. Under some circumstances the use of "$i\epsilon$" in the exact representation (3.1.20) may give additional information concerning the global nature of the states.

## 5mm . 3.2. Effective action for the quantum matter fields in curved distributional space-time 2mm

As in classical case one can obtain Colombeau generalized quantity $(W_\varepsilon)_\varepsilon$, called the effective action for the quantum matter fields in curved distributional spcetime, which, when functionally differentiated,yields

$$\left(\frac{2}{(-g(\varepsilon))^{\frac{1}{2}}}\frac{\delta W_\varepsilon}{\delta g^{\mu\nu}(\varepsilon)}\right)_\varepsilon = (\langle\mathbf{T}_{\mu\nu}(\varepsilon)\rangle)_\varepsilon \quad (3.2.1)$$

Note that the generating functional

$$(Z_\varepsilon[\mathbf{J}_\varepsilon])_\varepsilon = \left(\int D[\varphi_\varepsilon]\exp\left\{iS_{\mathbf{m}}(\varepsilon)+i\int\mathbf{J}_\varepsilon(x)\varphi_\varepsilon(x)d^nx\right\}\right)_\varepsilon \quad (3.2.2)$$

was interpreted physically as the vacuum persistence amplitude $(\langle\mathbf{out}_\varepsilon,0|0,\mathbf{in}_\varepsilon\rangle)_\varepsilon$. The presence of the external distributional current density $(\mathbf{J}_\varepsilon)_\varepsilon$ can cause the initial vacuum state $(|0,\mathbf{in}_\varepsilon\rangle)_\varepsilon$ to be unstable, i.e., it can bring about the production of particles.

Following canonical calculation one obtains cite: FoukzonPotapovMenkova16Vickers12 [19]

$$(Z_\varepsilon^{\pm}[0])_\varepsilon \propto \left([\det(-G_\varepsilon^{\pm}(x,x'))]^{\frac{1}{2}}\right)_\varepsilon \quad (3.2.3)$$

where the proportionality constant is metric-independent and can be ignored. Thus we obtain

$$(W_\varepsilon^\pm)_\varepsilon = -i(\ln Z_\varepsilon^\pm[0])_\varepsilon = -\frac{i}{2}\left(\mathbf{tr}\left[\ln\left(-\hat{G}_\varepsilon^\pm\right)\right]\right)_\varepsilon. \quad (3.2.4)$$

In (3.2.4) $\left(\hat{G}_\varepsilon^\pm\right)_\varepsilon$ is to be interpreted as an Colombeau generalized operator which acts on an linear space $\mathfrak{I}$ of generalized vectors $|x,\varepsilon\rangle, \varepsilon \in (0,1]$ normalized by

$$(\langle x,\varepsilon|x',\varepsilon\rangle)_\varepsilon = \delta(x-x')\left([-g^\pm(x,\varepsilon)]^{-\frac{1}{2}}\right)_\varepsilon \quad (3.2.5)$$

in such a way that

$$(G_\varepsilon^\pm(x,x'))_\varepsilon = \left(\langle x,\varepsilon|\hat{G}_\varepsilon^\pm|x',\varepsilon\rangle\right)_\varepsilon. \quad (3.2.6)$$

**Remark 3.2.1.** Note that the trace $(\mathbf{tr}[\bullet])_\varepsilon$ of an Colombeau generalized operator $(\mathfrak{R}_\varepsilon)_\varepsilon$ which acts on a linear space $\mathfrak{I}$, is defined by

$$(\mathbf{tr}[\mathfrak{R}_\varepsilon])_\varepsilon = \left(\int d^nx[-g^\pm(x,\varepsilon)]^{\frac{1}{2}}\mathfrak{R}_{xx;\varepsilon}\right)_\varepsilon = \left(\int d^nx[-g^\pm(x,\varepsilon)]^{\frac{1}{2}}\langle x|\mathfrak{R}_{xx;\varepsilon}|x'\rangle\right)_\varepsilon. \quad (3.2.7)$$

Writing now the Colombeau generalized operator $\left(\hat{G}_\varepsilon^\pm\right)_\varepsilon$ as

$$\left(\hat{G}_\varepsilon^\pm\right)_\varepsilon = -(\mathcal{F}_\varepsilon^{\pm-1})_\varepsilon = -i\left(\int_0^\infty ds \exp[-s\mathcal{F}_\varepsilon^\pm]\right)_\varepsilon, \quad (3.2.8)$$

by Eq.(3.1.20) we obtain

$$(\langle x|\exp[-s\mathcal{F}_\varepsilon^\pm]|x'\rangle) = i(4\pi)^{-n/2}\left[\left(\Delta_\pm^{1/2}(x,x';\varepsilon)\right)_\varepsilon\right]\exp\left[-im^2s + \frac{\sigma(x,x')}{2is}\right]\mathcal{F}_\varepsilon^\pm(x,x';is)(is)^{-n/2}. \quad (3.2.9)$$

Proceeding in standard manner we get cite: FoukzonPotapovMenkova16Vickers12 [19]

$$(W_\varepsilon^\pm)_\varepsilon = \frac{i}{2}\left[\left(\int d^nx[-g^\pm(x,\varepsilon)]^{\frac{1}{2}}\right)_\varepsilon\right]\left(\lim_{x\to x'}\int_{m^2}^\infty G_\varepsilon^\pm(x,x';m^2)dm^2\right)_\varepsilon. \quad (3.2.10)$$

Interchanging now the order of integration and taking the limit $x \to x'$ one obtains

$$(W_\varepsilon^\pm)_\varepsilon = \frac{i}{2}\left(\int_{m^2}^\infty dm^2\int d^nx[-g^\pm(x,\varepsilon)]^{\frac{1}{2}}G_\varepsilon^\pm(x,x;m^2)\right)_\varepsilon. \quad (3.2.11)$$

Colombeau generalized quantity $(W_\varepsilon^\pm)_\varepsilon$ is colled as the one-loop effective action. In the case of fermion effective actions, there would be a remaining trace over spinorial indices. From Eq.(3.2.11) we may define an effective Lagrangian density $\left(L_{\varepsilon;\mathbf{eff}}^\pm(x)\right)_\varepsilon$ by

$$(W_\varepsilon^\pm)_\varepsilon = \left(\int d^nx[-g^\pm(x,\varepsilon)]^{\frac{1}{2}}L_{\varepsilon;\mathbf{eff}}^\pm(x)\right)_\varepsilon \quad (3.2.12)$$

whence one get

$$(L_\varepsilon^\pm(x))_\varepsilon = \left([-g^\pm(x,\varepsilon)]^{\frac{1}{2}}\mathcal{L}_{\varepsilon;\mathbf{eff}}^\pm(x)\right)_\varepsilon = \frac{i}{2}\left(\lim_{x\to x'}\int_{m^2}^\infty dm^2G_\varepsilon^\pm(x,x';m^2)\right)_\varepsilon. \quad (3.2.13)$$

## 5mm . 3.3.Stress-tensor renormalization 2mm

Note that $(L_\varepsilon^\pm(x))_\varepsilon$ diverges at the lower end of the $s$ integral because the $\sigma/2s$ damping factor in the exponent vanishes in the limit $x \to x'$. (Convergence at the upper end is guaranteed by the $-i\epsilon$ that is implicitly added to $m^2$ in the De Witt-Schwinger representation of $(L_\varepsilon^\pm(x))_\varepsilon$. In four dimensions, the potentially divergent terms in the DeWitt- Schwinger expansion of $(L_\varepsilon^\pm(x))_\varepsilon$ are

$$(L^{\pm}_{\varepsilon;\mathrm{div}}(x))_{\varepsilon} =$$

$$-(32\pi^2)^{-1}\left(\lim_{x\to x'}\left[\left(\Delta_{\pm}^{1/2}(x,x';\varepsilon)\right)_{\varepsilon}\right]\int_0^{\infty}\frac{ds}{s^3}\exp\left[-im^2s+\frac{\sigma(x,x')}{2is}\right]\times \right. \tag{3.3.1}$$

$$\left.\times\left[a_0^{\pm}(x,x';\varepsilon)+isa_1^{\pm}(x,x';\varepsilon)+(is)^2a_2^{\pm}(x,x';\varepsilon)\right]\right)_{\varepsilon}$$

where the coefficients $a_0^{\pm}$, $a_1^{\pm}$ and $a_2^{\pm}$ are given by Eq.(3.1.17).The remaining terms in this asymptotic expansion, involving $a_3^{\pm}$ and higher, are finite in the limit $x\to x'$.

Let us determine now the precise form of the geometrical $(L^{\pm}_{\varepsilon;\mathrm{div}}(x))_{\varepsilon}$ terms, to compare them with the distributional generalization of the gravitational Lagrangian that appears in cite: FoukzonPotapovMenkova16Vickers12 [19]. This is a delicate matter because (3.3.1) is, of course, infinite. What we require is to display the divergent terms in the form $\infty\times$ [geometrical object]. This can be done in a variety of ways. For example, in $n$ dimensions, the asymptotic (adiabatic) expansion of $\left(L^{\pm}_{\varepsilon;\mathbf{eff}}(x)\right)_{\varepsilon}$ is

$$\left(L^{\pm}_{\varepsilon;\mathbf{eff}}(x)\right)_{\varepsilon} \asymp$$

$$2^{-1}(4\pi)^{-n/2}\left(\lim_{x\to x'}\left[\left(\Delta_{\pm}^{1/2}(x,x';\varepsilon)\right)_{\varepsilon}\right]\sum_{j=0}^{\infty}a_j(x,x';\varepsilon)\times\right. \tag{3.3.2}$$

$$\left.\times\int_0^{\infty}ids(is)^{j-1-n/2}\exp\left[-im^2s+\frac{\sigma(x,x')}{2is}\right]\right)_{\varepsilon}$$

of which the first $n/2+1$ terms are divergent as $\sigma\to 0.$ If $n$ is treated as a variable which can be analytically continued throughout the complex plane, then we may take the $x\to x'$ limit

$$\left(L^{\pm}_{\varepsilon;\mathbf{eff}}(x)\right)_{\varepsilon}\asymp 2^{-1}(4\pi)^{-n/2}\left(\sum_{j=0}^{\infty}a_j(x;\varepsilon)\int_0^{\infty}ids(is)^{j-1-n/2}\exp[-im^2s]\right)_{\varepsilon} = \tag{3.3.3}$$

$$2^{-1}(4\pi)^{-n/2}\sum_{j=0}^{\infty}a_j(x;\varepsilon)(m^2)^{n/2-j}\Gamma\left(j-\frac{n}{2}\right), a_j(x;\varepsilon)=a_j(x,x;\varepsilon).$$

From Eq.(3.3.3) follows we shall wish to retain the units of $L^{\pm}_{\varepsilon;\mathbf{eff}}(x)$ as (length)$^{-4}$, even when $n\neq 4.$ It is therefore necessary to introduce an arbitrary mass scale $\mu$ and to rewrite Eq.(3.3.3) as

$$\left(L^{\pm}_{\varepsilon;\mathbf{eff}}(x)\right)_{\varepsilon}\asymp 2^{-1}(4\pi)^{-n/2}\left(\frac{m}{\mu}\right)^{n-4}\left(\sum_{j=0}^{\infty}a_j(x;\varepsilon)(m^2)^{4-2j}\Gamma\left(j-\frac{n}{2}\right)\right)_{\varepsilon}. \tag{3.3.4}$$

If $n\to 4,$ the first three terms of Eq.(3.3.4) diverge because of poles in the $\Gamma$- functions:

$$\Gamma\left(-\frac{n}{4}\right)=\frac{4}{n(n-2)}\left(\frac{2}{4-n}-\gamma\right)+O(n-4),$$
$$\Gamma\left(1-\frac{n}{2}\right)=\frac{4}{(2-n)}\left(\frac{2}{4-n}-\gamma\right)+O(n-4), \tag{3.3.5}$$
$$\Gamma\left(2-\frac{n}{2}\right)=\frac{2}{4-n}-\gamma+O(n-4).$$

Denoting these first three terms by $(L^{\pm}_{\varepsilon;\mathbf{div}}(x))_{\varepsilon},$ we have

$$(L^{\pm}_{\varepsilon;\mathbf{div}}(x))_{\varepsilon} = (4\pi)^{-n/2}\left\{\frac{1}{n-4}+\frac{1}{2}\left[\gamma+\ln\left(\frac{m^2}{\mu^2}\right)\right]\right\}\times \left(\left[\frac{4m^4a_0(x;\varepsilon)}{n(n-2)}-\frac{2m^2a_1(x;\varepsilon)}{n-2}+a_2(x;\varepsilon)\right]\right)_{\varepsilon}. \tag{3.3.6}$$

The functions $a_0(x;\varepsilon), a_1(x;\varepsilon)$ and $a_2(x;\varepsilon)$ are given by taking the coincidence limits of (3.1.17)

$$\begin{gathered}(a^{\pm}_0(x;\varepsilon))_{\varepsilon} = 1, (a^{\pm}_1(x;\varepsilon))_{\varepsilon} = \left(\frac{1}{6}-\xi\right)(\mathbf{R}^{\pm}(\varepsilon))_{\varepsilon},\\ (a^{\pm}_2(x;\varepsilon))_{\varepsilon} = \frac{1}{180}(\mathbf{R}^{\pm}_{\alpha\beta\gamma\delta}(x,\varepsilon)\mathbf{R}^{\pm\alpha\beta\gamma\delta}(x,\varepsilon))_{\varepsilon} - \frac{1}{180}(\mathbf{R}^{\pm\alpha\beta}(x,\varepsilon)\mathbf{R}^{\pm}_{\alpha\beta}(x,\varepsilon))_{\varepsilon} - \\ -\frac{1}{6}\left(\frac{1}{5}-\xi\right)(\Box_{\varepsilon,x}\mathbf{R}^{\pm}(x,\varepsilon))_{\varepsilon} + \frac{1}{2}\left(\frac{1}{6}-\xi\right)(\mathbf{R}^{\pm 2}(x,\varepsilon))_{\varepsilon}.\end{gathered} \tag{3.3.7}$$

Finally one obtains cite: FoukzonPotapovMenkova16Vickers12[19]

$$(L^{\pm}_{\varepsilon;\mathbf{ren}}(x))_{\varepsilon} \asymp -\frac{1}{64\pi^2}\left(\int_0^{\infty} ids\ln(is)\frac{\partial^3}{\partial(is)^3}\left[\mathcal{F}^{\pm}_{\varepsilon}(x,x;is)e^{-ism^2}\right]\right)_{\varepsilon}. \tag{3.3.8}$$

**Remark 3.3.1**.All the higher order $(j > 2)$ terms in the DeWitt-Schwinger expansion of the effective Lagrangian (3.3.4) are infrared divergent at $n = 4$ as $m \to 0,$ we can still use this expansion to yield the ultraviolet divergent terms arising from $j = 0,1,$ and 2 in the four-dimensional case. We may put $m = 0$ immediately in the $j = 0$ and 1 terms in the expansion, because they are of positive power for $n \sim 4.$ These terms therefore vanish. The only nonvanishing potentially ultraviolet divergent term is therefore $j = 2$ :

$$2^{-1}(4\pi)^{-n/2}\left(\frac{m}{\mu}\right)^{n-4}a_2(x,\varepsilon)\Gamma\left(2-\frac{n}{2}\right), \tag{3.3.9}$$

which must be handled carefully. Substituting for $a_2(x)$ with $\xi = \xi(n)$ from (3.3.7), and rearranging terms, we may write the divergent term in the effective action arising from (3.3.9) as follows

$$\begin{gathered}(W^{\pm}_{\varepsilon,\mathbf{div}})_{\varepsilon} = 2^{-1}(4\pi)^{-n/2}\left(\frac{m}{\mu}\right)^{n-4}\Gamma\left(2-\frac{n}{2}\right)\left(\int d^nx[-g^{\pm}(x,\varepsilon)]^{\frac{1}{2}}a_2(x,\varepsilon)\right)_{\varepsilon} = \\ 2^{-1}(4\pi)^{-n/2}\left(\frac{m}{\mu}\right)^{n-4}\Gamma\left(2-\frac{n}{2}\right)\times \\ \left(\int d^nx[-g^{\pm}(x,\varepsilon)]^{\frac{1}{2}}\left[\widetilde{\alpha}\mathsf{F}^{\pm}_{\varepsilon}(x)+\widetilde{\beta}G^{\pm}_{\varepsilon}(x)\right]\right)_{\varepsilon} + O(n-4),\end{gathered} \tag{3.3.10}$$

where

$$\begin{gathered}(\mathsf{F}_{\varepsilon}(x))_{\varepsilon} = \\ (\mathbf{R}^{\pm\alpha\beta\gamma\delta}(x,\varepsilon)\mathbf{R}^{\pm}_{\alpha\beta\gamma\delta}(x,\varepsilon))_{\varepsilon} - 2(\mathbf{R}^{\pm\alpha\beta}(x,\varepsilon)\mathbf{R}^{\pm}_{\alpha\beta}(x,\varepsilon))_{\varepsilon} + \frac{1}{3}(\mathbf{R}^{\pm 2}(x,\varepsilon))_{\varepsilon}, \\ (G^{\pm}_{\varepsilon}(x))_{\varepsilon} = (\mathbf{R}^{\pm\alpha\beta\gamma\delta}(x,\varepsilon)\mathbf{R}^{\pm}_{\alpha\beta\gamma\delta}(x,\varepsilon))_{\varepsilon}, \\ \widetilde{\alpha} = \frac{1}{120}, \widetilde{\beta} = -\frac{1}{360}.\end{gathered} \tag{3.3.11}$$

Finally we obtain cite: FoukzonPotapovMenkova16Vickers12 [19]

$$\begin{gathered}(\langle T^{\mu}_{\mu}(x,\varepsilon)\rangle_{\mathbf{ren}})_{\varepsilon} = -(1/2880\pi^2)\left[\widetilde{\alpha}\left(\mathsf{F}_{\varepsilon}(x)-\tfrac{2}{3}\Box_{\varepsilon,x}\mathbf{R}^{\pm}(x,\varepsilon)\right)_{\varepsilon}+\widetilde{\beta}(G^{\pm}_{\varepsilon}(x))_{\varepsilon}\right] = \\ -(1/2880\pi^2)\times \\ \left[(\mathbf{R}^{\pm}_{\alpha\beta\gamma\delta}(x,\varepsilon)\mathbf{R}^{\pm\alpha\beta\gamma\delta}(x,\varepsilon))_{\varepsilon} - (\mathbf{R}^{\pm}_{\alpha\beta}(x,\varepsilon)\mathbf{R}^{\pm\alpha\beta}(x,\varepsilon))_{\varepsilon} - (\Box_{\varepsilon,x}\mathbf{R}^{\pm}(x,\varepsilon))_{\varepsilon}\right].\end{gathered} \tag{3.3.12}$$

Therefore for the case of the distributional Schwarzchild spesetime using Eq.(2.2.4) and

Eq.(3.3.12) for $r \to 2m$ we obtain

$$\begin{aligned}\left(\langle T_{\mu}^{\mu}(r,\varepsilon)\rangle_{\mathbf{ren}}\right)_{\varepsilon} &\asymp -(2880\cdot\pi^2)^{-1}\left[\left(\left[16^{-1}m^2(r-2m)^2+\varepsilon^2\right]^{-1}\right)_{\varepsilon}+\ldots\right]\approx \\ &\approx -O(1)(2880\cdot 16\cdot\pi^2)^{-1}m^{-2}(r-2m)^{-2}.\end{aligned} \tag{3.3.13}$$

**Remark 3.3.2.** Thus QFT in ditributional curved spacetime predict that the infalling observer burns up at the BH horizon.

**Remark 3.3.3.** In order avoid singularity at horizon $r = 2m$ in Eq.(3.3.13) one have applied the Loop Quantum Gravity approach cite: Olmedo16GambiniOlmedoPullin14Mavromatos09
cite: Rivasseau12BarriosGambiniPullin15GambiniPullin14 [41]-[46].The first one concerns the requirement of selfadjointness to the metric components. For instance, the classical quantity

$$g_{tx} = -\frac{(E^x)'K_\varphi}{2\sqrt{E^x}\sqrt{1+K_\varphi^2-\frac{2Gm}{\sqrt{E^x}}}}, \tag{3.3.14}$$

defined as an evolving constant (i.e. a Dirac observable), must correspond to a selfadjoint operator at the quantum level. Classically, $K_\varphi$ and $E^x$ are pure gauge, and $g_{tx}$ is just a function of the observable $m$. In the interior of the horizon, if $\widehat{g}_{tx}$ is a selfadjoint operator, a necessary condition will be cite: Olmedo16GambiniOlmedoPullin14Mavromatos09
cite: Rivasseau12BarriosGambiniPullin15GambiniPullin14 [41]-[46]

$$1+K_\varphi^2-\frac{2Gm}{l_{\mathbf{P}}\sqrt{k_j}} \geq 0. \tag{3.3.15}$$

At the singularity, i.e. $j = 1$, and owing to the bounded nature of $K_\varphi^2 < \infty$,

$$\sqrt{k_1} \geq \frac{2Gm}{l_{\mathbf{P}}(1+K_\varphi^2)} > 0. \tag{3.3.16}$$

Therefore, this argument strongly suggests that the classical singularity will be resolved at the quantum level since $k_1$ must be a non-vanishing integer.

**Remark 3.3.4.** Let $\langle T_\nu^\mu(r)\rangle_{\mathbf{ren}}^H$ be $\langle 0^H|T_\nu^\mu(r)|0^H\rangle_{\mathbf{ren}}$, where $|0^H\rangle$ is the Hartle-Hawking vacuum state cite: FrolovNovikov98 [37]. Notice that the main feature of the tensor $\widehat{T}_\nu^\mu(r) \triangleq \langle T_\nu^\mu(r)\rangle_{\mathbf{ren}}^H$ formally calculated in classical literature (see, for example, cite: FrolovNovikov98 [37] chapter 11.3) is that its components are finite on the event horizon $r_+$. An observer at rest at a point $r$ close to the event horizon records the local energy density $\epsilon = -\widehat{T}_t^t(r)$. This quantity remains finite as $r \to r_+$. On the other hand, the temperature measured by the such observer is

$$\Theta_{\mathbf{loc}}(r) = \frac{k}{2\pi}\left(1+\frac{r_+}{r}\right)^{-1/2}, \tag{3.3.17}$$

grows infinitely near the horizon cite: FrolovNovikov98 [37]. The local temperature can be measured by using a two-level system as a thermometer. Transitions between levels are caused by the absorption and emission of quanta of the fields (photons). After a sufficiently long exposure, the probability for a system to occupy the upper level will be less than that for the lower level by a factor $\exp(\Delta E/\Theta_{\mathbf{loc}}(r))$, where $\Delta E$ is the energy difference between the levels. It is well known that the temperature in the vicinity of $r_+$ is $\Theta_{\mathbf{loc}} \approx \theta_a = a/2\pi$, where $a$ is the observer's acceleration cite: FrolovNovikov98 [37]; as $r \to r_+, \Theta_{\mathbf{loc}}(r) \to \infty$. The radiation energy density $\epsilon$ in the neighborhood of such a point is cite: FrolovNovikov98 [37]

$$\epsilon \simeq \sigma(k/2\pi)^4 \ll \sigma\theta_a^4. \tag{3.3.18}$$

Therefore Stefan-Boltzmann law under formal calculation by using classical Schwarzschild geometry is evidently violated. Let us remind that the acceleration $a = \sqrt{a_i a^i}$ of free fall of a body which is initially at rest in the Schwarzschild reference frame is cite: FrolovNovikov98[37]

$$a(r) = \frac{m}{r^2(1-\frac{2m}{r})^{1/2}}. \tag{3.3.19}$$

The acceleration points along the radius and is directed toward the center; as $r \asymp 2m$ :

$$a(r) = \frac{\sqrt{2m}}{4m(r-2m)^{1/2}}. \quad (3.3.19)$$

From Eq. (3.3.19) and Eq. (3.3 13) finally we obtain as $\varepsilon \asymp 0, r \asymp 2m$ :

$$\sigma\theta_a^4 = \frac{\sigma}{64m^2(r-2m)^2} \asymp -(\langle T_\mu^\mu(r,\varepsilon)\rangle_{\mathbf{ren}})_\varepsilon. \quad (3.3.2$$

Therefore Stefan-Boltzmann law under rigorous calculation by using distributional Schwarzschild geometry evidently is not violated.

5mm . **3.4.Unruh effect revisited** 2mm

We remind now that a black holes have an approximate Rindler region near the Schwarzschild horizon. For the the distributional Schwarzschild solution (2.1.8) by coordinate transformation

$$r = 2m(1+(\delta^2+\epsilon^2)), \epsilon \in (0,1], \quad (3.4.1)$$

where $\varepsilon \ll \epsilon$, we obtain

$$(ds_\epsilon^2)_\epsilon = -\left((\delta^2+\epsilon^2)_\epsilon\right)dt^2 + 16m^2 d\delta^2 + 4m^2 d\Omega_2^2 + O(\varepsilon^2/\epsilon^2)\ldots \quad (3.4.2)$$

The $(t,\delta)$ piece of this metric (3.4.2) is Rindler space (we can rescale $t$ ,$\delta$ and $\epsilon$ to make it look exactly like (2.2.10) for $\varepsilon^2/\epsilon^2 \to 0$. Thus from (3.3.13) using (3.4.1) we obtain directly for $\delta \asymp 0$

$$(\langle T_\mu^\mu(\delta,\varepsilon)\rangle_{\mathbf{ren}})_\varepsilon \asymp \delta^{-4}. \quad (3.4.3)$$

Therefore sufficiently strongly accelerated observer burns up near the Rindler horizon. Thus Polchinski's account doesn't violation of the Einstein equivalence principle.

**Remark 3.4.1.**Note that by using Eq.(A.8) and Eq.(A.9) (see appendix A) one obtains Eq. (3.4.3) directly from distributionel Rindler metric (2.2.10).

$$(\langle T_\mu^\mu(\delta,\varepsilon)\rangle_{\mathbf{ren}})_\varepsilon \asymp \frac{-4g^4 O(1)}{\left[(a+gx)^2+\varepsilon^2\right]^2}. \quad (3.4.4)$$

# 4.1.The Colombeau distributional Kerr spacetime in Boyer- Lindquist form.

The classical Kerr metric in Boyer-Lindquist form reads

$$\begin{gathered} ds^2 = -\Xi(r,\theta)dt^2 - \frac{4mra\sin^2\theta}{\rho^2}dtd\phi + \frac{\rho^2}{\Delta_a}dr^2 + \rho^2 d\theta^2 + \\ \left(r^2+a^2+\frac{2mra^2\sin^2\theta}{\rho^2}\right)\sin^2\theta d\phi^2, \end{gathered} \quad (4.1.1)$$

where

$$\begin{gathered} \rho^2 = r^2 + a^2\cos^2\theta, \Delta_a = r^2 - 2mr + a^2, \\ \Xi(r,\theta) = 1 - \frac{2mr}{\rho^2} = \frac{r^2-2mr+a^2\cos^2\theta}{\rho^2}. \end{gathered} \quad (4.1.2)$$

**Remark 4.2.1.**Note that For small values of the parameter $a$, where we may neglect terms of the order of $a^2$, we get from (4.1.1) the Lense-Thirring metric with $J_z = ma$

$$ds^2 = -\left(1-\frac{2m}{r}\right)dt^2 + \left(1-\frac{2m}{r}\right)^{-1}dr^2 + r^2(d\theta^2+\sin^2\theta d\phi^2) - \frac{4ma\sin^2\theta}{r}dtd\phi. \quad (4.1.3)$$

**I.Slow Kerr gravitational singularity**: $a < m$.

Note that

$$\Xi(r,\theta) = \frac{r^2-2mr+a^2\cos^2\theta}{\rho^2} = \frac{(r-r_{E_+}(\theta))(r-r_{E_-}(\theta))}{\rho^2}, \quad (4.1.4)$$

where $r_{E_{\pm}}(\theta) = m \pm \sqrt{m^2 - a^2\cos^2\theta}$ and $\Delta_a = r^2 - 2mr + a^2 = (r - r_+)(r - r_-)$, where $r_{\pm} = m \pm \sqrt{m^2 - a^2}$.

**Remark 4.1.2.**Let $\Xi(\theta)$ be a submanifold given by equation $\phi = \text{const}$, then metric (4.1.1) restricted on submanifold $\cup_\theta \Xi(\theta)$ reads

$$ds^2 = -\Xi(r,\theta)dt^2 + \frac{\rho^2}{\Delta_a}dr^2 + \rho^2 d\theta^2. \tag{4.1.5}$$

Note that: (i) the metric (4.1.5) is degenerate on outer ergosurfaces $r = r_{E_+}(\theta)$ and inner ergosurfaces $r - r_{E_-}(\theta)$,(ii) the metric (4.1.5) is singular on horizon $r = r_+$,(iii) the metric (4.1.5) is singular on submanifold given by equation $r = r_-$.

**Remark 4.1.3.**Note that we will be consider the distributional Kerr spacetime not as full distributionel BH with Colombeau generalized metric (4.1.7), but only as gravitational singularity located on submanifold $\cup_\theta \Xi(\theta)$ which coincide with outer ergosurface of classical Kerr spacetime. In accordance with Eq.(4.1.11), Eq.(4.1.19) and Eq.(4.1.20) submanifold $\cup_\theta \Xi(\theta)$ presented the singular boundary of distributional spacetime presented by Colombeau generalized metric (4.1.7).

We introduce now the following regularized (above ergosurface $r = r_{E_+}(\theta)$) quantity

$$\Xi^+_\varepsilon(r_\varepsilon,\theta) = \frac{(r_\varepsilon - r_{E_-}(\theta))\sqrt{(r_\varepsilon - r_{E_+}(\theta))^2 + \varepsilon^2}}{\rho^2_\varepsilon(r_\varepsilon)}, \qquad \Delta_{a,\varepsilon} = r^2_\varepsilon - 2mr_\varepsilon + a^2, \tag{4.1.6}$$

where $\rho^2_\varepsilon = \rho^2_\varepsilon(r_\varepsilon) = r^2_\varepsilon + a^2\cos^2\theta, \varepsilon \in (0,\delta],\ r_\varepsilon \geq r_{E_+}(\theta) \geq r_+$. Thus Colombeau generalized metric (above ergosurface $r = r_{E_+}(\theta)$) corresponding to classical Kerr metric (4.1.1) reads

$$\begin{gathered}(ds^{+2}_\varepsilon)_\varepsilon = -[(\Xi_\varepsilon(r_\varepsilon,\theta))_\varepsilon]dt^2 - \left[\left(\frac{4mr_\varepsilon a\sin^2\theta}{\rho^2_\varepsilon}\right)_\varepsilon\right]dtd\phi + \\ \left[\left(\frac{\rho^2_\varepsilon}{\Delta_{a,\varepsilon}}\right)_\varepsilon\right][(dr^2_\varepsilon)_\varepsilon] + (\rho^2_\varepsilon)_\varepsilon d\theta^2 + \\ \left(r^2_\varepsilon + a^2 + \frac{2mra^2\sin^2\theta}{\rho^2_\varepsilon}\right)_\varepsilon \sin^2\theta d\phi^2.\end{gathered} \tag{4.1.7}$$

**Remark 4.1.4.**Let $\Xi(\theta)$ be a submanifold given by equation $\phi = \text{const}$, then Colombeau generalized metric (7.2.7) restricted on $\Xi(\theta)$ reads

$$(ds^{+2}_\varepsilon)_\varepsilon = -[(\Xi_\varepsilon(r_\varepsilon,\theta))_\varepsilon]dt^2 + \left[\left(\frac{\rho^2_\varepsilon}{\Delta_{a,\varepsilon}}\right)_\varepsilon\right][(dr^2_\varepsilon)_\varepsilon] + [(\rho^2_\varepsilon)_\varepsilon]d\theta^2. \tag{4.1.8}$$

Note that Colombeau generalized metric (4.1.7) nondegenerate on outer ergosurfaces $(r_\varepsilon)_\varepsilon = r_{E_+}(\theta)$, see Pic.4.1.

**Remark 4.1.4.**Note that for small values of the parameter $a$, where we may neglect terms of the order of $a^2$, we get from (4.1.7) effectively the following Colombeau generalized metric

$$\begin{gathered}(ds^{+2}_\varepsilon)_\varepsilon = -\left(1 - \frac{r_{E_+}(\theta)}{(r_\varepsilon)_\varepsilon}\right)dt^2 + \left(1 - \frac{r_{E_+}(\theta)}{(r_\varepsilon)_\varepsilon}\right)^{-1}[(dr^2_\varepsilon)_\varepsilon] + \\ [(r^2_\varepsilon)_\varepsilon](d\theta^2 + \sin^2\theta d\phi^2) - \frac{4ma\sin^2\theta}{(r_\varepsilon)_\varepsilon}dtd\phi,\end{gathered} \tag{4.1.9}$$

where $r_{E_+}(\theta) = m + \sqrt{m^2 - a^2\cos^2\theta}$.

**Remark 4.1.5.** Note that Colombeau generalized metric (7.2.9) restricted on $\Xi(\theta)$ reads

$$(ds_{\varepsilon}^{+2})_{\varepsilon} = -\left(1 - \frac{r_{E_+}(\theta)}{(r_{\varepsilon})_{\varepsilon}}\right)dt^2 + \left(1 - \frac{r_{E_+}(\theta)}{(r_{\varepsilon})_{\varepsilon}}\right)^{-1}[(dr_{\varepsilon}^2)_{\varepsilon}] + [(r_{\varepsilon}^2)_{\varepsilon}]d\theta^2 \qquad (4.1.10)$$

(**I**)Let $(\mathbf{R}^{a\ll 1}(r_{\varepsilon},\varepsilon))_{\varepsilon}$ be Colombeau generalized curvature scalar $(\mathbf{R}(r_{\varepsilon},\varepsilon))_{\varepsilon}$ corresponding to the metric (4.1.10) with $a \ll 1$. Main singular part $\mathbf{sing}\left[(\mathbf{R}^{a\ll 1}(r_{\varepsilon},\varepsilon))_{\varepsilon}\right]$ of the Colombeau generalized curvature scalar $(\mathbf{R}^{a\ll 1}(r_{\varepsilon},\varepsilon))_{\varepsilon}$corresponding to the metric (4.1.10) reads

$$\mathbf{sing}\left[(\mathbf{R}^{a\ll 1}(r_{\varepsilon},\varepsilon))_{\varepsilon}\right] =_{\widetilde{\mathbb{R}}} \left(\frac{\varepsilon^2}{r_{E_+}(\theta)\left[(r_{\varepsilon} - r_{E_+}(\theta))^2 + \varepsilon^2\right]^{3/2}}\right)_{\varepsilon}, \qquad (4.1.11)$$

where $\mathbf{cl}[(r_{\varepsilon})_{\varepsilon}] = r_{E_+}(\theta)$.

(**II**) Let $\left(\mathbf{R}^{\mu\nu(a\ll 1)}(r_{\varepsilon},\varepsilon)\mathbf{R}_{\mu\nu}^{(a\ll 1)}(r_{\varepsilon},\varepsilon)\right)_{\varepsilon}$ be Colombeau generalized quadratic scalar $(\mathbf{R}^{\mu\nu}(r_{\varepsilon},\varepsilon)\mathbf{R}_{\mu\nu}(r_{\varepsilon},\varepsilon))_{\varepsilon}$ corresponding to the metric (4.1.10) with $a \ll 1$. From Eq.(4.1.10) and Eq.(A.1)-Eq.(A.2) one obtains that singular part

$$\mathbf{sing}\left[\left(\mathbf{R}^{\mu\nu(a\ll 1)}(r_{\varepsilon},\varepsilon)\mathbf{R}_{\mu\nu}^{(a\ll 1)}(r_{\varepsilon},\varepsilon)\right)_{\varepsilon}\right]$$

of the quadratic scalar $\left(\mathbf{R}^{\mu\nu(a\ll 1)}(r_{\varepsilon},\varepsilon)\mathbf{R}_{\mu\nu}^{(a\ll 1)}(r_{\varepsilon},\varepsilon)\right)_{\varepsilon}$ reads:

$$\mathbf{sing}\left[\left(\mathbf{R}^{\mu\nu(a\ll 1)}(r_{\varepsilon},\varepsilon)\mathbf{R}_{\mu\nu}^{(a\ll 1)}(r_{\varepsilon},\varepsilon)\right)_{\varepsilon}\right] =_{\widetilde{\mathbb{R}}} \left(\frac{\varepsilon^4}{4(r_{E_+}(\theta))^4\left[\varepsilon^2 + (r_{\varepsilon} - 2m)^2\right]^3}\right)_{\varepsilon}. \qquad (4.1.12)$$

(**III**) Let $\left(\mathbf{R}^{\rho\sigma\mu\nu(a\ll 1)}(r_{\varepsilon},\varepsilon)\mathbf{R}_{\rho\sigma\mu\nu}^{(a\ll 1)}(r_{\varepsilon},\varepsilon)\right)_{\varepsilon}$ be Colombeau generalized quadratic scalar $(\mathbf{R}^{\rho\sigma\mu\nu}(r_{\varepsilon},\varepsilon)\mathbf{R}_{\rho\sigma\mu\nu}(r_{\varepsilon},\varepsilon))_{\varepsilon}$ corresponding to the metric (7.2.10) with $a \ll 1$. From Eq.(7.2.10) and Eq.(A.1)-Eq.(A.2) one obtains that main singular part

$$\mathbf{sing}\left[\left(\mathbf{R}^{\rho\sigma\mu\nu(a\ll 1)}(r_{\varepsilon},\varepsilon)\mathbf{R}_{\rho\sigma\mu\nu}^{(a\ll 1)}(r_{\varepsilon},\varepsilon)\right)_{\varepsilon}\right]$$

of the Colombeau generalized quadratic scalar $\left(\mathbf{R}^{\rho\sigma\mu\nu(a\ll 1)}(r_{\varepsilon},\varepsilon)\mathbf{R}_{\rho\sigma\mu\nu}^{(a\ll 1)}(r_{\varepsilon},\varepsilon)\right)_{\varepsilon}$ reads

$$\mathbf{sing}\left[\left(\mathbf{R}^{\rho\sigma\mu\nu(a\ll 1)}(r_{\varepsilon},\varepsilon)\mathbf{R}_{\rho\sigma\mu\nu}^{(a\ll 1)}(r_{\varepsilon},\varepsilon)\right)_{\varepsilon}\right] =_{\widetilde{\mathbb{R}}} \left(\frac{\varepsilon^4}{4(r_{E_+}(\theta))^4\left[\varepsilon^2 + (r_{\varepsilon} - 2m)^2\right]^3}\right)_{\varepsilon}. \qquad (4.1.13)$$

**Remark 4.1.6**.Note that from Eq.(4.1.11)-Eq.(4.1.13) at outer ergosurfaces $(r_{\varepsilon})_{\varepsilon} = r_{E_+}(\theta)$, (see Pic.4.1) follows that

$$\mathbf{sing}\left[(\mathbf{R}^{a\ll 1}(r_{\varepsilon},\varepsilon))_{\varepsilon}\right]\Big|_{(r_{\varepsilon})_{\varepsilon}=r_{E+}(\theta)} =_{\widetilde{\mathbb{R}}} r_{E_+}^{-1}(\theta)(\varepsilon^{-1})_{\varepsilon} \approx_{\widetilde{\mathbb{R}}} \infty. \qquad (4.1.14)$$

and

$$\mathbf{sing}\left[\left(\mathbf{R}^{\mu\nu(a\ll 1)}(r_{\varepsilon},\varepsilon)\mathbf{R}_{\mu\nu}^{(a\ll 1)}(r_{\varepsilon},\varepsilon)\right)_{\varepsilon}\right]\Big|_{(r_{\varepsilon})_{\varepsilon}=r_{E+}(\theta)} =_{\widetilde{\mathbb{R}}} r_{E_+}^{-4}(\theta)(\varepsilon^{-2})_{\varepsilon} \approx_{\widetilde{\mathbb{R}}} \infty,$$
$$\mathbf{sing}\left[\left(\mathbf{R}^{\rho\sigma\mu\nu(a\ll 1)}(r_{\varepsilon},\varepsilon)\mathbf{R}_{\rho\sigma\mu\nu}^{(a\ll 1)}(r_{\varepsilon},\varepsilon)\right)_{\varepsilon}\right]\Big|_{(r_{\varepsilon})_{\varepsilon}=r_{E+}(\theta)} =_{\widetilde{\mathbb{R}}} r_{E_+}^{-4}(\theta)(\varepsilon^{-2})_{\varepsilon} \approx_{\widetilde{\mathbb{R}}} \infty. \qquad (4.1.15)$$

Let $(\mathbf{R}^{a<m}(r_{\varepsilon},\varepsilon))_{\varepsilon}$ be Colombeau generalized curvature scalar $(\mathbf{R}^{a<m}(r_{\varepsilon},\varepsilon))_{\varepsilon}$ corresponding to the metric (4.1.8) with $a < m$. We let now that

$$\Delta_{\varepsilon} = A_{\varepsilon}(B_{\varepsilon} + C_{\varepsilon}) = A_{\varepsilon}\rho_{\varepsilon}^2\Delta_{a,\varepsilon}^{-1}, B_{\varepsilon}r_{\varepsilon}^2 = \rho_{\varepsilon}^2,$$
$$B_{\varepsilon} + C_{\varepsilon} = \rho_{\varepsilon}^2\Delta_{a,\varepsilon}^{-1}, A_{\varepsilon} = \Xi_{\varepsilon}^{+}(r,\theta), D_{\varepsilon} = 0. \qquad (4.1.16)$$

From Eq.(4.1.8),Eq.(4.1.16) and Eq.(A.1)-Eq.(A.2) we obtain

$$(\mathbf{R}^{a<m}(r_\varepsilon,\varepsilon))_\varepsilon = \left(\frac{A_\varepsilon}{\Delta_\varepsilon}\left[\frac{2}{r_\varepsilon}\left(-2\frac{A'_\varepsilon}{A_\varepsilon}-3\frac{B'_\varepsilon}{B_\varepsilon}+\frac{\Delta'_\varepsilon}{\Delta_\varepsilon}\right)+\frac{2}{r^2}\frac{C_\varepsilon}{B_\varepsilon}-\frac{A''_\varepsilon}{A_\varepsilon}-2\frac{B''_\varepsilon}{B_\varepsilon}\right.\right.$$
$$\left.\left.+\frac{1}{2}\left(\frac{B'_\varepsilon}{B_\varepsilon}\right)^2-2\frac{A'_\varepsilon B'_\varepsilon}{A_\varepsilon B_\varepsilon}+\left(\frac{1}{2}\frac{A'_\varepsilon}{A_\varepsilon}+\frac{B'_\varepsilon}{B_\varepsilon}\right)\frac{\Delta'_\varepsilon}{\Delta_\varepsilon}\right]\right)_\varepsilon =$$
$$\left(\frac{\Delta_{a,\varepsilon}}{\rho_\varepsilon^2}\left[\frac{2}{r_\varepsilon}\left(-2\frac{A'_\varepsilon}{A_\varepsilon}-3\frac{B'_\varepsilon}{B_\varepsilon}+\frac{\Delta'_\varepsilon}{\Delta_\varepsilon}\right)+\frac{2}{r^2}\frac{C_\varepsilon}{B_\varepsilon}-\frac{A''_\varepsilon}{A_\varepsilon}-2\frac{B''_\varepsilon}{B_\varepsilon}\right.\right.$$
$$\left.\left.+\frac{1}{2}\left(\frac{B'_\varepsilon}{B_\varepsilon}\right)^2-2\frac{A'_\varepsilon B'_\varepsilon}{A_\varepsilon B_\varepsilon}+\left(\frac{1}{2}\frac{A'_\varepsilon}{A_\varepsilon}+\frac{B'_\varepsilon}{B_\varepsilon}\right)\frac{\Delta'_\varepsilon}{\Delta_\varepsilon}\right]\right)_\varepsilon . \tag{4.1.17}$$

Note that

$$\frac{\partial}{\partial r_\varepsilon}\sqrt{(r_\varepsilon - r_{E_+}(\theta))^2+\varepsilon^2} = \frac{r_\varepsilon - r_{E_+}(\theta)}{\sqrt{(r_\varepsilon - r_{E_+}(\theta))^2+\varepsilon^2}}$$
$$\frac{\partial^2}{\partial r_\varepsilon^2}\sqrt{(r_\varepsilon - r_{E_+})^2+\varepsilon^2} = \frac{\varepsilon^2}{\left[(r_\varepsilon - r_{E_+})^2+\varepsilon^2\right]^{\frac{3}{2}}}. \tag{4.1.18}$$

From Eq.(4.1.17) and Eq.(4.1.18) one obtains that main singular part $\mathbf{sing}[(\mathbf{R}^{a<m}(r_\varepsilon,\varepsilon))_\varepsilon]$ of the Colombeau generalized curvature scalar $(\mathbf{R}^{a<m}(r_\varepsilon,\varepsilon))_\varepsilon$ corresponding to the metric (4.1.8) (mod nonsingular multiplier) reads

$$\mathbf{sing}[(\mathbf{R}^{a<m}(r_\varepsilon,\varepsilon))_\varepsilon] =_{\widetilde{\mathbb{R}}} \left(\frac{\varepsilon^2}{\left[(r_\varepsilon - r_{E_+}(\theta))^2+\varepsilon^2\right]^2}\right)_\varepsilon . \tag{4.1.19}$$

**Remark 4.1.7.(I)** Let $\left(\mathbf{R}^{\mu\nu(a<m)}(r_\varepsilon,\varepsilon)\mathbf{R}^{(a<m)}_{\mu\nu}(r_\varepsilon,\varepsilon)\right)_\varepsilon$ be Colombeau generalized quadratic scalar $(\mathbf{R}^{\mu\nu}(r_\varepsilon,\varepsilon)\mathbf{R}_{\mu\nu}(r_\varepsilon,\varepsilon))_\varepsilon$ corresponding to the metric (4.1.8) with $a < m$. From Eq.(4.1.8)-Eq.(4.1.16) and Eq.(A.1)-Eq.(A.2) one obtains that main singular part

$$\mathbf{sing}\left[\left(\mathbf{R}^{\mu\nu(a<m)}(r_\varepsilon,\varepsilon)\mathbf{R}^{(a<m)}_{\mu\nu}(r_\varepsilon,\varepsilon)\right)_\varepsilon\right]$$

of the Colombeau generalized quadratic scalar $\left(\mathbf{R}^{\mu\nu(a<m)}(r_\varepsilon,\varepsilon)\mathbf{R}^{(a<m)}_{\mu\nu}(r_\varepsilon,\varepsilon)\right)_\varepsilon$ reads

$$\mathbf{sing}\left[\left(\mathbf{R}^{\mu\nu(a<m)}(r_\varepsilon,\varepsilon)\mathbf{R}^{(a<m)}_{\mu\nu}(r_\varepsilon,\varepsilon)\right)_\varepsilon\right] =_{\widetilde{\mathbb{R}}} \left(\frac{\varepsilon^4}{4(r_{E_+}(\theta))^4\left[\varepsilon^2+(r_\varepsilon-2m)^2\right]^3}\right)_\varepsilon . \tag{4.1.20}$$

**(II)** Let $\left(\mathbf{R}^{\rho\sigma\mu\nu(a<m)}(r_\varepsilon,\varepsilon)\mathbf{R}^{(a<m)}_{\rho\sigma\mu\nu}(r_\varepsilon,\varepsilon)\right)_\varepsilon$ be Colombeau generalized quadratic scalar $(\mathbf{R}^{\rho\sigma\mu\nu}(r_\varepsilon,\varepsilon)\mathbf{R}_{\rho\sigma\mu\nu}(r_\varepsilon,\varepsilon))_\varepsilon$ corresponding to the metric (4.1.8) with $a < m$. From Eq.(4.1.8), Eq.(4.1.16) and Eq.(A.1)-Eq.(A.2) one obtains that main singular part

$$\mathbf{sing}\left[\left(\mathbf{R}^{\rho\sigma\mu\nu(a<m)}(r_\varepsilon,\varepsilon)\mathbf{R}^{(a<m)}_{\rho\sigma\mu\nu}(r_\varepsilon,\varepsilon)\right)_\varepsilon\right]$$

of the Colombeau generalized quadratic scalar $\left(\mathbf{R}^{\rho\sigma\mu\nu(a<m)}(r_\varepsilon,\varepsilon)\mathbf{R}^{(a<m)}_{\rho\sigma\mu\nu}(r_\varepsilon,\varepsilon)\right)_\varepsilon$ reads

$$\mathbf{sing}\left[\left(\mathbf{R}^{\rho\sigma\mu\nu(a<m)}(r_\varepsilon,\varepsilon)\mathbf{R}^{(a<m)}_{\rho\sigma\mu\nu}(r_\varepsilon,\varepsilon)\right)_\varepsilon\right] =_{\widetilde{\mathbb{R}}} \left(\frac{\varepsilon^4}{4(r_{E_+}(\theta))^4\left[\varepsilon^2+(r_\varepsilon-2m)^2\right]^3}\right)_\varepsilon . \tag{4.1.21}$$

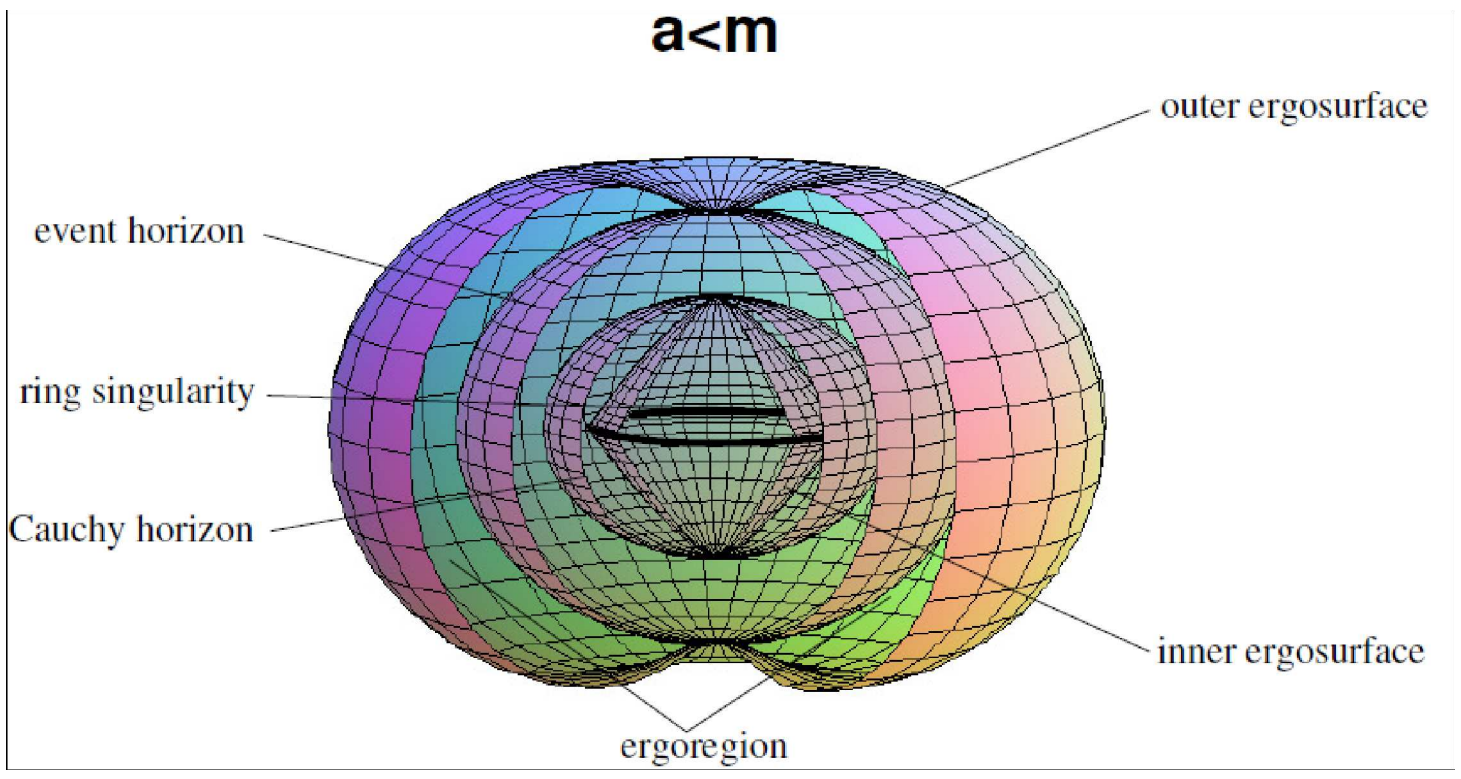

Pic.4.1.1.Ergosurface,horizon,and singularity for slow
Kerr black hole.

**II.Critical Kerr gravitational singularity**: $a = m$.

Note that in contrast with full distributional Kerr spacetime the case of the critical Kerr gravitational singularity considered in this subsection (see Remark 4.1.3) not principal different in comparison with a case of the slow Kerr gravitational singularity considered above. In particular the Eqs.(4.1.19)-(4.1.21) holds with $r_{E_+}(\theta)$ given by Eq.(4.1.22)

$$r_{E_+}(\theta) = m + \sqrt{m^2 - a^2\cos^2\theta} = m(1 + \sin^2\theta). \qquad (4.1.22)$$

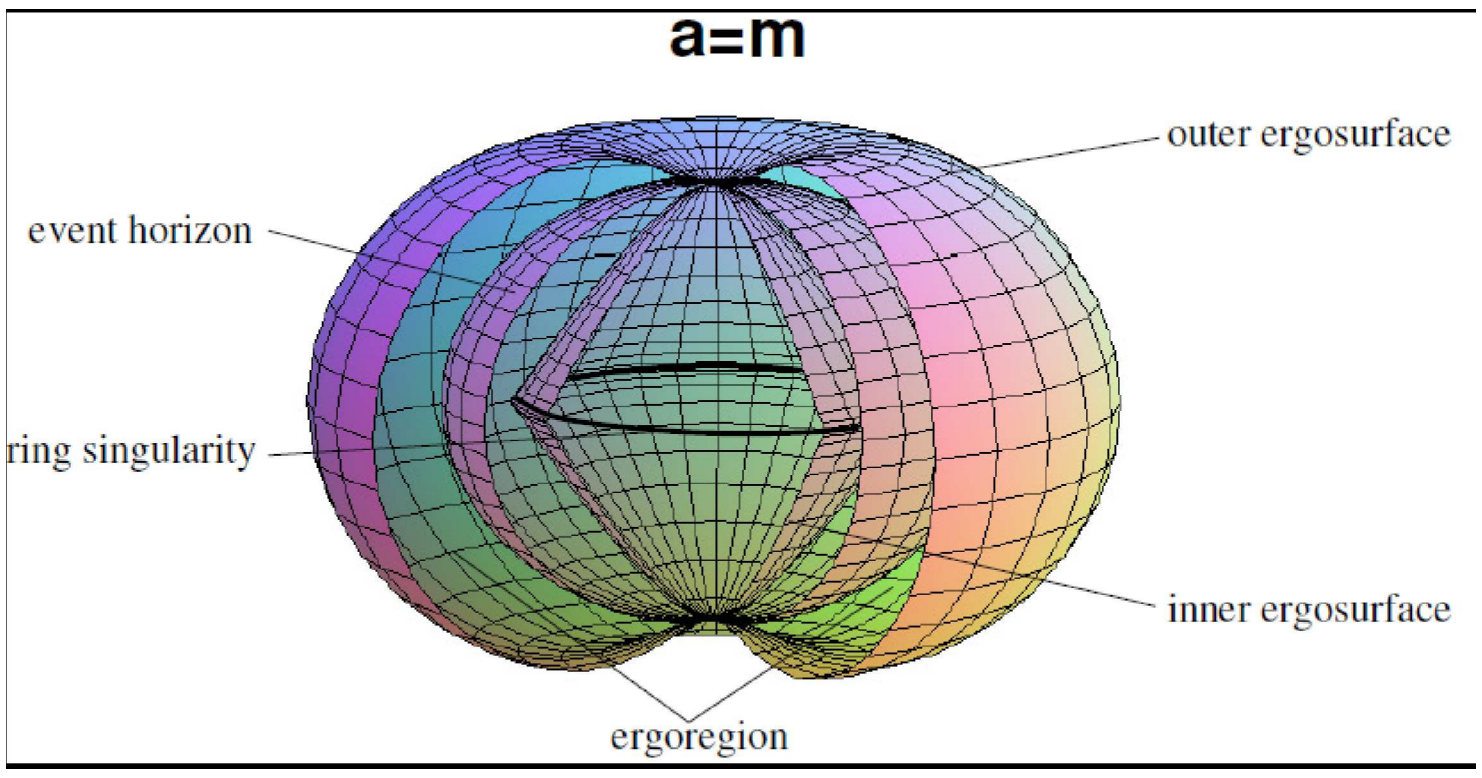

Pic.4.1.2.Ergosurface,horizon,and singularity for
critical Kerr black hole.

**III.Fast Kerr gravitational singularity**: $a > m$.

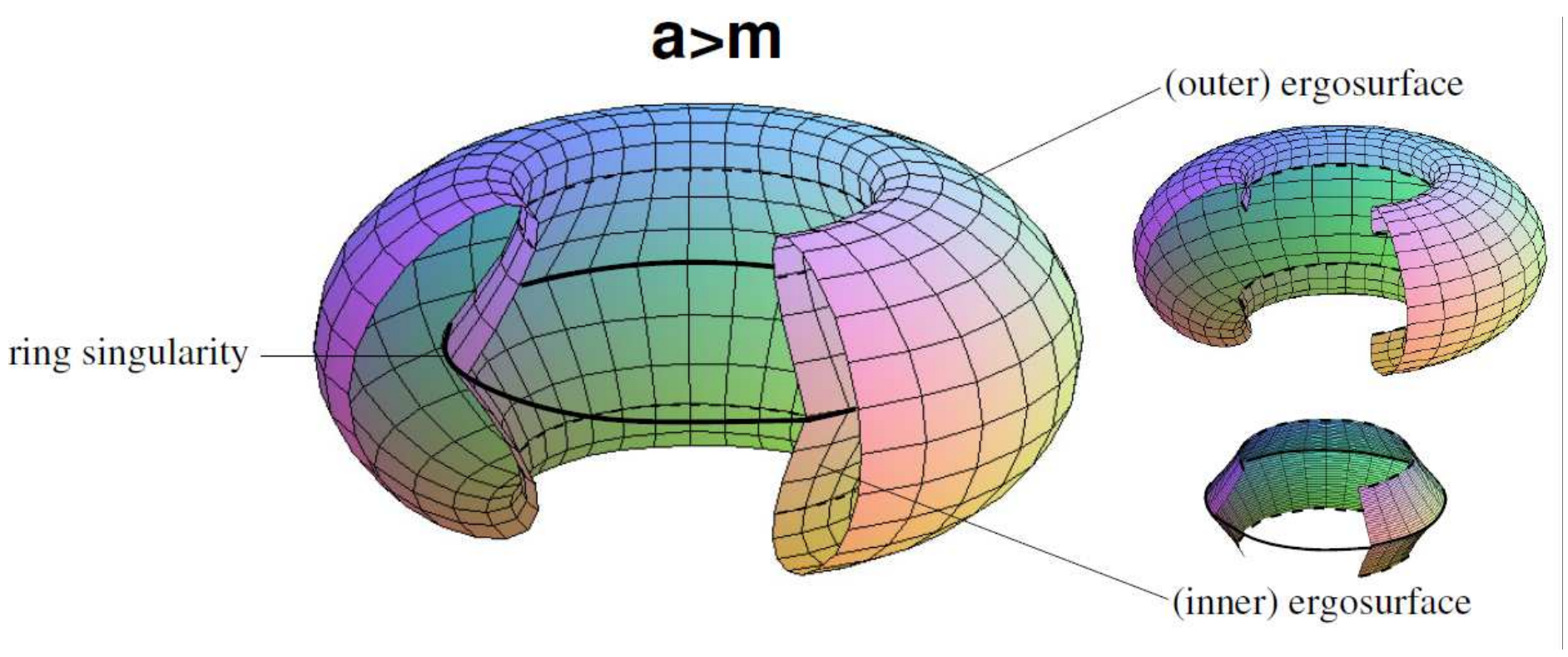

Pic.4.1.3.Ergosurface,horizon,and singularity for fast

Kerr black hole.

Let $\Xi_\eta(\theta)$ be a submanifold given by equations (i) $\phi$ = const and (ii) $m^2 - a^2\cos^2\theta \geq 0$, i.e.

$$\cos^2\theta \leq \frac{m^2}{a^2} = \eta. \tag{7.2.23}$$

Let $\Theta_\eta$ be a set $\Theta_\eta = \{\theta|\cos^2\theta \leq \eta\}$ and let $\chi(\theta,\eta)$ be the indicator function of the set $\Theta_\eta$,i.e. $\chi(\theta,\eta)$ is the function defined to be identically 1 on $\Theta_\eta$, and is 0 elsewhere.
We introduce now the following regularized (above ergosurface $r = r_{E_+}(\theta), \theta \in \Theta_\eta$) quantity

$$\Xi_\varepsilon^+(r_\varepsilon,\theta,\eta) = \frac{\chi(\theta,\eta)(r_\varepsilon - r_{E_-}(\theta))\sqrt{(r_\varepsilon - r_{E_+}(\theta))^2 + \varepsilon^2}}{\rho_\varepsilon^2(r_\varepsilon)}, \tag{4.1.25}$$

where $\rho_\varepsilon^2 = \rho_\varepsilon^2(r_\varepsilon) = r_\varepsilon^2 + a^2\cos^2\theta, \varepsilon \in (0,\delta],\ r_\varepsilon \geq r_{E_+}(\theta) > 0$. Thus Colombeau generalized metric (above ergosurface $r = r_{E_+}(\theta)$) corresponding to classical Kerr metric (4.1.1) reads

$$\begin{gathered}(ds_\varepsilon^{+2})_\varepsilon = -\chi(\theta,\eta)[(\Xi_\varepsilon(r_\varepsilon,\theta,\eta))_\varepsilon]dt^2 - \left[\left(\frac{4mr_\varepsilon a\sin^2\theta}{\rho_\varepsilon^2}\right)_\varepsilon\right]dtd\phi + \\ \left[\left(\frac{\rho_\varepsilon^2}{\Delta_{a,\varepsilon}}\right)_\varepsilon\right][(dr_\varepsilon^2)_\varepsilon] + (\rho_\varepsilon^2)_\varepsilon d\theta^2 + \\ \left(r_\varepsilon^2 + a^2 + \frac{2mra^2\sin^2\theta}{\rho_\varepsilon^2}\right)_\varepsilon \sin^2\theta d\phi^2.\end{gathered} \tag{4.1.26}$$

**Remark 4.1.8**.Note that we will be consider the distributional Kerr spacetime not as full distributionel BH with Colombeau generalized metric (4.1.7), but only as gravitational singularity located on submanifold $\cup_{\theta\in\Theta_\eta}\Xi_\eta(\theta)$ which coincide with an part of the outer ergosurface of classical Kerr spacetime. In accordance with Eq.(4.1.11),Eq.(4.1.19) and Eq.(4.1.20) submanifold $\cup_{\theta\in\Theta_\eta}\Xi_\eta(\theta)$ presented the singular boundary of distributional spacetime with Colombeau generalized metric (4.1.26).

**Remark 4.1.9**.Let $\tilde{\tilde{\Xi}}_\eta(\theta)$ be a submanifold given by equations (i) $\phi$ = const and (ii) $\cos^2\theta \leq \eta$, then Colombeau generalized metric (4.1.26) restricted on submanifold $\cup_{\theta\in\Theta_\eta}\tilde{\tilde{\Xi}}_\eta(\theta)$ reads

$$(ds_\varepsilon^{+2})_\varepsilon = -\chi(\theta,\eta)[(\Xi_\varepsilon(r_\varepsilon,\theta,\eta))_\varepsilon]dt^2 + \left[\left(\frac{\rho_\varepsilon^2}{\Delta_{a,\varepsilon}}\right)_\varepsilon\right][(dr_\varepsilon^2)_\varepsilon] + [(\rho_\varepsilon^2)_\varepsilon]d\theta^2. \tag{4.1.27}$$

Note that Colombeau generalized metric (7.2.27) nondegenerate on outer ergosurfaces $(r_\varepsilon)_\varepsilon = r_{E_+}(\theta)$, see Pic.4.1.3.
From Eq.(4.1.27) and and Eq.(A.1)-Eq.(A.2) one obtains that main singular part $\mathbf{sing}[(\mathbf{R}^{a>m}(r_\varepsilon,\varepsilon))_\varepsilon]$ of the Colombeau generalized curvature scalar $(\mathbf{R}^{a>m}(r_\varepsilon,\varepsilon))_\varepsilon$

corresponding to the metric (4.1.27) (mod nonsingular multiplier) reads

$$\mathbf{sing}[(\mathbf{R}^{a>m}(r_\varepsilon,\varepsilon))_\varepsilon] =_{\widetilde{\mathbb{R}}} \left(\frac{\chi(\theta,\eta)\varepsilon^2}{\left[(r_\varepsilon - r_{E_+}(\theta))^2 + \varepsilon^2\right]^2}\right)_\varepsilon. \qquad (4.1.28)$$

**Remark 4.1.10.(I)** Let $\left(\mathbf{R}^{\mu\nu(a>m)}(r_\varepsilon,\varepsilon)\mathbf{R}^{(a>m)}_{\mu\nu}(r_\varepsilon,\varepsilon)\right)_\varepsilon$ be Colombeau generalized quadratic scalar $(\mathbf{R}^{\mu\nu}(r_\varepsilon,\varepsilon)\mathbf{R}_{\mu\nu}(r_\varepsilon,\varepsilon))_\varepsilon$ corresponding to the metric (4.1.27) with $a > m$. From Eq.(4.1.27) and Eq.(A.1)-Eq.(A.2) one obtains singular part

$$\mathbf{sing}\left[\left(\mathbf{R}^{\mu\nu(a>m)}(r_\varepsilon,\varepsilon)\mathbf{R}^{(a>m)}_{\mu\nu}(r_\varepsilon,\varepsilon)\right)_\varepsilon\right]$$

of the Colombeau generalized quadratic scalar $\left(\mathbf{R}^{\mu\nu(a>m)}(r_\varepsilon,\varepsilon)\mathbf{R}^{(a>m)}_{\mu\nu}(r_\varepsilon,\varepsilon)\right)_\varepsilon$ reads

$$\mathbf{sing}\left[\left(\mathbf{R}^{\mu\nu(a>m)}(r_\varepsilon,\varepsilon)\mathbf{R}^{(a>m)}_{\mu\nu}(r_\varepsilon,\varepsilon)\right)_\varepsilon\right] =_{\widetilde{\mathbb{R}}} \left(\frac{\chi(\theta,\eta)\varepsilon^4}{4(r_{E_+}(\theta))^4\left[\varepsilon^2 + (r_\varepsilon - 2m)^2\right]^3}\right)_\varepsilon. \qquad (4.1.29)$$

**(II)** Let $\left(\mathbf{R}^{\rho\sigma\mu\nu(a>m)}(r_\varepsilon,\varepsilon)\mathbf{R}^{(a>m)}_{\rho\sigma\mu\nu}(r_\varepsilon,\varepsilon)\right)_\varepsilon$ be Colombeau generalized quadratic scalar $(\mathbf{R}^{\rho\sigma\mu\nu}(r_\varepsilon,\varepsilon)\mathbf{R}_{\rho\sigma\mu\nu}(r_\varepsilon,\varepsilon))_\varepsilon$ corresponding to the metric (4.1.27) with $a > m$. From Eq.(4.1.27) and Eq.(A.1)-Eq.(A.2) one obtains singular part

$$\mathbf{sing}\left[\left(\mathbf{R}^{\rho\sigma\mu\nu(a>m)}(r_\varepsilon,\varepsilon)\mathbf{R}^{(a>m)}_{\rho\sigma\mu\nu}(r_\varepsilon,\varepsilon)\right)_\varepsilon\right]$$

of the Colombeau generalized quadratic scalar $\left(\mathbf{R}^{\rho\sigma\mu\nu(a>m)}(r_\varepsilon,\varepsilon)\mathbf{R}^{(a>m)}_{\rho\sigma\mu\nu}(r_\varepsilon,\varepsilon)\right)_\varepsilon$ reads

$$\mathbf{sing}\left[\left(\mathbf{R}^{\rho\sigma\mu\nu(a>m)}(r_\varepsilon,\varepsilon)\mathbf{R}^{(a>m)}_{\rho\sigma\mu\nu}(r_\varepsilon,\varepsilon)\right)_\varepsilon\right] =_{\widetilde{\mathbb{R}}} \left(\frac{\chi(\theta,\eta)\varepsilon^4}{4(r_{E_+}(\theta))^4\left[\varepsilon^2 + (r_\varepsilon - 2m)^2\right]^3}\right)_\varepsilon. \qquad (4.1.30)$$

.

# 5.Novel explanation of the Active Galactic Nuclei.The Power Source of Quasars as a result of vacuum polarization by the gravitational singularities on BHs horizon.

Accretion of gas onto the supermassive Kerr black holes lurking at the center of active galactic nuclei (AGN) gives rise to powerful relativistic jets.

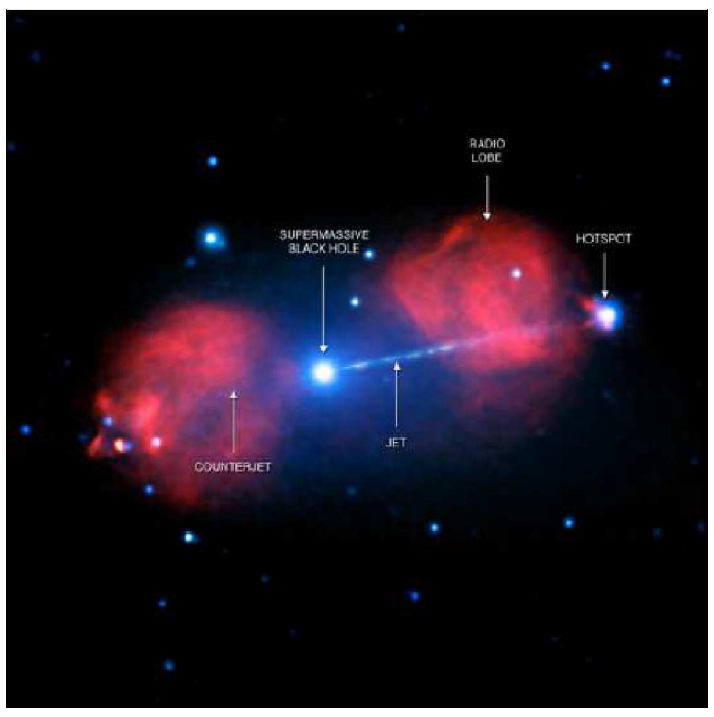

Fig.5.1.Jet from Black Hole in a Galaxy Pictor A
The active galaxy Pictor A lies nearly 500 million
light-years from Earth and contains a
supermassive black hole at its centre.
This is a composite radio and X-ray image.

We remind that in the standard theory of an accretion disk around a black hole it is assumed that there is no coupling between the disk and the central black hole [51]. However, in the presence of a magnetic field, a magnetic coupling between the disk and the black hole could exist and play an important role in the balance and transportation of energy and angular momentum.In the absence of the magnetic coupling, the energy source for the radiation of the disk is the gravitational energy of the disk (i.e., the gravitational binding energy between the disk and the black hole). But, if the magnetic coupling exists and the black hole is rotating, the rotational energy of the black hole provides an additional energy source for the radiation of the disk.With the magnetic coupling, the black hole exerts a torque on the disk, which transfers energy and angular momentum between the black hole and the disk. If the black hole rotates faster than the disk, energy and angular momentum are extracted from the black hole and transfered to the disk. The energy deposited into the disk is eventually radiated away by the disk,which will increase the efficiency of the disk and make the disk brighter than usual. If the black hole rotates slower than the disk, energy and angular momentum are transfered from the disk to the black hole, which will lower the efficiency of the disk and make the disk dimmer than usual. Therefore, the magnetic coupling between the black hole and the disk has important effects on the radiation properties of the disk [52]-[53].

The current paradigm for AGN is that their radio emission is explained by synchrotron radiation from relativistic electrons that are Doppler boosted through bulk motion. In this model, the intrinsic brightness temperatures cannot exceed $10^{11}$ to $10^{12}$ K [55]. Typical Doppler boosting is expected to be able to raise this temperature by a factor of 10.

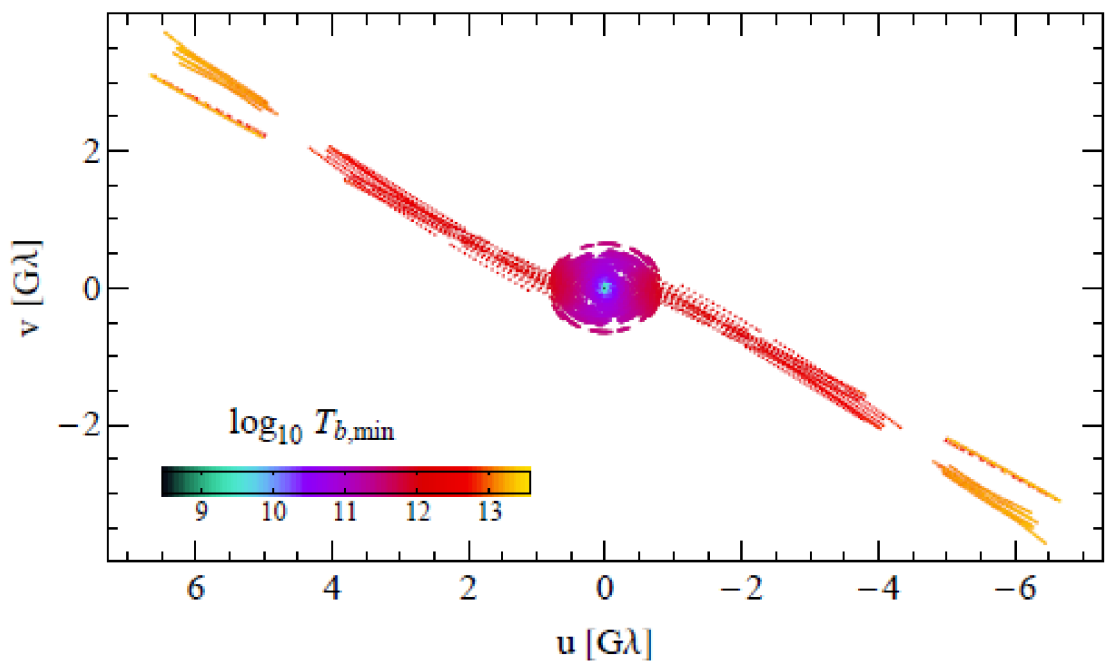

Fig.5.2.Fourier coverage (uv-coverage) of the fringe fitted data (i.e.,reliable fringe detections) of the Radio Astron observations of BL Lac on 2013 November 10-11at 22 GHz. Color marks the lower limit of observed brightness temperature obtained from visibility amplitudes. Adapted from [54].

The observed brightness temperature of the most compact structures in BL Lac, constrained by baselines longer than $5.3G\lambda$, must indeed exceed $2 \times 10^{13}$K and can reach as high as ~ $3 \times 10^{14}$K [55]. As follows from Fig. 7.2, these visibilities correspond to the structural scales of $30 - 40$ $\mu as$ oriented along position angles of $25^\circ - 30^\circ$. These values are indeed close to the width of the inner jet and the normal to its direction.The observed, $T_{b,obs}$, and intrinsic, $T_{b,int}$, brightness temperatures are related by

$$T_{b,obs} = \delta(1+z)^{-1}T_{b,int} \tag{5.1}$$

with $\delta = 7.2$. The estimeted by (4.1.1) a lower limit of the intrinsic brightness temperature in the core component of our Radio Astron observations of $T_{b,int} > 2.910^{12}$ K. It is commonly considered that inverse Compton losses limit the intrinsic brightness temperature for incoherent synchrotron sources, such as AGN, to about $10^{12}$K [53].In case of a strong flare, the "Compton catastrophe" is calculated to take about one day to drive the brightness temperature below $10^{12}$K [53]. The estimated lower limit for the intrinsic brightness temperature of the core in the Radio Astron image of $T_{b,int} > 2.910^{12}$K is therefore more than an order of magnitude larger than the equipartition brightness temperature limit established in [55] and at least several times larger than the limit established by inverse Compton cooling.

**Remark 5.2.1**. Note that if the estimate of the maximum brightness temperature given in [53], is closer to actual values, it would imply $T_{b;int}5 \times 10^{13}$K. This is difficult to reconcile with current incoherent synchrotron emission models from relativistic electrons, requiring alternative models such as emission from relativistic protons.

**Remark 5.2.2**. However the proton, as we know, is 1836 times heavier than an electron and absolutely huge energy is required to accelerated it to sublight speed.

**Remark 5.2.3**.These alternative models such as emission from relativistic protons can be suported by semiclassical gravity effect finds its roots in the singular behavior of quantum fields on curved distributional spacetimes presented by rotating gravitational singularities.

## 6mm . **Conclusion** 3mm

On a Riemannian or a semi-Riemannian manifold, the metric determines invariants like the Levi-Civita connection and the Riemann curvature. If the metric becomes degenerate (as in singular semi-Riemannian geometry), these constructions no longer work, because they are based on the inverse of the metric, and on related operations like the contraction between covariant indices. In order to avoid these difficultnes distribytional geometry by using Colombeau generalized functions
cite: Colombeau84Parker79VickersWilson98VickersWilson99Vickers99GerochTraschen87
cite: BalasinNachbagauer93BalasinNachbagauer94 [3]-[10].In authors papers cite: Foukzon15
cite: FoukzonPotapovMenkova16Vickers12[18]-[19] appropriate generalization of classical GR based on Colombeau generalized functions is proposed.

Such generalization of classical GR based on appropriate generalization of the Einstein equivalence principle (GEEP) mentioned above in subsection 2.3. Using Rindler distributional geometry Unruh effect revisited. We pointed out that GEEP avoid the contradiction which was mentioned by Z.Merali in paper cite: Merali13[47], and therefore Polchinski's account cite: AlmheiriMarolfPolchinskiSully13[1] doesn't violate of the Einstein equivalence principle.

# 6mm . Appendix 3mm

## 5mm . Appendix A1 ref: lem:conv_K-M_ESS 2mm

Let us introduce now Colombeau generalized metric which has the form

$$\begin{cases} (ds_\varepsilon^2)_\varepsilon = -(A_\varepsilon(r)(dx^0)^2)_\varepsilon - 2(D_\varepsilon(r)dx^0dr)_\varepsilon + ((B_\varepsilon(r)+C_\varepsilon(r))(dr)^2)_\varepsilon \\ +(B_\varepsilon(r)r^2[(d\theta)^2+\sin^2\theta(d\phi)^2])_\varepsilon, r\in\mathbb{R}. \end{cases} \quad (A1.1)$$

The Colombeau scalars $(\mathbf{R}(r,\varepsilon))_\varepsilon, (\mathbf{R}^{\mu\nu}(r,\varepsilon)\mathbf{R}_{\mu\nu}(r,\varepsilon))_\varepsilon$ and $(\mathbf{R}^{\rho\sigma\mu\nu}(r,\varepsilon)\mathbf{R}_{\rho\sigma\mu\nu}(r,\varepsilon))_\varepsilon$, in terms of Colombeau generalized functions $(A_\varepsilon(r))_\varepsilon, (B_\varepsilon(r))_\varepsilon, (C_\varepsilon(r))_\varepsilon, (D_\varepsilon(r))_\varepsilon$ are expressed as

$$\begin{gathered}
(\mathbf{R}(r,\varepsilon))_\varepsilon = \left(\frac{A_\varepsilon}{\Delta_\varepsilon}\left[\frac{2}{r}\left(-2\frac{A'_\varepsilon}{A_\varepsilon} - 3\frac{B'_\varepsilon}{B_\varepsilon} + \frac{\Delta'_\varepsilon}{\Delta_\varepsilon}\right) + \frac{2}{r^2}\frac{A_\varepsilon C_\varepsilon + D_\varepsilon^2}{A_\varepsilon B_\varepsilon} - \frac{A''_\varepsilon}{A_\varepsilon} - 2\frac{B''_\varepsilon}{B_\varepsilon}\right.\right. \\
\left.\left. +\frac{1}{2}\left(\frac{B'_\varepsilon}{B_\varepsilon}\right)^2 - 2\frac{A'_\varepsilon B'_\varepsilon}{A_\varepsilon B_\varepsilon} + \left(\frac{1}{2}\frac{A'_\varepsilon}{A_\varepsilon} + \frac{B'_\varepsilon}{B_\varepsilon}\right)\frac{\Delta'_\varepsilon}{\Delta_\varepsilon}\right]\right)_\varepsilon, \\
(\mathbf{R}^{\mu\nu}(r,\varepsilon)\mathbf{R}_{\mu\nu}(r,\varepsilon))_\varepsilon = \left(\frac{A_\varepsilon^2}{\Delta_\varepsilon^2}\left(\frac{1}{2}\frac{A''_\varepsilon}{A_\varepsilon} - \frac{1}{4}\frac{A'_\varepsilon\Delta'_\varepsilon}{A_\varepsilon\Delta_\varepsilon} + \frac{1}{2}\frac{A'_\varepsilon B'_\varepsilon}{A_\varepsilon B_\varepsilon} + \frac{1}{r}\frac{A'_\varepsilon}{A_\varepsilon}\right)^2\right)_\varepsilon + \\
+2\left(\frac{A_\varepsilon^2}{\Delta_\varepsilon^2}\left[\frac{1}{r}\left(\frac{1}{2}\frac{\Delta'_\varepsilon}{\Delta_\varepsilon} - \frac{A'_\varepsilon}{A_\varepsilon} - 2\frac{B'_\varepsilon}{B_\varepsilon}\right) + \frac{1}{r^2}\frac{A_\varepsilon C_\varepsilon + D_\varepsilon^2}{A_\varepsilon B_\varepsilon} - \frac{1}{2}\frac{A'_\varepsilon B'_\varepsilon}{A_\varepsilon B_\varepsilon} - \right.\right. \\
\left.\left. -\frac{1}{2}\frac{B''_\varepsilon}{B_\varepsilon} + \frac{1}{4}\frac{B'_\varepsilon\Delta'_\varepsilon}{B_\varepsilon\Delta_\varepsilon}\right]^2\right)_\varepsilon + \\
\left(\frac{A_\varepsilon^2}{\Delta_\varepsilon^2}\left[\frac{1}{2}\frac{A''_\varepsilon}{A_\varepsilon} - \frac{1}{4}\frac{A'_\varepsilon\Delta'_\varepsilon}{A_\varepsilon\Delta_\varepsilon} + \frac{1}{2}\frac{A'_\varepsilon B'_\varepsilon}{A_\varepsilon B_\varepsilon} + \frac{B''_\varepsilon}{B_\varepsilon} - \frac{1}{2}\left(\frac{B'_\varepsilon}{B_\varepsilon}\right)^2\right.\right. \\
\left.\left. -\frac{1}{2}\frac{B'_\varepsilon\Delta'_\varepsilon}{B_\varepsilon\Delta_\varepsilon} + \frac{1}{r}\left(\frac{A'_\varepsilon}{A_\varepsilon} - \frac{\Delta'_\varepsilon}{\Delta_\varepsilon} + 2\frac{B'_\varepsilon}{B_\varepsilon}\right)\right]^2\right)_\varepsilon, \qquad (A1.2) \\
(\mathbf{R}^{\rho\sigma\mu\nu}(r,\varepsilon)\mathbf{R}_{\rho\sigma\mu\nu}(r,\varepsilon))_\varepsilon = \\
\left(\frac{A_\varepsilon^2}{\Delta_\varepsilon^2}\left(\frac{A''_\varepsilon}{A_\varepsilon} - \frac{1}{2}\frac{A'_\varepsilon\Delta'_\varepsilon}{A_\varepsilon\Delta_\varepsilon}\right)^2 + 2\frac{A_\varepsilon^2}{\Delta_\varepsilon^2}\left(\frac{1}{r}\frac{A'_\varepsilon}{A_\varepsilon} + \frac{1}{2}\frac{A'_\varepsilon B'_\varepsilon}{A_\varepsilon B_\varepsilon}\right)^2\right. \\
+4\frac{A_\varepsilon^2}{\Delta_\varepsilon^2}\left[\frac{1}{r}\frac{B'_\varepsilon}{B_\varepsilon} - \frac{1}{r^2}\frac{A_\varepsilon C_\varepsilon + D_\varepsilon^2}{A_\varepsilon B_\varepsilon} + \frac{1}{4}\left(\frac{B'_\varepsilon}{B_\varepsilon}\right)^2\right]^2 + \\
+2\frac{A_\varepsilon^2}{\Delta_\varepsilon^2}\left[\frac{1}{r}\left(\frac{A'_\varepsilon}{A_\varepsilon} + 2\frac{B'_\varepsilon}{B_\varepsilon} - \frac{\Delta'_\varepsilon}{\Delta_\varepsilon}\right) + \frac{1}{2}\frac{A'_\varepsilon B'_\varepsilon}{A_\varepsilon B_\varepsilon} + \frac{B''_\varepsilon}{B_\varepsilon}\right. \\
\left.\left. -\frac{1}{2}\left(\frac{B'_\varepsilon}{B_\varepsilon}\right)^2 - \frac{1}{2}\frac{B'_\varepsilon\Delta'_\varepsilon}{B_\varepsilon\Delta_\varepsilon}\right]^2\right)_\varepsilon. \\
\Delta_\varepsilon = A_\varepsilon(B_\varepsilon + C_\varepsilon) + D_\varepsilon^2
\end{gathered}$$

**Remark A.1**. Note that the Colombeau scalars $(\mathbf{R}(r,\varepsilon))_\varepsilon, (\mathbf{R}^{\mu\nu}(r,\varepsilon)\mathbf{R}_{\mu\nu}(r,\varepsilon))_\varepsilon$ and $(\mathbf{R}^{\rho\sigma\mu\nu}(r,\varepsilon)\mathbf{R}_{\rho\sigma\mu\nu}(r,\varepsilon))_\varepsilon$ can be extended on Colombeau generalized numbers $\mathbf{r} = [(r_\varepsilon)_\varepsilon] \in \widetilde{\mathbb{R}}$ as corresponding generalized point value (see Definition 1.5.4) by formulas:

$$\begin{gathered}(\mathbf{i})[(\mathbf{R}(\mathbf{r},\varepsilon))_\varepsilon] \triangleq [(\mathbf{R}(r_\varepsilon,\varepsilon))_\varepsilon],\\ (\mathbf{ii})[(\mathbf{R}^{\mu\nu}(\mathbf{r},\varepsilon)\mathbf{R}_{\mu\nu}(\mathbf{r},\varepsilon))_\varepsilon] \triangleq [(\mathbf{R}^{\mu\nu}(r_\varepsilon,\varepsilon)\mathbf{R}_{\mu\nu}(r_\varepsilon,\varepsilon))_\varepsilon]\\ (\mathbf{iii})[(\mathbf{R}^{\rho\sigma\mu\nu}(\mathbf{r},\varepsilon)\mathbf{R}_{\rho\sigma\mu\nu}(\mathbf{r},\varepsilon))_\varepsilon] \triangleq [(\mathbf{R}^{\rho\sigma\mu\nu}(r_\varepsilon,\varepsilon)\mathbf{R}_{\rho\sigma\mu\nu}(r_\varepsilon,\varepsilon))_\varepsilon].\end{gathered} \quad (A1.3)$$

The distributional Mǿller's metric is

$$\begin{gathered}(d_\varepsilon s^2)_\varepsilon = -(A_\varepsilon(x)dt^2)_\varepsilon + dx^2 + dy^2 + dz^2,\\ A_\varepsilon(x) = \left[(a+gx)^2 + \varepsilon^2\right], \varepsilon \in (0,1].\end{gathered} \quad (A1.4)$$

In order to aply formulas (A1.2) directly we chose now $g = g(\theta,\varphi)$, where angles $\theta,\varphi$ correspond to spherical coordinates: $x = r\sin\theta\cos\varphi, y = r\sin\theta\sin\varphi, z = r\cos\theta$. In spherical coordinates we get

$$\begin{gathered}(d_\varepsilon s^2)_\varepsilon = -(A_\varepsilon(r\eta(\theta,\varphi))dt^2)_\varepsilon + dr^2 + r^2 d\Omega^2,\\ d\Omega^2 = d\theta^2 + \sin\theta d\varphi^2, \eta(\theta,\varphi) = \sin\theta\cos\varphi,\\ A_\varepsilon(r\eta) = \left[(a+g\eta r)^2 + \varepsilon^2\right], A'_\varepsilon(r\eta) = 2g\eta(a+g\eta r), A''_\varepsilon(r) = 2g^2\eta^2,\\ \eta = \eta(\theta,\varphi).\end{gathered} \quad (A1.5)$$

We choose now in the formulae (A1.2): $B_\varepsilon(r) = 1, C_\varepsilon(r) = 0$ and $g\eta = g^* = const$, $\sin\theta\cos\varphi \neq 0$ and rewrite Eq.(A1.5) in the following equivalent form

$$\begin{gathered}(d_\varepsilon s^2)_\varepsilon = -(A_\varepsilon(r)dt^2)_\varepsilon + dr^2 + r^2 d\Omega^2,\\ A_\varepsilon(r) = (a+g^* r)^2 + \varepsilon^2.\end{gathered} \quad (A1.6)$$

Note that

$$\Delta_\varepsilon = A_\varepsilon(B_\varepsilon + C_\varepsilon) = A_\varepsilon. \quad (A1.7)$$

From Eq.(A1.5)-Eq.(A1.7) by formulae (A1.2) we get

$$
\begin{gathered}
(\mathbf{R}(r,\varepsilon))_{\varepsilon} = \\
\left(\frac{A_{\varepsilon}}{\Delta_{\varepsilon}}\left[\frac{2}{r}\left(-2\frac{A'_{\varepsilon}}{A_{\varepsilon}} - 3\frac{B'_{\varepsilon}}{B_{\varepsilon}} + \frac{\Delta'_{\varepsilon}}{\Delta_{\varepsilon}}\right) + \frac{2}{r^2}\frac{A_{\varepsilon}C_{\varepsilon} + D^2_{\varepsilon}}{A_{\varepsilon}B_{\varepsilon}} - \frac{A''_{\varepsilon}}{A_{\varepsilon}} - 2\frac{B''_{\varepsilon}}{B_{\varepsilon}}\right.\right. \\
\left.\left. + \frac{1}{2}\left(\frac{B'_{\varepsilon}}{B_{\varepsilon}}\right)^2 - 2\frac{A'_{\varepsilon}B'_{\varepsilon}}{A_{\varepsilon}B_{\varepsilon}} + \left(\frac{1}{2}\frac{A'_{\varepsilon}}{A_{\varepsilon}} + \frac{B'_{\varepsilon}}{B_{\varepsilon}}\right)\frac{\Delta'_{\varepsilon}}{\Delta_{\varepsilon}}\right]\right)_{\varepsilon} = \\
\left(-\frac{2}{r}\frac{A'_{\varepsilon}}{A_{\varepsilon}} - \frac{A''_{\varepsilon}}{A_{\varepsilon}} + \frac{1}{2}\frac{A'^2_{\varepsilon}}{A^2_{\varepsilon}}\right)_{\varepsilon} = \\
-\left(\frac{4g^*(a+g^*r)}{r\left[(a+g^*r)^2+\varepsilon^2\right]} - \frac{2g^{*2}}{(a+g^*r)^2+\varepsilon^2} + \frac{2\widetilde{g}^2(a+g^*r)^2}{\left[(a+g^*r)^2+\varepsilon^2\right]^2}\right)_{\varepsilon} = \\
\left(2\widetilde{g}^2\frac{(a+g^*r)^2+\varepsilon^2-\varepsilon^2}{\left[(a+g^*r)^2+\varepsilon^2\right]^2} - \frac{4g^*(a+g^*r)}{r\left[(a+g^*r)^2+\varepsilon^2\right]}\right. \qquad (A1.8) \\
\left. - \frac{2\widetilde{g}^2}{(a+\widetilde{g}r)^2+\varepsilon^2}\right)_{\varepsilon} = \\
\left(\frac{2\widetilde{g}^2}{(a+\widetilde{g}r)^2+\varepsilon^2} + \frac{-2\widetilde{g}^2\varepsilon^2}{\left[(a+\widetilde{g}r)^2+\varepsilon^2\right]^2} - \right. \\
\left.\frac{4\widetilde{g}(a+\widetilde{g}r)}{r\left[(a+\widetilde{g}r)^2+\varepsilon^2\right]} - \frac{2\widetilde{g}^2}{(a+\widetilde{g}r)^2+\varepsilon^2}\right)_{\varepsilon} = \\
\left(\frac{-2\widetilde{g}^2\varepsilon^2}{\left[(a+\widetilde{g}r)^2+\varepsilon^2\right]^2} - \frac{4\widetilde{g}(a+\widetilde{g}r)}{r\left[(a+\widetilde{g}r)^2+\varepsilon^2\right]}\right)_{\varepsilon}.
\end{gathered}
$$

From Eq.(A1.8) in the limit $a+g^*r \to 0$ we get

$$
(\mathbf{R}(r,\varepsilon))_{\varepsilon} \asymp \left(\frac{-2g^{*2}\varepsilon^2}{\left[(a+g^*r)^2+\varepsilon^2\right]^2}\right)_{\varepsilon}, a+g^*r \asymp 0. \qquad (A1.9)
$$

**Remark A1.3**. Note that: (1) Eq.(1.2.14) in a nice agriment with Eq.(A1.9), see Remark 1.2.2-Remark 1.2.4. (2) For $r$ located beyond horizon, i.e. $a+g^*r \neq 0$ one obtains classical result

$$
(\mathbf{R}(r,\varepsilon))_{\varepsilon} \approx_{\widetilde{\mathbb{R}}} O\left(g^{*2}\right)\left(\varepsilon^2\right)_{\varepsilon} \approx_{\widetilde{\mathbb{R}}} 0, \qquad (A1.10)
$$

see Definition 1.5.2.(i).(3) At horizon $r_{hor} : a+g^*r_{hor} = 0$ from Eq.(A1.9) one obtains nonclassical result

$$
(\mathbf{R}(r_{hor},\varepsilon))_{\varepsilon} \approx_{\widetilde{\mathbb{R}}} O\left(g^{*2}\right)\left(\varepsilon^{-2}\right)_{\varepsilon} \approx_{\widetilde{\mathbb{R}}} \infty, \qquad (A1.11)
$$

see Definition 1.5.2.(ii).

**Remark A1.4**. Let $[(a+g^*r_{\varepsilon})_{\varepsilon}] \approx_{\widetilde{\mathbb{R}}} 0,$ then from Eq.(A1.3) and Eq.(A1.9) we obtain

$$
[(\mathbf{R}(r_{\varepsilon},\varepsilon))_{\varepsilon}] \approx_{\widetilde{\mathbb{R}}} \left[\left(\frac{-2g^{*2}\varepsilon^2}{\left[(a+g^*r_{\varepsilon})^2+\varepsilon^2\right]^2}\right)_{\varepsilon}\right]. \qquad (A1.12)
$$

From Eq.(A1.5)-Eq.(A1.7) by formulae (A1.2) we get

$$
\begin{aligned}
&\left(\mathbf{R}^{\mu\nu}(r,\varepsilon)\mathbf{R}_{\mu\nu}(r,\varepsilon)\right)_{\varepsilon} = \left(\frac{A_{\varepsilon}^{2}}{\Delta_{\varepsilon}^{2}}\left(\frac{1}{2}\frac{A_{\varepsilon}''}{A_{\varepsilon}} - \frac{1}{4}\frac{A_{\varepsilon}'\Delta_{\varepsilon}'}{A_{\varepsilon}\Delta_{\varepsilon}} + \frac{1}{2}\frac{A_{\varepsilon}'B_{\varepsilon}'}{A_{\varepsilon}B_{\varepsilon}} + \frac{1}{r}\frac{A_{\varepsilon}'}{A_{\varepsilon}}\right)^{2}\right)_{\varepsilon} + \\
&+2\left(\frac{A_{\varepsilon}^{2}}{\Delta_{\varepsilon}^{2}}\left[\frac{1}{r}\left(\frac{1}{2}\frac{\Delta_{\varepsilon}'}{\Delta_{\varepsilon}} - \frac{A_{\varepsilon}'}{A_{\varepsilon}} - 2\frac{B_{\varepsilon}'}{B_{\varepsilon}}\right) + \frac{1}{r^{2}}\frac{A_{\varepsilon}C_{\varepsilon} + D_{\epsilon}^{2}}{A_{\varepsilon}B_{\varepsilon}} - \frac{1}{2}\frac{A_{\varepsilon}'B_{\varepsilon}'}{A_{\varepsilon}B_{\varepsilon}} - \right.\right. \\
&\left.\left. -\frac{1}{2}\frac{B_{\varepsilon}''}{B_{\varepsilon}} + \frac{1}{4}\frac{B_{\varepsilon}'\Delta_{\varepsilon}'}{B_{\varepsilon}\Delta_{\varepsilon}}\right]^{2}\right)_{\varepsilon} + \\
&\left(\frac{A_{\varepsilon}^{2}}{\Delta_{\varepsilon}^{2}}\left[\frac{1}{2}\frac{A_{\varepsilon}''}{A_{\varepsilon}} - \frac{1}{4}\frac{A_{\varepsilon}'\Delta_{\varepsilon}'}{A_{\varepsilon}\Delta_{\varepsilon}} + \frac{1}{2}\frac{A_{\varepsilon}'B_{\varepsilon}'}{A_{\varepsilon}B_{\varepsilon}} + \frac{B_{\varepsilon}''}{B_{\varepsilon}} - \frac{1}{2}\left(\frac{B_{\varepsilon}'}{B_{\varepsilon}}\right)^{2}\right.\right. \\
&\left.\left. -\frac{1}{2}\frac{B_{\varepsilon}'\Delta_{\varepsilon}'}{B_{\varepsilon}\Delta_{\varepsilon}} + \frac{1}{r}\left(\frac{A_{\varepsilon}'}{A_{\varepsilon}} - \frac{\Delta_{\varepsilon}'}{\Delta_{\varepsilon}} + 2\frac{B_{\varepsilon}'}{B_{\varepsilon}}\right)\right]^{2}\right)_{\varepsilon} = \\
&\left(\left(\frac{1}{2}\frac{A_{\varepsilon}''}{A_{\varepsilon}} - \frac{1}{4}\frac{A_{\varepsilon}'\Delta_{\varepsilon}'}{A_{\varepsilon}\Delta_{\varepsilon}} + \frac{1}{r}\frac{A_{\varepsilon}'}{A_{\varepsilon}}\right)^{2}\right)_{\varepsilon} + 2\left(\left[\frac{1}{r}\left(\frac{1}{2}\frac{\Delta_{\varepsilon}'}{\Delta_{\varepsilon}} - \frac{A_{\varepsilon}'}{A_{\varepsilon}}\right)\right]^{2}\right)_{\varepsilon} + \\
&\left(\left[\frac{1}{2}\frac{A_{\varepsilon}''}{A_{\varepsilon}} - \frac{1}{4}\frac{A_{\varepsilon}'\Delta_{\varepsilon}'}{A_{\varepsilon}\Delta_{\varepsilon}} + \frac{1}{r}\left(\frac{A_{\varepsilon}'}{A_{\varepsilon}} - \frac{\Delta_{\varepsilon}'}{\Delta_{\varepsilon}}\right)\right]^{2}\right)_{\varepsilon} = \\
&\left(\left[\frac{1}{2}\frac{A_{\varepsilon}''}{A_{\varepsilon}} - \frac{1}{4}\frac{A_{\varepsilon}'^{2}}{A_{\varepsilon}^{2}} + \frac{1}{r}\frac{A_{\varepsilon}'}{A_{\varepsilon}}\right]^{2}\right)_{\varepsilon} + 2\left(\left[-\frac{1}{2r}\frac{A_{\varepsilon}'}{A_{\varepsilon}}\right]^{2}\right)_{\varepsilon} + \\
&+\left(\left[\frac{1}{2}\frac{A_{\varepsilon}''}{A_{\varepsilon}} - \frac{1}{4}\frac{A_{\varepsilon}'^{2}}{A_{\varepsilon}^{2}}\right]^{2}\right)_{\varepsilon} = \\
&\left(\left[\frac{1}{2}\frac{2g^{*2}}{(a+g^{*}r)^{2}+\varepsilon^{2}} - \frac{1}{4}\frac{4g^{*2}(a+g^{*}r)^{2}}{\left[(a+g^{*}r)^{2}+\varepsilon^{2}\right]^{2}}\right.\right. \\
&\left.\left. +\frac{1}{r}\frac{2g^{*}(a+g^{*}r)}{\left[(a+g^{*}r)^{2}+\varepsilon^{2}\right]}\right]^{2}\right)_{\varepsilon} + \\
&+2\left(\left[-\frac{1}{2r}\frac{2g^{*}(a+g^{*}r)}{(a+g^{*}r)^{2}+\varepsilon^{2}} - \frac{1}{r^{2}}\right]^{2}\right)_{\varepsilon} + \\
&\left(\left[\frac{1}{2}\frac{2g^{*2}}{(a+g^{*}r)^{2}+\varepsilon^{2}} - \frac{1}{4}\frac{4g^{2}(a+g^{*}r)^{2}}{\left[(a+g^{*}r)^{2}+\varepsilon^{2}\right]^{2}}\right]^{2}\right)_{\varepsilon} = \qquad (A1.13) \\
&\left(\left[\frac{g^{*2}}{(a+\widetilde{g}r)^{2}+\varepsilon^{2}} - \frac{g^{*2}\left[(a+g^{*}r)^{2}+\varepsilon^{2}-\varepsilon^{2}\right]}{\left[(a+g^{*}r)^{2}+\varepsilon^{2}\right]^{2}}\right.\right. \\
&\left.\left. +\frac{1}{r}\frac{2g^{*}(a+g^{*}r)}{\left[(a+g^{*}r)^{2}+\varepsilon^{2}\right]}\right]\right)_{\varepsilon} + \\
&+2\left(\left[-\frac{1}{2r}\frac{2g^{*}(a+g^{*}r)}{(a+g^{*}r)^{2}+\varepsilon^{2}} - \frac{1}{r^{2}}\right]^{2}\right)_{\varepsilon} + \\
&\left(\left[\frac{g^{*2}}{(a+g^{*}r)^{2}+\varepsilon^{2}} - \frac{g^{*2}\left[(a+g^{*}r)^{2}+\varepsilon^{2}-\varepsilon^{2}\right]}{\left[(a+g^{*}r)^{2}+\varepsilon^{2}\right]^{2}}\right]^{2}\right)_{\varepsilon} = \\
&\left(\left[\frac{g^{*2}}{(a+\widetilde{g}r)^{2}+\varepsilon^{2}} - \frac{g^{*2}\left[(a+g^{*}r)^{2}+\varepsilon^{2}\right]}{\left[(a+g^{*}r)^{2}+\varepsilon^{2}\right]^{2}} + \frac{g^{*2}\varepsilon^{2}}{\left[(a+g^{*}r)^{2}+\varepsilon^{2}\right]^{2}}\right.\right.
\end{aligned}
$$

From Eq.(A1.13) in the limit $a + g^{*}r \to 0$ we get

$$(\mathbf{R}^{\mu\nu}(r,\varepsilon)\mathbf{R}_{\mu\nu}(r,\varepsilon))_{\varepsilon} \asymp \left( \frac{4g^{*4}O(1)\varepsilon^{4}}{\left[(a+g^{*}r)^{2}+\varepsilon^{2}\right]^{4}} \right)_{\varepsilon}, a + g^{*}r \asymp 0. \qquad (A1.14)$$

**Remark A1.5**. At horizon $r_{hor} : a + \widetilde{g}r_{hor} = 0$ from Eq.(A1.14) one obtains nonclassical result

$$(\mathbf{R}^{\mu\nu}(r_{hor},\varepsilon)\mathbf{R}_{\mu\nu}(r_{hor},\varepsilon))_{\varepsilon} \approx_{\widetilde{\mathbb{R}}} O\left(g^{*2}\right)\left(\varepsilon^{-4}\right)_{\varepsilon} \approx_{\widetilde{\mathbb{R}}} \infty, \qquad (A1.15)$$

see Definition 1.5.2.(ii).

**Remark A1.6**. Let $[(a + g^{*}r_{\varepsilon})_{\varepsilon}] \approx_{\widetilde{\mathbb{R}}} 0,$ then from Eq.(A1.3) and Eq.(A1.14) we obtain

$$[(\mathbf{R}^{\mu\nu}(r_{\varepsilon},\varepsilon)\mathbf{R}_{\mu\nu}(r_{\varepsilon},\varepsilon))_{\varepsilon}] \approx_{\widetilde{\mathbb{R}}} \left[ \left( \frac{4g^{*4}O(1)\varepsilon^{4}}{\left[(a+g^{*}r_{\varepsilon})^{2}+\varepsilon^{2}\right]^{4}} \right)_{\varepsilon} \right]. \qquad (A1.16)$$

From Eq.(A1.4)-Eq.(A1.6) by formulae (A1.2) we get

$$(\mathbf{R}^{\rho\sigma\mu\nu}(r,\varepsilon)\mathbf{R}_{\rho\sigma\mu\nu}(r,\varepsilon))_{\varepsilon} =$$

$$\left(\frac{A_{\varepsilon}^{2}}{\Delta_{\varepsilon}^{2}}\left(\frac{A_{\varepsilon}''}{A_{\varepsilon}} - \frac{1}{2}\frac{A_{\varepsilon}'\Delta_{\varepsilon}'}{A_{\varepsilon}\Delta_{\varepsilon}}\right)^{2} + 2\frac{A_{\varepsilon}^{2}}{\Delta_{\varepsilon}^{2}}\left(\frac{1}{r}\frac{A_{\varepsilon}'}{A_{\varepsilon}} + \frac{1}{2}\frac{A_{\varepsilon}'B_{\varepsilon}'}{A_{\varepsilon}B_{\varepsilon}}\right)^{2}\right.$$

$$+4\frac{A_{\varepsilon}^{2}}{\Delta_{\varepsilon}^{2}}\left[\frac{1}{r}\frac{B_{\varepsilon}'}{B_{\varepsilon}} - \frac{1}{r^{2}}\frac{A_{\varepsilon}C_{\varepsilon} + D_{\varepsilon}^{2}}{A_{\varepsilon}B_{\varepsilon}} + \frac{1}{4}\left(\frac{B_{\varepsilon}'}{B_{\varepsilon}}\right)^{2}\right]^{2} +$$

$$+2\frac{A_{\varepsilon}^{2}}{\Delta_{\varepsilon}^{2}}\left[\frac{1}{r}\left(\frac{A_{\varepsilon}'}{A_{\varepsilon}} + 2\frac{B_{\varepsilon}'}{B_{\varepsilon}} - \frac{\Delta_{\varepsilon}'}{\Delta_{\varepsilon}}\right) + \frac{1}{2}\frac{A_{\varepsilon}'B_{\varepsilon}'}{A_{\varepsilon}B_{\varepsilon}} + \frac{B_{\varepsilon}''}{B_{\varepsilon}}\right.$$

$$\left.\left.-\frac{1}{2}\left(\frac{B_{\varepsilon}'}{B_{\varepsilon}}\right)^{2} - \frac{1}{2}\frac{B_{\varepsilon}'\Delta_{\varepsilon}'}{B_{\varepsilon}\Delta_{\varepsilon}}\right]^{2}\right)_{\varepsilon} =$$

$$\left(\left(\frac{A_{\varepsilon}''}{A_{\varepsilon}} - \frac{1}{2}\frac{A_{\varepsilon}'^{2}}{A_{\varepsilon}^{2}}\right)^{2}\right)_{\varepsilon} + 2\frac{1}{r}\left(\frac{A_{\varepsilon}'^{2}}{A_{\varepsilon}^{2}}\right)_{\varepsilon} =$$

$$\left(\left(\frac{2\tilde{g}^{2}}{(a+g^{*}r)^{2}+\varepsilon^{2}} - \frac{2\tilde{g}^{2}(a+\tilde{g}r)^{2}}{\left[(a+g^{*}r)^{2}+\varepsilon^{2}\right]^{2}}\right)^{2}\right)_{\varepsilon} +$$

$$\frac{1}{r}\left(\frac{8g^{*2}(a+g^{*}r)^{2}}{\left[(a+g^{*}r)^{2}+\varepsilon^{2}\right]^{2}}\right)_{\varepsilon} = \qquad (A1.17)$$

$$\left(\left(\frac{2g^{*2}}{(a+g^{*}r)^{2}+\varepsilon^{2}} - \frac{2g^{*2}\left[(a+g^{*}r)^{2}+\varepsilon^{2}-\varepsilon^{2}\right]}{\left[(a+g^{*}r)^{2}+\varepsilon^{2}\right]^{2}}\right)^{2}\right)_{\varepsilon}$$

$$+\frac{1}{r}\left(\frac{8g^{*2}(a+g^{*}r)^{2}}{\left[(a+g^{*}r)^{2}+\varepsilon^{2}\right]^{2}}\right)_{\varepsilon} =$$

$$\left(\left(\frac{2\tilde{g}^{2}}{(a+\tilde{g}r)^{2}+\varepsilon^{2}} - \frac{2\tilde{g}^{2}\left[(a+\tilde{g}r)^{2}+\varepsilon^{2}\right]}{\left[(a+\tilde{g}r)^{2}+\varepsilon^{2}\right]^{2}} + \frac{2\tilde{g}^{2}\varepsilon^{2}}{\left[(a+\tilde{g}r)^{2}+\varepsilon^{2}\right]^{2}}\right)^{2}\right)_{\varepsilon} +$$

$$\frac{1}{r}\left(\frac{8\tilde{g}^{2}(a+\tilde{g}r)^{2}}{\left[(a+\tilde{g}r)^{2}+\varepsilon^{2}\right]^{2}}\right)_{\varepsilon} = \left(\frac{4\tilde{g}^{4}\varepsilon^{4}}{\left[(a+\tilde{g}r)^{2}+\varepsilon^{2}\right]^{4}}\right)_{\varepsilon}$$

$$+\frac{1}{r}\left(\frac{8\tilde{g}^{2}(a+\tilde{g}r)^{2}}{\left[(a+\tilde{g}r)^{2}+\varepsilon^{2}\right]^{2}}\right)_{\varepsilon}$$

In the limit $a + g^{*}r \to 0$ from (A1.12) we get

$$(\mathbf{R}^{\rho\sigma\mu\nu}(r,\varepsilon)\mathbf{R}_{\rho\sigma\mu\nu}(r,\varepsilon))_{\varepsilon} \asymp \left(\frac{4g^{*4}\varepsilon^{4}}{\left[(a+g^{*}r)^{2}+\varepsilon^{2}\right]^{4}}\right)_{\varepsilon}, a + g^{*}r \asymp 0. \qquad (A1.18)$$

**Remark A1.7**.At horizon $r_{hor} : a + \tilde{g}r_{hor} = 0$ from Eq.(A1.18) one obtains nonclassical result

$$(\mathbf{R}^{\rho\sigma\mu\nu}(r,\varepsilon)\mathbf{R}_{\rho\sigma\mu\nu}(r,\varepsilon))_{\varepsilon} \approx_{\widetilde{\mathbb{R}}} O\left(g^{*2}\right)\left(\varepsilon^{-4}\right)_{\varepsilon} \approx_{\widetilde{\mathbb{R}}} \infty, \qquad (A1.19)$$

see Definition 1.5.2.(ii).

**Remark A1.8**. Let $[(a + g^{*}r_{\varepsilon})_{\varepsilon}] \approx_{\widetilde{\mathbb{R}}} 0$, then from Eq.(A1.3) and Eq.(A1.18) we obtain

$$[(\mathbf{R}^{\rho\sigma\mu\nu}(r_\varepsilon,\varepsilon)\mathbf{R}_{\rho\sigma\mu\nu}(r_\varepsilon,\varepsilon))_\varepsilon] \approx_{\widetilde{\mathbb{R}}} \left[ \left( \frac{4g^{*4}\varepsilon^4}{\left[(a+g^*r_\varepsilon)^2+\varepsilon^2\right]^4} \right)_\varepsilon \right]. \qquad (A1.20)$$

5mm . **Appendix A2 ref: lem:conv_K-M_ESS** 2mm

Let us consider now distributional Colombeau metric given by Eq.(1.3.30) with $c = 1$

$$(ds_\varepsilon^2)_\varepsilon = -\frac{\left((r-\overline{\rho})^2+\varepsilon^2\right)_\varepsilon}{(r+\overline{\rho})^2}dt^2 + \left(1+\frac{\overline{\rho}}{r}\right)^4[dr^2+r^2(d\theta^2+\sin^2\theta d\varphi)], \qquad (A2.1)$$

where $\varepsilon \in (0,1]$, $\overline{\rho} = r_s/4$, $r_s$ is a schwarzschild radius.

We choose now $D_\varepsilon(r) = C_\varepsilon(r) = 0$, and rewrite Eq.(A2.1) in the following equivalent form

$$(d_\varepsilon s^2)_\varepsilon = -(A_\varepsilon(r)dt^2)_\varepsilon + B(r)[dr^2+r^2d\Omega^2],$$

$$A_\varepsilon(r) = \frac{(r-\overline{\rho})^2+\varepsilon^2}{(r+\overline{\rho})^2}, B(r) = \left(1+\frac{\overline{\rho}}{r}\right)^4,$$

$$\Delta_\varepsilon = A_\varepsilon B_\varepsilon, \frac{A_\varepsilon}{\Delta_\varepsilon} = \frac{1}{B_\varepsilon}, \Delta'_\varepsilon = A'_\varepsilon B_\varepsilon + A_\varepsilon B'_\varepsilon, \frac{\Delta'_\varepsilon}{\Delta_\varepsilon} = \frac{A'_\varepsilon B_\varepsilon + A_\varepsilon B'_\varepsilon}{A_\varepsilon B_\varepsilon} = \frac{A'_\varepsilon}{A_\varepsilon} + \frac{B'_\varepsilon}{B_\varepsilon}.$$

$$\frac{A'_\varepsilon}{A_\varepsilon} = -\frac{2(\varepsilon^2+2\overline{\rho}^2-2r\overline{\rho})}{(r+\overline{\rho})\left((r-\overline{\rho})^2+\varepsilon^2\right)},$$

$$\left(\frac{A'_\varepsilon}{A_\varepsilon}\right)^2 = \frac{4(\varepsilon^2+2\overline{\rho}^2-2r\overline{\rho})^2}{(r+\overline{\rho})^2\left((r-\overline{\rho})^2+\varepsilon^2\right)^2} = \frac{4\varepsilon^4+16\overline{\rho}(\overline{\rho}-r)+16\overline{\rho}^2(r-\overline{\rho})^2}{(r+\overline{\rho})^2\left((r-\overline{\rho})^2+\varepsilon^2\right)^2} =$$

$$\frac{4\varepsilon^4+16\overline{\rho}(\overline{\rho}-r)}{(r+\overline{\rho})^2\left((r-\overline{\rho})^2+\varepsilon^2\right)^2} + \frac{16\overline{\rho}^2\left[(r-\overline{\rho})^2+\varepsilon^2-\varepsilon^2\right]}{(r+\overline{\rho})^2\left((r-\overline{\rho})^2+\varepsilon^2\right)^2} = \qquad (A2.2)$$

$$\frac{4\varepsilon^4+16\overline{\rho}(\overline{\rho}-r)}{(r+\overline{\rho})^2\left((r-\overline{\rho})^2+\varepsilon^2\right)^2} + \frac{16\overline{\rho}^2}{(r+\overline{\rho})^2\left((r-\overline{\rho})^2+\varepsilon^2\right)} -$$

$$-\frac{16\overline{\rho}^2\varepsilon^2}{(r+\overline{\rho})^2\left((r-\overline{\rho})^2+\varepsilon^2\right)^2},$$

$$\frac{A''_\varepsilon}{A_\varepsilon} = \frac{2(3\varepsilon^2+8\overline{\rho}^2-4r\overline{\rho})}{(r+\overline{\rho})^2\left((r-\overline{\rho})^2+\varepsilon^2\right)}.$$

We assume now that $r \asymp \overline{\rho}$, then from Eqs.(A2.2) we obtain

$$\frac{A'_\varepsilon}{A_\varepsilon} \asymp -\frac{2\varepsilon^2}{\overline{\rho}\left((r-\overline{\rho})^2+\varepsilon^2\right)}$$

$$\left(\frac{A'_\varepsilon}{A_\varepsilon}\right)^2 \asymp \frac{\varepsilon^4}{\overline{\rho}^2\left((r-\overline{\rho})^2+\varepsilon^2\right)^2} + \frac{4}{\left[(r-\overline{\rho})^2+\varepsilon^2\right]} - \frac{4\varepsilon^2}{\left((r-\overline{\rho})^2+\varepsilon^2\right)^2},$$

$$\frac{A''_\varepsilon}{A_\varepsilon} \asymp \frac{(3\varepsilon^2+4\overline{\rho}^2)}{2\overline{\rho}^2\left((r-\overline{\rho})^2+\varepsilon^2\right)} \asymp \frac{2}{(r-\overline{\rho})^2+\varepsilon^2}, \tag{A2.3}$$

$$-\frac{A''_\varepsilon}{A_\varepsilon} + \frac{1}{2}\left(\frac{A'_\varepsilon}{A_\varepsilon}\right)^2 \asymp -\frac{2}{(r-\overline{\rho})^2+\varepsilon^2} + \frac{2}{(r-\overline{\rho})^2+\varepsilon^2} - \frac{2\varepsilon^2}{\left[(r-\overline{\rho})^2+\varepsilon^2\right]^2}$$

$$= -\frac{2\varepsilon^2}{\left[(r-\overline{\rho})^2+\varepsilon^2\right]^2}.$$

From Eq.(A2.3) by formulae (A1.2) we get

$$\begin{gathered}
(\mathbf{R}(r,\varepsilon))_\varepsilon = \left(\frac{A_\varepsilon}{\Delta_\varepsilon}\left[\frac{2}{r}\left(-2\frac{A'_\varepsilon}{A_\varepsilon} - 3\frac{B'_\varepsilon}{B_\varepsilon} + \frac{\Delta'_\varepsilon}{\Delta_\varepsilon}\right) - \frac{A''_\varepsilon}{A_\varepsilon} - 2\frac{B''_\varepsilon}{B_\varepsilon}\right.\right. \\
\left.\left. + \frac{1}{2}\left(\frac{B'_\varepsilon}{B_\varepsilon}\right)^2 - 2\frac{A'_\varepsilon B'_\varepsilon}{A_\varepsilon B_\varepsilon} + \left(\frac{1}{2}\frac{A'_\varepsilon}{A_\varepsilon} + \frac{B'_\varepsilon}{B_\varepsilon}\right)\frac{\Delta'_\varepsilon}{\Delta_\varepsilon}\right]\right)_\varepsilon = \\
\left(\frac{1}{B_\varepsilon}\left[\frac{2}{r}\left(-\frac{A'_\varepsilon}{A_\varepsilon} - 2\frac{B'_\varepsilon}{B_\varepsilon}\right) - \frac{A''_\varepsilon}{A_\varepsilon} - 2\frac{B''_\varepsilon}{B_\varepsilon} + \right.\right. \\
\left.\left. + \frac{1}{2}\left(\frac{B'_\varepsilon}{B_\varepsilon}\right)^2 - 2\frac{A'_\varepsilon B'_\varepsilon}{A_\varepsilon B_\varepsilon} + \left(\frac{1}{2}\frac{A'_\varepsilon}{A_\varepsilon} + \frac{B'_\varepsilon}{B_\varepsilon}\right)\left(\frac{A'_\varepsilon}{A_\varepsilon} + \frac{B'_\varepsilon}{B_\varepsilon}\right)\right]\right)_\varepsilon = \\
\left(\frac{1}{B_\varepsilon}\left[\frac{2}{r}\left(-\frac{A'_\varepsilon}{A_\varepsilon} - 2\frac{B'_\varepsilon}{B_\varepsilon}\right) - \frac{A''_\varepsilon}{A_\varepsilon} - 2\frac{B''_\varepsilon}{B_\varepsilon} + \right.\right. \\
\left.\left. + \frac{1}{2}\frac{(A'_\varepsilon)^2}{A_\varepsilon^2} + \frac{3}{2}\frac{(B'_\varepsilon)^2}{B_\varepsilon^2} - \frac{1}{2}\frac{A'_\varepsilon B'_\varepsilon}{A_\varepsilon B_\varepsilon}\right]\right)_\varepsilon.
\end{gathered} \tag{A2.4}$$

From Eq.(A2.4) in the limit $r \asymp \overline{\rho}$ by formulae (A2.3) we get

$$(\mathbf{R}(r,\varepsilon))_\varepsilon \asymp -\frac{2\varepsilon^2}{\left[(r-\overline{\rho})^2+\varepsilon^2\right]^2}. \tag{A2.5}$$

**Remark A2.1**. Note that: (1) Eq.(A2.5) in a nice agriment with Eq.(A1.9). For $r$ located beyond horizon, i.e. $r-\overline{\rho} \neq 0$ one obtains classical result

$$(\mathbf{R}(r,\varepsilon))_\varepsilon \approx_{\widetilde{\mathbb{R}}} O\left(1\right)\left(\varepsilon^2\right)_\varepsilon \approx_{\widetilde{\mathbb{R}}} 0, \tag{A2.6}$$

see Definition 1.5.2.(i).(3) At horizon $r_{hor}$ : $r_{hor} - \overline{\rho} = 0$ from Eq.(A2.5) one obtains nonclassical result

$$(\mathbf{R}(r_{hor},\varepsilon))_\varepsilon \approx_{\widetilde{\mathbb{R}}} O\left(1\right)\left(\varepsilon^{-2}\right)_\varepsilon \approx_{\widetilde{\mathbb{R}}} \infty, \tag{A2.7}$$

see Definition 1.5.2.(ii).

**Remark A2.2**. Let $[(r_\varepsilon - \overline{\rho})_\varepsilon] \approx_{\widetilde{\mathbb{R}}} 0,$ then from Eq.(A1.3) and Eq.(A2.5) we obtain

$$[(\mathbf{R}(r_\varepsilon,\varepsilon))_\varepsilon] \approx_{\widetilde{\mathbb{R}}} \left[\left(-\frac{2\varepsilon^2}{\left[(r_\varepsilon-\overline{\rho})^2+\varepsilon^2\right]^2}\right)_\varepsilon\right]. \tag{A2.8}$$

From Eq.(A2.3) by formulae (A1.2) we get

$$(\mathbf{R}^{\mu\nu}(r,\varepsilon)\mathbf{R}_{\mu\nu}(r,\varepsilon))_{\varepsilon} = \left(\frac{A_{\varepsilon}^2}{\Delta_{\varepsilon}^2}\left(\frac{1}{2}\frac{A''_{\varepsilon}}{A_{\varepsilon}} - \frac{1}{4}\frac{A'_{\varepsilon}\Delta'_{\varepsilon}}{A_{\varepsilon}\Delta_{\varepsilon}} + \frac{1}{2}\frac{A'_{\varepsilon}B'_{\varepsilon}}{A_{\varepsilon}B_{\varepsilon}} + \frac{1}{r}\frac{A'_{\varepsilon}}{A_{\varepsilon}}\right)^2\right)_{\varepsilon} +$$

$$+2\left(\frac{A_{\varepsilon}^2}{\Delta_{\varepsilon}^2}\left[\frac{1}{r}\left(\frac{1}{2}\frac{\Delta'_{\varepsilon}}{\Delta_{\varepsilon}} - \frac{A'_{\varepsilon}}{A_{\varepsilon}} - 2\frac{B'_{\varepsilon}}{B_{\varepsilon}}\right) - \frac{1}{2}\frac{A'_{\varepsilon}B'_{\varepsilon}}{A_{\varepsilon}B_{\varepsilon}} -\right.\right.$$

$$\left.\left.-\frac{1}{2}\frac{B''_{\varepsilon}}{B_{\varepsilon}} + \frac{1}{4}\frac{B'_{\varepsilon}\Delta'_{\varepsilon}}{B_{\varepsilon}\Delta_{\varepsilon}}\right]^2\right)_{\varepsilon} +$$

$$\left(\frac{A_{\varepsilon}^2}{\Delta_{\varepsilon}^2}\left[\frac{1}{2}\frac{A''_{\varepsilon}}{A_{\varepsilon}} - \frac{1}{4}\frac{A'_{\varepsilon}\Delta'_{\varepsilon}}{A_{\varepsilon}\Delta_{\varepsilon}} + \frac{1}{2}\frac{A'_{\varepsilon}B'_{\varepsilon}}{A_{\varepsilon}B_{\varepsilon}} + \frac{B''_{\varepsilon}}{B_{\varepsilon}} - \frac{1}{2}\left(\frac{B'_{\varepsilon}}{B_{\varepsilon}}\right)^2\right.\right.$$

$$\left.\left.-\frac{1}{2}\frac{B'_{\varepsilon}\Delta'_{\varepsilon}}{B_{\varepsilon}\Delta_{\varepsilon}} + \frac{1}{r}\left(\frac{A'_{\varepsilon}}{A_{\varepsilon}} - \frac{\Delta'_{\varepsilon}}{\Delta_{\varepsilon}} + 2\frac{B'_{\varepsilon}}{B_{\varepsilon}}\right)\right]^2\right)_{\varepsilon} =$$

$$\left(\frac{1}{B_{\varepsilon}^2}\left(\frac{1}{2}\frac{A''_{\varepsilon}}{A_{\varepsilon}} - \frac{1}{4}\frac{A'_{\varepsilon}}{A_{\varepsilon}}\left(\frac{A'_{\varepsilon}}{A_{\varepsilon}} + \frac{B'_{\varepsilon}}{B_{\varepsilon}}\right) + \frac{1}{2}\frac{A'_{\varepsilon}B'_{\varepsilon}}{A_{\varepsilon}B_{\varepsilon}} + \frac{1}{r}\frac{A'_{\varepsilon}}{A_{\varepsilon}}\right)^2\right)_{\varepsilon} + \qquad (A2.9)$$

$$+2\left(\frac{1}{B_{\varepsilon}^2}\left[\frac{1}{r}\left(\frac{1}{2}\left(\frac{A'_{\varepsilon}}{A_{\varepsilon}} + \frac{B'_{\varepsilon}}{B_{\varepsilon}}\right) - \frac{A'_{\varepsilon}}{A_{\varepsilon}} - 2\frac{B'_{\varepsilon}}{B_{\varepsilon}}\right) - \frac{1}{2}\frac{A'_{\varepsilon}B'_{\varepsilon}}{A_{\varepsilon}B_{\varepsilon}} -\right.\right.$$

$$\left.\left.-\frac{1}{2}\frac{B''_{\varepsilon}}{B_{\varepsilon}} + \frac{1}{4}\frac{B'_{\varepsilon}}{B_{\varepsilon}}\left(\frac{A'_{\varepsilon}}{A_{\varepsilon}} + \frac{B'_{\varepsilon}}{B_{\varepsilon}}\right)\right]^2\right)_{\varepsilon} +$$

$$\left(\frac{A_{\varepsilon}^2}{\Delta_{\varepsilon}^2}\left[\frac{1}{2}\frac{A''_{\varepsilon}}{A_{\varepsilon}} - \frac{1}{4}\frac{A'_{\varepsilon}}{A_{\varepsilon}}\left(\frac{A'_{\varepsilon}}{A_{\varepsilon}} + \frac{B'_{\varepsilon}}{B_{\varepsilon}}\right) + \frac{1}{2}\frac{A'_{\varepsilon}B'_{\varepsilon}}{A_{\varepsilon}B_{\varepsilon}} + \frac{B''_{\varepsilon}}{B_{\varepsilon}} - \frac{1}{2}\left(\frac{B'_{\varepsilon}}{B_{\varepsilon}}\right)^2\right.\right.$$

$$\left.\left.-\frac{1}{2}\frac{B'_{\varepsilon}\Delta'_{\varepsilon}}{B_{\varepsilon}\Delta_{\varepsilon}} + \frac{1}{r}\frac{B'_{\varepsilon}}{B_{\varepsilon}}\right]^2\right)_{\varepsilon}.$$

From Eq.(A2.9) in the limit $r \asymp \overline{\rho}$ by formulae (A2.3) we get

$$(\mathbf{R}^{\mu\nu}(r,\varepsilon)\mathbf{R}_{\mu\nu}(r,\varepsilon))_{\varepsilon} \asymp \left(\frac{\varepsilon^4}{\left[(r-\overline{\rho})^2+\varepsilon^2\right]^4}\right)_{\varepsilon}. \qquad (A2.10)$$

**Remark A2.3**. Note that: (1) For $r$ located beyond horizon, i.e. $r - \overline{\rho} \neq 0$ one obtains classical result

$$(\mathbf{R}^{\mu\nu}(r,\varepsilon)\mathbf{R}_{\mu\nu}(r,\varepsilon))_{\varepsilon} \approx_{\widetilde{\mathbb{R}}} O(1)\left(\varepsilon^2\right)_{\varepsilon} \approx_{\widetilde{\mathbb{R}}} 0, \qquad (A2.11)$$

see Definition 1.5.2.(i).(2) At horizon $r_{hor} : r_{hor} - \overline{\rho} = 0$ from Eq.(A2.10) one obtains nonclassical result

$$(\mathbf{R}^{\mu\nu}(r_{hor},\varepsilon)\mathbf{R}_{\mu\nu}(r_{hor},\varepsilon))_{\varepsilon} \approx_{\widetilde{\mathbb{R}}} O(1)\left(\varepsilon^{-2}\right)_{\varepsilon} \approx_{\widetilde{\mathbb{R}}} \infty, \qquad (A2.12)$$

see Definition 1.5.2.(ii).

**Remark A2.3**. Let $[(r_{\varepsilon} - \overline{\rho})_{\varepsilon}] \approx_{\widetilde{\mathbb{R}}} 0,$ then from Eq.(A1.3) and Eq.(A2.10) we obtain

$$[(\mathbf{R}^{\mu\nu}(r_{\varepsilon},\varepsilon)\mathbf{R}_{\mu\nu}(r_{\varepsilon},\varepsilon))_{\varepsilon}] \approx_{\widetilde{\mathbb{R}}} \left[\left(\frac{\varepsilon^4}{\left[(r_{\varepsilon}-\overline{\rho})^2+\varepsilon^2\right]^4}\right)_{\varepsilon}\right]. \qquad (A2.13)$$

From Eq.(A2.3) by formulae (A1.2) we get

$$
\begin{aligned}
&(\mathbf{R}^{\rho\sigma\mu\nu}(r,\varepsilon)\mathbf{R}_{\rho\sigma\mu\nu}(r,\varepsilon))_{\varepsilon} = \\
&\left(\frac{A_{\varepsilon}^{2}}{\Delta_{\varepsilon}^{2}}\left(\frac{A_{\varepsilon}''}{A_{\varepsilon}}-\frac{1}{2}\frac{A_{\varepsilon}'\Delta_{\varepsilon}'}{A_{\varepsilon}\Delta_{\varepsilon}}\right)^{2}+2\frac{A_{\varepsilon}^{2}}{\Delta_{\varepsilon}^{2}}\left(\frac{1}{r}\frac{A_{\varepsilon}'}{A_{\varepsilon}}+\frac{1}{2}\frac{A_{\varepsilon}'B_{\varepsilon}'}{A_{\varepsilon}B_{\varepsilon}}\right)^{2}\right. \\
&+4\frac{A_{\varepsilon}^{2}}{\Delta_{\varepsilon}^{2}}\left[\frac{1}{r}\frac{B_{\varepsilon}'}{B_{\varepsilon}}+\frac{1}{4}\left(\frac{B_{\varepsilon}'}{B_{\varepsilon}}\right)^{2}\right]^{2}+ \\
&+2\frac{A_{\varepsilon}^{2}}{\Delta_{\varepsilon}^{2}}\left[\frac{1}{r}\left(\frac{A_{\varepsilon}'}{A_{\varepsilon}}+2\frac{B_{\varepsilon}'}{B_{\varepsilon}}-\frac{\Delta_{\varepsilon}'}{\Delta_{\varepsilon}}\right)+\frac{1}{2}\frac{A_{\varepsilon}'B_{\varepsilon}'}{A_{\varepsilon}B_{\varepsilon}}+\frac{B_{\varepsilon}''}{B_{\varepsilon}}\right. \\
&\left.\left.-\frac{1}{2}\left(\frac{B_{\varepsilon}'}{B_{\varepsilon}}\right)^{2}-\frac{1}{2}\frac{B_{\varepsilon}'\Delta_{\varepsilon}'}{B_{\varepsilon}\Delta_{\varepsilon}}\right]^{2}\right)_{\varepsilon} = \\
&\left(\frac{1}{B_{\varepsilon}^{2}}\left(\frac{A_{\varepsilon}''}{A_{\varepsilon}}-\frac{1}{2}\frac{A_{\varepsilon}'}{A_{\varepsilon}}\left(\frac{A_{\varepsilon}'}{A_{\varepsilon}}+\frac{B_{\varepsilon}'}{B_{\varepsilon}}\right)\right)^{2}+2\frac{1}{B_{\varepsilon}^{2}}\left(\frac{1}{r}\frac{A_{\varepsilon}'}{A_{\varepsilon}}+\frac{1}{2}\frac{A_{\varepsilon}'B_{\varepsilon}'}{A_{\varepsilon}B_{\varepsilon}}\right)^{2}\right. \\
&+4\frac{1}{B_{\varepsilon}^{2}}\left[\frac{1}{r}\frac{B_{\varepsilon}'}{B_{\varepsilon}}+\frac{1}{4}\left(\frac{B_{\varepsilon}'}{B_{\varepsilon}}\right)^{2}\right]^{2}+ \\
&+2\frac{1}{B_{\varepsilon}^{2}}\left[\frac{1}{r}\left(\frac{A_{\varepsilon}'}{A_{\varepsilon}}+2\frac{B_{\varepsilon}'}{B_{\varepsilon}}-\left(\left(\frac{A_{\varepsilon}'}{A_{\varepsilon}}+\frac{B_{\varepsilon}'}{B_{\varepsilon}}\right)\right)\right)+\frac{1}{2}\frac{A_{\varepsilon}'B_{\varepsilon}'}{A_{\varepsilon}B_{\varepsilon}}+\frac{B_{\varepsilon}''}{B_{\varepsilon}}\right. \\
&\left.\left.-\frac{1}{2}\left(\frac{B_{\varepsilon}'}{B_{\varepsilon}}\right)^{2}-\frac{1}{2}\frac{B_{\varepsilon}'}{B_{\varepsilon}}\left(\left(\frac{A_{\varepsilon}'}{A_{\varepsilon}}+\frac{B_{\varepsilon}'}{B_{\varepsilon}}\right)\right)\right]^{2}\right)_{\varepsilon}.
\end{aligned}
\tag{A2.14}
$$

From Eq.(A2.14) in the limit $r \to \overline{\rho}$ by formulae (A2.3) we get

$$(\mathbf{R}^{\rho\sigma\mu\nu}(r,\varepsilon)\mathbf{R}_{\rho\sigma\mu\nu}(r,\varepsilon))_{\varepsilon} \approx_{\widetilde{\mathbb{R}}} \widetilde{K(r)} + \left(\frac{\varepsilon^{4}}{\left[(r-\overline{\rho})^{2}+\varepsilon^{2}\right]^{4}}\right)_{\varepsilon}, \tag{A2.15}$$

where $K(r)$ is a Kretschman scalar: $K(r) = 3 \cdot 4^{13} r^{6} r_{s}^{2}(4r + r_{s})^{-12}$.

**Remark A2.4**. Note that: (1) For $r$ located beyond horizon, i.e. $r - \overline{\rho} \neq 0$ one obtains classical result

$$(\mathbf{R}^{\rho\sigma\mu\nu}(r,\varepsilon)\mathbf{R}_{\rho\sigma\mu\nu}(r,\varepsilon))_{\varepsilon} \approx_{\widetilde{\mathbb{R}}} \widetilde{K(\overline{\rho})}, \tag{A2.16}$$

see Definition 1.5.2.(i).(2) At horizon $r_{hor} : r_{hor} - \overline{\rho} = 0$ from Eq.(A2.15) one obtains nonclassical result

$$(\mathbf{R}^{\rho\sigma\mu\nu}(r,\varepsilon)\mathbf{R}_{\rho\sigma\mu\nu}(r,\varepsilon))_{\varepsilon} \approx_{\widetilde{\mathbb{R}}} O(1)\left(\varepsilon^{-2}\right)_{\varepsilon} \approx_{\widetilde{\mathbb{R}}} \infty, \tag{A2.17}$$

see Definition 1.5.2.(ii).

**Remark A2.5**. Let $[(r_{\varepsilon} - \overline{\rho})_{\varepsilon}] \approx_{\widetilde{\mathbb{R}}} 0,$ then from Eq.(A1.3) and Eq.(A2.15) we obtain

$$[(\mathbf{R}^{\rho\sigma\mu\nu}(r_{\varepsilon},\varepsilon)\mathbf{R}_{\rho\sigma\mu\nu}(r_{\varepsilon},\varepsilon))_{\varepsilon}] \approx_{\widetilde{\mathbb{R}}} \widetilde{K(\overline{\rho})} + \left[\left(\frac{\varepsilon^{4}}{\left[(r_{\varepsilon}-\overline{\rho})^{2}+\varepsilon^{2}\right]^{4}}\right)_{\varepsilon}\right]. \tag{A2.18}$$

**Remark A2.6**. We assume now there exist an fundamental generalized lengh

$$(l_{\varepsilon})_{\varepsilon\in(0,\eta]} = a(\varepsilon)_{\varepsilon\in(0,\eta]}, \eta \ll 1 \tag{A2.19}$$

such that $|(r_{\varepsilon} - \overline{\rho})_{\varepsilon}| \geq (l_{\varepsilon})_{\varepsilon} = a(\varepsilon)_{\varepsilon}$ It meant there exist a thickness of horizon.
By using (A2.19) we get the estimate

$$\left(\frac{\varepsilon^4}{\left[(r_\varepsilon-\overline{\rho})^2+\varepsilon^2\right]^4}\right)_{\varepsilon\in(0,\eta]} =$$

$$\left(\frac{1}{\left[(r_\varepsilon-\overline{\rho})^2+\varepsilon^2\right]^2}\right)_{\varepsilon\in(0,\eta]} \times \left(\frac{\varepsilon^2}{\left[(r_\varepsilon-\overline{\rho})^2+\varepsilon^2\right]}\right)_{\varepsilon\in(0,\eta]} \times$$

$$\times\left(\frac{\varepsilon^2}{\left[(r_\varepsilon-\overline{\rho})^2+\varepsilon^2\right]}\right)_{\varepsilon\in(0,\eta]} \leq \qquad (A2.20)$$

$$\leq \left(\frac{1}{\left[(r_\varepsilon-\overline{\rho})^2+\varepsilon^2\right]^2}\right)_{\varepsilon\in(0,\eta]} \left(\left(\frac{\varepsilon^2}{[a^2\varepsilon^2+\varepsilon^2]}\right)_{\varepsilon\in(0,\eta]}\right)^2 =$$

$$\frac{1}{[a^2+1]^2}\left(\frac{1}{\left[(r_\varepsilon-\overline{\rho})^2+\varepsilon^2\right]^2}\right)_{\varepsilon\in(0,\eta]}.$$

5mm . **Appendix B ref: thm:conv_lambda0mle** 2mm

We calculate now the distributional curvature at Schwarzschild horizon. In the usual Schwarzschild coordinates $(t, r > 0, \theta, \phi), r \neq 2m$ the metric is

$$\begin{cases} ds^2 = h(r)dt^2 - h(r)^{-1}dr^2 + r^2 d\Omega^2, \\ h(r) = -1 + \frac{2m}{r}. \end{cases} \qquad (B.1)$$

Metric takes the form above horizon $r > 2m$ and below horizon $r < 2m$ correspondingly

$$\begin{cases} \text{above horizon } r > 2m : \\ ds^{+2} = h^+(r)dt^2 - [h^+(r)]^{-1}dr^2 + r^2 d\Omega^2, \\ h^+(r) = -1 + \frac{2m}{r} = -\frac{r-2m}{r} \\ \text{below horizon } r < 2m : \\ ds^{-2} = h^-(r)dt^2 - h^-(r)^{-1}dr^2 + r^2 d\Omega^2, \\ h^-(r) = -1 + \frac{2m}{r} = \frac{2m-r}{r} \end{cases} \qquad (B.2)$$

**Remark B.1**. Following the above discussion we consider the metric coefficients $h^+(r), [h^+(r)]^{-1}$ $h^-(r)$, and $[h^-(r)]^{-1}$ as an element of $D'(\mathbb{R}^3)$ and embed it into $\mathbf{D}\left(\widetilde{\mathbb{R}}^3\right)$ by replacement above horizon $r \geqslant 2m$ and below horizon $r \leqslant 2m$ correspondingly

$$\begin{aligned} r \geqslant 2m &: r - 2m \mapsto \sqrt{(r-2m)^2+\epsilon^2}\,, \\ r \leqslant 2m &: 2m - r \mapsto \sqrt{(2m-r)^2+\epsilon^2}\,. \end{aligned} \qquad (B.3)$$

Note that, accordingly, we have fixed the differentiable structure of the manifold: the Cartesian coordinates associated with the spherical Schwarzschild coordinates in (B.1) are extended through the origin. We have above $r \geqslant 2m$ (below $(r \leqslant 2m)$) horizon

$$h(r)=\left\{\begin{array}{c}-\frac{r-2m}{r}\text{ if } r\geqslant 2m\\ 0\text{ if } r\leqslant 2m\end{array}\right\}\mapsto (h_\epsilon^+(r))_\epsilon=\left(-\frac{\sqrt{(r-2m)^2+\epsilon^2}}{r}\right)_\epsilon,$$

where $(h_\epsilon^+(r))_\epsilon\in\mathbf{G}(\mathbb{R}^3,B^+(2m,R)),B^+(2m,R)=\{x\in\mathbb{R}^3|2m\leqslant\|x\|\leqslant R\}$.

$$h^{-1}(r)=\left\{\begin{array}{c}-\frac{r}{r-2m},r>2m\\ \infty,r=2m\end{array}\right\}\mapsto (h_\epsilon^+)^{-1}(r)=$$

$$h^-(r)=\left\{\begin{array}{c}-\frac{r-2m}{r}\text{ if } r\leqslant 2m\\ 0\text{ if } r\geq 2m\end{array}\right\}\mapsto h_\epsilon^-(r)= \qquad (B.4)$$

$$=\left(\frac{\sqrt{(2m-r)^2+\epsilon^2}}{r}\right)_\epsilon\in\mathbf{G}(\mathbb{R}^3,B^-(0,2m)),$$

where $B^-(0,2m)=\{x\in\mathbb{R}^3|0<\|x\|\leqslant 2m\}$

$$\left\{\begin{array}{c}-\frac{r}{r-2m},r<2m\\ \infty,r=2m\end{array}\right\}\mapsto (h_\epsilon^-)^{-1}(r)=$$

$$=\left(\frac{r}{\sqrt{(r-2m)^2+\epsilon^2}}\right)_\epsilon\in\mathbf{G}(\mathbb{R}^3,B^-(0,2m))$$

Inserting (B.4) into (B.2) we obtain a generalized object modeling the singular Schwarzschild metric above (below) gorizon, i.e.,

$$\begin{array}{c}(ds_\epsilon^{+2})_\epsilon=(h_\epsilon^+(r)dt^2)_\epsilon-\left([h_\epsilon^+(r)]^{-1}dr^2\right)_\epsilon+r^2d\Omega^2,\\ (ds_\epsilon^{-2})_\epsilon=(h_\epsilon^-(r)dt^2)_\epsilon-\left([h_\epsilon^-(r)]^{-1}dr^2\right)_\epsilon+r^2d\Omega^2\end{array} \qquad (B.5)$$

The generalized Ricci tensor above horizon $[\mathbf{R}^+]_\alpha^\beta$ may now be calculated componentwise using the classical formulae

$$\begin{array}{c}\left([\mathbf{R}_\epsilon^+]_0^0\right)_\epsilon=\left([\mathbf{R}_\epsilon^+]_1^1\right)_\epsilon=\frac{1}{2}\left((h_\epsilon^{+\prime\prime})_\epsilon+\frac{2}{r}(h_\epsilon^{+\prime})_\epsilon\right)\\ \left([\mathbf{R}_\epsilon^+]_2^2\right)_\epsilon=\left([\mathbf{R}_\epsilon^+]_3^3\right)_\epsilon=\frac{(h_\epsilon^{+\prime})_\epsilon}{r}+\frac{1+(h_\epsilon^+)_\epsilon}{r^2}.\end{array} \qquad (B.6)$$

From (B.4) by differentiation we obtain

$$h_\epsilon^{+'}(r) = -\frac{r-2m}{r\left[(r-2m)^2+\epsilon^2\right]^{1/2}} + \frac{\left[(r-2m)^2+\epsilon^2\right]^{1/2}}{r^2},$$

$$r(h_\epsilon^{+'})_\epsilon + 1 + (h_\epsilon^{+})_\epsilon =$$

$$r\left\{-\frac{r-2m}{r\left[(r-2m)^2+\epsilon^2\right]^{1/2}} + \frac{\left[(r-2m)^2+\epsilon^2\right]^{1/2}}{r^2}\right\} + 1 - \frac{\sqrt{(r-2m)^2+\epsilon^2}}{r} =$$

$$-\frac{r-2m}{\left[(r-2m)^2+\epsilon^2\right]^{1/2}} + \frac{\left[(r-2m)^2+\epsilon^2\right]^{1/2}}{r} + 1 - \frac{\sqrt{(r-2m)^2+\epsilon^2}}{r} =$$

$$-\frac{r-2m}{\left[(r-2m)^2+\epsilon^2\right]^{1/2}} + 1.$$

$$h_\epsilon''(r) = -\left(\frac{r-2m}{r\left[(r-2m)^2+\epsilon^2\right]^{1/2}}\right)' + \left(\frac{\left[(r-2m)^2+\epsilon^2\right]^{1/2}}{r^2}\right)' =$$

$$= -\frac{1}{r\left[(r-2m)^2+\epsilon^2\right]^{1/2}} + \frac{(r-2m)^2}{r\left[(r-2m)^2+\epsilon^2\right]^{3/2}} + \frac{r-2m}{r^2\left[(r-2m)^2+\epsilon^2\right]^{1/2}} +$$

$$+\frac{r-2m}{r^2\left[(r-2m)^2+\epsilon^2\right]^{1/2}} - \frac{2\left[(r-2m)^2+\epsilon^2\right]^{1/2}}{r^3}.$$

$$r^2(h_\epsilon^{+''})_\epsilon + 2r(h_\epsilon^{+'})_\epsilon = \qquad (B.7)$$

$$r^2\left\{-\frac{1}{r\left[(r-2m)^2+\epsilon^2\right]^{1/2}} + \frac{(r-2m)^2}{r\left[(r-2m)^2+\epsilon^2\right]^{3/2}} + \frac{r-2m}{r^2\left[(r-2m)^2+\epsilon^2\right]^{1/2}} +\right.$$

$$\left.+\frac{r-2m}{r^2\left[(r-2m)^2+\epsilon^2\right]^{1/2}} - \frac{2\left[(r-2m)^2+\epsilon^2\right]^{1/2}}{r^3}\right\} +$$

$$+2r\left\{-\frac{r-2m}{r\left[(r-2m)^2+\epsilon^2\right]^{1/2}} + \frac{\left[(r-2m)^2+\epsilon^2\right]^{1/2}}{r^2}\right\} =$$

$$-\frac{r}{\left[(r-2m)^2+\epsilon^2\right]^{1/2}} + \frac{r(r-2m)^2}{\left[(r-2m)^2+\epsilon^2\right]^{3/2}} + \frac{r-2m}{\left[(r-2m)^2+\epsilon^2\right]^{1/2}} +$$

$$+\frac{r-2m}{\left[(r-2m)^2+\epsilon^2\right]^{1/2}} - \frac{2\left[(r-2m)^2+\epsilon^2\right]^{1/2}}{r} +$$

$$-\frac{2(r-2m)}{\left[(r-2m)^2+\epsilon^2\right]^{1/2}} + \frac{2\left[(r-2m)^2+\epsilon^2\right]^{1/2}}{r} =$$

$$-\frac{r}{\left[(r-2m)^2+\epsilon^2\right]^{1/2}} + \frac{r(r-2m)^2}{\left[(r-2m)^2+\epsilon^2\right]^{3/2}}.$$

angular components of the Ricci tensor (using the abbreviation

$$\tilde{\Phi}(r) = \int_0^\pi \sin\theta d\theta \int_0^{2\pi} d\phi \Phi(\vec{x}) \qquad (B.8)$$

and let $\Phi(\vec{x})$ be the function $\Phi(\vec{x}) \in S_{2m}(\mathbb{R}^3, B^{+}(2m,R_0))$, where by $S_{2m}(\mathbb{R}^3, B^{+}(2m,R_0))$ we denote the class of the functions $\Phi(\vec{x})$ with compact support such that:

(i) $\mathbf{supp}(\Phi(\vec{x})) \subset B^{+}(2m,R_0) \subset \{\vec{x} | R_0 \geq \|\vec{x}\| \geq 2m\}$ (ii) $\tilde{\Phi}(r) \in C^{\infty}(\mathbb{R})$.

Then for any function $\Phi(\vec{x}) \in S_{2m}(\mathbb{R}^3, B^{+}(2m,R_0))$ we get:

$$\begin{aligned}
&\int_K \left([\mathbf{R}_\epsilon^{+}]_2^2\right)_\epsilon \Phi(\vec{x}) d^3x = \int_K \left([\mathbf{R}_\epsilon^{+}]_3^3\right)_\epsilon \Phi(\vec{x}) d^3x = \\
&= \int_{2m}^{R} \left( r(h_\epsilon^{+\prime})_\epsilon + 1 + (h_\epsilon^{+})_\epsilon \right) \tilde{\Phi}(r) dr = \\
&= \int_{2m}^{R} \left\{ -\frac{r-2m}{\left[(r-2m)^2+\epsilon^2\right]^{1/2}} \right\} \tilde{\Phi}(r) dr + \int_{2m}^{R} \tilde{\Phi}(r) dr.
\end{aligned} \tag{B.9}$$

By replacement $r - 2m = u$, from (B.9) we obtain

$$\begin{aligned}
&\int_K \left([\mathbf{R}_\epsilon^{+}]_2^2\right)_\epsilon \Phi(x) d^3x = \int_K \left([\mathbf{R}_\epsilon^{+}]_3^3\right)_\epsilon \Phi(x) d^3x = \\
&= -\int_0^{R-2m} \frac{u\tilde{\Phi}(u+2m)du}{(u^2+\epsilon^2)^{1/2}} + \int_0^{R-2m} \tilde{\Phi}(u+2m)du.
\end{aligned} \tag{B.10}$$

By replacement $u = \epsilon\eta$, from (B.10) we obtain the expression

$$\begin{aligned}
&\mathbf{I}_3^{+}(\epsilon) = \int_K \left([\mathbf{R}_\epsilon^{+}]_3^3\right)_\epsilon \Phi(x) d^3x = \mathbf{I}_2^{+}(\epsilon) = \int_K \left([\mathbf{R}_\epsilon^{+}]_2^2\right)_\epsilon \Phi(\vec{x}) d^3x = \\
&= -\epsilon \times \left( \int_0^{\frac{R-2m}{\epsilon}} \frac{\eta\tilde{\Phi}(\epsilon\eta+2)d\eta}{(\eta^2+1)^{1/2}} - \int_0^{\frac{R-2m}{\epsilon}} \tilde{\Phi}(\epsilon\eta+2m)d\eta \right).
\end{aligned} \tag{B.11}$$

From Eq.(B.11) we get

$$\begin{aligned}
&\mathbf{I}_3^{+}(\epsilon) = \mathbf{I}_2^{+}(\epsilon) = -\epsilon \frac{\tilde{\Phi}(2m)}{0!} \int_0^{\frac{R-2m}{\epsilon}} \left[ \frac{\eta}{(\eta^2+1)^{1/2}} - 1 \right] d\eta + \\
&-\frac{\epsilon^2}{1!} \int_0^{\frac{R-2m}{\epsilon}} \left[ \frac{\eta}{(\eta^2+1)^{1/2}} - 1 \right] \tilde{\Phi}^{(1)}(\xi)\eta d\eta = \\
&-\epsilon\tilde{\Phi}(2m) \left[ \sqrt{\left(\frac{R-2m}{\epsilon}\right)^2 + 1} - 1 - \frac{R-2m}{\epsilon} \right] + \\
&-\frac{\epsilon^2}{1} \int_0^{\frac{R-2m}{\epsilon}} \left[ \frac{\eta}{(\eta^2+1)^{1/2}} - 1 \right] \tilde{\Phi}^{(1)}(\xi)\eta d\eta,
\end{aligned} \tag{B.12}$$

where we have expressed the function $\tilde{\Phi}(\epsilon\eta + 2m)$ as

$$\widetilde{\Phi}(\epsilon\eta+2m) = \sum_{l=0}^{n-1}\frac{\Phi^{(l)}(2m)}{l!}(\epsilon\eta)^l + \frac{1}{n!}(\epsilon\eta)^n\Phi^{(n)}(\xi)\ , \qquad (B.13)$$
$$\xi \triangleq \theta\epsilon\eta+2m\ ,\quad 1>\theta>0\ ,\quad n=1$$

with $\tilde{\Phi}^{(l)}(\xi) \triangleq d^l\tilde{\Phi}/d\xi^l$.

Equations (B.12)-(3.13) give

$$\begin{gathered}\lim_{\epsilon\to 0}\mathbf{I}_3^+(\epsilon) = \lim_{\epsilon\to 0}\mathbf{I}_2^+(\epsilon) = \\ \lim_{\epsilon\to 0}\left\{-\epsilon\tilde{\Phi}(2m)\left[\sqrt{\left(\frac{R-2m}{\epsilon}\right)^2+1}-1-\frac{R-2m}{\epsilon}\right]\right\} + \\ +\lim_{\epsilon\to 0}\left\{-\frac{\epsilon^2}{1}\int\limits_0^{\frac{R-2m}{\epsilon}}\left[\frac{\eta}{(\eta^2+1)^{1/2}}-1\right]\tilde{\Phi}^{(1)}(\xi)\eta d\eta\right\} = 0.\end{gathered} \qquad (B.14)$$

Since $S'_{2m}(B^+(2m,R)) \subset D'(\mathbb{R}^3)$, where $B^+(2m,R) = \{x\in\mathbb{R}^3 | 2m \leqslant \|x\| \leqslant R\}$ from Eq.(B.14) we get: $\mathcal{S}\mathcal{S}'_{2m}(\mathbb{R}^3) \subset \mathcal{D}'(\mathbb{R}^3)$,

$$\begin{gathered} w\text{-}\lim_{\epsilon\to 0}[\mathbf{R}_\epsilon^+]_3^3 = \lim_{\epsilon\to 0}\mathbf{I}_3^+(\epsilon) = 0, \\ w\text{-}\lim_{\epsilon\to 0}[\mathbf{R}_\epsilon^+]_2^2 = \lim_{\epsilon\to 0}\mathbf{I}_2^+(\epsilon) = 0.\end{gathered} \qquad (B.15)$$

For $\left([\mathbf{R}_\epsilon^+]_1^1\right)_\epsilon, \left([\mathbf{R}_\epsilon^+]_0^0\right)_\epsilon$ we get:

$$\begin{gathered} 2\int_K\left([\mathbf{R}_\epsilon^+]_1^1\right)_\epsilon\Phi(x)d^3x = 2\int_K\left([\mathbf{R}_\epsilon^+]_0^0\right)_\epsilon\Phi(x)d^3x = \\ \int\limits_{2m}^{R}\left(r^2(h_\epsilon^{+\prime\prime})_\epsilon + 2r(h_\epsilon^{+\prime})_\epsilon\right)\tilde{\Phi}(r)dr = \\ = \int\limits_{2m}^{R}\left\{-\frac{r}{\left[(r-2m)^2+\epsilon^2\right]^{1/2}} + \frac{r(r-2m)^2}{\left[(r-2m)^2+\epsilon^2\right]^{3/2}}\right\}\tilde{\Phi}(r)dr.\end{gathered} \qquad (B.16)$$

where use is made of the relation

$$\lim_{s\to\infty}\left[\int\limits_0^s\frac{\eta^2 d\eta}{(\eta^2+1)^{3/2}} - \int\limits_0^s\frac{d\eta}{(u^2+1)^{1/2}}\right] = -1 \qquad (B.17)$$

Finally we obtain

$$w\text{-}\lim_{\epsilon\to 0}[\mathbf{R}_\epsilon^+]_1^1 = w\text{-}\lim_{\epsilon\to 0}[\mathbf{R}_\epsilon^+]_0^0 = -m\tilde{\Phi}(2m). \qquad (B.18)$$

The Colombeau generalized Ricci tensor below horizon $[\mathbf{R}_\epsilon^-]_\alpha^\beta = [\mathbf{R}_\epsilon^-]_\alpha^\beta$ may now be calculated componentwise using the classical formulae

$$\begin{cases}\left([\mathbf{R}_\epsilon^-]_0^0\right)_\epsilon = \left([\mathbf{R}_\epsilon^-]_1^1\right)_\epsilon = \frac{1}{2}\left((h_\epsilon^{-\prime\prime})_\epsilon + \frac{2}{r}(h_\epsilon^{-\prime})_\epsilon\right), \\ \left([\mathbf{R}_\epsilon^-]_2^2\right)_\epsilon = \left([\mathbf{R}_\epsilon^-]_3^3\right)_\epsilon = \frac{(h_\epsilon^{-\prime})_\epsilon}{r} + \frac{1+(h_\epsilon^-)_\epsilon}{r^2}.\end{cases} \qquad (B.19)$$

From (B.4) we obtain

$$h_\epsilon^-(r) = -\frac{r-2m}{r} \mapsto h_\epsilon^-(r) = \left(\frac{\sqrt{(2m-r)^2+\epsilon^2}}{r}\right) = -h_\epsilon^+(r), r < 2m.$$

$$h_\epsilon^{-\prime}(r) = -h_\epsilon^{+\prime}(r) = \frac{r-2m}{r\left[(r-2m)^2+\epsilon^2\right]^{1/2}} - \frac{\left[(r-2m)^2+\epsilon^2\right]^{1/2}}{r^2},$$

$$r(h_\epsilon^{-\prime})_\epsilon + 1 + (h_\epsilon^-)_\epsilon = -r(h_\epsilon^{+\prime})_\epsilon + 1 - (h_\epsilon^+)_\epsilon =$$

$$\frac{r-2m}{\left[(r-2m)^2+\epsilon^2\right]^{1/2}} + 1. \qquad (B.20)$$

$$h_\epsilon^{-\prime\prime}(r) = -h_\epsilon^{+\prime\prime}(r) =$$

$$-\frac{r-2m}{r^2\left[(r-2m)^2+\epsilon^2\right]^{1/2}} + \frac{2\left[(r-2m)^2+\epsilon^2\right]^{1/2}}{r^3}.$$

$$r^2(h_\epsilon^{-\prime\prime})_\epsilon + 2r(h_\epsilon^{-\prime})_\epsilon = -r^2(h_\epsilon^{+\prime\prime})_\epsilon - 2r(h_\epsilon^{+\prime})_\epsilon =$$

$$\frac{r}{\left[(r-2m)^2+\epsilon^2\right]^{1/2}} - \frac{r(r-2m)^2}{\left[(r-2m)^2+\epsilon^2\right]^{3/2}}.$$

Investigating the weak limit of the angular components of the Ricci tensor (using the abbreviation $\tilde{\Phi}(r) = \int_0^\pi \sin\theta d\theta \int_0^{2\pi} d\phi \Phi(\vec{x})$ and let $\Phi(\vec{x})$ be the function $\Phi(\vec{x}) \in S_{2m}^-(\mathbb{R}^3, B^-(0,2m))$, where by $S_{2m}^-(\mathbb{R}^3, B^-(0,2m))$ we denote the class of the functions $\Phi(\vec{x})$ with compact support $K \subset B^-(0,2m), B^-(0,2m) = \{\vec{x}|0 \leq \|\vec{x}\| \leq 2m\}$ such that:

(i) $\mathbf{supp}(\Phi(\vec{x})) \subset \{\vec{x}|0 \leq \|\vec{x}\| \leq 2m\}$ (ii) $\tilde{\Phi}(r) \in C^\infty(\mathbb{R})$.

Then for any function $\Phi(\vec{x}) \in S_{2m}^-(\mathbb{R}^3, B^-(0,2m))$ we get

$$\int_K \left([\mathbf{R}_\epsilon^-]_2^2\right)_\epsilon \Phi(\vec{x}) d^3x = \int_K \left([\mathbf{R}_\epsilon^-]_3^3\right)_\epsilon \Phi(\vec{x}) d^3x =$$

$$\int_0^{2m} \left(r(h_\epsilon^{-\prime})_\epsilon + 1 + (h_\epsilon^-)_\epsilon\right) \tilde{\Phi}(r) dr = \qquad (B.21)$$

$$\int_0^{2m} \left\{\frac{r-2m}{\left[(r-2m)^2+\epsilon^2\right]^{1/2}}\right\} \tilde{\Phi}(r) dr + \int_0^{2m} \tilde{\Phi}(r) dr.$$

By replacement $r - 2m = u,$ from Eq.(B.21) we obtain

$$\int_K \left([\mathbf{R}_\epsilon^-]_2^2\right)_\epsilon \Phi(x) d^3x = \int_K \left([\mathbf{R}_\epsilon^-]_3^3\right)_\epsilon \Phi(x) d^3x =$$

$$\int_{-2m}^0 \frac{u\tilde{\Phi}(u+2m)du}{(u^2+\epsilon^2)^{1/2}} + \int_{-2m}^0 \tilde{\Phi}(u+2m)du. \qquad (B.22)$$

By replacement $u = \epsilon\eta,$ from (B.22) we obtain

$$\mathbf{I}_3^-(\epsilon) = \int_K \left([\mathbf{R}_\epsilon^-]_3^3\right)_\epsilon \Phi(x) d^3x = \mathbf{I}_2^-(\epsilon) = \int_K \left([\mathbf{R}_\epsilon^-]_2^2\right)_\epsilon \Phi(\vec{x}) d^3x =$$

$$\epsilon \times \left( \int_{-\frac{2m}{\epsilon}}^{0} \frac{\eta\tilde{\Phi}(\epsilon\eta + 2m)d\eta}{(\eta^2+1)^{1/2}} + \int_{-\frac{2m}{\epsilon}}^{0} \tilde{\Phi}(\epsilon\eta + 2m)d\eta \right), \qquad (B.23)$$

which is calculated to give

$$\begin{aligned}
\mathbf{I}_3^-(\epsilon) = \mathbf{I}_2^-(\epsilon) = \epsilon \frac{\tilde{\Phi}(2m)}{0!} \int_{-\frac{2m}{\epsilon}}^{0} \left[ \frac{\eta}{(\eta^2+1)^{1/2}} + 1 \right] d\eta + \\
+ \frac{\epsilon^2}{1!} \int_{-\frac{2m}{\epsilon}}^{0} \left[ \frac{\eta}{(\eta^2+1)^{1/2}} + 1 \right] \tilde{\Phi}^{(1)}(\xi)\eta d\eta = \\
\epsilon\tilde{\Phi}(2m)\left[ 1 - \sqrt{\left(\frac{2m}{\epsilon}\right)^2 + 1} + \frac{2m}{\epsilon} \right] + \\
+ \frac{\epsilon^2}{1} \int_{-\frac{2m}{\epsilon}}^{0} \left[ \frac{\eta}{(\eta^2+1)^{1/2}} + 1 \right] \tilde{\Phi}^{(1)}(\xi)\eta d\eta,
\end{aligned} \qquad (B.24)$$

where we have expressed the function $\widetilde{\Phi}(\epsilon\eta + 2m)$ as

$$\begin{cases} \widetilde{\Phi}(\epsilon\eta + 2m) = \sum_{l=0}^{n-1} \frac{\Phi^{(l)}(2m)}{l!} (\epsilon\eta)^l + \frac{1}{n!} (\epsilon\eta)^n \Phi^{(n)}(\xi) \ , \\ \xi \triangleq \theta\epsilon\eta + 2m \ , \quad 1 > \theta > 0 \ , \quad n = 1 \end{cases} \qquad (B.25)$$

with $\tilde{\Phi}^{(l)} \triangleq d^l\tilde{\Phi}/dr^l$. Equation (B.25) gives

$$\begin{cases} \lim_{\epsilon\to 0} \mathbf{I}_3^-(\epsilon) = \lim_{\epsilon\to 0} \mathbf{I}_2^-(\epsilon) = \\ \lim_{\epsilon\to 0} \left\{ \epsilon\tilde{\Phi}(2m)\left[ 1 - \sqrt{\left(\frac{2m}{\epsilon}\right)^2 + 1} + \frac{2m}{\epsilon} \right] \right\} + \\ + \lim_{\epsilon\to 0} \left\{ \frac{\epsilon^2}{2} \int_{-\frac{2m}{\epsilon}}^{0} \left[ \frac{\eta}{(\eta^2+1)^{1/2}} + 1 \right] \tilde{\Phi}^{(1)}(\xi)\eta d\eta \right\} = 0. \end{cases} \qquad (B.26)$$

Since $S'_{2m}(B^-(0,2m)) \subset D'(\mathbb{R}^3)$, where $B^-(0,2m) = \{x \in \mathbb{R}^3 | 0 \leqslant \|x\| \leqslant 2m\}$ from Eq.(B.26) we obtain $\mathcal{S}'_{2m}(B_R^-(2m)) \subset \mathcal{S}'_{2m}(\mathbb{R}^3)$,

$$\begin{cases} w\text{-}\lim_{\epsilon\to 0} [\mathbf{R}_\epsilon^-]_3^3 = \lim_{\epsilon\to 0} \mathbf{I}_3^-(\epsilon) = 0. \\ w\text{-}\lim_{\epsilon\to 0} [\mathbf{R}_\epsilon^-]_2^2 = \lim_{\epsilon\to 0} \mathbf{I}_2^-(\epsilon) = 0. \end{cases} \qquad (B.27)$$

For $\left([\mathbf{R}_\epsilon^-]_1^1\right)_\epsilon, \left([\mathbf{R}_\epsilon^-]_0^0\right)_\epsilon$ we get:

$$2\int_K \left([\mathbf{R}_\epsilon^-]_1^1\right)_\epsilon \Phi(x)d^3x = 2\int_K \left([\mathbf{R}_\epsilon^-]_0^0\right)_\epsilon \Phi(x)d^3x =$$

$$\int_0^{2m} \left(r^2(h_\epsilon^{-''})_\epsilon + 2r(h_\epsilon^{-'})_\epsilon\right)\tilde{\Phi}(r)dr = \qquad (B.28)$$

$$= \int_0^{2m} \left\{ \frac{r}{\left[(r-2m)^2+\epsilon^2\right]^{1/2}} - \frac{r(r-2m)^2}{\left[(r-2m)^2+\epsilon^2\right]^{3/2}} \right\}\tilde{\Phi}(r)dr.$$

By replacement $r - 2m = u,$ from (B.28) we obtain

$$I_1^\dagger(\epsilon) = 2\int \left([\mathbf{R}_\epsilon^-]_1^1\right)_\epsilon \Phi(x)d^3x = I_2^\dagger(\epsilon) = 2\int \left([\mathbf{R}_\epsilon^-]_0^0\right)_\epsilon \Phi(x)d^3x$$

$$= \int_0^{2m} \left(r^2(h_\epsilon^{-''})_\epsilon + 2r(h_\epsilon^{-'})_\epsilon\right)\tilde{\Phi}(r)dr = \qquad (B.29)$$

$$= \int_{-2m}^{0} \left\{ \frac{u+2m}{(u^2+\epsilon^2)^{1/2}} - \frac{u^2(u+2m)}{(u^2+\epsilon^2)^{3/2}} \right\}\tilde{\Phi}(u+2m)du.$$

By replacement $u = \epsilon\eta,$ from (B.29) we obtain

$$2\int_K \left([\mathbf{R}_\epsilon^-]_1^1\right)_\epsilon \Phi(x)d^3x = 2\int_K \left([\mathbf{R}_\epsilon^-]_0^0\right)_\epsilon \Phi(x)d^3x =$$

$$\int_{-\frac{2m}{\epsilon}}^{0} \left(r^2(h_\epsilon^{-''})_\epsilon + 2r(h_\epsilon^{-'})_\epsilon\right)\tilde{\Phi}(r)dr =$$

$$= \epsilon \int_{-\frac{2m}{\epsilon}}^{0} \left\{ \frac{\epsilon\eta+2m}{(\epsilon^2\eta^2+\epsilon^2)^{1/2}} - \frac{\epsilon^2\eta^2(\epsilon\eta+2m)}{(\epsilon^2\eta^2+\epsilon^2)^{3/2}} \right\}\tilde{\Phi}(\epsilon\eta+2m)d\eta =$$

$$\int_{-\frac{2m}{\epsilon}}^{0} \frac{\epsilon^2\eta\tilde{\Phi}(\epsilon\eta+2m)d\eta}{(\epsilon^2\eta^2+\epsilon^2)^{1/2}} + 2m\int_{-\frac{2m}{\epsilon}}^{0} \frac{\epsilon\tilde{\Phi}(\epsilon\eta+2m)d\eta}{(\epsilon^2\eta^2+\epsilon^2)^{1/2}} - \qquad (B.30)$$

$$-\int_{-\frac{2m}{\epsilon}}^{0} \frac{\epsilon^4\eta^3\tilde{\Phi}(\epsilon\eta+2m)d\eta}{(\epsilon^2\eta^2+\epsilon^2)^{3/2}} - 2m\int_{-\frac{2m}{\epsilon}}^{0} \frac{\epsilon^3\eta^2\tilde{\Phi}(\epsilon\eta+2m)d\eta}{(\epsilon^2\eta^2+\epsilon^2)^{3/2}} =$$

$$\epsilon\int_{-\frac{2m}{\epsilon}}^{0} \frac{\eta\tilde{\Phi}(\epsilon\eta+2m)d\eta}{(\eta^2+1)^{1/2}} - \int_{-\frac{2m}{\epsilon}}^{0} \frac{\eta^3\tilde{\Phi}(\epsilon\eta+2m)d\eta}{(\eta^2+1)^{3/2}} +$$

$$+2m\left[\int_{-\frac{2m}{\epsilon}}^{0} \frac{\tilde{\Phi}(\epsilon\eta+2m)d\eta}{(\eta^2+1)^{1/2}} - \int_{-\frac{2m}{\epsilon}}^{0} \frac{\eta^2\tilde{\Phi}(\epsilon\eta+2m)d\eta}{(\eta^2+1)^{3/2}}\right].$$

which is calculated to give

$$\mathbf{I}_0^-(\epsilon)=\mathbf{I}_1^-(\epsilon)=2m\frac{\tilde{\Phi}(2m)}{0!}\epsilon^l\int\limits_{-\frac{2m}{\epsilon}}^{0}\left[\frac{1}{(\eta^2+1)^{1/2}}-\frac{\eta^2}{(\eta^2+1)^{3/2}}\right]d\eta+$$

$$+\frac{\epsilon}{1!}\int\limits_{0}^{\frac{2m}{\epsilon}}\tilde{\Phi}^{(1)}(\xi)\left[\frac{1}{(\eta^2+1)^{1/2}}-\frac{\eta^2}{(\eta^2+1)^{3/2}}\right]\eta d\eta+O(\epsilon^2). \qquad (B.31)$$

where we have expressed the function $\widetilde{\Phi}(\epsilon\eta+2m)$ as

$$\widetilde{\Phi}(\epsilon\eta+2m)=\sum_{l=0}^{n-1}\frac{\Phi^{\alpha\beta(l)}(2m)}{l!}(\epsilon\eta)^l+\frac{1}{n!}(\epsilon\eta)^n\Phi^{\alpha\beta(n)}(\xi)\ , \qquad (B.32)$$

$$\xi\triangleq\theta\epsilon\eta+2m\ ,\quad 1>\theta>0\ ,\quad n=1$$

with $\tilde{\Phi}^{(l)}(\xi)\triangleq d^l\tilde{\Phi}/d\xi^l$.Equation (B.32) gives

$$\lim_{\epsilon\to0}\mathbf{I}_0^-(\epsilon)=\lim_{\epsilon\to0}\mathbf{I}_1^-(\epsilon)=$$

$$2m\lim_{\epsilon\to0}\left\{\frac{\tilde{\Phi}(2m)}{0!}\int\limits_{-\frac{2m}{\epsilon}}^{0}\left[\frac{1}{(\eta^2+1)^{1/2}}-\frac{\eta^2}{(\eta^2+1)^{3/2}}\right]d\eta\right\}=$$

$$2m\tilde{\Phi}(2m)\lim_{s\to0}\left[\int_{-s}^{0}\frac{d\eta}{(\eta^2+1)^{1/2}}-\int_{-s}^{0}\frac{\eta^2d\eta}{(\eta^2+1)^{3/2}}\right]= \qquad (B.33)$$

$$=2m\tilde{\Phi}(2m).$$

where use is made of the relation

$$\lim_{s\to\infty}\left[\int\limits_{-s}^{0}\frac{d\eta}{(u^2+1)^{1/2}}-\int\limits_{-s}^{0}\frac{\eta^2d\eta}{(\eta^2+1)^{3/2}}\right]=1. \qquad (B.34)$$

Thus

$$w\text{-}\lim_{\epsilon\to0}[\mathbf{R}_\epsilon^-]_1^1=w\text{-}\lim_{\epsilon\to0}[\mathbf{R}_\epsilon^-]_0^0=m\tilde{\Phi}(2m). \qquad (B.35)$$

## 5mm . **Appendix C ref: thm:conv_lambda0mle** 2mm

We calculate now the distributional Colombeau scalars $(\mathbf{R}(r,\varepsilon))_\varepsilon,(\mathbf{R}^{\mu\nu}(r,\varepsilon)\mathbf{R}_{\mu\nu}(r,\varepsilon))_\varepsilon$ and $(\mathbf{R}^{\rho\sigma\mu\nu}(r,\varepsilon)\mathbf{R}_{\rho\sigma\mu\nu}(r,\varepsilon))_\varepsilon$, in terms of Colombeau generalized functions $(A_\varepsilon(r))_\varepsilon,(B_\varepsilon(r))_\varepsilon$, $(C_\varepsilon(r))_\varepsilon,(D_\varepsilon(r))_\varepsilon$ is given above in Appendix B at Schwarzschild horizon.We choose now

$$B_\varepsilon(r)=1,C_\varepsilon(r)=-1+A_\varepsilon^{-1}(r),D_\varepsilon(r)=0, \qquad (C.1)$$

and rewrite Eq.(A.1) in the following equivalent form

$$(ds_\varepsilon^2)_\varepsilon=-(A_\varepsilon(r)dt^2)_\varepsilon+(A_\varepsilon^{-1}(r)dr^2)_\varepsilon+r^2d\Omega^2, \qquad (C.2)$$

where $A_\varepsilon(r)$ is given above by using Eqs.(B.2)-(B.4). Thus we obtain

$$A_\varepsilon(r) = -r^{-1}\sqrt{(r-2m)^2+\varepsilon^2}, \Delta_\varepsilon = A_\varepsilon(B_\varepsilon + C_\varepsilon) + D_\varepsilon^2 = 1, \Delta'_\varepsilon = 0,$$

$$A'_\varepsilon(r) = \frac{-2m(r-2m)}{r^2\sqrt{(r-2m)^2+\varepsilon^2}}, \qquad (C.3)$$

$$A''_\varepsilon(r) = \frac{2m(-16m^3+24m^2r-12mr^2-4m\varepsilon^2+2r^3+r\varepsilon^2)}{r^3\left[(r-2m)^2+\varepsilon^2\right]^{3/2}}$$

From Eqs.(A.2) and Eq.(C.3) we obtain

$$\begin{aligned}(\mathbf{R}(r,\varepsilon))_\varepsilon &= \left(\frac{A_\varepsilon}{\Delta_\varepsilon}\left[\frac{2}{r}\left(-2\frac{A'_\varepsilon}{A_\varepsilon} - 3\frac{B'_\varepsilon}{B_\varepsilon} + \frac{\Delta'_\varepsilon}{\Delta_\varepsilon}\right) + \frac{2}{r^2}\frac{A_\varepsilon C_\varepsilon + D_\varepsilon^2}{A_\varepsilon B_\varepsilon} - \frac{A''_\varepsilon}{A_\varepsilon} - 2\frac{B''_\varepsilon}{B_\varepsilon}\right.\right. \\ &\left.\left. + \frac{1}{2}\left(\frac{B'_\varepsilon}{B_\varepsilon}\right)^2 - 2\frac{A'_\varepsilon B'_\varepsilon}{A_\varepsilon B_\varepsilon} + \left(\frac{1}{2}\frac{A'_\varepsilon}{A_\varepsilon} + \frac{B'_\varepsilon}{B_\varepsilon}\right)\frac{\Delta'_\varepsilon}{\Delta_\varepsilon}\right]\right)_\varepsilon = \\ &\left(A_\varepsilon\left[\frac{2}{r}\left(-2\frac{A'_\varepsilon}{A_\varepsilon}\right) - \frac{2A_\varepsilon(1-A_\varepsilon^{-1})}{r^2} - \frac{A''_\varepsilon}{A_\varepsilon}\right]\right)_\varepsilon = \\ &\left(-\frac{4A'_\varepsilon}{r} - \frac{2A_\varepsilon}{r^2} + \frac{2}{r^2} - A''_\varepsilon\right) = \qquad (C.4) \\ &\left(\frac{8m(r-2m)}{r^3\sqrt{(r-2m)^2+\varepsilon^2}}\right)_\varepsilon + 2r^{-3}\left(\sqrt{(r-2m)^2+\varepsilon^2}\right)_\varepsilon - \frac{2}{r^2} - \\ &-\left(\frac{2m(-16m^3+24m^2r-12mr^2-4m\varepsilon^2+2r^3+r\varepsilon^2)}{r^3\left[(r-2m)^2+\varepsilon^2\right]^{3/2}}\right)_\varepsilon .\end{aligned}$$

Finally we obtain the following expression for the distributional Colombeau scalar $(\mathbf{R}(r,\varepsilon))_\varepsilon$

$$\begin{aligned}(\mathbf{R}(r,\varepsilon))_\varepsilon &= \left(\frac{8m(r-2m)}{r^3\sqrt{(r-2m)^2+\varepsilon^2}}\right)_\varepsilon + 2r^{-3}\left(\sqrt{(r-2m)^2+\varepsilon^2}\right)_\varepsilon - \frac{2}{r^2} - \\ &-\left(\frac{2m(-16m^3+24m^2r-12mr^2-4m\varepsilon^2+2r^3+r\varepsilon^2)}{r^3\left[(r-2m)^2+\varepsilon^2\right]^{3/2}}\right)_\varepsilon .\end{aligned} \qquad (C.5)$$

**Remark C.1**. Note that from Eq.(C.5) follows that: $r \neq 2m \Rightarrow (\mathbf{R}(r,\varepsilon))_\varepsilon \approx_{\widetilde{\mathbb{R}}} 0$, see Definition 1.5.2.(i).

We assume now that $r \asymp 2m$ and therefore from Eq.(C.4) we obtain

$$(\mathbf{R}(r,\varepsilon))_\varepsilon \asymp \left(\frac{4m^2\varepsilon^2}{8m^3\left[(r-2m)^2+\varepsilon^2\right]^{3/2}}\right)_\varepsilon . \qquad (C.6)$$

**Remark C.2**. Note that from Eq.(C.5) at horizon $r = 2m$ follows that:

$$(\mathbf{R}(r,\varepsilon))_\varepsilon = \left(\frac{4m^2\varepsilon^2}{8m^3[\varepsilon^2]^{3/2}}\right)_\varepsilon = (4m)^{-1}(\varepsilon^{-1})_\varepsilon \approx_{\widetilde{\mathbb{R}}} \infty, \qquad (C.7)$$

see Definition 1.5.2.(ii).

**Remark C.3**. Note that from Eq.(C.5) follows that:

$$w\text{-}\lim_{\varepsilon\to 0} \mathbf{R}(r,\varepsilon) \sim \delta(r-2m). \qquad (C.8)$$

From Eqs.(A.2) and Eq.(C.3) we obtain

$$(\mathbf{R}^{\mu\nu}(r,\varepsilon)\mathbf{R}_{\mu\nu}(r,\varepsilon))_{\varepsilon} =$$

$$+2\left(\left[\frac{1}{r}A'_{\varepsilon} + \frac{-A_{\varepsilon}+1}{r^2}\right]^2\right)_{\varepsilon} + 2\left(\left[\frac{1}{2}A''_{\varepsilon} \quad + \frac{1}{r}A'_{\varepsilon}\right]^2\right)_{\varepsilon} =$$

$$2\left(\left[\left(\frac{1}{r^3}\sqrt{\varepsilon^2+(r-2m)^2} + \frac{1}{r^2}\right) + 2\frac{m}{r^3}\frac{r-2m}{\sqrt{\varepsilon^2+(r-2m)^2}}\right]^2\right)_{\varepsilon} + \qquad (C.9)$$

$$2\left(\frac{m(-16m^3+24m^2r-12mr^2-4m\varepsilon^2+2r^3+r\varepsilon^2)}{r^3\left(\varepsilon^2+(r-2m)^2\right)^{\frac{3}{2}}} - 2\frac{m}{r^3}\frac{r-2m}{\sqrt{\varepsilon^2+(r-2m)^2}}\right)^2.$$

**Remark C.4**. Note that from Eq.(C.9) follows that:

$$r \neq 2m \Rightarrow (\mathbf{R}^{\mu\nu}(r,\varepsilon)\mathbf{R}_{\mu\nu}(r,\varepsilon))_{\varepsilon} \approx_{\widetilde{\mathbb{R}}} 0. \qquad (C.10)$$

We assume now that $r \asymp 2m$ and therefore from Eq.(C.9) we obtain

$$(\mathbf{R}^{\mu\nu}(r,\varepsilon)\mathbf{R}_{\mu\nu}(r,\varepsilon))_{\varepsilon} \asymp \frac{\varepsilon^4}{4m^4\left(\varepsilon^2+(r-2m)^2\right)^3} \qquad (C.10)$$

**Remark C.5**. Note that from Eq.(C.9) at horizon $r = 2m$ follows that:

$$(\mathbf{R}^{\mu\nu}(r,\varepsilon)\mathbf{R}_{\mu\nu}(r,\varepsilon))_{\varepsilon} = \left(\frac{1}{4m^4\varepsilon^2}\right)_{\varepsilon} \approx_{\widetilde{\mathbb{R}}} \infty. \qquad (C.11)$$

From Eqs.(A.2) and Eq.(C.3) we obtain

$$(\mathbf{R}^{\rho\sigma\mu\nu}(r,\varepsilon)\mathbf{R}_{\rho\sigma\mu\nu}(r,\varepsilon))_{\varepsilon} =$$

$$\left((A''_{\varepsilon})^2 + 2\left(\frac{A'_{\varepsilon}}{r}\right)^2\right)_{\varepsilon} + 4\left(\left[\frac{1}{r^2}(1-A_{\varepsilon})\right]^2\right)_{\varepsilon} + 2\left(\left[\frac{A'_{\varepsilon}}{r}\right]^2\right)_{\varepsilon} =$$

$$(C.12)$$

$$\left(A''^2_{\varepsilon} + 4\frac{A'^2_{\varepsilon}}{r^2}\right)_{\varepsilon} + 4\left(\frac{1}{r^4}(1-A_{\varepsilon})^2\right)_{\varepsilon} =$$

$$A''^2_{\varepsilon} + 4\frac{A'^2_{\varepsilon}}{r^2} + \frac{4}{r^4}(1-A_{\varepsilon})^2$$

6mm

# Acknowledgments

3mm

We thank the Editor and the referee for their comments.